  {}%
\documentclass[traditabstract, longauth]{aa}

\def\setsymbol#1#2{\expandafter\def\csname #1\endcsname{#2}}
\def\getsymbol#1{\csname #1\endcsname}

\def\Planck{\textit{Planck}}

\def\alltwentyfifteenresultspapers{\nocite{planck2014-a01, planck2014-a03, planck2014-a04, planck2014-a05, planck2014-a06, planck2014-a07, planck2014-a08, planck2014-a09, planck2014-a11, planck2014-a12, planck2014-a13, planck2014-a14, planck2014-a15, planck2014-a16, planck2014-a17, planck2014-a18, planck2014-a19, planck2014-a20, planck2014-a22, planck2014-a24, planck2014-a26, planck2014-a28, planck2014-a29, planck2014-a30, planck2014-a31, planck2014-a35, planck2014-a36, planck2014-a37, planck2014-ES}}

\newbox\tablebox    \newdimen\tablewidth
\def\leaderfil{\leaders\hbox to 5pt{\hss.\hss}\hfil}
\def\endPlancktable{\tablewidth=\columnwidth 
    $$\hss\copy\tablebox\hss$$
    \vskip-\lastskip\vskip -2pt}
\def\endPlancktablewide{\tablewidth=\textwidth 
    $$\hss\copy\tablebox\hss$$
    \vskip-\lastskip\vskip -2pt}
\def\tablenote#1 #2\par{\begingroup \parindent=0.8em
    \abovedisplayshortskip=0pt\belowdisplayshortskip=0pt
    \noindent
    $$\hss\vbox{\hsize\tablewidth \hangindent=\parindent \hangafter=1 \noindent
    \hbox to \parindent{$^#1$\hss}\strut#2\strut\par}\hss$$
    \endgroup}
\def\doubleline{\vskip 3pt\hrule \vskip 1.5pt \hrule \vskip 5pt}

\def\L2{\ifmmode L_2\else $L_2$\fi}

\def\DeltaT{\ifmmode \Delta T\else $\Delta T$\fi}
\def\deltat{\ifmmode \Delta t\else $\Delta t$\fi}
\def\fknee{\ifmmode f_{\rm knee}\else $f_{\rm knee}$\fi}
\def\Fmax{\ifmmode F_{\rm max}\else $F_{\rm max}$\fi}
\def\solar{\ifmmode{\rm M}_{\mathord\odot}\else${\rm M}_{\mathord\odot}$\fi}
\def\Msolar{\ifmmode{\rm M}_{\mathord\odot}\else${\rm M}_{\mathord\odot}$\fi}
\def\Lsolar{\ifmmode{\rm L}_{\mathord\odot}\else${\rm L}_{\mathord\odot}$\fi}
\def\inv{\ifmmode^{-1}\else$^{-1}$\fi}
\def\mo{\ifmmode^{-1}\else$^{-1}$\fi}
\def\sup#1{\ifmmode ^{\rm #1}\else $^{\rm #1}$\fi}
\def\expo#1{\ifmmode \times 10^{#1}\else $\times 10^{#1}$\fi}
\def\,{\thinspace}
\def\lsim{\mathrel{\raise .4ex\hbox{\rlap{$<$}\lower 1.2ex\hbox{$\sim$}}}}
\def\gsim{\mathrel{\raise .4ex\hbox{\rlap{$>$}\lower 1.2ex\hbox{$\sim$}}}}

\def\simprop{\mathrel{\raise .4ex\hbox{\rlap{$\propto$}\lower 1.2ex\hbox{$\sim$}}}}
\def\deg{\ifmmode^\circ\else$^\circ$\fi}
\def\pdeg{\ifmmode $\setbox0=\hbox{$^{\circ}$}\rlap{\hskip.11\wd0 .}$^{\circ}
          \else \setbox0=\hbox{$^{\circ}$}\rlap{\hskip.11\wd0 .}$^{\circ}$\fi}
\def\arcs{\ifmmode {^{\scriptstyle\prime\prime}}
          \else $^{\scriptstyle\prime\prime}$\fi}
\def\arcm{\ifmmode {^{\scriptstyle\prime}}
          \else $^{\scriptstyle\prime}$\fi}
\newdimen\sa  \newdimen\sb
\def\parcs{\sa=.07em \sb=.03em
     \ifmmode \hbox{\rlap{.}}^{\scriptstyle\prime\kern -\sb\prime}\hbox{\kern -\sa}
     \else \rlap{.}$^{\scriptstyle\prime\kern -\sb\prime}$\kern -\sa\fi}
\def\parcm{\sa=.08em \sb=.03em
     \ifmmode \hbox{\rlap{.}\kern\sa}^{\scriptstyle\prime}\hbox{\kern-\sb}
     \else \rlap{.}\kern\sa$^{\scriptstyle\prime}$\kern-\sb\fi}
\def\ra[#1 #2 #3.#4]{#1\sup{h}#2\sup{m}#3\sup{s}\llap.#4}
\def\dec[#1 #2 #3.#4]{#1\deg#2\arcm#3\arcs\llap.#4}
\def\deco[#1 #2 #3]{#1\deg#2\arcm#3\arcs}
\def\rra[#1 #2]{#1\sup{h}#2\sup{m}}

\def\dots{\relax\ifmmode \ldots\else $\ldots$\fi}
\def\WHzsr{\ifmmode $W\,Hz\mo\,sr\mo$\else W\,Hz\mo\,sr\mo\fi}
\def\mHz{\ifmmode $\,mHz$\else \,mHz\fi}
\def\GHz{\ifmmode $\,GHz$\else \,GHz\fi}
\def\mKs{\ifmmode $\,mK\,s$^{1/2}\else \,mK\,s$^{1/2}$\fi}
\def\muKs{\ifmmode \,\mu$K\,s$^{1/2}\else \,$\mu$K\,s$^{1/2}$\fi}
\def\muKRJs{\ifmmode \,\mu$K$_{\rm RJ}$\,s$^{1/2}\else \,$\mu$K$_{\rm RJ}$\,s$^{1/2}$\fi}
\def\muKHz{\ifmmode \,\mu$K\,Hz$^{-1/2}\else \,$\mu$K\,Hz$^{-1/2}$\fi}
\def\MJysr{\ifmmode \,$MJy\,sr\mo$\else \,MJy\,sr\mo\fi}
\def\MJysrmK{\ifmmode \,$MJy\,sr\mo$\,mK$_{\rm CMB}\mo\else \,MJy\,sr\mo\,mK$_{\rm CMB}\mo$\fi}
\def\microns{\ifmmode \,\mu$m$\else \,$\mu$m\fi}

\def\muK{\ifmmode \,\mu$K$\else \,$\mu$\hbox{K}\fi}
\def\microK{\ifmmode \,\mu$K$\else \,$\mu$\hbox{K}\fi}
\def\muW{\ifmmode \,\mu$W$\else \,$\mu$\hbox{W}\fi}
\def\kms{\ifmmode $\,km\,s$^{-1}\else \,km\,s$^{-1}$\fi}
\def\kmsMpc{\ifmmode $\,\kms\,Mpc\mo$\else \,\kms\,Mpc\mo\fi}
\providecommand{\sorthelp}[1]{}

\usepackage{graphicx}
\usepackage{color}
\usepackage{natbib}
\usepackage[breaklinks,colorlinks, citecolor=blue]{hyperref}
\usepackage{breakurl}
\usepackage{amssymb,,amsmath}
\usepackage{longtable}
\usepackage{multirow}
\usepackage{txfonts}
\usepackage{array}
\usepackage{bmpsize}
\usepackage{rotating}
\usepackage{ifthen}
\usepackage{caption}
\usepackage{fixltx2e}
\usepackage{lscape}

\bibpunct{(}{)}{;}{a}{}{,}

\newcommand{\hatn}{\vec{\hat{n\,}}}
\def\lsim{\mathrel{\raise .4ex\hbox{\rlap{$<$}\lower 1.2ex\hbox{$\sim$}}}}
\def\gsim{\mathrel{\raise .4ex\hbox{\rlap{$>$}\lower 1.2ex\hbox{$\sim$}}}}
\def\arcm{\ifmmode {^{\scriptscriptstyle\prime}}
          \else $^{\scriptscriptstyle\prime}$\fi}
\def\muKRJ{\ifmmode \,\mu$K$_{\rm RJ}$\else \,$\mu$\hbox{K}$_{\rm RJ}$\fi}    
\def\muKCMB{\ifmmode \,\mu$K$_{\rm CMB}$\else \,$\mu$\hbox{K}$_{\rm CMB}$\fi} 
\def\Planck{\textit{Planck}}
\def\leaderfil{\leaders\hbox to 5pt{\hss.\hss}\hfil}

\DeclareGraphicsExtensions{.jpg, .png, .pdf, .eps}

\begin{document}

\title{\textit{Planck} 2015 results. III. LFI systematic uncertainties}

\titlerunning{LFI systematic uncertainties}
\authorrunning{Planck Collaboration}
\author{\small
Planck Collaboration: P.~A.~R.~Ade\inst{77}
\and
J.~Aumont\inst{53}
\and
C.~Baccigalupi\inst{75}
\and
A.~J.~Banday\inst{82, 8}
\and
R.~B.~Barreiro\inst{58}
\and
N.~Bartolo\inst{27, 59}
\and
S.~Basak\inst{75}
\and
P.~Battaglia\inst{29, 31}
\and
E.~Battaner\inst{83, 84}
\and
K.~Benabed\inst{54, 81}
% ...
}
\institute{\small
APC, AstroParticule et Cosmologie, Universit\'{e} Paris Diderot, CNRS/IN2P3, CEA/lrfu, Observatoire de Paris, Sorbonne Paris Cit\'{e}, 10, rue Alice Domon et L\'{e}onie Duquet, 75205 Paris Cedex 13, France\goodbreak
\and
African Institute for Mathematical Sciences, 6-8 Melrose Road, Muizenberg, Cape Town, South Africa\goodbreak
\and
Agenzia Spaziale Italiana Science Data Center, Via del Politecnico snc, 00133, Roma, Italy\goodbreak
\and
Aix Marseille Universit\'{e}, CNRS, LAM (Laboratoire d'Astrophysique de Marseille) UMR 7326, 13388, Marseille, France\goodbreak
\and
Astrophysics Group, Cavendish Laboratory, University of Cambridge, J J Thomson Avenue, Cambridge CB3 0HE, U.K.\goodbreak
\and
CGEE, SCS Qd 9, Lote C, Torre C, 4$^{\circ}$ andar, Ed. Parque Cidade Corporate, CEP 70308-200, Bras\'{i}lia, DF, Brazil\goodbreak
\and
CITA, University of Toronto, 60 St. George St., Toronto, ON M5S 3H8, Canada\goodbreak
\and
CNRS, IRAP, 9 Av. colonel Roche, BP 44346, F-31028 Toulouse cedex 4, France\goodbreak
\and
California Institute of Technology, Pasadena, California, U.S.A.\goodbreak
\and
Centro de Estudios de F\'{i}sica del Cosmos de Arag\'{o}n (CEFCA), Plaza San Juan, 1, planta 2, E-44001, Teruel, Spain\goodbreak
\and
Computational Cosmology Center, Lawrence Berkeley National Laboratory, Berkeley, California, U.S.A.\goodbreak
\and
DSM/Irfu/SPP, CEA-Saclay, F-91191 Gif-sur-Yvette Cedex, France\goodbreak
\and
DTU Space, National Space Institute, Technical University of Denmark, Elektrovej 327, DK-2800 Kgs. Lyngby, Denmark\goodbreak
\and
D\'{e}partement de Physique Th\'{e}orique, Universit\'{e} de Gen\`{e}ve, 24, Quai E. Ansermet,1211 Gen\`{e}ve 4, Switzerland\goodbreak
\and
Departamento de Astrof\'{i}sica, Universidad de La Laguna (ULL), E-38206 La Laguna, Tenerife, Spain\goodbreak
\and
Departamento de F\'{\i}sica, Universidad de Oviedo, Avda. Calvo Sotelo s/n, Oviedo, Spain\goodbreak
\and
Departamento de Matem\'{a}ticas, Estad\'{\i}stica y Computaci\'{o}n, Universidad de Cantabria, Avda. de los Castros s/n, Santander, Spain\goodbreak
\and
Department of Astrophysics/IMAPP, Radboud University Nijmegen, P.O. Box 9010, 6500 GL Nijmegen, The Netherlands\goodbreak
\and
Department of Physics \& Astronomy, University of British Columbia, 6224 Agricultural Road, Vancouver, British Columbia, Canada\goodbreak
\and
Department of Physics and Astronomy, Dana and David Dornsife College of Letter, Arts and Sciences, University of Southern California, Los Angeles, CA 90089, U.S.A.\goodbreak
\and
Department of Physics and Astronomy, University College London, London WC1E 6BT, U.K.\goodbreak
\and
Department of Physics, Florida State University, Keen Physics Building, 77 Chieftan Way, Tallahassee, Florida, U.S.A.\goodbreak
\and
Department of Physics, Gustaf H\"{a}llstr\"{o}min katu 2a, University of Helsinki, Helsinki, Finland\goodbreak
\and
Department of Physics, Princeton University, Princeton, New Jersey, U.S.A.\goodbreak
\and
Department of Physics, University of California, Santa Barbara, California, U.S.A.\goodbreak
\and
Department of Physics, University of Illinois at Urbana-Champaign, 1110 West Green Street, Urbana, Illinois, U.S.A.\goodbreak
\and
Dipartimento di Fisica e Astronomia G. Galilei, Universit\`{a} degli Studi di Padova, via Marzolo 8, 35131 Padova, Italy\goodbreak
\and
Dipartimento di Fisica e Scienze della Terra, Universit\`{a} di Ferrara, Via Saragat 1, 44122 Ferrara, Italy\goodbreak
\and
Dipartimento di Fisica, Universit\`{a} La Sapienza, P. le A. Moro 2, Roma, Italy\goodbreak
\and
Dipartimento di Fisica, Universit\`{a} degli Studi di Milano, Via Celoria, 16, Milano, Italy\goodbreak
\and
Dipartimento di Fisica, Universit\`{a} degli Studi di Trieste, via A. Valerio 2, Trieste, Italy\goodbreak
\and
Dipartimento di Matematica, Universit\`{a} di Roma Tor Vergata, Via della Ricerca Scientifica, 1, Roma, Italy\goodbreak
\and
Discovery Center, Niels Bohr Institute, Blegdamsvej 17, Copenhagen, Denmark\goodbreak
\and
European Space Agency, ESAC, Planck Science Office, Camino bajo del Castillo, s/n, Urbanizaci\'{o}n Villafranca del Castillo, Villanueva de la Ca\~{n}ada, Madrid, Spain\goodbreak
\and
European Space Agency, ESTEC, Keplerlaan 1, 2201 AZ Noordwijk, The Netherlands\goodbreak
\and
Gran Sasso Science Institute, INFN, viale F. Crispi 7, 67100 L'Aquila, Italy\goodbreak
\and
HGSFP and University of Heidelberg, Theoretical Physics Department, Philosophenweg 16, 69120, Heidelberg, Germany\goodbreak
\and
Haverford College Astronomy Department, 370 Lancaster Avenue, Haverford, Pennsylvania, U.S.A.\goodbreak
\and
Helsinki Institute of Physics, Gustaf H\"{a}llstr\"{o}min katu 2, University of Helsinki, Helsinki, Finland\goodbreak
\and
INAF - Osservatorio Astrofisico di Catania, Via S. Sofia 78, Catania, Italy\goodbreak
\and
INAF - Osservatorio Astronomico di Padova, Vicolo dell'Osservatorio 5, Padova, Italy\goodbreak
\and
INAF - Osservatorio Astronomico di Roma, via di Frascati 33, Monte Porzio Catone, Italy\goodbreak
\and
INAF - Osservatorio Astronomico di Trieste, Via G.B. Tiepolo 11, Trieste, Italy\goodbreak
\and
INAF/IASF Bologna, Via Gobetti 101, Bologna, Italy\goodbreak
\and
INAF/IASF Milano, Via E. Bassini 15, Milano, Italy\goodbreak
\and
INFN, Sezione di Bologna, Via Irnerio 46, I-40126, Bologna, Italy\goodbreak
\and
INFN, Sezione di Roma 1, Universit\`{a} di Roma Sapienza, Piazzale Aldo Moro 2, 00185, Roma, Italy\goodbreak
\and
INFN, Sezione di Roma 2, Universit\`{a} di Roma Tor Vergata, Via della Ricerca Scientifica, 1, Roma, Italy\goodbreak
\and
INFN/National Institute for Nuclear Physics, Via Valerio 2, I-34127 Trieste, Italy\goodbreak
\and
IUCAA, Post Bag 4, Ganeshkhind, Pune University Campus, Pune 411 007, India\goodbreak
\and
Imperial College London, Astrophysics group, Blackett Laboratory, Prince Consort Road, London, SW7 2AZ, U.K.\goodbreak
\and
Infrared Processing and Analysis Center, California Institute of Technology, Pasadena, CA 91125, U.S.A.\goodbreak
\and
Institut d'Astrophysique Spatiale, CNRS (UMR8617) Universit\'{e} Paris-Sud 11, B\^{a}timent 121, Orsay, France\goodbreak
\and
Institut d'Astrophysique de Paris, CNRS (UMR7095), 98 bis Boulevard Arago, F-75014, Paris, France\goodbreak
\and
Institute of Astronomy, University of Cambridge, Madingley Road, Cambridge CB3 0HA, U.K.\goodbreak
\and
Institute of Theoretical Astrophysics, University of Oslo, Blindern, Oslo, Norway\goodbreak
\and
Instituto de Astrof\'{\i}sica de Canarias, C/V\'{\i}a L\'{a}ctea s/n, La Laguna, Tenerife, Spain\goodbreak
\and
Instituto de F\'{\i}sica de Cantabria (CSIC-Universidad de Cantabria), Avda. de los Castros s/n, Santander, Spain\goodbreak
\and
Istituto Nazionale di Fisica Nucleare, Sezione di Padova, via Marzolo 8, I-35131 Padova, Italy\goodbreak
\and
Jet Propulsion Laboratory, California Institute of Technology, 4800 Oak Grove Drive, Pasadena, California, U.S.A.\goodbreak
\and
Jodrell Bank Centre for Astrophysics, Alan Turing Building, School of Physics and Astronomy, The University of Manchester, Oxford Road, Manchester, M13 9PL, U.K.\goodbreak
\and
Kavli Institute for Cosmological Physics, University of Chicago, Chicago, IL 60637, USA\goodbreak
\and
Kavli Institute for Cosmology Cambridge, Madingley Road, Cambridge, CB3 0HA, U.K.\goodbreak
\and
Kazan Federal University, 18 Kremlyovskaya St., Kazan, 420008, Russia\goodbreak
\and
LAL, Universit\'{e} Paris-Sud, CNRS/IN2P3, Orsay, France\goodbreak
\and
LERMA, CNRS, Observatoire de Paris, 61 Avenue de l'Observatoire, Paris, France\goodbreak
\and
Laboratoire AIM, IRFU/Service d'Astrophysique - CEA/DSM - CNRS - Universit\'{e} Paris Diderot, B\^{a}t. 709, CEA-Saclay, F-91191 Gif-sur-Yvette Cedex, France\goodbreak
\and
Laboratoire de Physique Subatomique et Cosmologie, Universit\'{e} Grenoble-Alpes, CNRS/IN2P3, 53, rue des Martyrs, 38026 Grenoble Cedex, France\goodbreak
\and
Laboratoire de Physique Th\'{e}orique, Universit\'{e} Paris-Sud 11 \& CNRS, B\^{a}timent 210, 91405 Orsay, France\goodbreak
\and
Lawrence Berkeley National Laboratory, Berkeley, California, U.S.A.\goodbreak
\and
Max-Planck-Institut f\"{u}r Astrophysik, Karl-Schwarzschild-Str. 1, 85741 Garching, Germany\goodbreak
\and
National University of Ireland, Department of Experimental Physics, Maynooth, Co. Kildare, Ireland\goodbreak
\and
Nicolaus Copernicus Astronomical Center, Bartycka 18, 00-716 Warsaw, Poland\goodbreak
\and
Niels Bohr Institute, Blegdamsvej 17, Copenhagen, Denmark\goodbreak
\and
SISSA, Astrophysics Sector, via Bonomea 265, 34136, Trieste, Italy\goodbreak
\and
SMARTEST Research Centre, Universit\`{a} degli Studi e-Campus, Via Isimbardi 10, Novedrate (CO), 22060, Italy\goodbreak
\and
School of Physics and Astronomy, Cardiff University, Queens Buildings, The Parade, Cardiff, CF24 3AA, U.K.\goodbreak
\and
Space Sciences Laboratory, University of California, Berkeley, California, U.S.A.\goodbreak
\and
Special Astrophysical Observatory, Russian Academy of Sciences, Nizhnij Arkhyz, Zelenchukskiy region, Karachai-Cherkessian Republic, 369167, Russia\goodbreak
\and
Sub-Department of Astrophysics, University of Oxford, Keble Road, Oxford OX1 3RH, U.K.\goodbreak
\and
UPMC Univ Paris 06, UMR7095, 98 bis Boulevard Arago, F-75014, Paris, France\goodbreak
\and
Universit\'{e} de Toulouse, UPS-OMP, IRAP, F-31028 Toulouse cedex 4, France\goodbreak
\and
University of Granada, Departamento de F\'{\i}sica Te\'{o}rica y del Cosmos, Facultad de Ciencias, Granada, Spain\goodbreak
\and
University of Granada, Instituto Carlos I de F\'{\i}sica Te\'{o}rica y Computacional, Granada, Spain\goodbreak
% ...
}

  \abstract{We present the current accounting of systematic effect uncertainties for the Low Frequency Instrument (LFI) that are relevant to the 2015 release of the \Planck\ cosmological results, showing the robustness and consistency of our data set, especially for polarization analysis.  We use two complementary approaches: (i) simulations based on measured data and physical models of the known systematic effects; and (ii) analysis of difference maps containing the same sky signal (``null-maps'').  The LFI temperature data are limited by instrumental noise.  At large angular scales the systematic effects are below the cosmic microwave background (CMB) temperature power spectrum by several orders of magnitude.  In polarization the systematic uncertainties are dominated by calibration uncertainties and compete with the CMB $E$-modes in the multipole range 10--20.  Based on our model of all known systematic effects, we show that these effects introduce a slight bias of around $0.2\,\sigma$ on the reionization optical depth derived from the 70\,GHz $EE$ spectrum using the 30 and 353\,GHz channels as foreground templates.  At 30\,GHz the systematic effects are smaller than the Galactic foreground at all scales in temperature and polarization, which allows us to consider this channel as a reliable template of synchrotron emission.  We assess the residual uncertainties due to LFI effects on CMB maps and power spectra after component separation and show that these effects are smaller than the CMB amplitude at all scales. We also assess the impact on non-Gaussianity studies and find it to be negligible.  Some residuals still appear in null maps from particular sky survey pairs, particularly at 30 GHz, suggesting possible straylight contamination due to an imperfect knowledge of the beam far sidelobes.}

\maketitle
\alltwentyfifteenresultspapers


   \section{Introduction}
\label{sec_introduction}

  This paper, one of a set associated with the 2015 release of data from the \Planck\footnote{\Planck\ (\url{http://www.esa.int/Planck}) is a project of the European Space Agency (ESA) with instruments provided by two scientific consortia funded by ESA member states and led by Principal Investigators from France and Italy, telescope reflectors provided through a collaboration between ESA and a scientific consortium led and funded by Denmark, and additional contributions from NASA (USA).} mission, describes the Low Frequency Instrument (LFI) systematic effects and their related uncertainties in cosmic microwave background (CMB) temperature and polarization scientific products. Systematic effects in the High Frequency Instrument data are discussed in \citet{planck2014-a08} and \citet{planck2014-a09}.

  The 2013 \Planck\ cosmological data release \citep{planck2013-p01} exploited data acquired during the first 14 months of the mission to produce the most accurate (to date) all-sky CMB temperature map and power spectrum in terms of sensitivity, angular resolution, and rejection of astrophysical and instrumental systematic effects. In \citet{planck2013-p02a} we showed that known and unknown systematic uncertainties are at least two orders of magnitude below the CMB temperature power spectrum, with residuals dominated by Galactic straylight and relative calibration uncertainty.

  The 2015 release \citep{planck2014-a01} is based on the entire mission (48 months for LFI and 29 months for HFI). For LFI, the sensitivity increase compared to the 2013 release is a approximately a factor of two on maps. This requires a thorough assessment of the level of systematic effects to demonstrate the robustness of the results and verify that the final uncertainties are noise-limited.

  We evaluate systematic uncertainties via two complementary approaches: (i) using null maps\footnote{A null map is the difference between maps over time periods in which the sky signal is the same. See Sect.~\ref{sec_assessment_null_tests}} to highlight potential residual signatures exceeding the white noise. We call this a ``top-down'' approach; (ii) simulating all the known systematic effects from time-ordered data to maps and power spectra. We call this a ``bottom-up'' approach.  This second strategy is particularly powerful, because it allows us to evaluate effects that are below the white noise level and do not show up in our null maps.  Furthermore, it allows us to assess the impact of residual effects on Gaussianity studies and component separation.

  In this paper we provide a comprehensive study of the instrumental systematic effects and the uncertainties that they cause on CMB maps and power spectra, in both temperature and polarization.  
    
  We give the details of the analyses leading to our results in  Sects.~\ref{sec_overview} and \ref{sec_assessment} . In Sect.~\ref{sec_overview} we discuss the instrumental effects that were not treated in the previous release.  Some of these effects are removed in the data processing pipeline according to algorithms described in \citet{planck2014-a03}. In Sect.~\ref{sec_assessment} we assess the residual systematic effect uncertainties according to two complementary ``top-down'' and ``bottom-up'' approaches. 

  We present the main results in Sect.~\ref{sec_summary_table},   which provides  an overview of all the main findings. We refer, in particular, to Tables~\ref{tab_summary_systematic_effects_maps_30}, \ref{tab_summary_systematic_effects_maps_44} and \ref{tab_summary_systematic_effects_maps_70} for residual uncertainties on maps and  Figs.~\ref{fig_systematic_effects_power_spectrum_30} through \ref{fig_component_separation_global_compsep} for the impact on power spectra.  These figures contain the power spectra of the systematic effects and are often referred to in the text, so we advise the reader to keep them at hand while going through the details in Sects.~\ref{sec_overview} and \ref{sec_assessment}. 

  This paper requires a general knowledge of the design of the LFI radiometers. For a detailed description we recommend reading section~3 of \citet{bersanelli2010}. Otherwise the reader can find a brief and simple description in section~2 of \citet{planck2011-1.4}. Throughout this paper we follow the naming convention described in appendix~A of \citet{mennella2010} and also available on-line in the Explanatory Supplement.\footnote{\burl{http://wiki.cosmos.esa.int/planckpla/index.php/Main_Page}}


  \section{LFI systematic effects affecting LFI data}
\label{sec_overview}

 In this section we describe the known systematic effects affecting the LFI data, and list them in Table~\ref{tab_list_systematic_effects}. 


  \begin{table*}[tmb]     
  \begingroup
  \newdimen\tblskip \tblskip=5pt
  \caption{List of known instrumental systematic effects in \Planck-LFI.}
  \label{tab_list_systematic_effects}
  \nointerlineskip
  \vskip -3mm
  \footnotesize
  \setbox\tablebox=\vbox{
    \newdimen\digitwidth 
    \setbox0=\hbox{\rm 0} 
    \digitwidth=\wd0 
    \catcode`*=\active 
    \def*{\kern\digitwidth}
    \newdimen\signwidth 
    \setbox0=\hbox{+} 
    \signwidth=\wd0 
    \catcode`!=\active 
    \def!{\kern\signwidth}

  {
  \halign{ 
  #\hfil\tabskip=1em& #\hfil& #\hfil&
  #\hfil\tabskip=0pt\cr      
  \noalign{\doubleline}
  \omit\hfil{\bf Effect}\hfil&
  \omit\hfil{\bf Source}\hfil&
  \omit\hfil{\bf Control/Removal}\hfil&
  \omit\hfil{\bf Reference}\hfil\cr      
  \noalign{\vskip -3pt}
  \noalign{\vskip 5pt\hrule\vskip 3pt}
  \noalign{\vskip 10pt}
  \multispan4\hfil{\bf Effects independent of the sky signal (temperature
  and polarization)}\hfil\cr 
  \noalign{\vskip 6pt}
  White noise correlation& 
  Phase switch imbalance& 
  Diode {weighting}& 
  \citet{planck2013-p02a}\cr
  \noalign{\vskip 6pt}
  1/$f$ noise&
  RF amplifiers&
  Pseudo-correlation and destriping&
  \citet{planck2013-p02a}\cr
  \noalign{\vskip 6pt}
  Bias fluctuations&
  RF amplifiers, back-end electronics&
  Pseudo-correlation and destriping&
  \ref{sec_assessment_simulations_bias}\cr
  \noalign{\vskip 6pt}
  Thermal fluctuations&
  4-K, 20-K and 300-K thermal stages&
  Calibration, destriping&
  \ref{sec_assessment_simulations_thermal}\cr
  \noalign{\vskip 6pt}
  1-Hz spikes&
  Back-end electronics&
  Template fitting and removal&
  \ref{sec_assessment_simulations_spikes}\cr 
  \noalign{\vskip 10pt}
  \multispan4\hfil{\bf Effects dependent on the sky signal (temperature and
  polarization)}\hfil\cr 
  \noalign{\vskip 6pt}
  {Main beam ellipticity}& {Main beams}&
  {Accounted for in window function}&
  \citet{planck2014-a04}\cr
  \noalign{\vskip 6pt}
  {Near sidelobe}& {Optical response at angles $<5^\circ$}&
  {Masking of Galaxy and point sources}&
  \citet{planck2014-a03},\cr
  \omit{pickup}\hfil&\omit {from the main beam}\hfil&\omit&\ref{sec_overview_near_sidelobes}, \ref{sec_assessment_simulations_optics_pointing}\cr
  \noalign{\vskip 6pt}
  {Far} sidelobe pickup& Main and sub-reflector spillover&
  Model sidelobes removed from timelines&
  \ref{sec_overview_far_sidelobes}, \ref{sec_assessment_simulations_optics_pointing}\cr
%
  \noalign{\vskip 6pt}
  Analogue-to-digital&
  Back-end analogue-to-digital&
  Template fitting and removal&
  \ref{sec_assessment_simulations_adc_non_linearity}\cr
  \omit converter nonlinearity\hfil&\omit  converter\hfill&\omit&\cr
  \noalign{\vskip 6pt}
  Imperfect photometric&
  Sidelobe pickup, radiometer noise&Adaptive smoothing algorithm using $4\pi$&\citet{planck2014-a03},\cr
  \omit calibration\hfil&\omit temperature changes, and other\hfil&\omit beam, 4-K reference load voltage output, \hfil&  
  \ref{sec_overview_imperfect_calibration}, \ref{sec_assessment_simulations_imperfect_calibration}\cr
  \omit&\omit non-idealities\hfil&\omit temperature sensor data\hfil&\cr
  \noalign{\vskip 6pt}
  Pointing&
  Uncertainties in pointing reconstru-&
  Negligible impact on anisotropy&
  \ref{sec_overview_effects_optics_pointing}, \ref{sec_assessment_simulations_optics_pointing}\cr
  \omit&\omit ction, thermal changes affecting\hfil&\omit measurements\hfil&\cr
  \omit&\omit focal plane geometry\hfil&\omit&\cr
  \noalign{\vskip 10pt}
  \multispan4\hfil{\bf Effects specifically impacting polarization}\hfil\cr 
  \noalign{\vskip 6pt}
  Bandpass asymmetries&
  Differential orthomode transducer&
  Spurious polarization removal&
  \ref{sec_overview_bandpass_mismatch}\cr
  \omit&\omit and receiver bandpass response\hfil&\omit&\cr
  \noalign{\vskip 6pt}
  Polarization angle&
  Uncertainty in the  polarization&
  Negligible impact&
  \ref{sec_overview_polarisation_angle}, \ref{sec_assessment_simulations_optics_pointing}\cr
  \omit uncertainty&\omit angle in-flight measurement\hfil&\omit&\cr
  \noalign{\vskip 6pt}
  Orthomode transducer&
  Imperfect polarization separation&
  Negligible impact&
  \citet{leahy2010}\cr
  \omit cross-polarization&\omit \hfil&\omit&\cr
  \noalign{\vskip 5pt\hrule\vskip 3pt}
  }
  }}
  \endPlancktablewide       
  \endgroup
  \end{table*}        


 Several of these effects were already discussed in the context of the 2013 release \citep{planck2013-p02a}, so we do not repeat the full description here. They are:

\begin{itemize}
 \item white noise correlation;
 \item 1/$f$ noise;
 \item bias fluctuations;
 \item thermal fluctuations (20-K front-end unit, 300-K back-end unit,
 4-K reference loads);
 \item so-called ``1-Hz'' spikes, caused by the housekeeping acquisition clock;
 \item analog-to-digital converter nonlinearity.
\end{itemize}

Here we describe effects that are either polarization-specific or that have been treated differently in this data release. These effects are:
  
\begin{itemize}
  \item near sidelobes pickup;
  \item far sidelobes pickup;
  \item imperfect photometric calibration;
  \item pointing uncertainties;
  \item bandpass mismatch;
  \item polarization angle uncertainties.
\end{itemize}

Two other effects that are listed in Table~\ref{tab_list_systematic_effects} but are not discussed in this paper are: (i) main beam ellipticity, and (ii) orthomode transducer cross-polarization. The first is discussed in sections~5 and 6 of \citet{planck2014-a05}. The second is negligible for LFI, as it is shown in section 4.1 of \citet{leahy2010}.

\subsection{Optics and pointing}
\label{sec_overview_effects_optics_pointing}

  \subsubsection{Far sidelobes}
\label{sec_overview_far_sidelobes}

  The far sidelobes are a source of systematic error because they pick up radiation far from the telescope line of sight and give rise to so-called ``straylight contamination.'' The LFI 30\,GHz channel is particularly sensitive to the straylight contamination, because the diffuse Galactic emission components are rather strong at this frequency, and the far-sidelobe level of the 30\,GHz beams is significantly higher compared to the other frequencies   \citep[for more details, see][]{sandri2010}. The simulated pattern shown in Fig.~\ref{fig_beam_18_x} provides an example of the far sidelobes of a 70\,GHz radiometer. The plot is a cut passing through the main reflector spillover of the \Planck\ telescope \footnote{For the definition of the main- and sub-reflector spillovers refer to figure~7 of \citet{planck2013-p02a}}.

  Straylight impacts the measurements in two ways: it directly contaminates the maps; and it affects the photometric calibration. In the latter case, the straylight could be a significant fraction of the measured signal that is compared with the calibrator itself (i.e., the Dipole), causing a systematic error in the recovered calibration constants. This error varies with time, depending on the orientation of the Galactic plane with respect to the line of sight.

  In the 2013 release we did not correct the LFI data for the straylight contamination and simply estimated the residual uncertainty in the final maps and power spectra \citep[see table~2 and figure~1 of][]{planck2013-p02a}. 

  In the CMB polarization analysis, instead, we accounted for this effect, both in the calibration phase and in the production of the calibrated timelines. This is particularly relevant at 30\,GHz, while at 44 and 70\,GHz the straylight spurious signal is small compared to the CMB, both in temperature and polarization (see the green dotted spectra in Figs.~\ref{fig_systematic_effects_power_spectrum_30}, \ref{fig_systematic_effects_power_spectrum_44} and \ref{fig_systematic_effects_power_spectrum_70}).

  We perform straylight correction in two steps: first, we calibrate the data, accounting for the straylight contamination in the sky signal; and then we remove it from the data themselves. To estimate the straylight signal, we assume a fiducial model of the sidelobes based on \texttt{GRASP} beams and radiometer band shapes, as well as a fiducial model of the sky emission based on simulated temperature and polarization maps. We discuss the details of these procedures in sections~7.1 and 7.4 of \citet{planck2014-a03} and section~2 of \citet{planck2014-a06}.
  
  \subsubsection{Near sidelobes}
\label{sec_overview_near_sidelobes}

\begin{figure}[h!]
    \begin{center}
      \includegraphics[width=88mm]{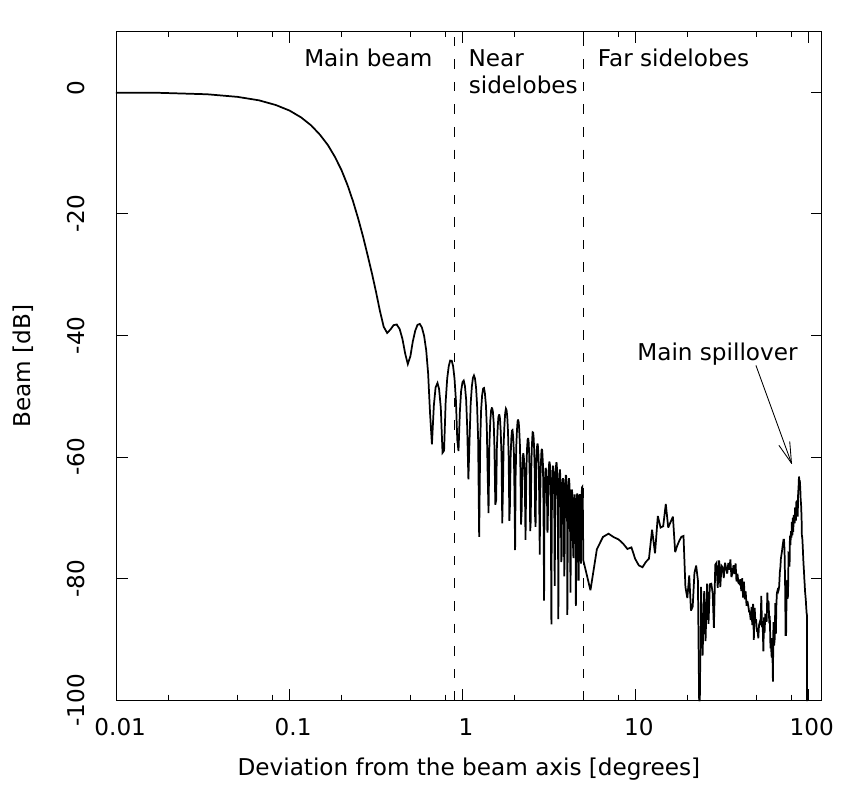}\\
    \end{center}
    \caption{Example of a cut of the simulated beam pattern of the 70\,GHz \texttt{LFI18-S} radiometer. The cut passes through the main reflector spillover of the \Planck\ telescope. The plot shows, in particular, the level and shape of the near sidelobes.}
  \label{fig_beam_18_x}
  \end{figure}

  The ``near sidelobes'' are defined as the lobes in the region of the beam pattern in the angular range extending between the main beam angular limit\footnote{The main beam is defined as extending to 1.9, 1.3, and 0.9\deg\ at 30, 44, and 70\,GHz, respectively.} and 5\deg  (see Fig.~\ref{fig_beam_18_x}) . We see that the power level of near sidelobes is about $-$40\,dB at 30\,GHz, and $-$50\,dB at 70\,GHz, with the shape of a typical diffraction pattern.
  
  Near sidelobes can be a source of systematic effects when the main beam scans the sky near the Galactic plane or in the proximity of bright sources. In the parts of the sky dominated by diffuse emission with little contrast in intensity, these lobes introduce a spurious signal of about $10^{-5}$ times the power entering the main beam. 
  
  We expect that the effect of near sidelobes on CMB measurements is small, provided that we properly mask the Galactic plane and the bright sources. For this reason we did not remove such an effect from the data and assessed its impact by generating simulated sky maps observed with and without the presence of near sidelobes in the beam and then taking the difference. We show and discuss such maps in Sect.~\ref{sec_assessment_simulations_optics_pointing} and the power spectra of this effect in Figs.~\ref{fig_systematic_effects_power_spectrum_30}, \ref{fig_systematic_effects_power_spectrum_44} and \ref{fig_systematic_effects_power_spectrum_70}.

\subsubsection{Polarization angle}
\label{sec_overview_polarisation_angle}

  We now discuss the systematic effect caused by the uncertainty in the orientation of the feed-horns in the focal plane. From thermo-elastic simulations we found this uncertainty to be about 0.2\deg\ \citep{lfi_alignment}. In this study we adopt a more conservative approach in which we set the uncertainties using measurements of the Crab Nebula. Then we perform a sensitivity study in which we consider a fiducial sky observed with a certain polarization angle for each feed-horn  and then reconstruct the sky with a slightly different polarization angle for each feed horn. The differences span the range of uncertainties in the polarization angle derived from measurements of the Crab Nebula. 
     
  In this section we first recall our definition of polarization angle and then we discuss the rationale we used to define the ``error bars'' used in our sensitivity study.

\begin{figure}[h!]
    \begin{center}
      \includegraphics[width=88mm]{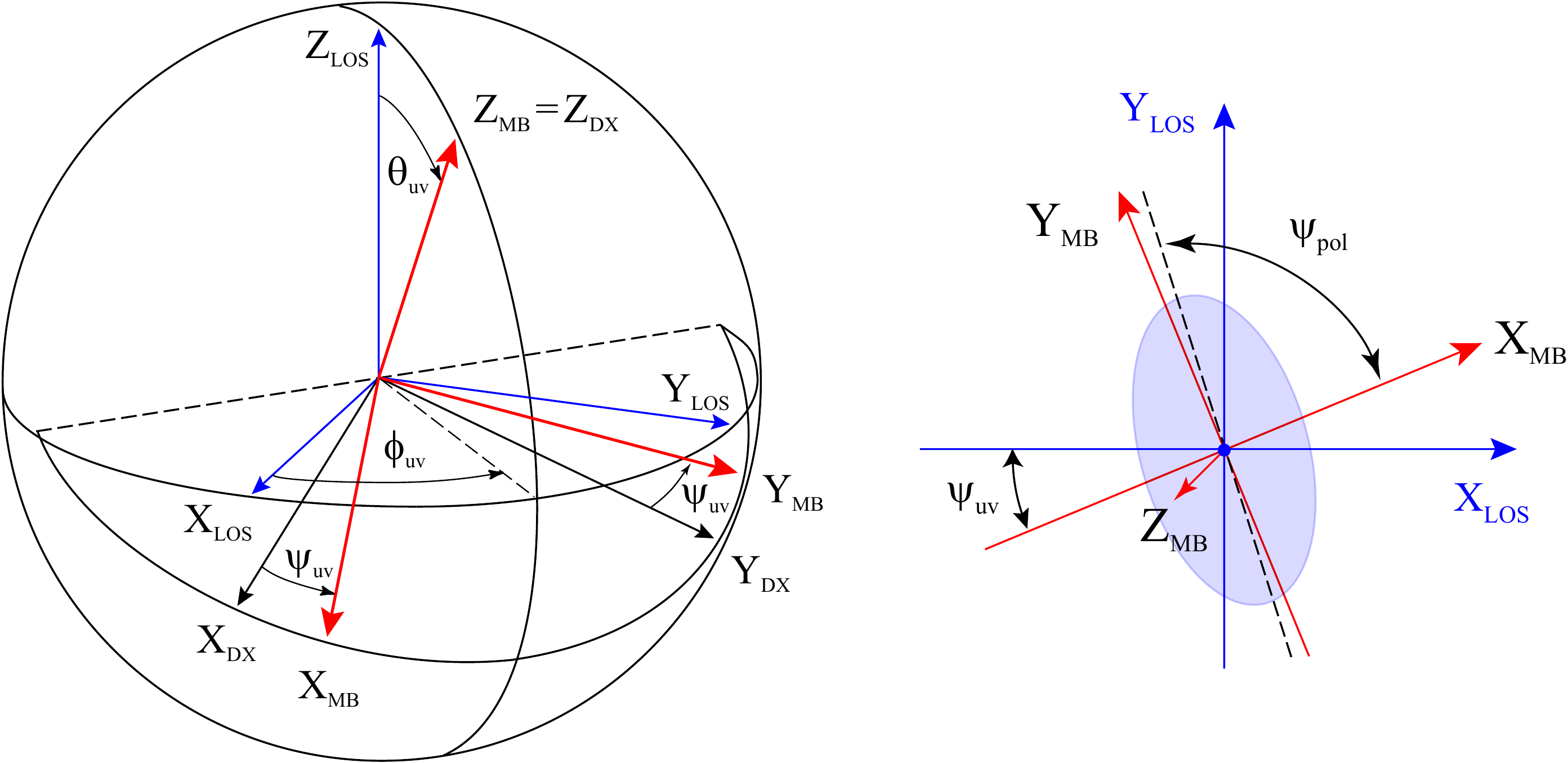}\\
    \end{center}
    
    \caption{Definition of polarization angle. \textit{Left}: the orientation of the main beam frame, (XYZ)$_\mathrm{MB}$, with respect to the line-of-sight (LOS) frame, (XYZ)$_\mathrm{LOS}$, is defined by the three angles $\theta_{uv}$, $\phi_{uv}$, and $\psi_{uv}$. The intermediate frame, (XYZ)$_\mathrm{DX}$, is the detector frame, defined by the two angles $\theta_{uv}$ and $\phi_{uv}$. \textit{Right}: the angle $\psi_\mathrm{pol}$ is defined with respect to X$_\mathrm{MB}$ and represents the orientation of the polarization ellipse along the beam line-of-sight. It is very close to 0 or 90 degrees for the \texttt{S} and \texttt{M} radiometers, respectively.}    \label{fig_polangle_definition}
  \end{figure}

  \paragraph{Definition of polarization angle.} Each LFI scanning beam\footnote{Here we refer to both the beams simulated with \texttt{GRASP} and to those reconstructed from Jupiter transits.} is defined in a reference frame specified by the three angles $\theta_\textrm{uv}$, $\phi_\textrm{uv}$, and $\psi_\textrm{uv}$, reported in table~5 of \citet{planck2014-a03} and shown in Fig.~\ref{fig_polangle_definition}. This choice implies that the power peak of the co-polar component lies along the main beam pointing direction, and a minimum in the cross-polar component appears in the same direction \citep{planck2014-a05}. In particular, the major axis of the polarization ellipse is along the $x$-axis for the radiometer side arm (\texttt{S}) and it is aligned with the $y$-axis for the radiometer main arm (\texttt{M}). 
  
  The main beam is essentially linearly polarized in directions close to the beam pointing. The $x$-axis of the main beam frame can be assumed to be the main beam polarization direction for the \texttt{S} radiometers and the $y$-axis of the main beam frame can be assumed to be the main beam polarization direction for the \texttt{M} radiometers.
      
  We define $\psi_\textrm{pol}$ to be the angle between the main beam polarization direction and the $x$-axis of the main beam frame, and define the main beam polarization angle, $\psi$, as $\psi$ = $\psi_\textrm{uv}$ + $\psi_\textrm{pol}$.  The angle $\psi_\textrm{pol}$ is nominally either 0\deg\ or 90\deg\ for the side and main arms, respectively.
      
  The values of $\psi_\textrm{pol}$ can be either determined from measured data using the Crab Nebula as a calibrator, or from optical simulations performed coupling the LFI feedhorns to the \Planck\ telescope, considering both the optical and radiometer bandpass response.\footnote{The polarization angle is defined as $\psi_\mathrm{pol} =\arctan(E_\mathrm{rhc} / E_\mathrm{lhc})$. Here $E_\mathrm{rhc}$ and $E_\mathrm{lhc}$ are the right- and left-hand circularly polarized components of the field, which can be defined in terms of the co- and cross-polar components, $E_\mathrm{co}$ and $E_\mathrm{cx}$, as $E_\mathrm{rhc(lhc)} = (E_\mathrm{co} -(+)\,E_\mathrm{cx}) / \sqrt{2}$ \citep{grasp_tech}.}

  For the current release our analysis uses values of $\psi_\textrm{pol}$ derived from simulations. Indeed, the optical model is well constrained by the main beam reconstruction carried out with seven Jupiter transits and provides us with more accurate estimates of the polarization angle compared to direct measurements.

    As an independent crosscheck, we also consider our measurements of the Crab nebula as a polarized calibrating source.  We use a least-squares fit of the time-ordered data measured during Crab scans to determine $I$, $Q$, and $U$ and, consequently, the polarization angle. Then we incorporate the instrument noise via the covariance matrix and we obtain the final error bars by adding in quadrature the uncertainties due to the bandpass mismatch correction \citep{planck2014-a03}.  
    
\begin{figure}[h!]
    \begin{center}
      \includegraphics[width=88mm]{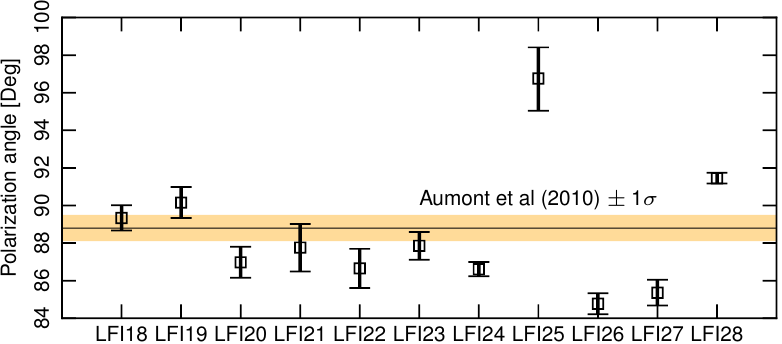}\\
    \end{center}
    
    \caption{Crab Nebula polarization angle measured by the various feed-horns in the focal plane. The straight horizontal line reports the value from \citet{Aumont2010} converted to Galactic coordinates, and the yellow area is the $\pm1\,\sigma$ uncertainty.}
    \label{fig_crab_polangle}
  \end{figure}

    While such a check is desirable, we find that the polarization angles derived from these data display systematic errors much larger than those expected from our noise and bandpass mismatch correction alone, especially at 30 and 44\,GHz (horns from \texttt{LFI24} through \texttt{LFI28}) (see Fig.~\ref{fig_crab_polangle}).  In particular, we find that the values obtained for the various horns in the focal plane display differences that are larger than our error estimates. 
    
    The horn with the largest apparent offset in angle, \texttt{LFI25}, is the solitary 44\,GHz horn on one side of the focal plane; in the next data release we will examine this discrepancy in more detail.

     An important difficulty is the determination of the relative gains of the individually polarized receivers, particularly during the Crab crossings, which appear near the minima of our principal temperature calibration.  Another souce of uncertainty missing from the Crab analysis is beam errors.  Of course, the LFI radiometer polarization angles are not changing over time, but the variability in the estimates limits our use of the Crab crossings for this purpose.

  \paragraph{Definition of error bars.} The $\psi_\textrm{pol}$ angle can differ from its nominal value because of small misalignments induced by the mechanical tolerances, thermo-mechanical effects during cooldown, and by uncertainties in the the optical and radiometer behavior across the band. If we consider the variation of $\psi_\textrm{pol}$ across the band in our simulations, for example, we find deviations from the nominal values that are, at most, 0.5\deg.

    To estimate the impact of imperfect knowledge of the polarization angle on CMB maps, we use the errors derived in the Crab analysis, which include the scan strategy and white noise and bandpass mismatch correction errors. While the errors derived this way are not designed to capture the time variation of the actual Crab measurements, we believe they provide a conservative upper bound to the errors in our knowledge of the instrument polarization angles.

\begin{figure}[h!]
    \begin{center}
      
\includegraphics[width=88mm]{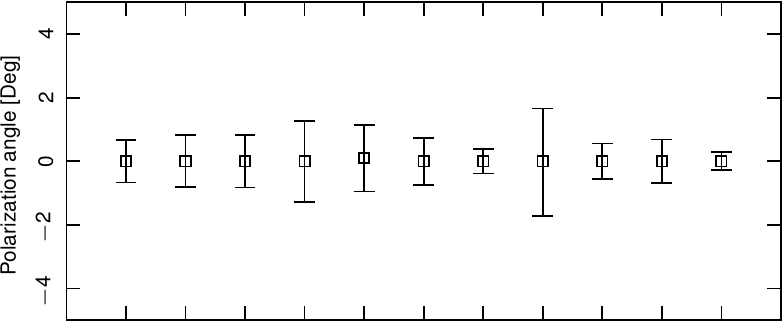}\\
      \vspace{1mm}
      \includegraphics[width=88mm]{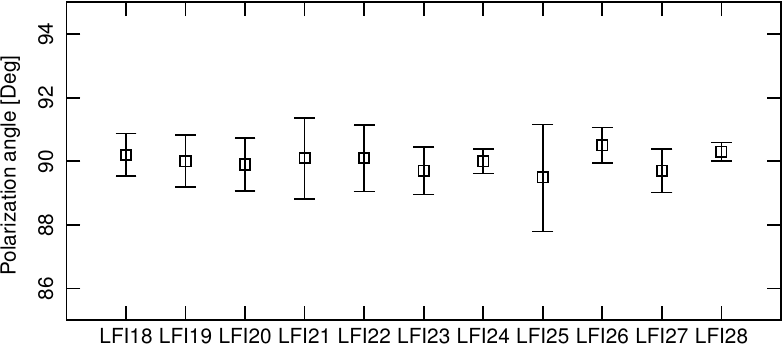}\\
      \vspace{2mm}
    \end{center}
    
    \caption{Simulated polarization angles and error bars from Crab measurements used in the analysis. \textit{Top}: radiometer main arm. \textit{Bottom}: radiometer side arm. The scatter of the plotted angles is much less than the error bars because the uncertainty in our simulations is much smaller than the error bars derived from Crab measurements.}
    \label{fig_polarization_angle_errorbars}
  \end{figure}

  The two panels in Fig.~\ref{fig_polarization_angle_errorbars} show the values of $\psi_\textrm{pol}$ derived from \texttt{GRASP} simulations (which are also used by the data analysis pipeline) and the error bars obtained from Crab observations. Notice that the scatter of the simulated angles is much less than the size of the error bars. This is consistent with uncertainty on the simulated angles that is much smaller than the error bars derived from Crab measurements. These data are the basis of the simulation exercise discussed in Sect.~\ref{sec_assessment_simulations_optics_pointing}.
      
    Our on-ground determination of radiometer polarization angles is more than sufficient for the CMB polarization.  As seen in the measurements of the Crab Nebula, the impact of gain errors among our polarized radiometers may be important, and we do include this effect in our gain error simulations.

\subsubsection{Pointing}
\label{sec_overview_pointing}

  Pointing reconstruction is performed in two steps. The first is the reconstruction of the satellite attitude, the second is the measurement of the orientation of the individual detectors with respect to the focal plane boresight (focal plane geometry reconstruction). In the first step we take into account all common-mode variations between the star camera and focal plane frames and assume the focal plane reconstruction, so that the focal plane geometry is essentially fixed over the entire mission.

  Planet scans indicate that the satellite attitude, reconstructed from the star camera data, contains slow timescale variations ($\ga 1$\,month) leading to total errors up to about 30\arcsec. The two major modes are a linear drift and a modulation that is heavily correlated with the Sun-Earth distance. To correct these fluctuations we fit a linear drift and a solar distance template to the planet position offsets, and include discontinuous steps at known disturbances of the thermal environment. Further details about the pointing reconstruction can be found in section~5.3 of \citet{planck2014-a01}.
  
  In this paper we evaluate the impact on the CMB maps and power spectra of residual uncertainties in the pointing reconstruction process. We perform the assessment using simulations in which the same sky is observed with two different pointing solutions that represent the uncertainty upper limit. We describe the approach and the results obtained in Sect.~\ref{sec_assessment_simulations_optics_pointing}.

\subsection{Imperfect calibration}
\label{sec_overview_imperfect_calibration}

  The analysis of the first data release showed that the uncertainty in the calibration is one of the main factors driving the systematic effects budget for \Planck-LFI. The accuracy of the retrieved calibration constant depends on the signal-to-noise ratio (S/N) between the dipole and instrumental noise along the scan directions, on effects causing gain variations (e.g., focal plane temperature fluctuations), and on the presence of Galactic straylight in the measured signal.
    
  In the analysis for the 2015 release we have substantially revised our calibration pipeline to account for these effects and to improve the accuracy of the calibration. The full details are provided in \citet{planck2014-a06}, and here we briefly list the most important changes: (i) we derive the Solar dipole parameters using LFI-only data, so that we no longer rely on parameters provided by \citet{hinshaw2009}; (ii) we take into account the shape of the beams over the full $4\pi$ sphere; (iii) we use an improved iterative calibration algorithm to estimate the calibration constant $K$ (measured in $\mathrm{V\,K}^{-1}$);\footnote{We have implemented such improvements into a new module named \texttt{DaCapo}, described in section~7.1 of \citet{planck2014-a03}.} and (iv) we use a new smoothing algorithm to reduce the statistical uncertainty in the estimates of $K$ and to account for gain changes caused by variations in the instrument environment.

  We nevertheless expect residual systematic effects in the calibration constants due to uncertainties in the following pipeline steps.

 \begin{enumerate}

  \item Solar dipole parameters derived from LFI data. This affects only the absolute calibration and impacts the overall dynamic range of the maps, as well as the power spectrum level. We discuss the absolute calibration accuracy in \citet{planck2014-a06} and do not address it further here.

  \item Optical model and radiometer bandpass response. This enters the computation of the $4\pi$ beams, which are used to account for Galactic straylight in the calibration.

  \item A number of effects (e.g., the impact of residual Galactic foregrounds)  that  might bias the estimates of the calibration constant $K$.

  \item The smoothing filter we use to reduce the scatter in the values of $K$ near periods of dipole minima might be too aggressive, removing features from the set of $K$ measurements that are not due to noise. This could cause systematic errors in the temperature and polarization data. 

 \end{enumerate} 
    
  We estimate the residual calibration uncertainties using simulations, as discussed in Sect.~\ref{sec_assessment_simulations_imperfect_calibration}. For this release we neglect effects caused by imperfect knowledge of the far sidelobes. In \citet{planck2014-a06} we provide an overall upper limit based on the consistency of power spectra derived from different radiometers.
    
  We are currently evaluating ways to improve this assessment in the context of the next \Planck\ release. One possibility would be to use Monte Carlo simulations to assess the impact on calibration of uncertainties in the beam far sidelobes.
 
 
    \subsection{Bandpass mismatch}
\label{sec_overview_bandpass_mismatch}

  Mismatch between the bandpasses of the two orthogonally-polarized arms of the LFI radiometers causes leakage of foreground total intensity into the polarization maps. The effect and our correction for it are described in section~11 of \citet{planck2014-a03} and references therein. A point to note is that the correction is only applied to at an angular resolution of 1\degr, although appendix C of \citet{planck2014-a35} describes a special procedure for correcting point source photometry derived from the full resolution maps.

  Residual discrepancies between the blind and model-driven estimates of the leakage are noted in \citet{planck2014-a03}, which imply that the small (typically $< 1$\,\%) mismatch corrections are not perfect.  The estimated fractional uncertainty in these corrections is $< 25$\,\% at 70\,GHz and $< 3$\,\% at 30\,GHz; the discrepancies are significant only because they are driven by the intense foreground  emission on the Galactic plane. 

  As detailed in Sect.~\ref{sec_masks}, our cosmological analysis of polarization data is restricted to 46\,\% of the sky with the weakest foreground emission. \citet{planck2014-a13} demonstrates that in this region the bandpass correction has a negligible effect on the angular power spectrum and cosmological parameters derived from it, the optical depth to reionization, $\tau$, and the power spectrum amplitude. The same applies to our upper limit on the  tensor-to-scalar  ratio,  Consequently the  impact of the {\em uncertainty} in the correction is also negligible for the cosmological results.


  \section{Assessing residual systematic effect uncertainties in maps and power spectra}
\label{sec_assessment}
      
	In this section we describe our assessment of systematic effects in the LFI data, which is based on a two-steps approach.
	
	The first is to simulate maps of each effect (see Table~\ref{tab_list_simulated_effects}) and combine them into a \textit{global} map that contains the sum of all the effects. We perform simulations for various time intervals, single surveys, individual years and full mission, and we use such simulations to produce a set of \textit{difference maps}. For example, we construct global systematic effects year-difference maps as the sum of all systematic effects for one year subtracted from from the sum of all systematics from another year. We also compare the pseudo-spectra computed on the full-mission maps with the expected sky signal to assess the impact of the various effects. This step is described in Sect.~\ref{sec_assessment_simulations}.
	
	The second is to calculate the same difference maps from flight data. We call these maps \textit{null maps}, because they should contain only white noise, as the sky observed in the time intervals of each pair of maps is the same. Here we compare the null maps pseudo-spectra with the pseudo-spectra of the \textit{global} systematic effects difference maps. Our objective, in this case, is to highlight any residuals in the pseudo-spectra obtained from flight data that are not accounted for by our simulations. This step is described in Sect.~\ref{sec_assessment_null_tests}.


\begin{table}[tmb]               
  \begingroup
  \newdimen\tblskip \tblskip=5pt
  \caption{List of the simulated systematic effects.}            
  \label{tab_list_simulated_effects}                     
  \nointerlineskip
  \vskip 2mm
  \footnotesize
  \setbox\tablebox=\vbox{
    \newdimen\digitwidth 
    \setbox0=\hbox{\rm 0} 
    \digitwidth=\wd0 
    \catcode`*=\active 
    \def*{\kern\digitwidth}
    \newdimen\signwidth 
    \setbox0=\hbox{+} 
    \signwidth=\wd0 
    \catcode`!=\active 
    \def!{\kern\signwidth}
  {\tabskip=0pt
  \halign{ 
  \hbox to 1.0in{#\leaderfil}\tabskip=0em& 
  \hfil#\cr                       
  \noalign{\doubleline}
  {Optical effects}& Near sidelobes\cr
  \omit                   & Pointing uncertainty\cr
  \omit                   &  Polarization angle uncertainty\cr
  \noalign{\vskip 6pt}
  {Thermal effects}& 4\,K stage temp. fluct.\cr
  \omit                   &  20\,K stage temp. fluct.\cr
  \omit                   & 300\,K stage temp. fluct.\cr
  \noalign{\vskip 6pt}
  {Calib. dependent}& ADC non-linearity\cr
  \omit                    &  Calibration uncertainty\cr
  \noalign{\vskip 6pt}
  {Electronics}     & 1-Hz spikes\cr
  \omit                    & Bias fluctuations\cr
    \noalign{\vskip 5pt\hrule\vskip 3pt}
  }
  }}
  \endPlancktable          

  \endgroup
\end{table}   


	In all cases we compute pseudo-spectra using the HEALPix \texttt{anafast} code and correct for the fraction of observed sky. In other words, in all the power spectra of this work we have $C_\ell = C_{\ell,\mathrm{anafast}} / f_\mathrm{sky}$, where $C_{\ell,\mathrm{anafast}}$ is the power spectrum as obtained by the \texttt{anafast} code and $f_\mathrm{sky}$ is the fraction of observed sky.
      
	In Sect.~\ref{sec_masks} we start by reviewing the masks applied in the calculation of the pseudo-spectra used in our assessment.

  
    \subsection{Masks}
\label{sec_masks}
\begin{figure}[!htpb]
    \begin{center}
	\includegraphics[width=88mm]{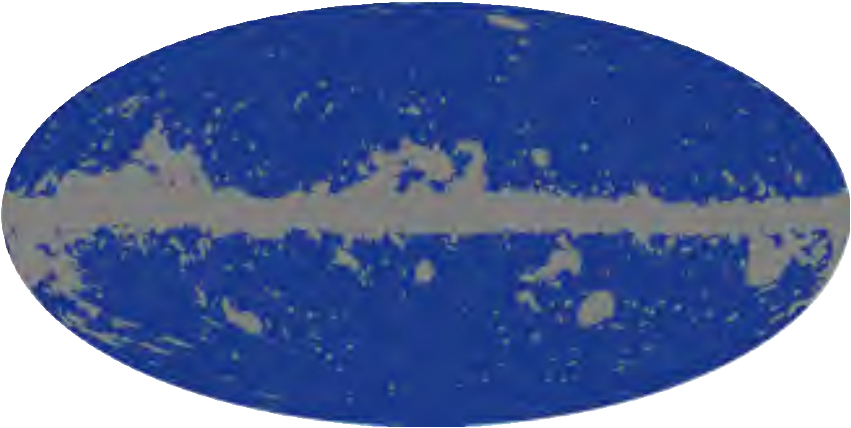}\\
	\vspace{.5cm}
	\includegraphics[width=88mm]{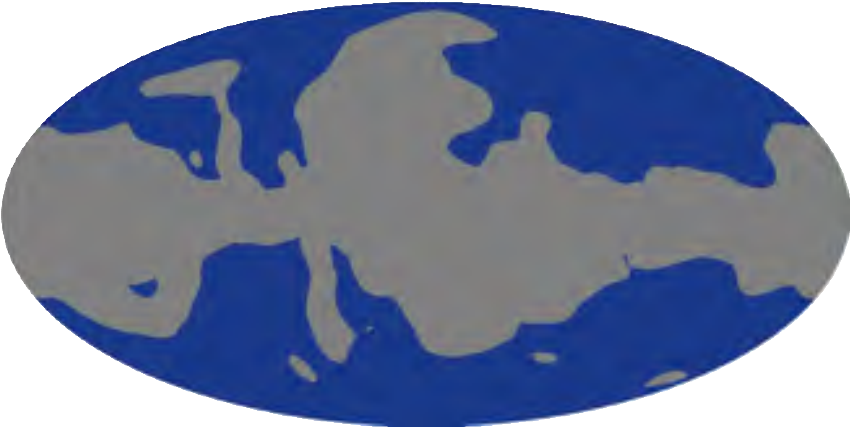}\\
	\vspace{.5cm}
	\includegraphics[width=88mm]{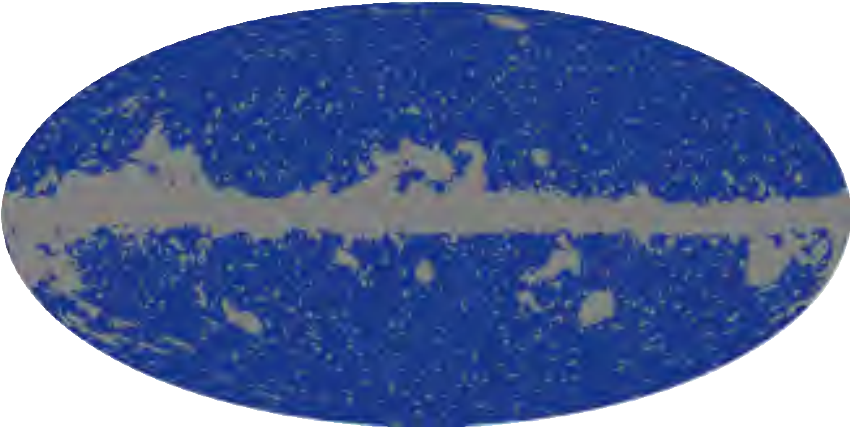}
    \end{center}
    \caption{Masks used in this work. \textit{Top:} mask used in the computation of power spectra of simulated systematic effect total intensity maps. \textit{Middle:} mask used in the computation of power spectra of $Q$ and $U$ systematic effect maps. \textit{Bottom:} mask used in the computation of power spectra of null maps.}
    \label{fig_masks}
  \end{figure}

  We have used three masks to compute the power spectra discussed in this paper , and we show them in the three panels of Fig.~\ref{fig_masks}.

  The first mask (top panel of Fig.~\ref{fig_masks}) is used for total intensity maps of the systematic effects. It removes the Galactic plane and point sources. It is the ``UT78'' mask described in section~4.1 of \citet{planck2014-a11}, obtained by combining the {\tt Commander}, {\tt SEVEM}, and {\tt SMICA} confidence masks.

  The second mask (middle panel of Fig.~\ref{fig_masks}) is used for $Q$ and $U$ maps of the systematic effects. It removes about 54\,\% of the sky, cutting out a large portion of the Galactic plane and the Northern and Southern Spurs. We adopted this mask in the low-$\ell$ likelihood used to extract the reionization optical depth parameter, $\tau$ \citep[see figure~3 in][]{planck2014-a13}. We chose to use the same mask in the assessment of systematic effect uncertainties in polarization.

  The third mask (bottom panel of Fig.~\ref{fig_masks}) is used in the null maps analysis at all frequencies both in temperature and polarization. We obtained this mask by combining the UPB77 30-GHz polarization mask \citep[right panel of figure~1 in][]{planck2014-a11} and the 30-GHz point source mask used for the 2013 release described in section~4 of \citet{planck2013-p06}\footnote{Because difference maps may contain unobserved pixels, in each null test we take the union between this mask and any set of unobserved pixels. For example, maps of single surveys do not cover the full sky, which requires us to combine the mask with the unobserved pixels in the null map.}.
   
      \subsection{Assessing systematic effects via simulations (``bottom-up'')}
  \label{sec_assessment_simulations}
  
    \subsubsection{Optics and pointing}
    \label{sec_assessment_simulations_optics_pointing}
  
        \paragraph{Far and near sidelobes.} To assess the effect of far and near sidelobes we simulate the residual signal by observing a fiducial sky with a beam pattern containing only the beam component being tested. The sky signal contains the CMB and foregrounds as observed by each radiometer, including the spurious polarization caused by the bandpass mismatch.  Further details regarding the the fiducial sky can be found in \citet{planck2014-a14}, a description of the simulation pipeline is in \citet{reinecke2006}, and the \texttt{Madam} mapmaker is described in \citet{kurki-suonio2009, keihanen2010}.

  We perform this assessment by projecting timelines of the sky convolved with the beam pattern into maps. First we use the spherical harmonic components of the beam \citep{planck2014-a05} to generate timelines.\footnote{In this step we use the {\tt conviqt\_v4} and {\tt multimod} routines \citep{reinecke2006} with parameters reported in table~6 of \citet{planck2014-a05}.}  Then we produce maps using the {\tt Madam} mapmaker with the same parameters used for the sky maps \citep{planck2014-a07}.
  
  We simulate far sidelobes with the {\tt GRASP} Mr-GTD\footnote{Multi reflector geometrical theory of diffraction \citep{grasp_mgtd}.} analysis, considering all the first-order contributions and two contributions at second order (reflections and diffractions from the sub-reflector, which are diffracted by the main reflector). This choice allows us to keep the computational time within the available CPU resources at the cost of a small fraction of power that is not accounted for in the sidelobe region of the beam pattern. 
  
  The power lost because of our approximation is $\la 0.5\,\%$ \citep[see table~1 in][]{planck2014-a05}. To estimate the level of uncertainty introduced by this lost power, we rescale the far-sidelobe spherical harmonic coefficients so that the total beam efficiency is equal to 100\,\%. Then we simulate maps of the effect from native and rescaled sidelobes and compare the two corresponding power spectra.
  
\begin{figure}[!htpb]
      \includegraphics[width=88mm]{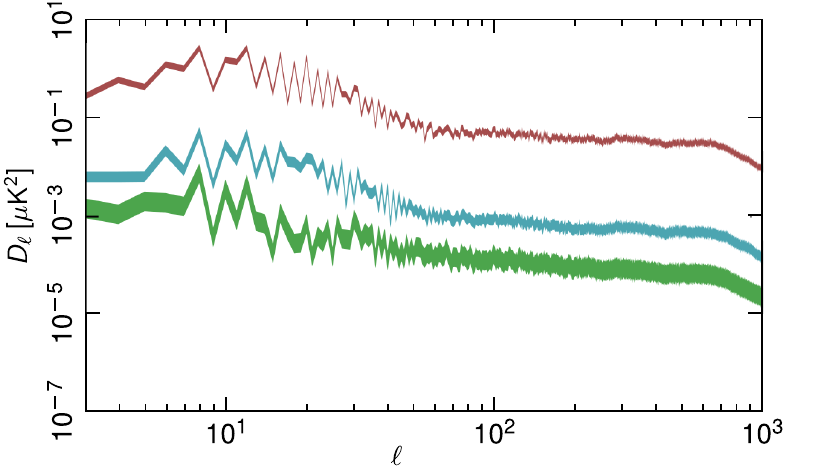} \\
      \includegraphics[width=88mm]{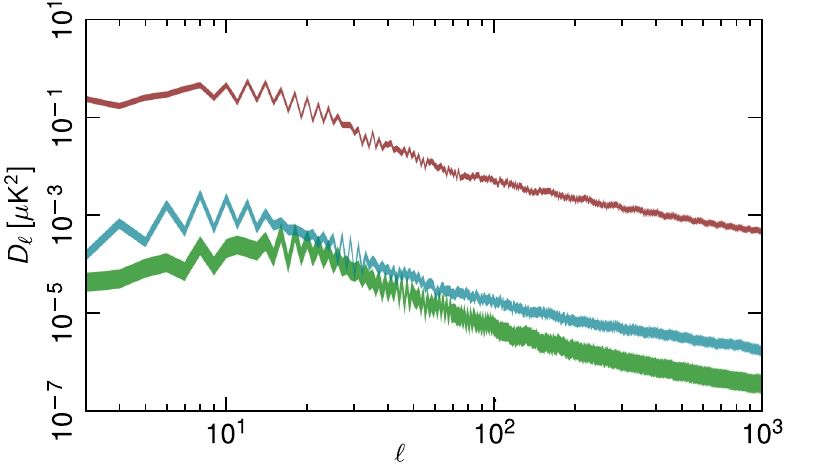} \\
      \includegraphics[width=85mm]{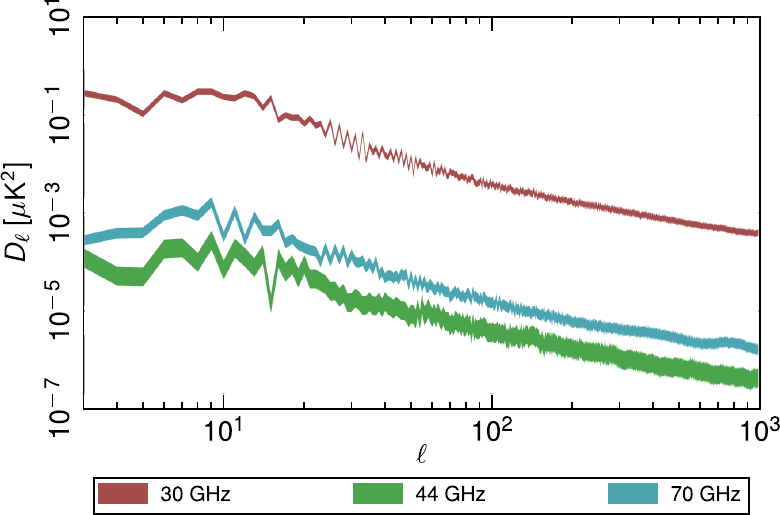}
    \caption{Uncertainty in the power spectra of the effect from far sidelobes
    introduced by the first-order approximation in {\tt GRASP} simulations.
    \textit{Top:} $TT$ spectrum. \textit{Middle:} $EE$ spectrum.
    \textit{Bottom:} $BB$ spectrum. For each frequency the coloured area is the
    region between the native power spectrum and the one rescaled to account
    for the missing power.}
  \label{fig_sidelobes_rescaling}
  \end{figure}

  Figure~\ref{fig_sidelobes_rescaling} shows the impact of this approximation. The coloured area is the region in $\ell$-space between the two power spectra and represents the uncertainty due to the first-order approximation in the {\tt GRASP} analysis. For the purpose of comparing the power spectrum of systematic effects with the sky signal, this uncertainty is small and we have neglected it.
  
\begin{figure*}[!htpb]
    \begin{center}
      \begin{tabular}{m{.25cm} m{5.6cm} m{5.6cm} m{5.6cm}}
	&\begin{center}$I$\end{center} &\begin{center}$Q$\end{center}
        &\begin{center}$U$\end{center}\\    
	30&
	\includegraphics[width=56mm]{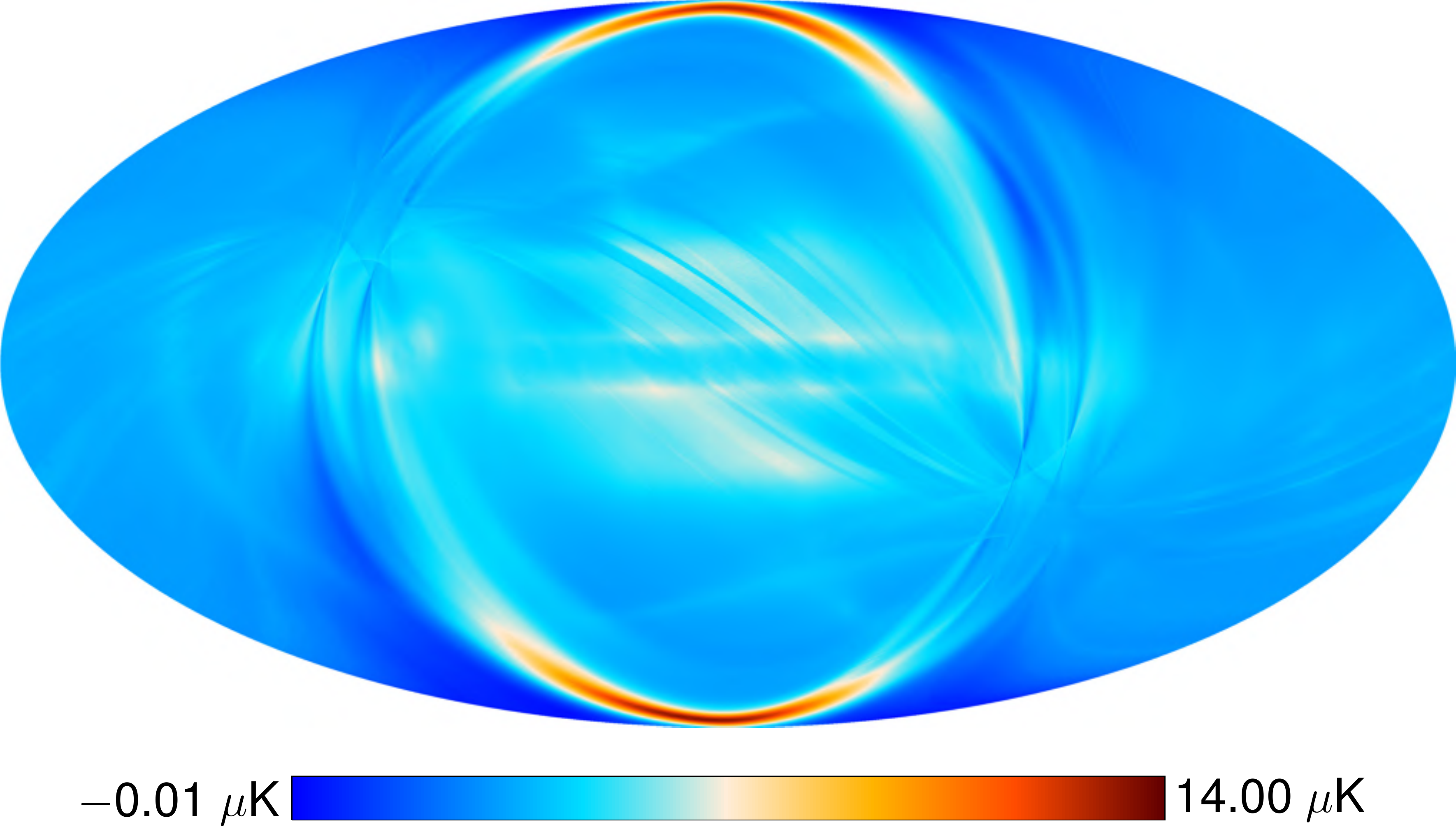}&
	\includegraphics[width=56mm]{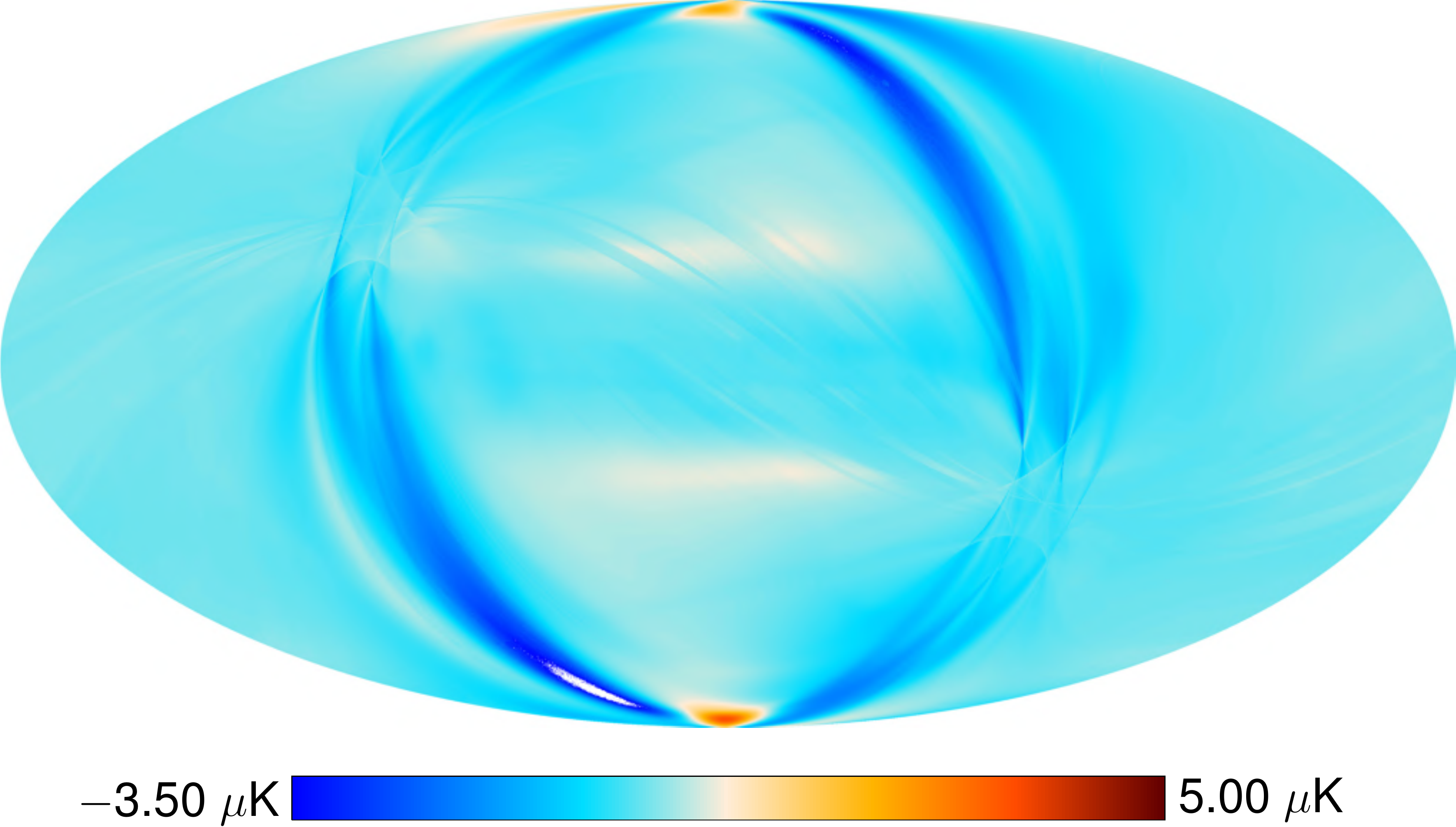}&
	\includegraphics[width=56mm]{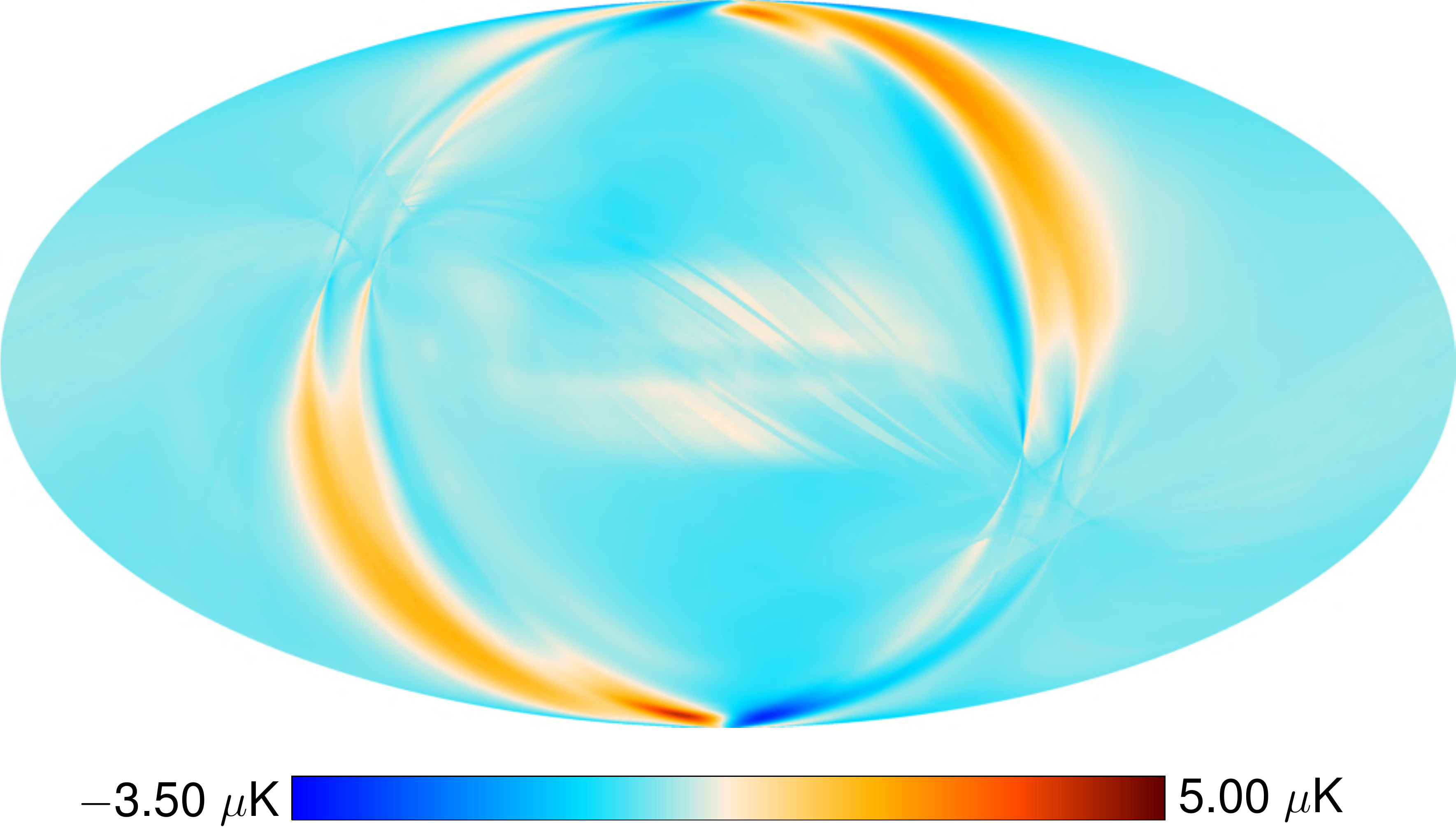}\\
	&&&\\
	44&
	\includegraphics[width=56mm]{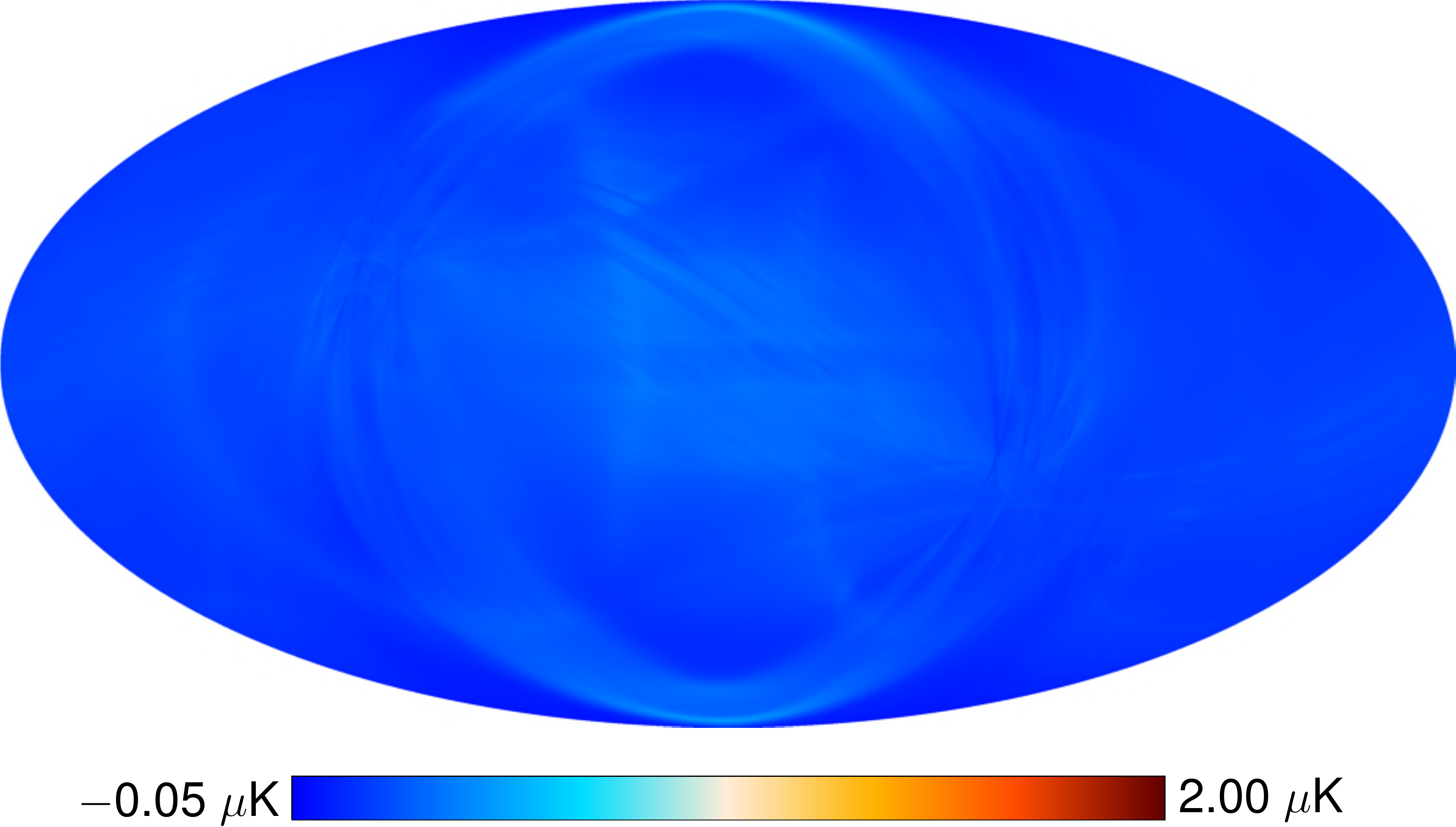}&
	\includegraphics[width=56mm]{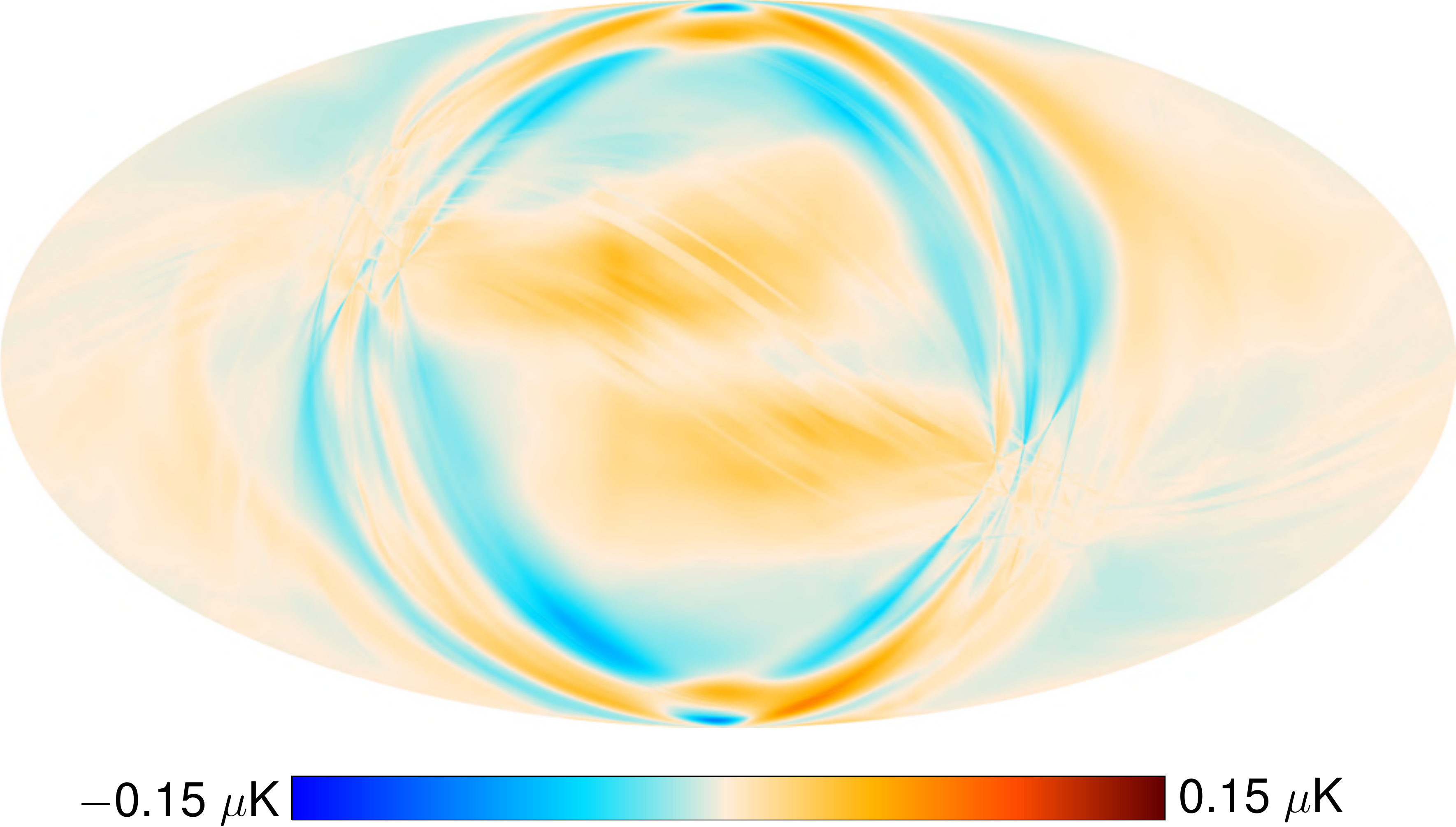}&
	\includegraphics[width=56mm]{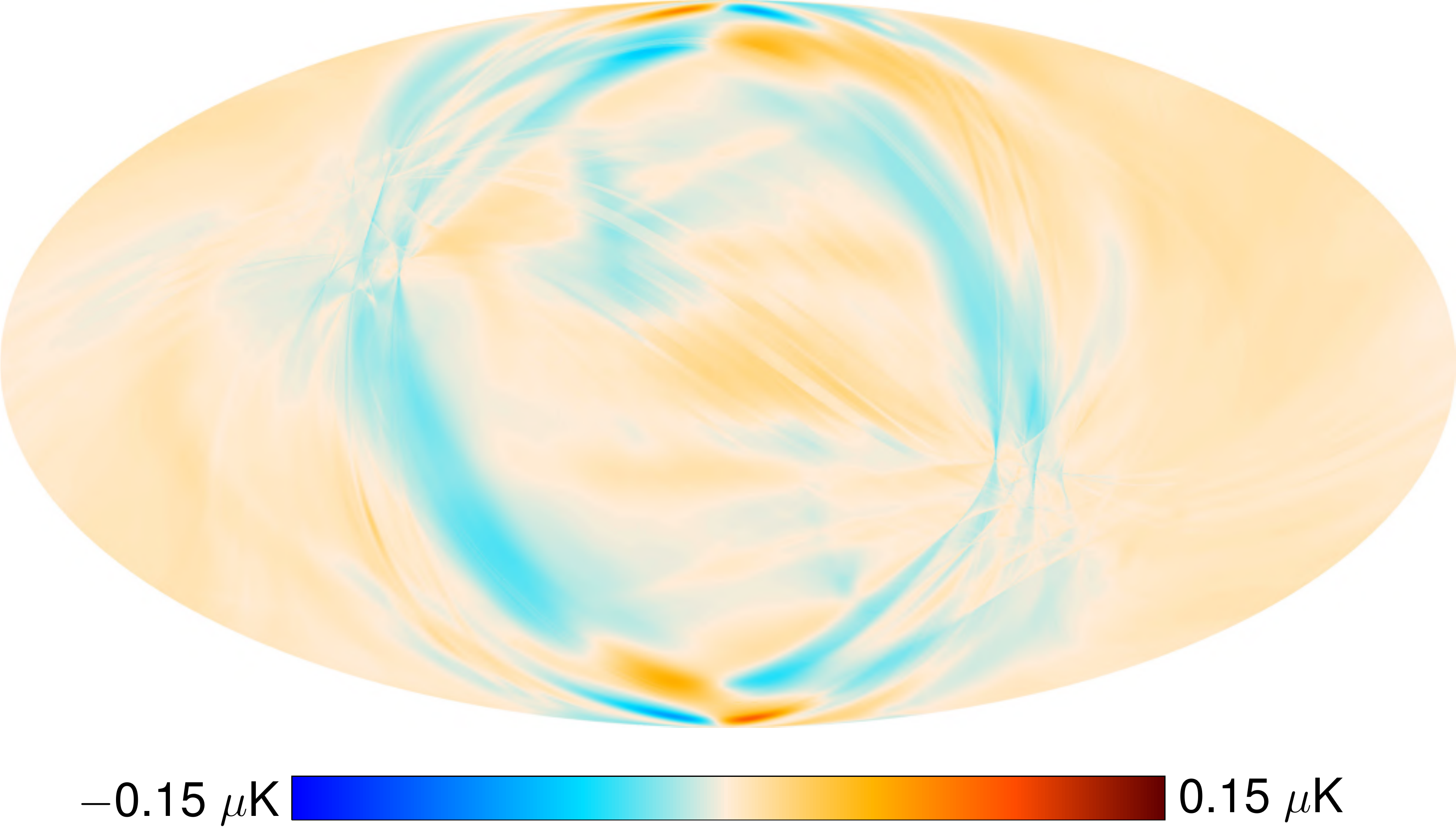}\\
	&&&\\
	70&
	\includegraphics[width=56mm]{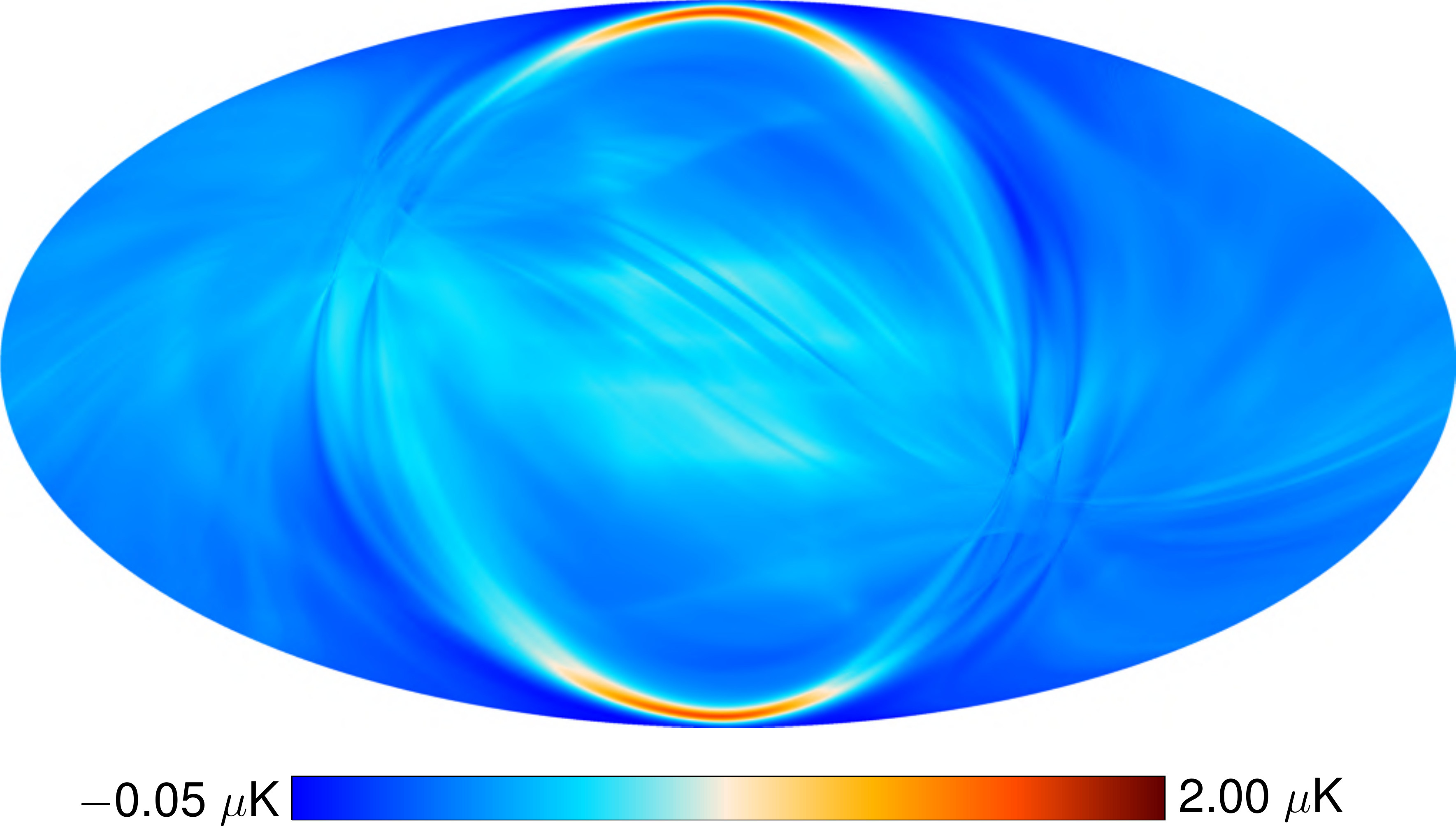}&
	\includegraphics[width=56mm]{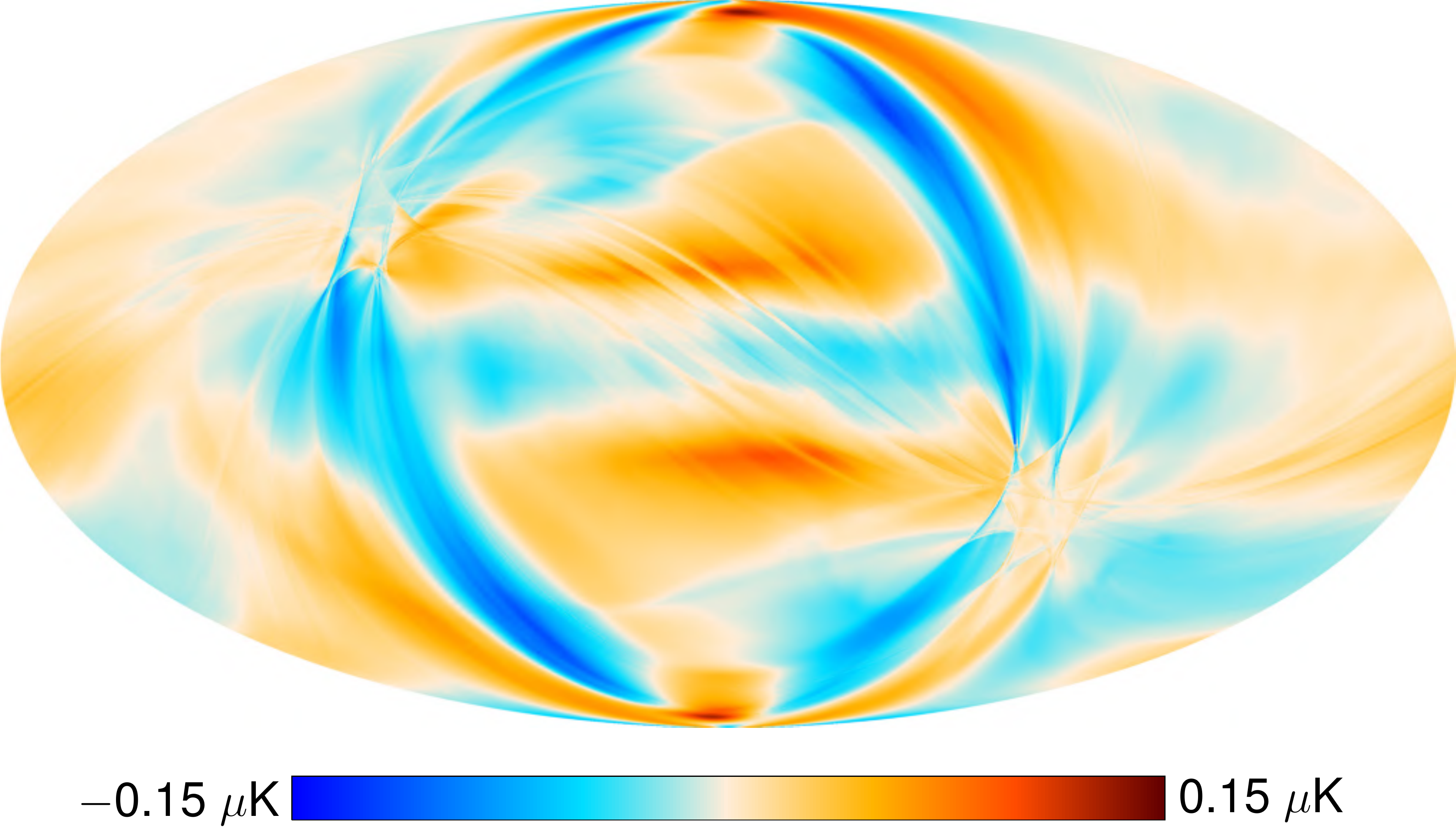}&
	\includegraphics[width=56mm]{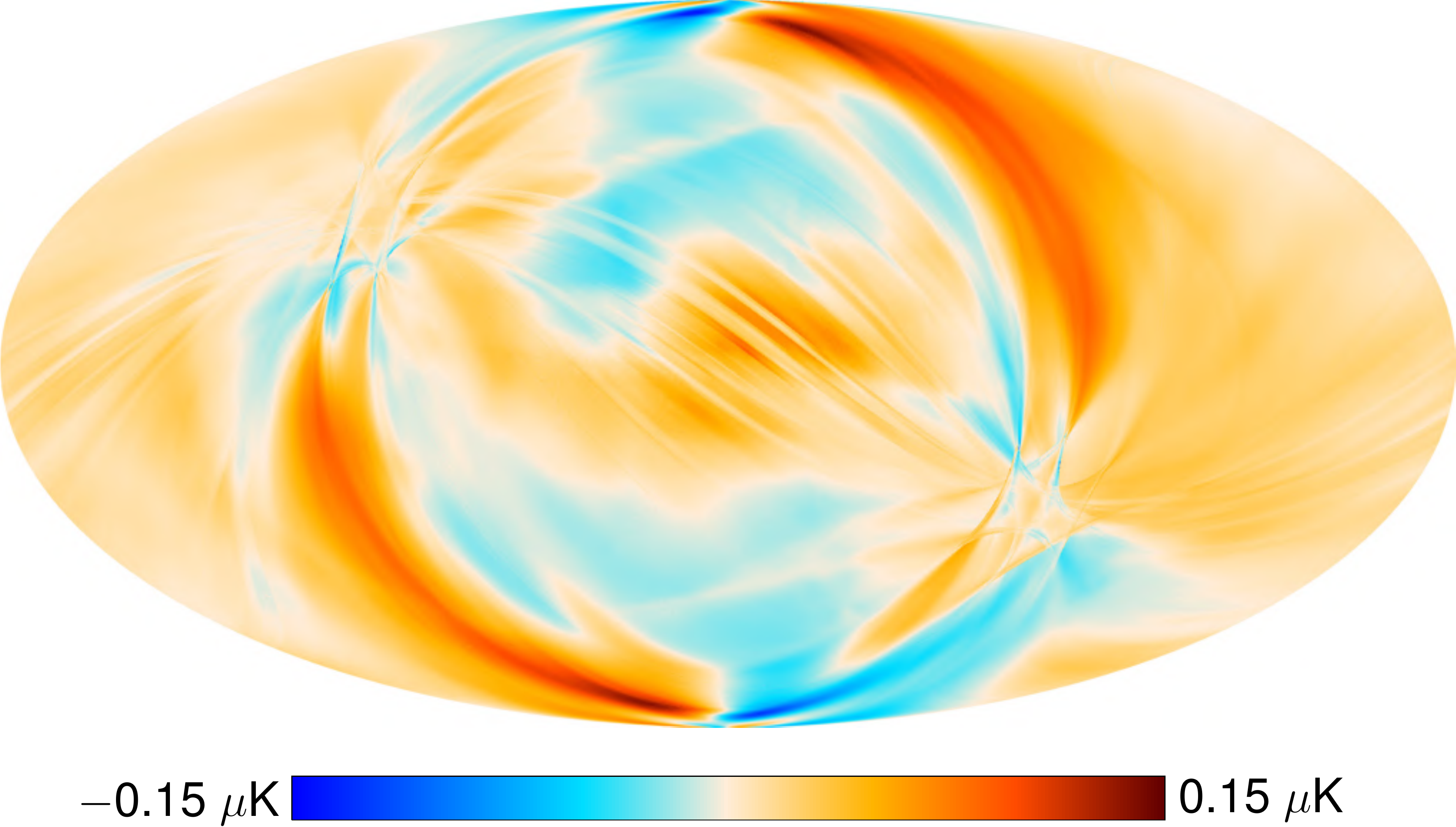}\\
	\end{tabular}
    \end{center}

    \caption{Maps of the effect from far sidelobes. Rows correspond to 30, 44, and 70\,GHz channels, while columns correspond to $I$, $Q$, and $U$. The straylight signal at 30\,GHz is larger compared to 44 and 70\,GHz. The 44\,GHz channel is the least contaminated by this effect. This behavior results from the combination of the beam far sidelobes and the intensity of the Galactic emissions at the LFI frequencies.}
  \label{fig_whole_mission_far_sidelobes_maps}
  \end{figure*}

  Figure~\ref{fig_whole_mission_far_sidelobes_maps} shows full-sky maps of the Galactic straylight detected by the far-sidelobe beam patterns. This signal is removed from the timelines, so we do not consider it in the budget of systematic uncertainties. Moreover, in Figs.~\ref{fig_systematic_effects_power_spectrum_30}, \ref{fig_systematic_effects_power_spectrum_44}, and \ref{fig_systematic_effects_power_spectrum_70} we plot the power spectrum of this effect and show that even if we did not remove it from the data, the effect would be at a level much lower than the sky signal.
      
  Our results show that the effect from Galactic straylight is significantly larger at 30\,GHz compared to 44 and 70\,GHz, for which the level of the spurious signal is similar. 
  
  To understand this result one must consider that this effect depends on two factors: (i) the level of the beam far sidelobes and (ii) the intensity of the Galactic signal. The level of the beam far sidelobes results from the coupling of the feed-horns beam pattern with the secondary mirror. 
  
  A larger feed-horn main beam \textit{illuminates} more effectively the secondary mirror. This determines a narrower main beam of the entire optical system and a higher level of far sidelobes. In \Planck-LFI the 44\,GHz feed-horns have narrower beams compared to the 30 and 70\,GHz horns and, for this reason, the power in the far sidelobes is significantly smaller. 
  
  If we also consider that the Galactic signal intensity decreases with frequency we understand our result, which is completely consistent with pre-launch optical simulations. The interested reader can find more details in \citet{Sandri2004}, \citet{Burigana2004} and \citet{sandri2010}.

  Ideally our analysis would assess the impact of the accuracy in our far-sidelobe model on the systematic effects analysis. The proper way to do this would be to identify the sources of uncertainty in the model and run Monte Carlo simulations, producing several far sidelobes with {\tt GRASP} and propagating the analysis to sky maps and power spectra.
    
  Such an analysis would require a considerable amount of computing time and we did not perform it for this release, but instead rely on null-test analyses. Null maps from consecutive surveys are quite sensitive to the pickup of straylight by the far sidelobes and can be used to assess the presence of straylight residuals in the data. We discussed this point in Sect.~\ref{sec_assessment_null_tests}. 

  We study the effect coming from the near sidelobes following the same procedure used for the far sidelobes. The main difference is that, in this case, we do not apply any correction to the data, so that our simulations estimate the systematic effect that we expect to be present in the data.

\begin{figure*}[!htpb]
    \begin{center}
      \begin{tabular}{m{.25cm} m{5.6cm} m{5.6cm} m{5.6cm}}
	& \begin{center}$I$\end{center} &\begin{center}$Q$\end{center}&\begin{center}$U$\end{center}\\    
	30&
	\includegraphics[width=56mm]{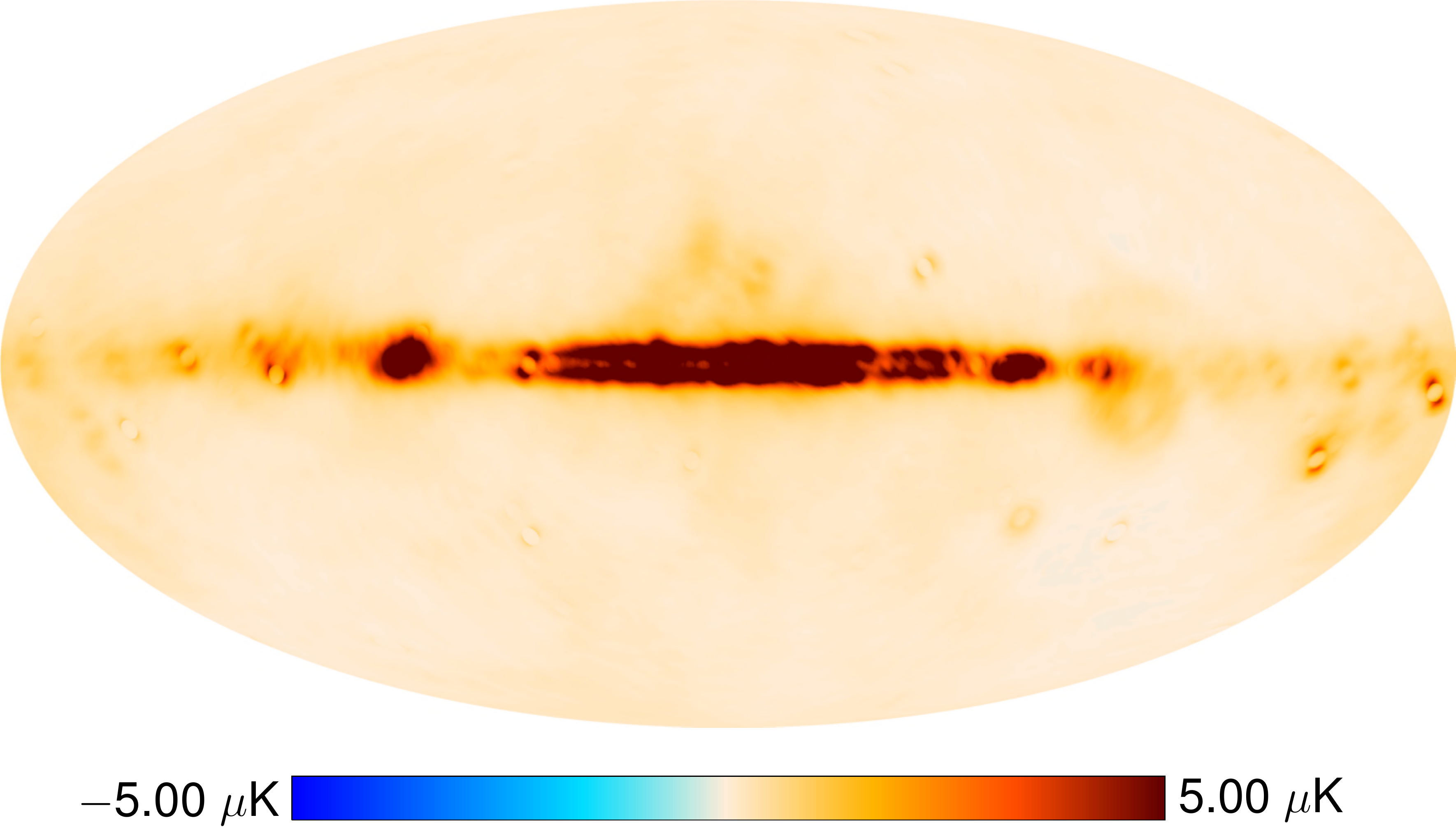}&
	\includegraphics[width=56mm]{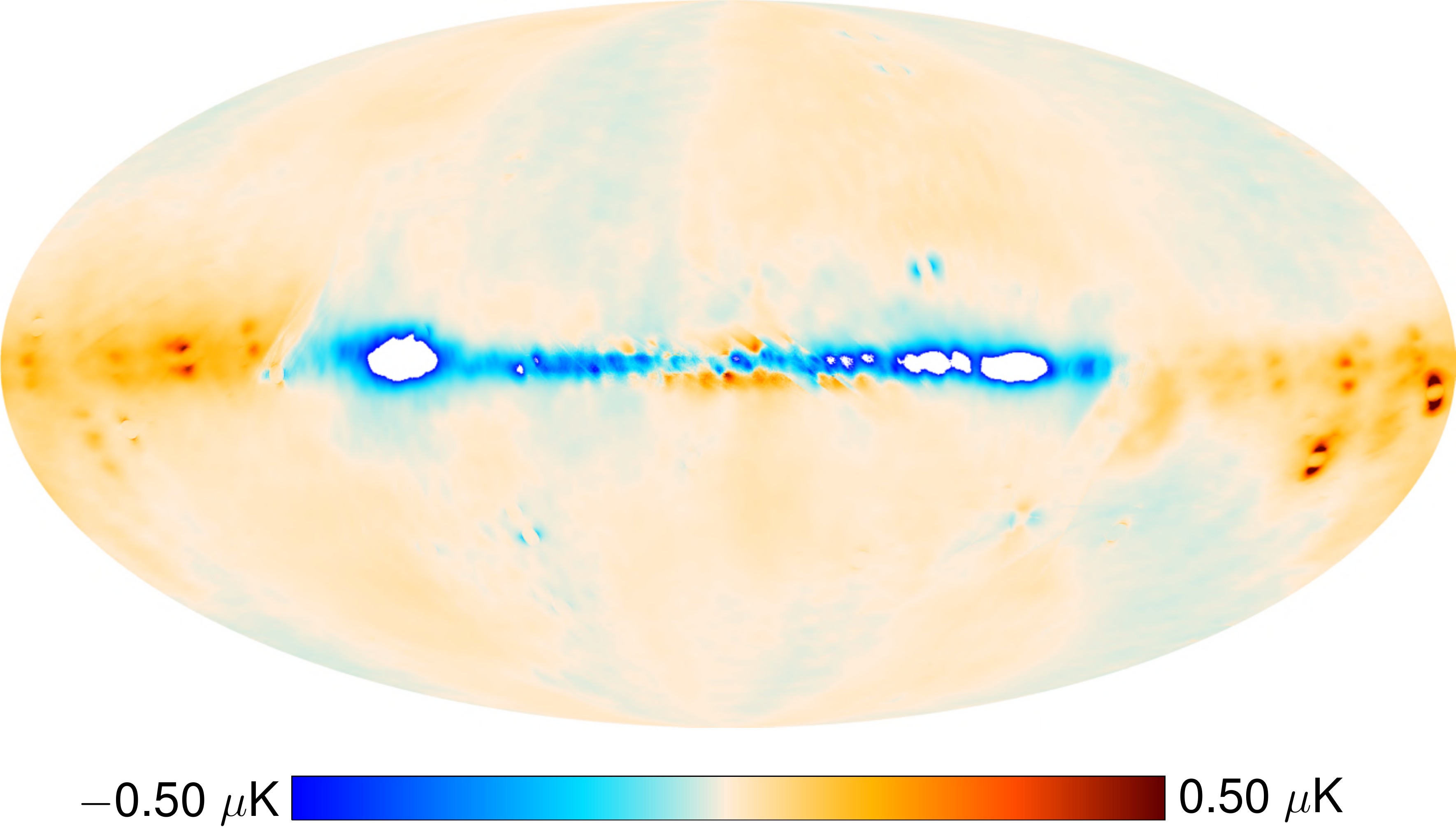}&
	\includegraphics[width=56mm]{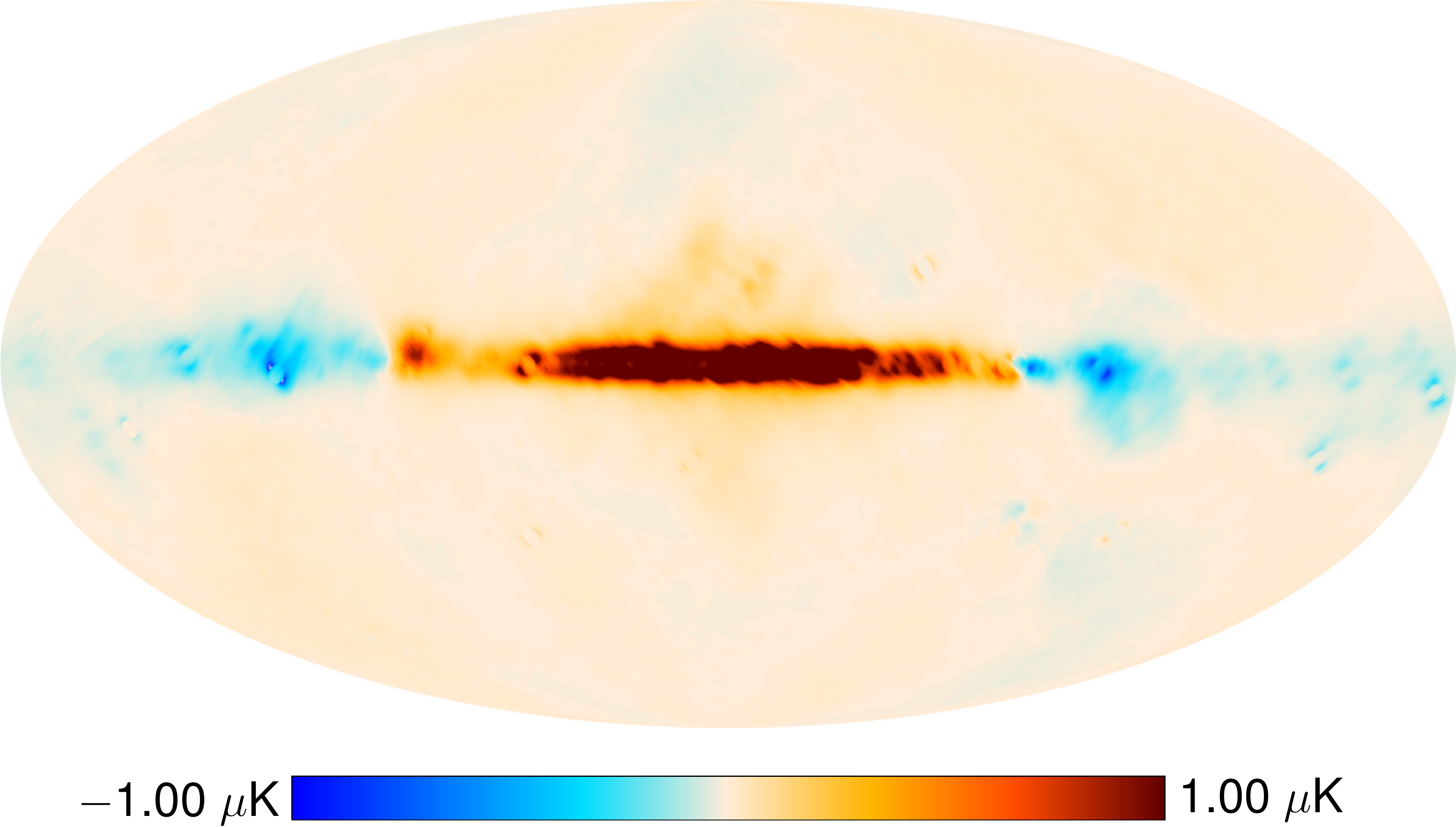}\\
	&&&\\
	44&
	\includegraphics[width=56mm]{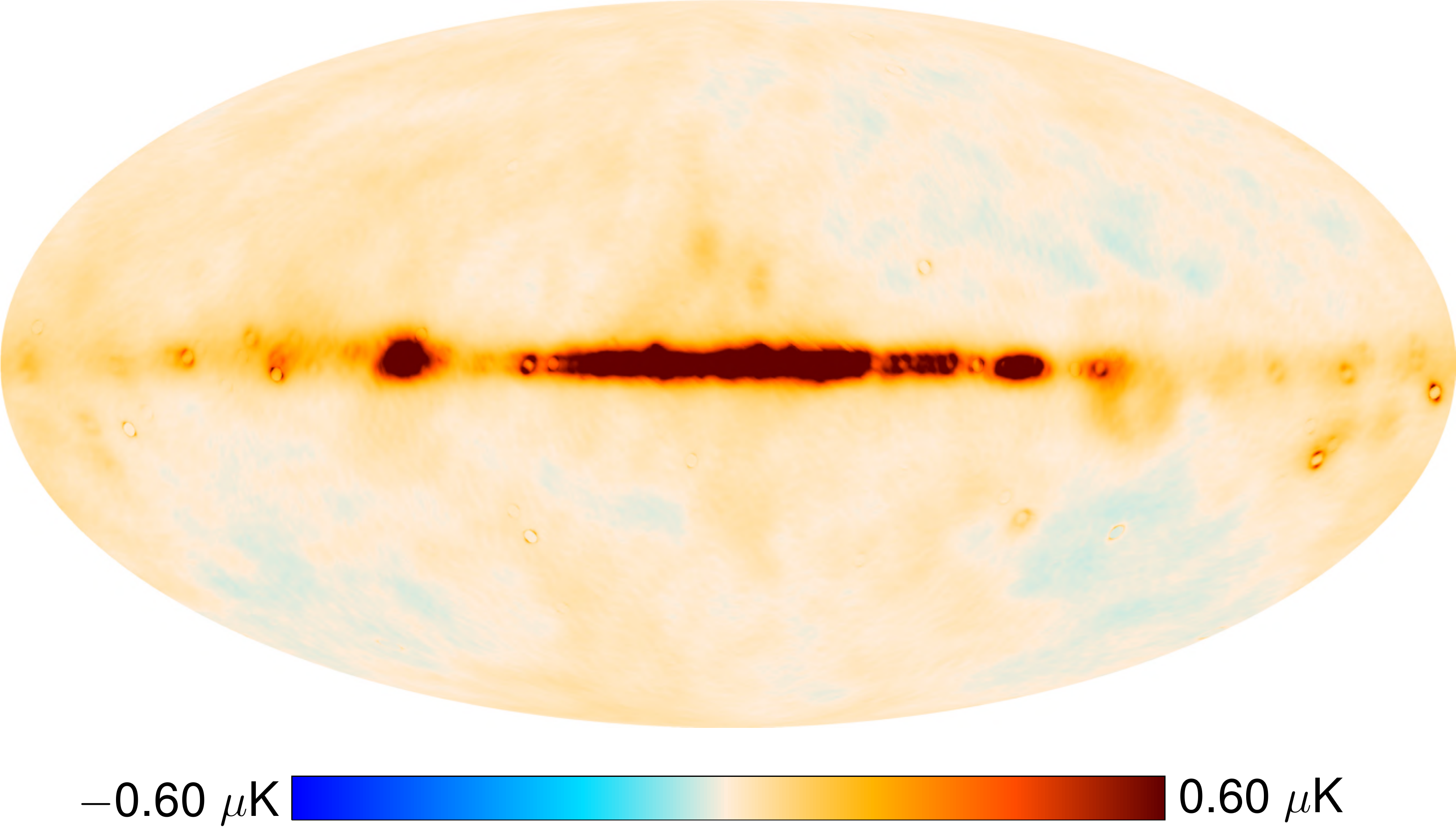}&
	\includegraphics[width=56mm]{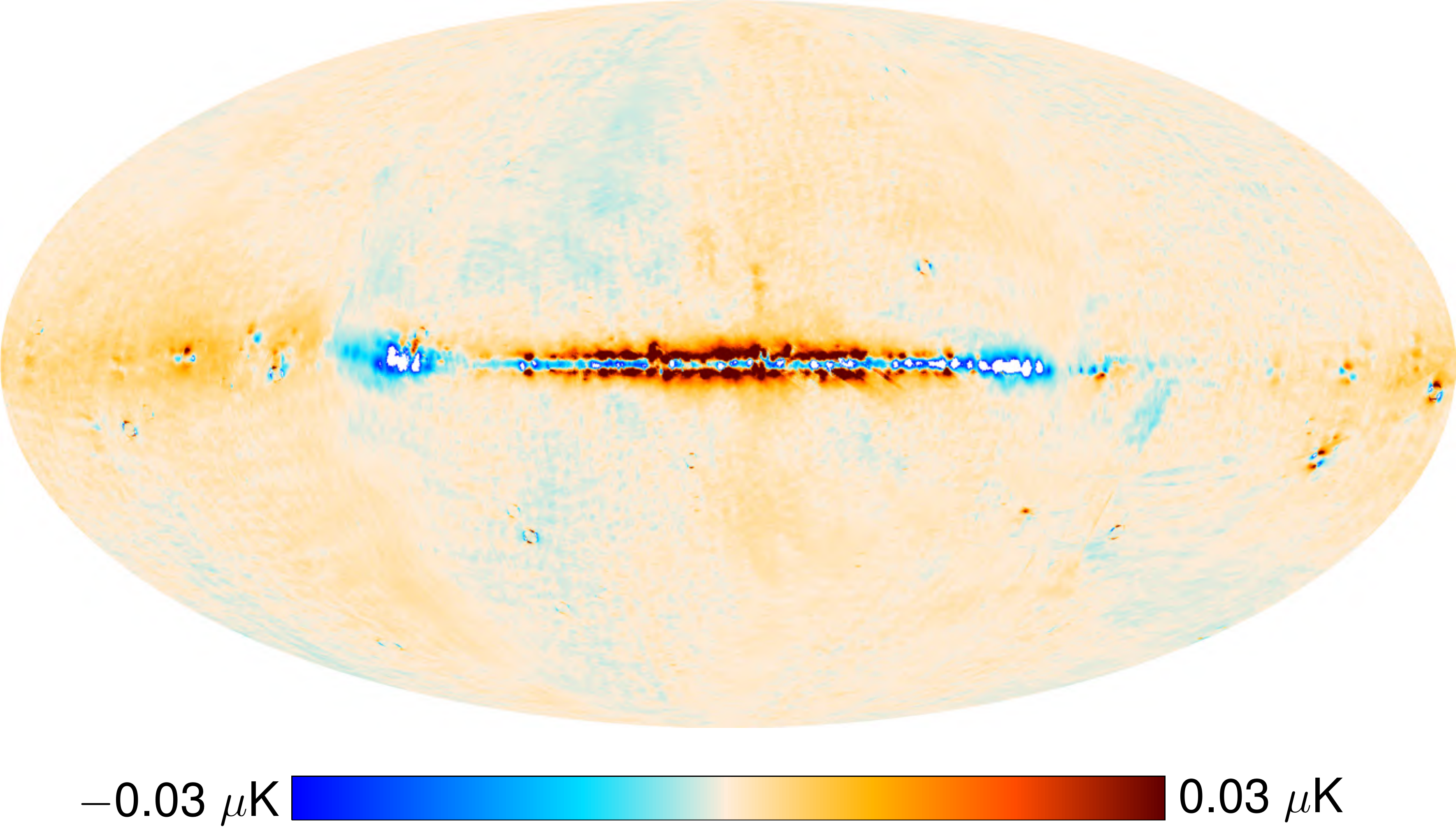}&
	\includegraphics[width=56mm]{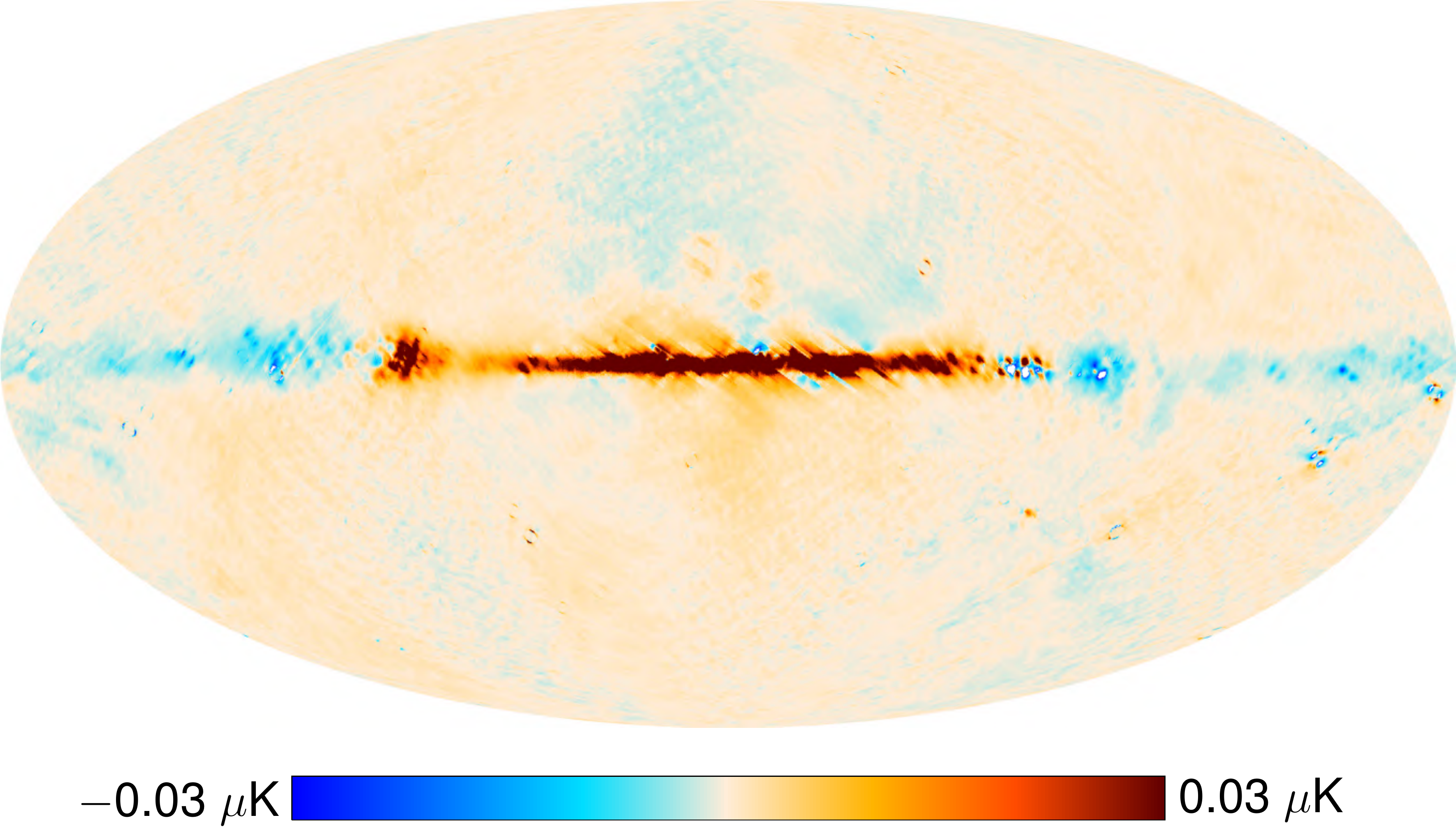}\\
	&&&\\
	70&
	\includegraphics[width=56mm]{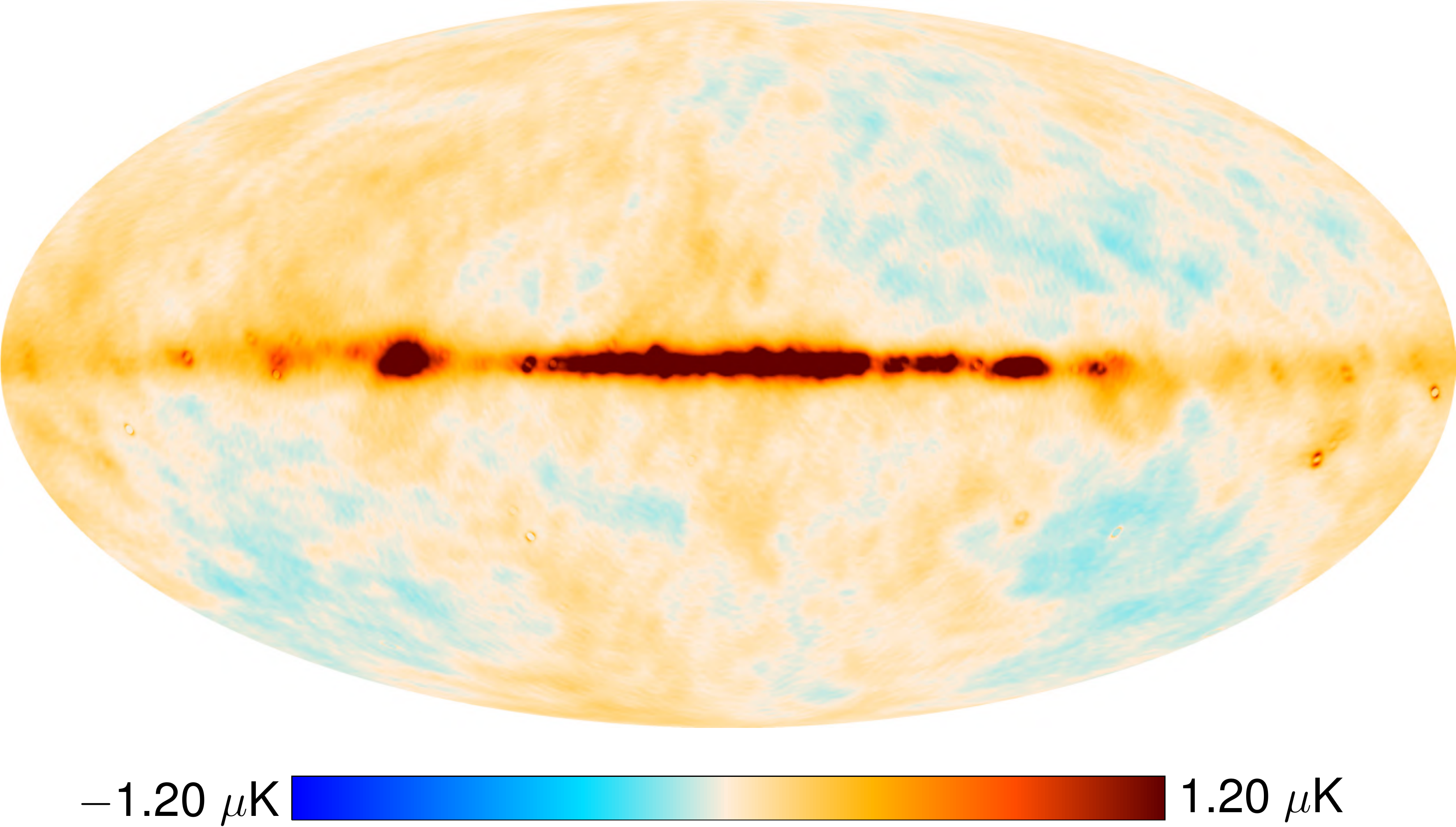}&
	\includegraphics[width=56mm]{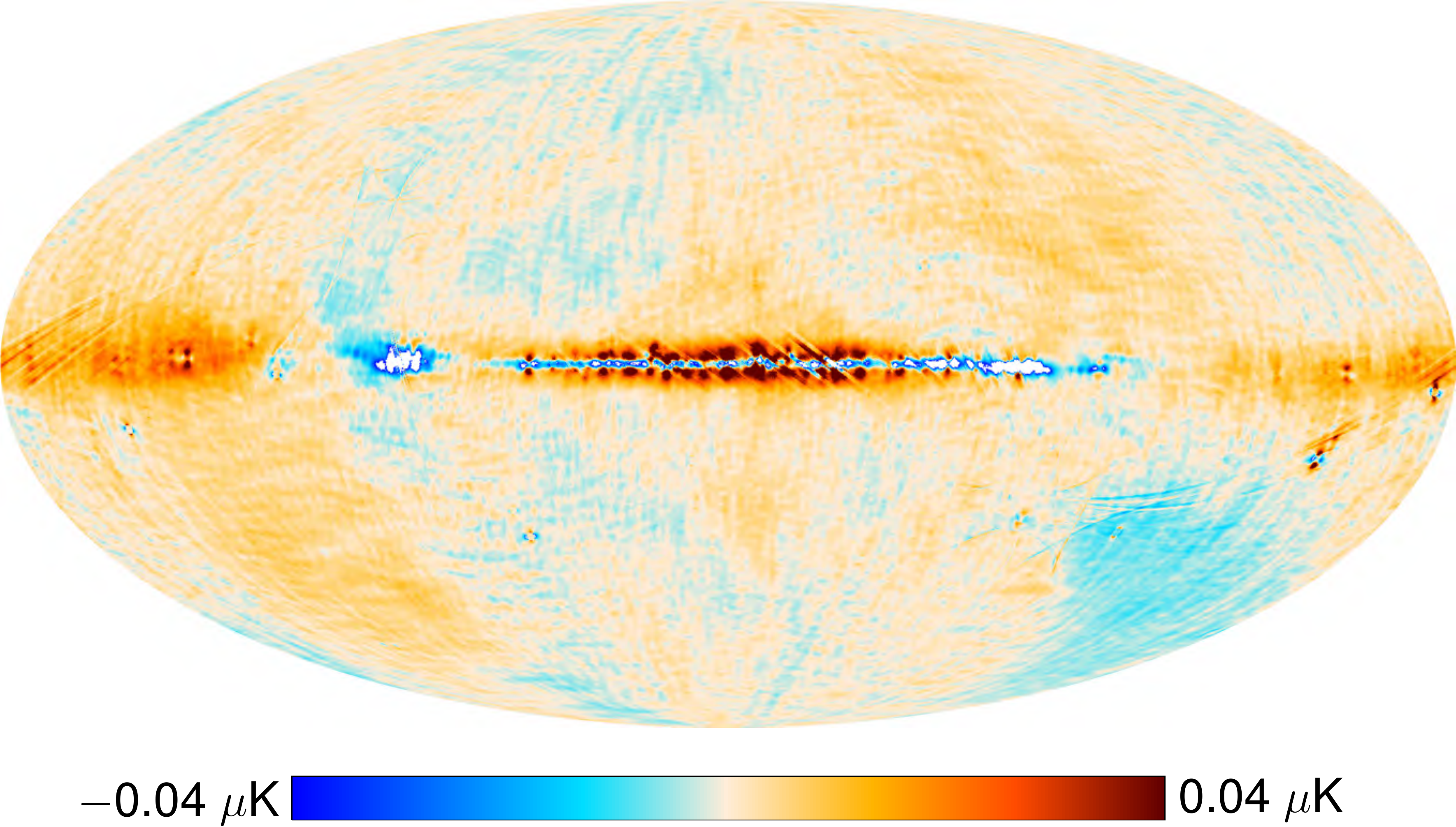}&
	\includegraphics[width=56mm]{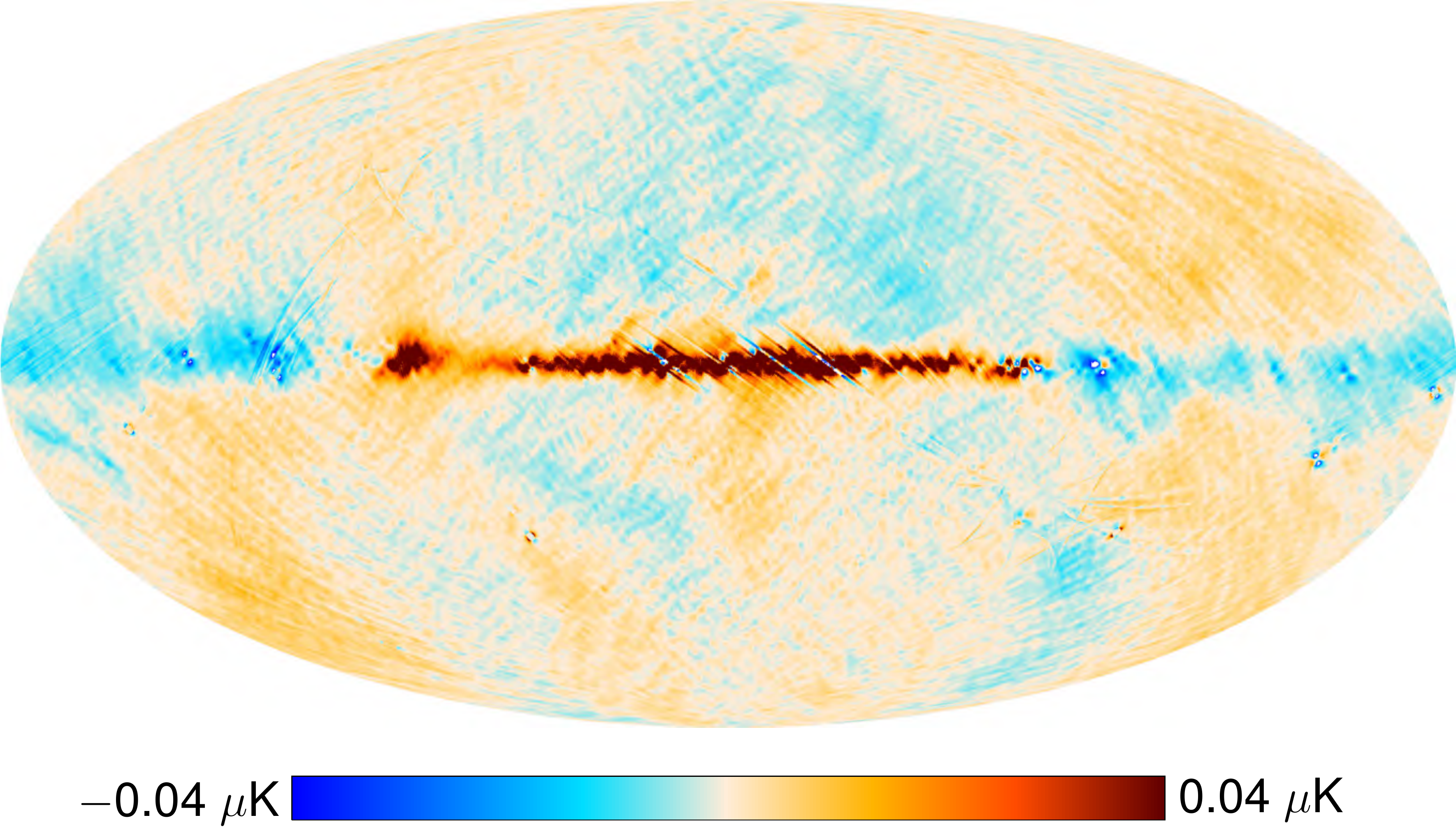}\\
	\end{tabular}
    \end{center}

    \caption{Maps of the effect from near sidelobes. Rows correspond to 30, 44,
    and 70\,GHz channels, while columns correspond to $I$, $Q$, and $U$.}
  \label{fig_whole_mission_intermediate_sidelobes_maps}
  \end{figure*}

  The maps in Fig.~\ref{fig_whole_mission_intermediate_sidelobes_maps} show that near sidelobes especially impact measurements close to the Galactic plane. This is expected, because this region of the beam pattern is close to the main beam and causes a spurious signal when the beam scans regions of the sky with large brightness variations over small angular scales. This implies that near sidelobes do not significantly impact the recovery of the CMB power spectrum if the Galactic plane is properly masked. We confirm this through the power spectra, as shown in Figs.~\ref{fig_systematic_effects_power_spectrum_30}, \ref{fig_systematic_effects_power_spectrum_44}, and \ref{fig_systematic_effects_power_spectrum_70}.
  
  \paragraph{Polarization angle.}  We study how the uncertainty in the polarization angle affects the recovered power spectra by means of a limited Monte Carlo exercise. We first produce a fiducial sky containing the CMB and foregrounds observed with the nominal polarization angles (Fig.~\ref{fig_polarization_angle_errorbars}) and then we generate five additional skies observed with a slightly different polarization angle for each feedhorn. Finally we compute the difference between each of the five maps and the fiducial sky.

  In each of the five cases we rotate the polarization angle of each feed-horn by an amount equal to either the maximum or the minimum of the error bars shown in Fig.~\ref{fig_polarization_angle_errorbars}. In this way we can explore, for a small number of cases, a range of variability in the polarization angle that is larger compared to the range expected from the focal plane thermo-mechanical analysis.

\begin{figure*}[!htpb]
	\begin{center}
	  \begin{tabular}{m{.25cm} m{5.6cm} m{5.6cm} m{5.6cm}}
	    & \begin{center}$I$\end{center} &\begin{center}$Q$\end{center}&\begin{center}$U$\end{center}\\    
	    30&
	    \includegraphics[width=56mm]{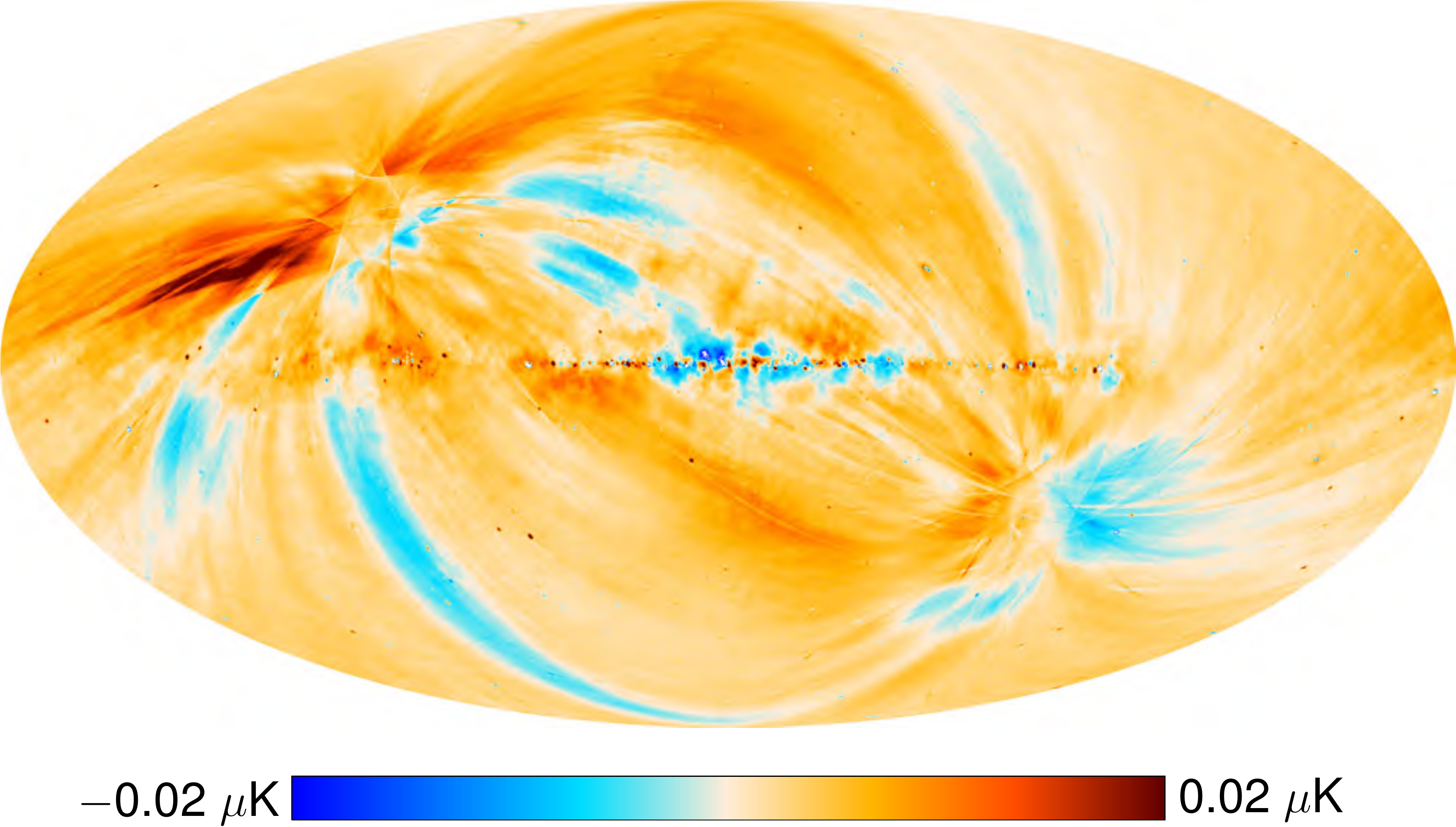}&
	    \includegraphics[width=56mm]{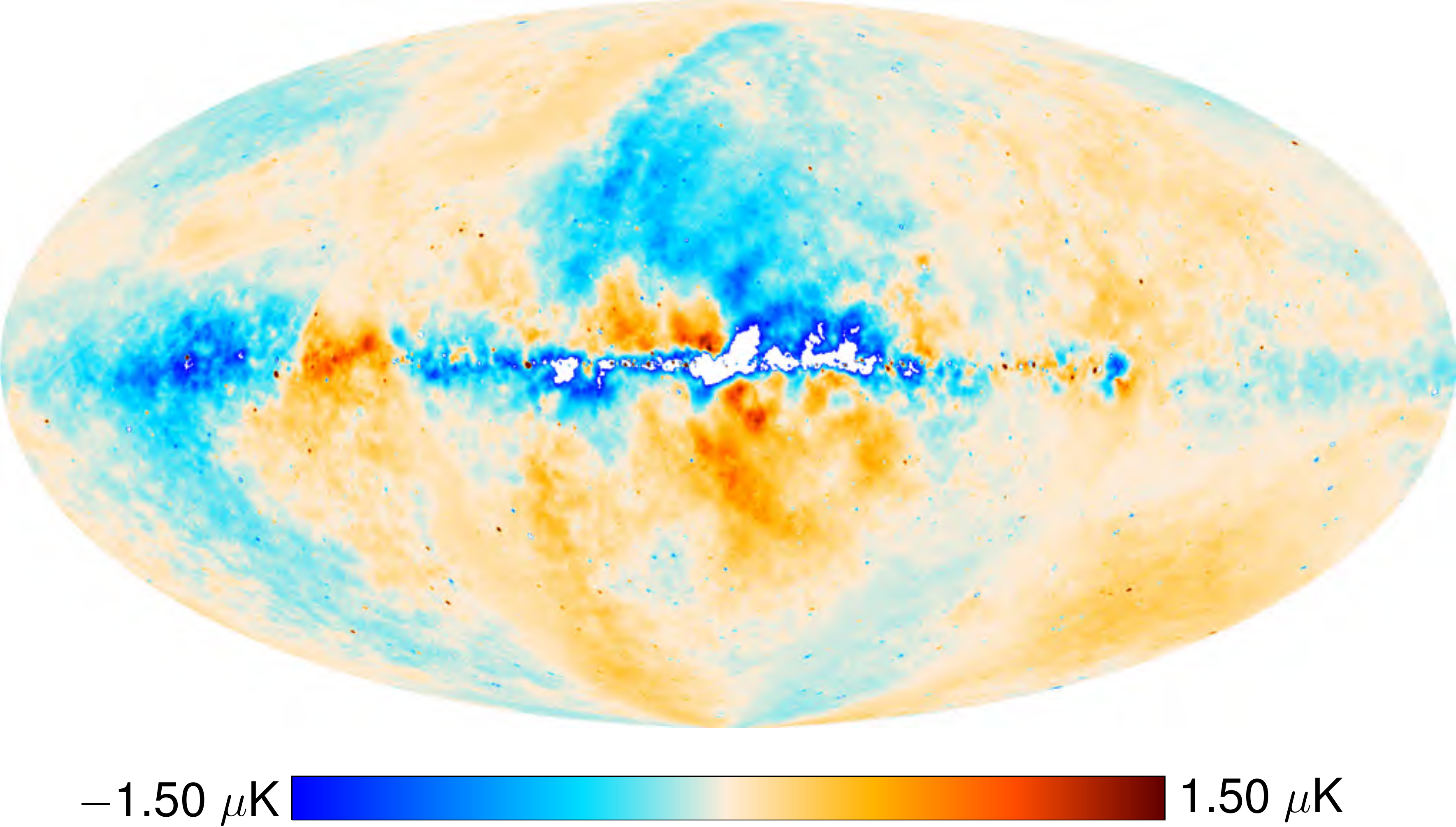}&
	    \includegraphics[width=56mm]{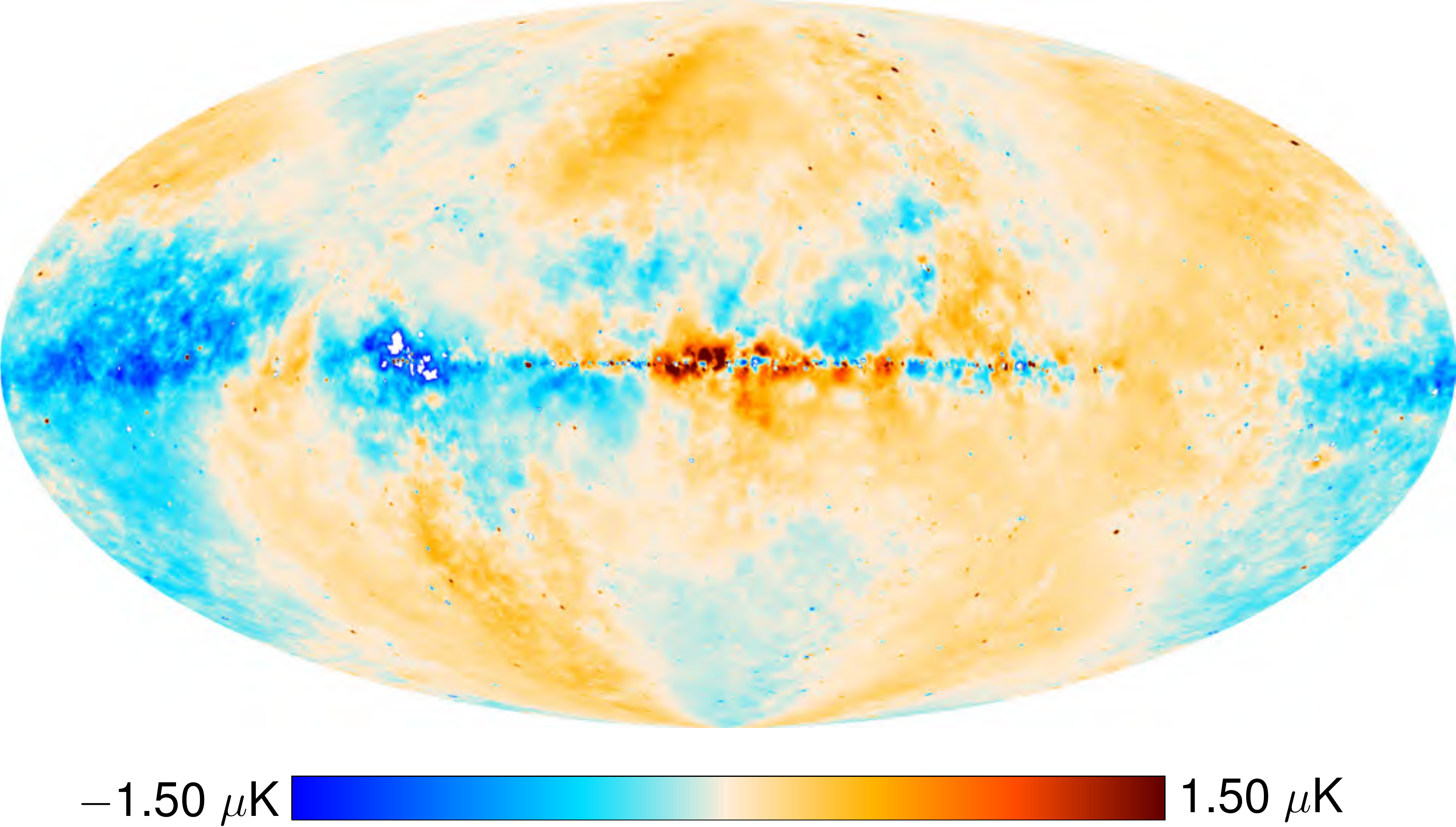}\\
	    &&&\\
	    44&
	    \includegraphics[width=56mm]{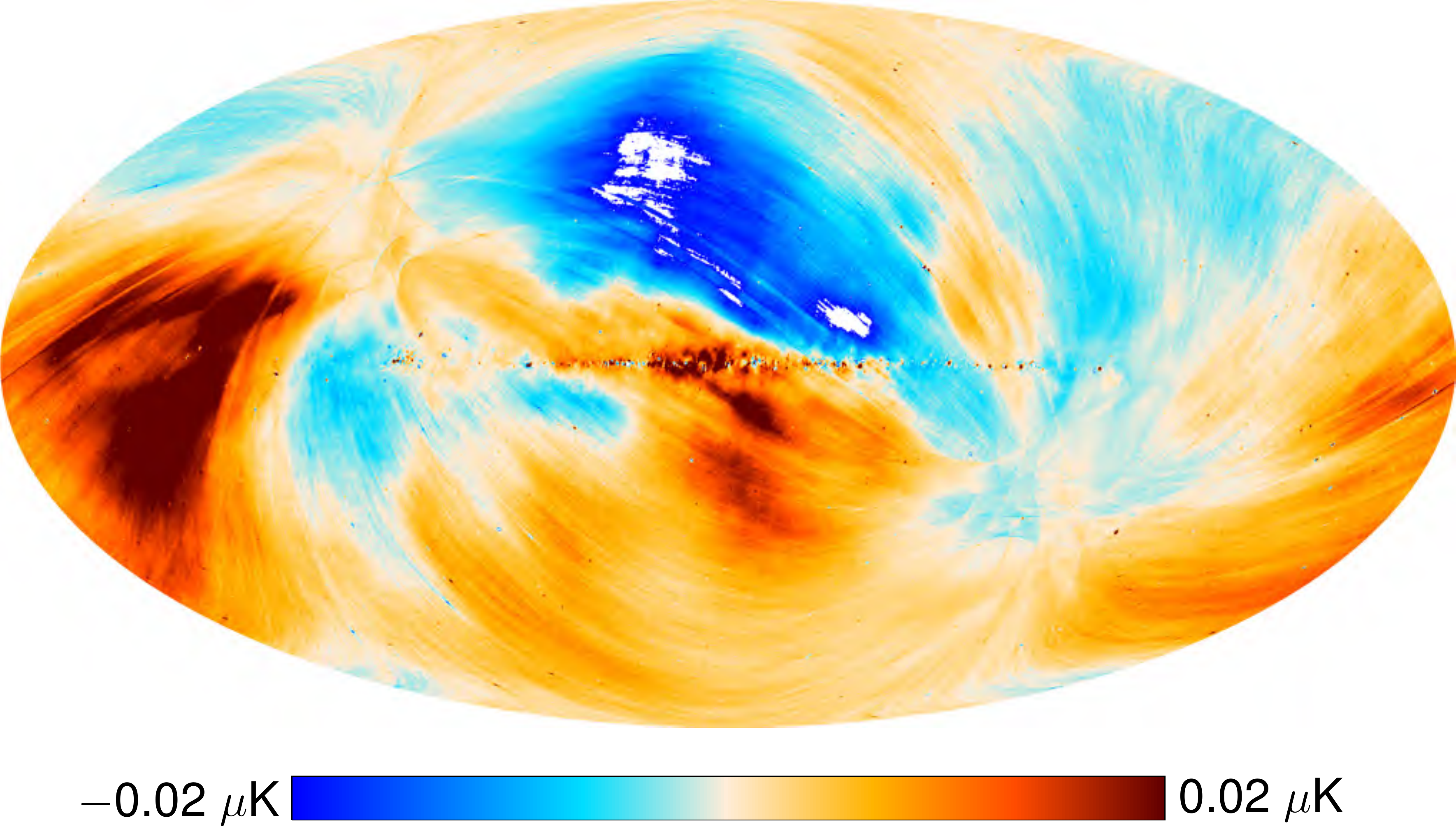}&
	    \includegraphics[width=56mm]{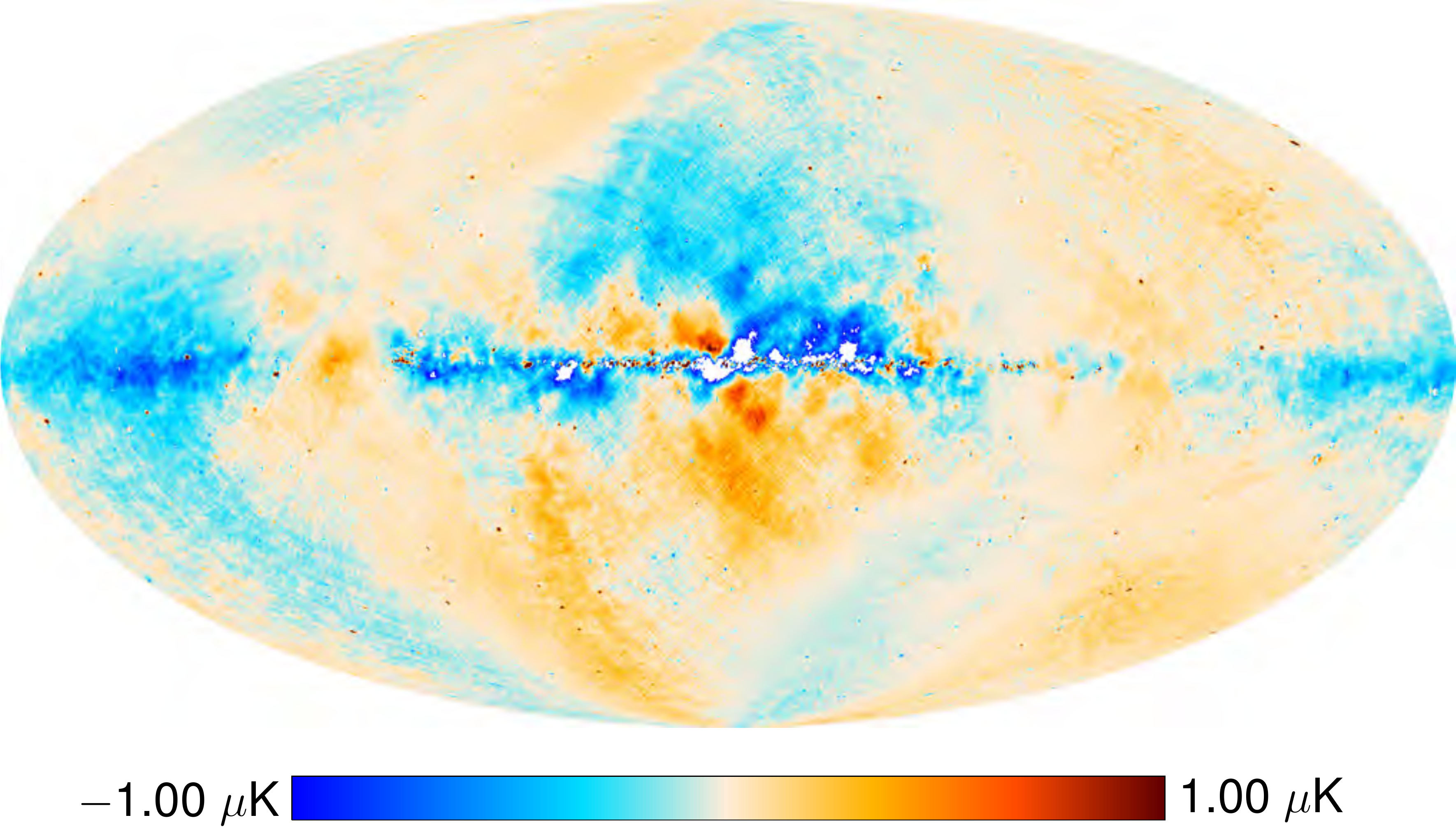}&
	    \includegraphics[width=56mm]{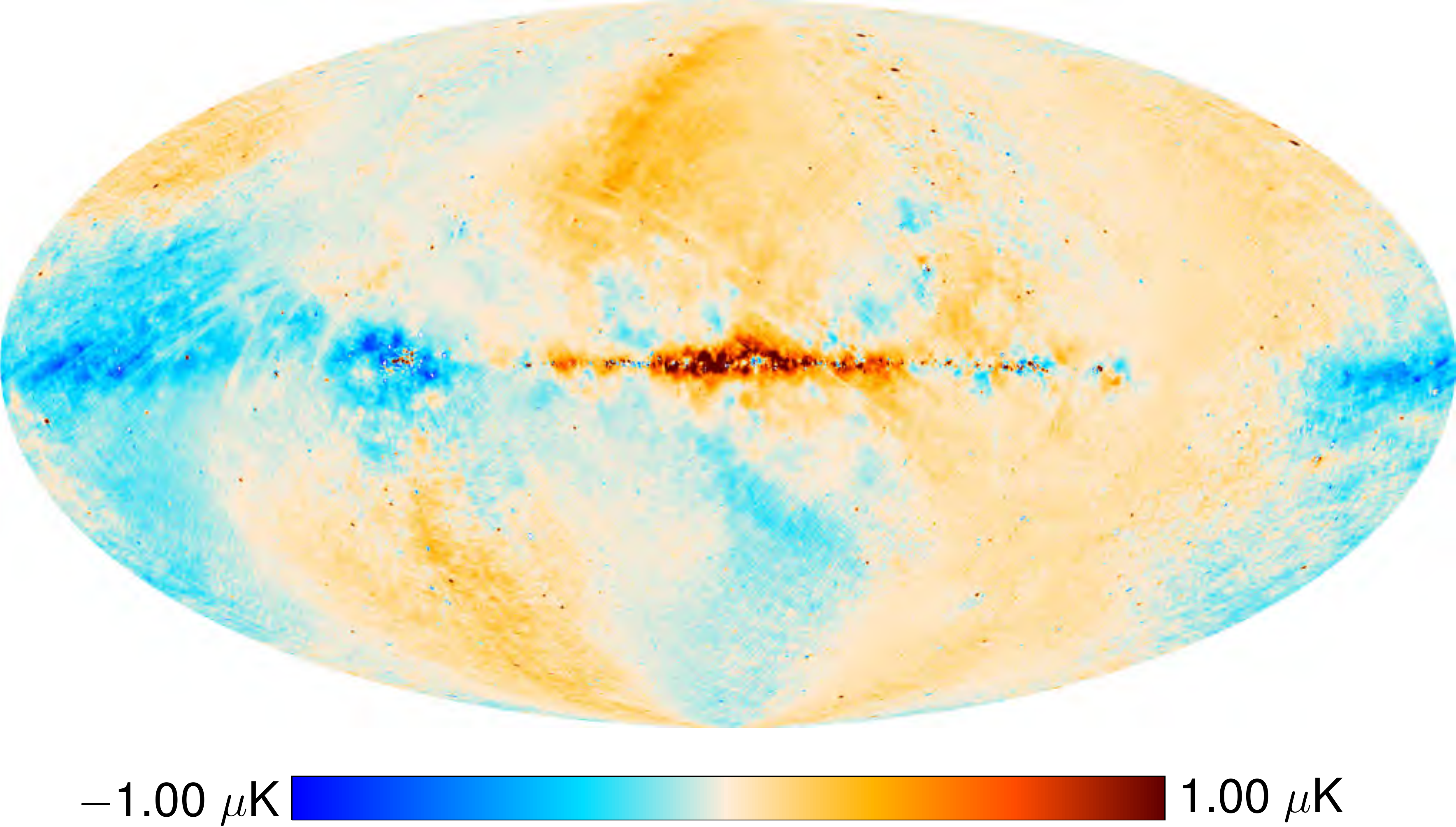}\\
	    &&&\\
	    70&
	    \includegraphics[width=56mm]{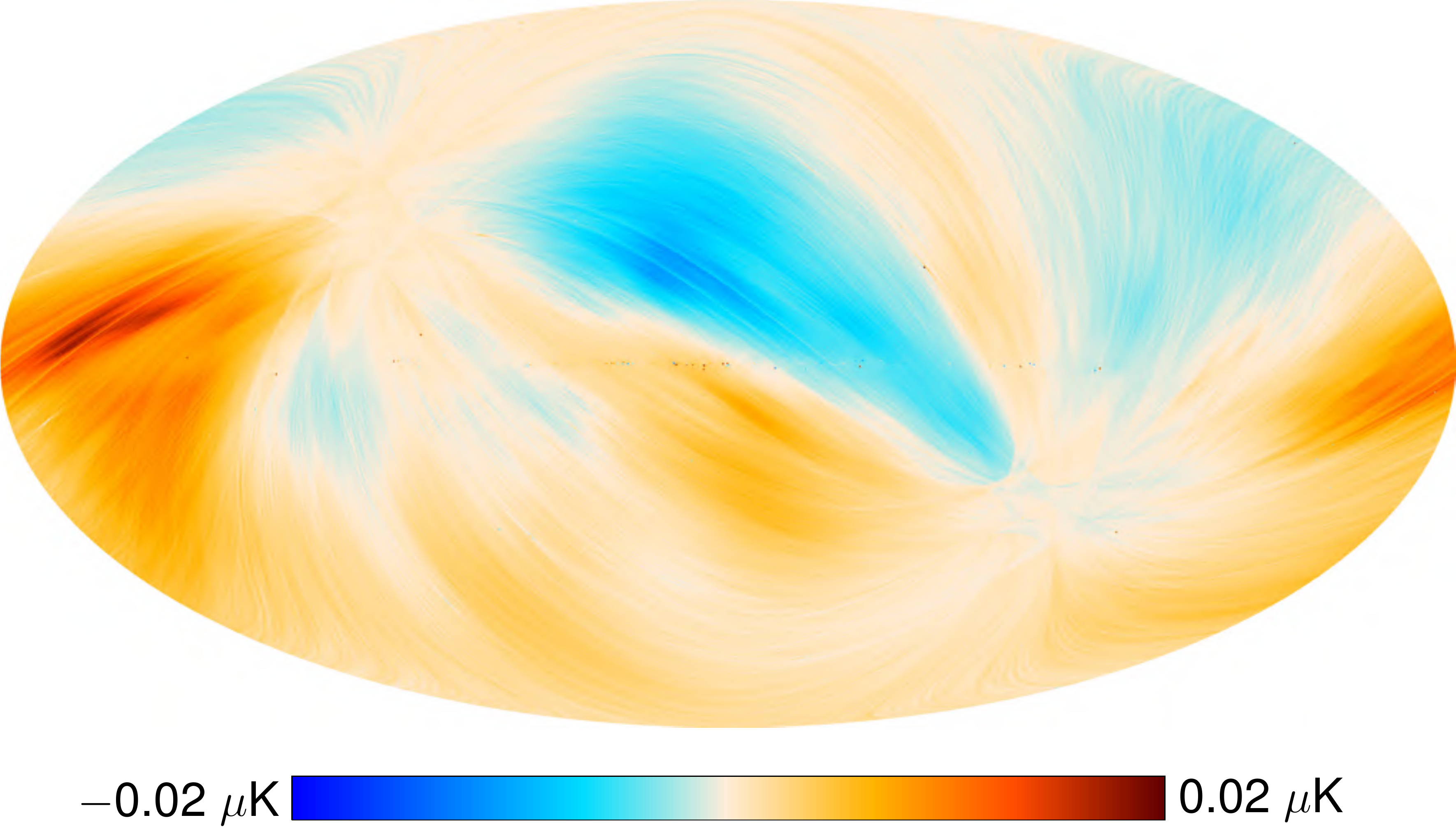}&
	    \includegraphics[width=56mm]{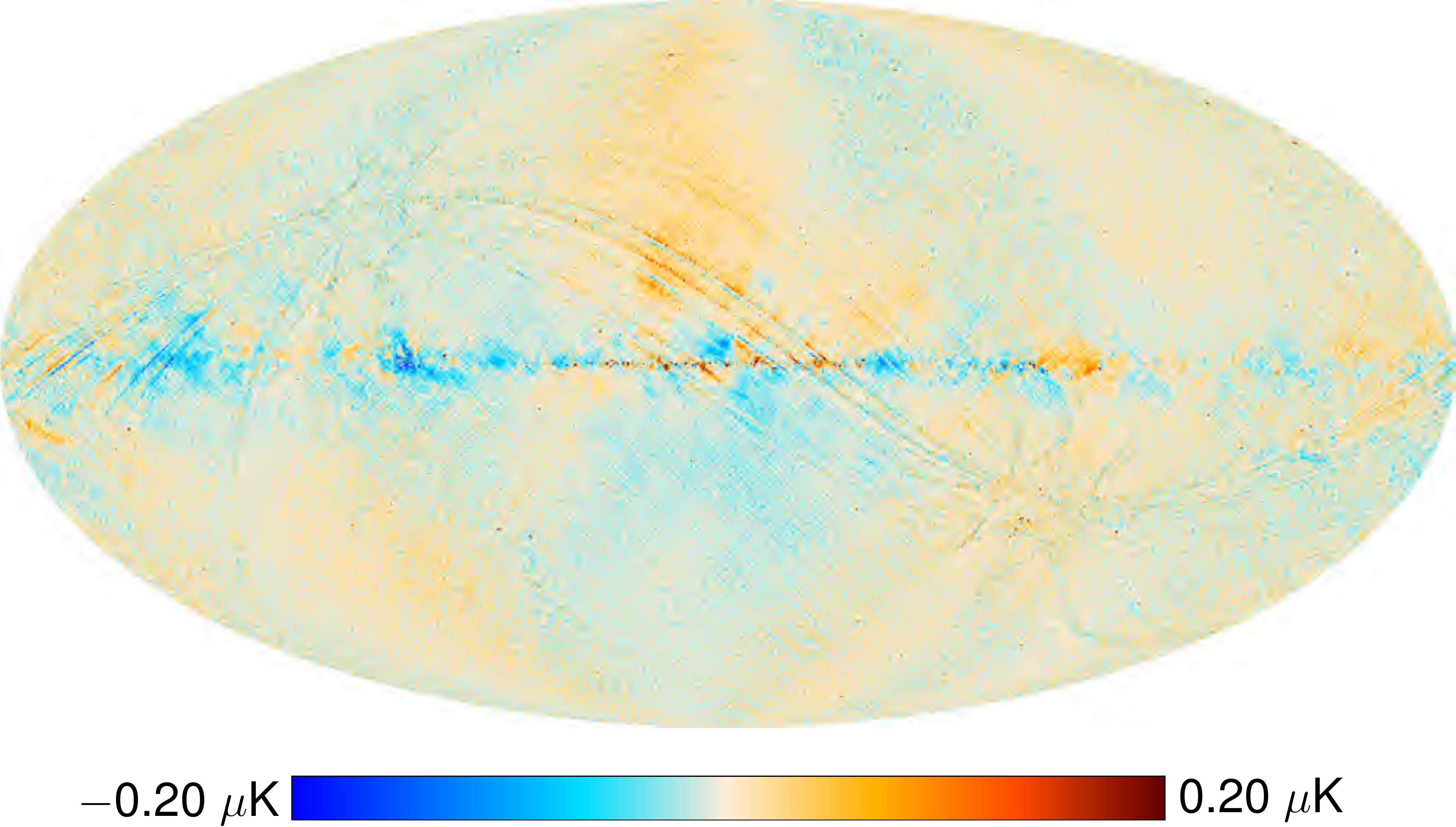}&
	    \includegraphics[width=56mm]{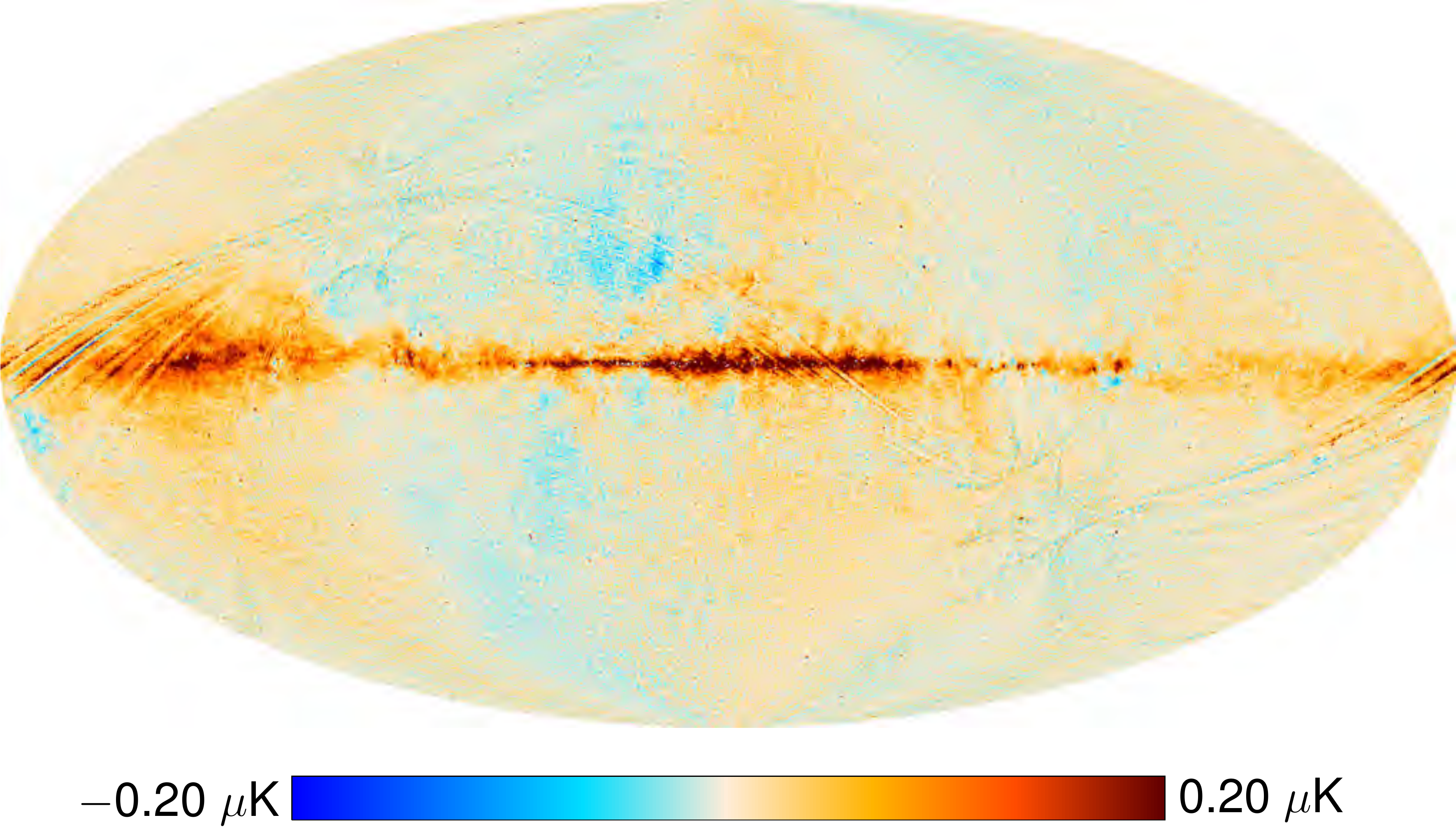}\\
	    \end{tabular}
	\end{center}
	
	\caption{Maps of the effect from uncertainty in the polarization angle. The maps shown are randomly selected from one of the five tested realizations. Rows correspond to 30, 44, and 70\,GHz channels, while columns correspond to $I$, $Q$, and $U$.}
      \label{fig_whole_mission_polangle_maps}
      \end{figure*}

  The difference maps in Fig.~\ref{fig_whole_mission_polangle_maps} show that the effect is negligible in temperature (as expected) and is less than 1\,$\mu$K at 70\,GHz in polarization. At 30 and 44\,GHz the maximum amplitude of the effect is around 2\,$\mu$K and $1\,\mu$K, respectively. The maps shown represent one of the five cases picked randomly from the set.
      
\begin{figure}[!htpb]
	\begin{center}
	    \includegraphics[width=88mm]{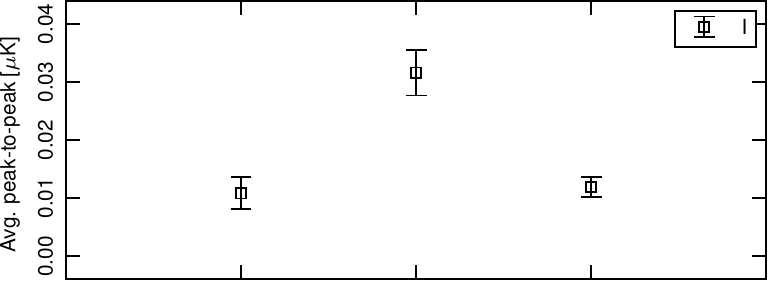}\\
	    \includegraphics[width=88mm]{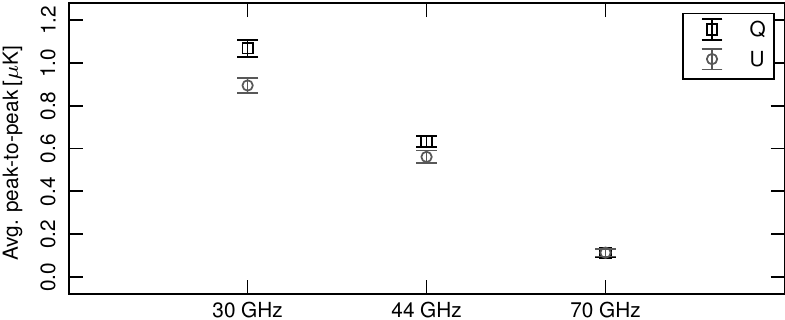}
	\end{center}
	
	\caption{Peak-to-peak of the effect from polarization angle uncertainty on LFI maps. The error bars represent the dispersion of the five cases chosen in the analysis. \textit{Top}: total intensity.
        \textit{Bottom}: $Q$ and $U$ Stokes parameters.}
      \label{fig_average_polangle_effect_on_map_p2p}
      \end{figure}

\begin{figure}[!htpb]
	\begin{center}
	    \includegraphics[width=88mm]{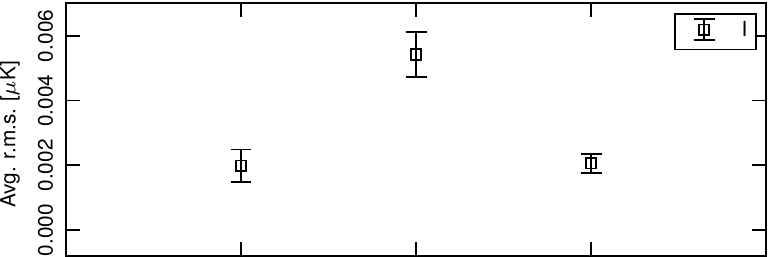}\\
	    \includegraphics[width=88mm]{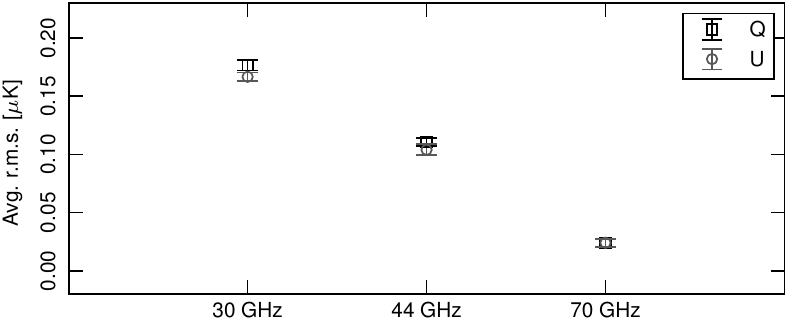}
	\end{center}
	
	\caption{Rms of the effect from polarization angle uncertainty on LFI
        maps. The error bars represent the dispersion of the five cases chosen
        in the analysis. \textit{Top}: total intensity. \textit{Bottom}:
        $Q$ and $U$ Stokes parameters.}
      \label{fig_average_polangle_effect_on_map_rms}
      \end{figure}

  In Figs.~\ref{fig_average_polangle_effect_on_map_p2p} and \ref{fig_average_polangle_effect_on_map_rms} we show the dispersion of the peak-to-peak and rms of this effect on maps, once we apply the masks in Fig.~\ref{fig_masks} (top one for total intensity and middle one for $Q$ and $U$ maps). The rms of the effect is smaller than 1\,$\mu$K and also the dispersion introduced by the five different cases is small. 
  
   We observe that the peak-to-peak and rms of the effect in the polarization map decrease with frequency (see the bottom panels of Figs.~\ref{fig_average_polangle_effect_on_map_p2p} and \ref{fig_average_polangle_effect_on_map_rms}). This correlates with the smaller contribution of polarized synchrotron emission in maps at higher frequency. 
  
  We also observe a higher residual at 44\,GHz in temperature maps compared to the 30 and 70\,GHz channels. We did not expect this behavior, and it is currently not understood. The effect in temperature, however, is much less than 0.1\,$\mu$K and, therefore, completely negligible.

\begin{figure*}[!htpb]
	  \hspace{3.6cm} TT \hspace{5.2cm} EE \hspace{5.1cm} BB\\
	  \begin{tabular}{m{.2cm} m{16cm}}
	  \begin{turn}{90}30\,GHz\end{turn}&\includegraphics[width=17cm]{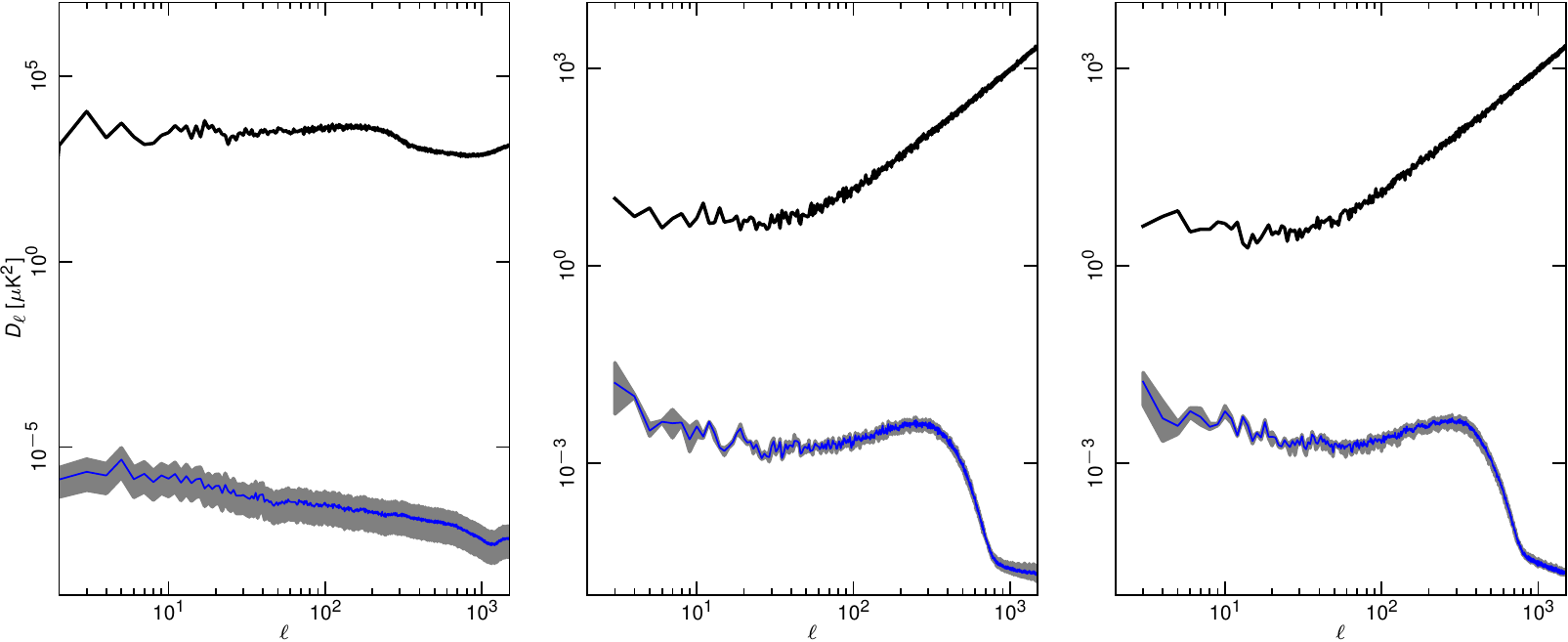}\\
	  &\mbox{ }\\
	  \begin{turn}{90}44\,GHz\end{turn}&\includegraphics[width=17cm]{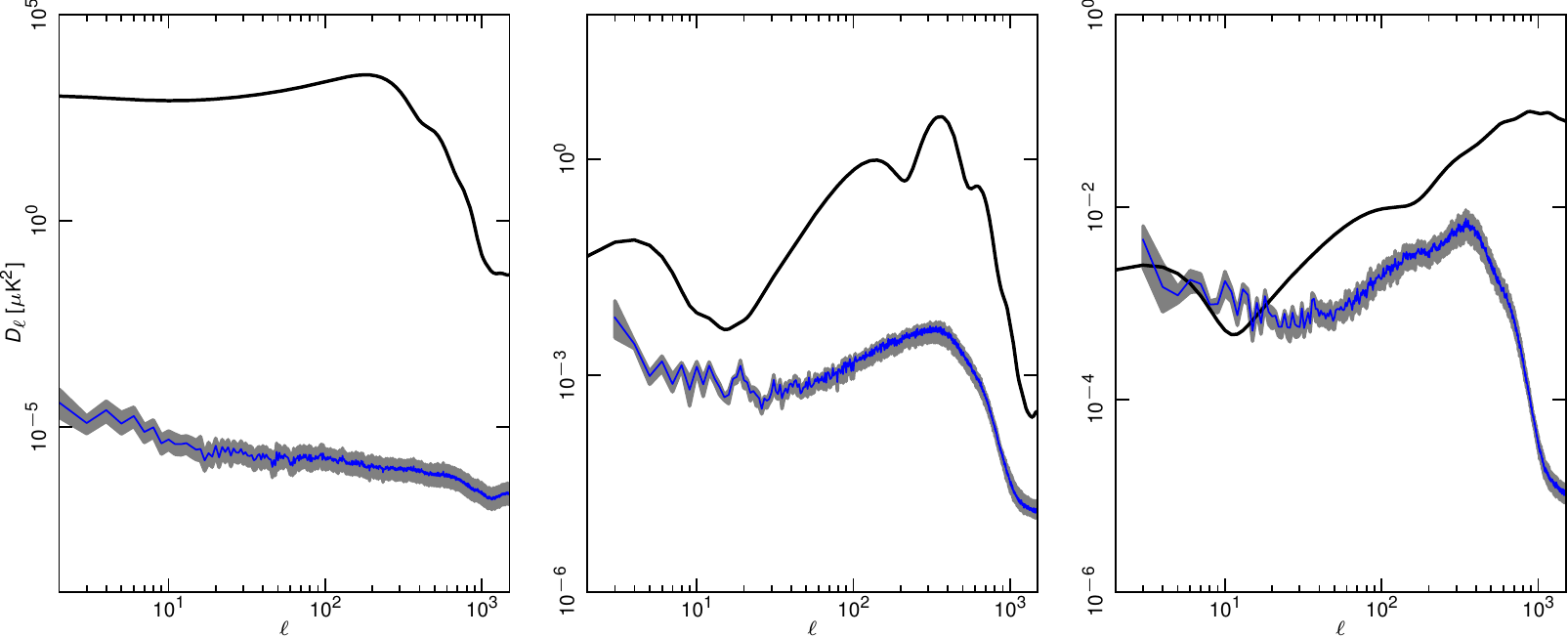}\\
	  &\mbox{ }\\
	  \begin{turn}{90}70\,GHz\end{turn}&\includegraphics[width=17cm]{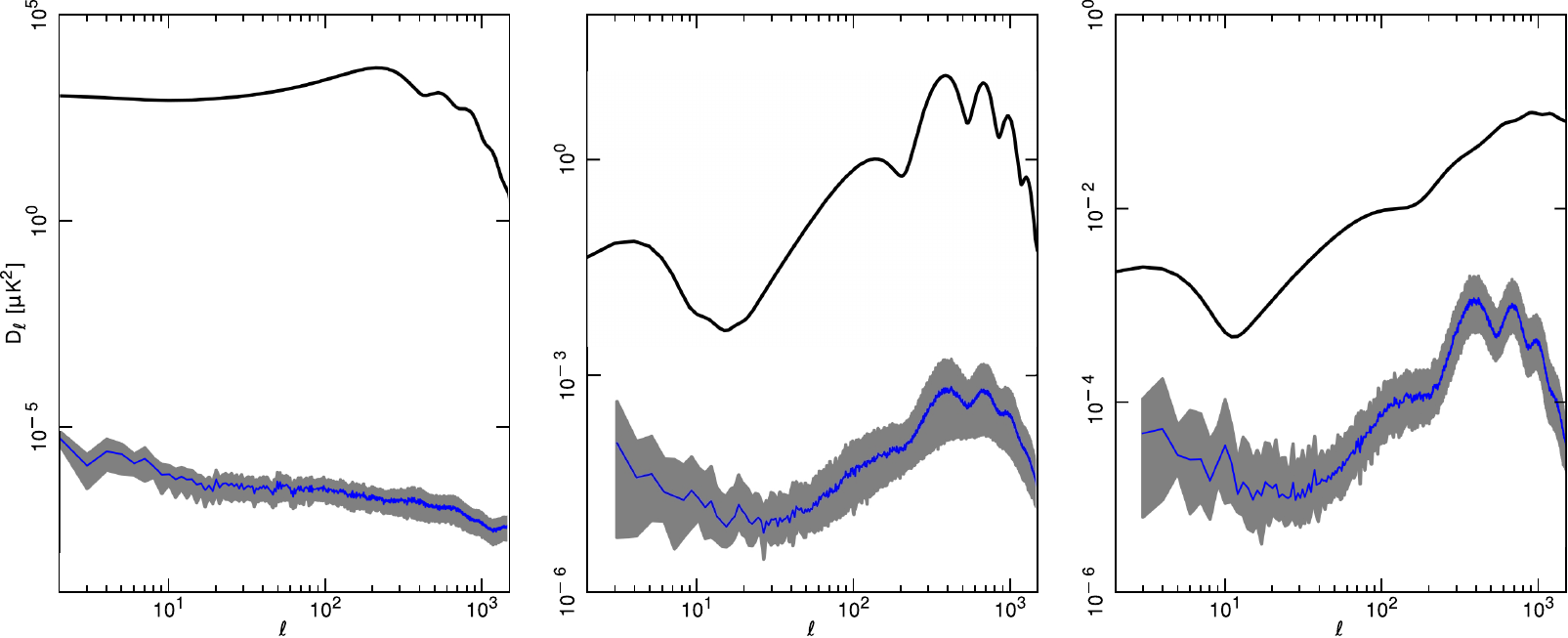}
	  \end{tabular}
	  \caption{
	  \label{fig_systematic_effects_power_spectrum_polangle}
	  Angular power spectra of the residual effect due to polarization
          angle uncertainty compared to the foreground spectra at 30\,GHz and
          to \Planck\ beam-filtered temperature and polarization spectra at 44
          and 70\,GHz. The blue curve represents the average spectrum, while
          the grey band is the envelope of all the power spectra calculated
          from the various realizations of the effect. The theoretical $B$-mode
          CMB spectrum assumes a tensor-to-scalar ratio $r = 0.1$, a tensor
          spectral index $n_\mathrm{T}=0$ and has not been beam-filtered.
          Rows are for 30, 44, and 70\,GHz spectra, while columns are for
          $TT$, $EE$ and $BB$ power spectra. }
      \end{figure*}

  From the five sets of difference maps we have computed power spectra and evaluated their dispersion. We show the results in Fig.~\ref{fig_systematic_effects_power_spectrum_polangle}, where the grey area represents the region containing all the spectra and the blue curve is the average of these five spectra. The blue curve corresponds to the spectrum that is also reported in Figs.~\ref{fig_systematic_effects_power_spectrum_30}, \ref{fig_systematic_effects_power_spectrum_44}, and \ref{fig_systematic_effects_power_spectrum_70}.
       
  \paragraph{Pointing.} We have simulated the effect caused by pointing uncertainty by adding a Gaussian noise realization independently to both co-scan and cross-scan bore sight pointing. The noise realization was drawn from a $1/f$ noise model with a smooth cutoff at 10\,mHz, which matches the single-planet transit analysis and multiple transit results over the entire mission. The rms net effect of the added pointing errors is an uncertainty of about 4.8\arcsec\ at timescales shorter than 10\,000\,s, and about 5.1\arcsec\ at timescales longer than 10\,000\,s. The overall uncertainty is approximately 7.0\arcsec.
      
\begin{figure*}[!htpb]
	\begin{center}
	  \begin{tabular}{m{.25cm} m{5.6cm} m{5.6cm} m{5.6cm}}
	    & \begin{center}$I$\end{center} &\begin{center}$Q$\end{center}&\begin{center}$U$\end{center}\\    
	    30&
	    \includegraphics[width=56mm]{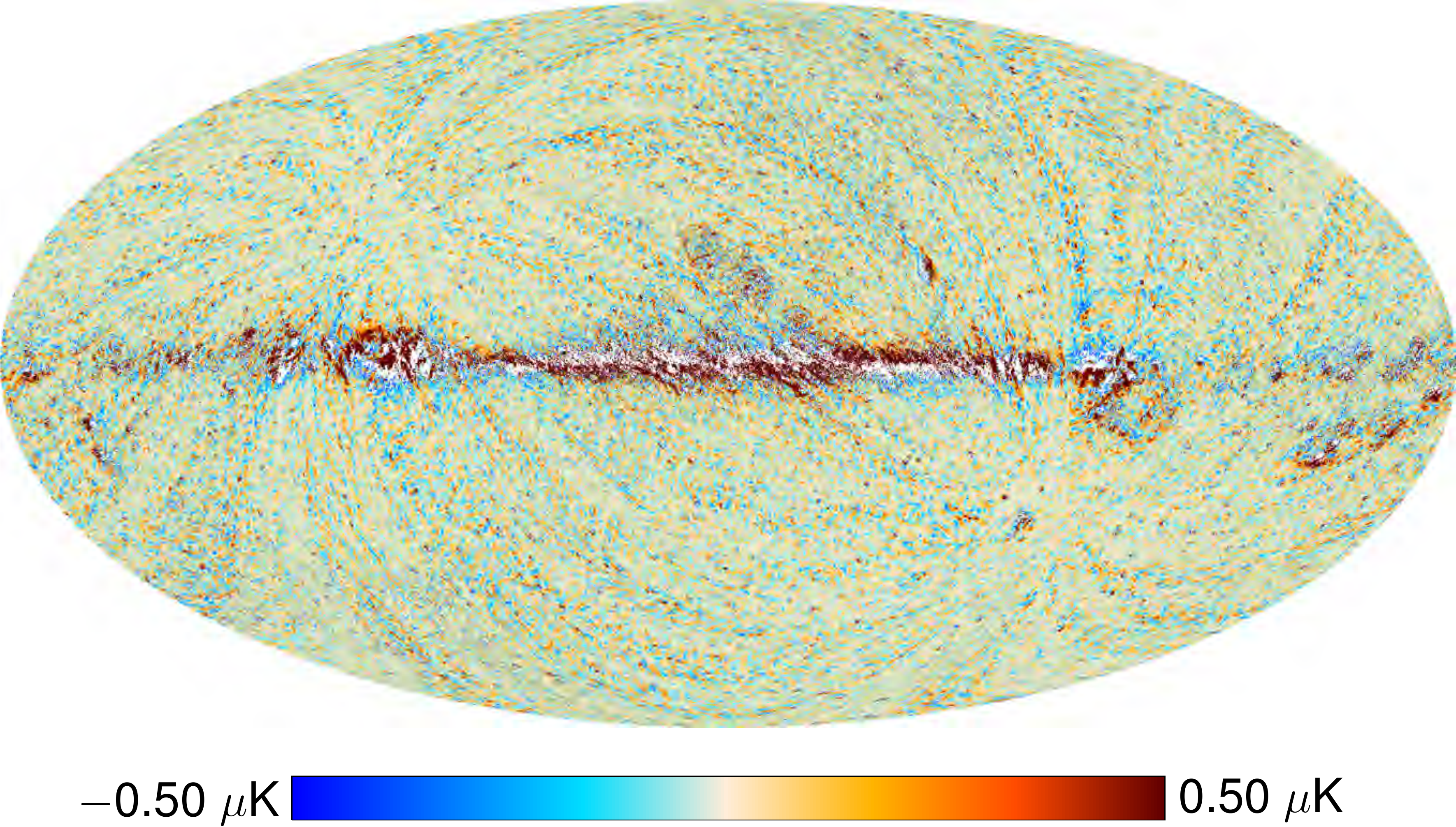}&
	    \includegraphics[width=56mm]{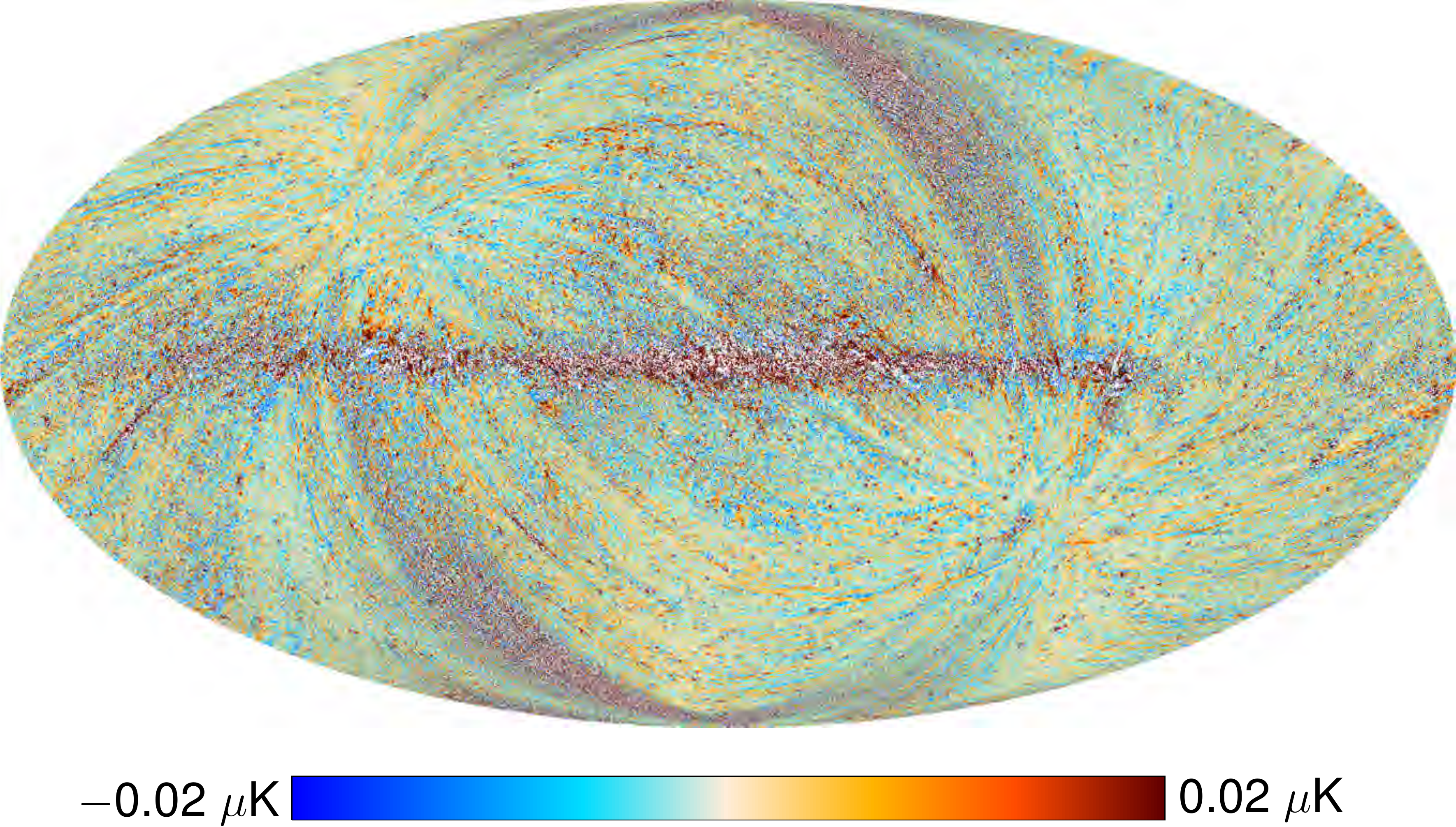}&
	    \includegraphics[width=56mm]{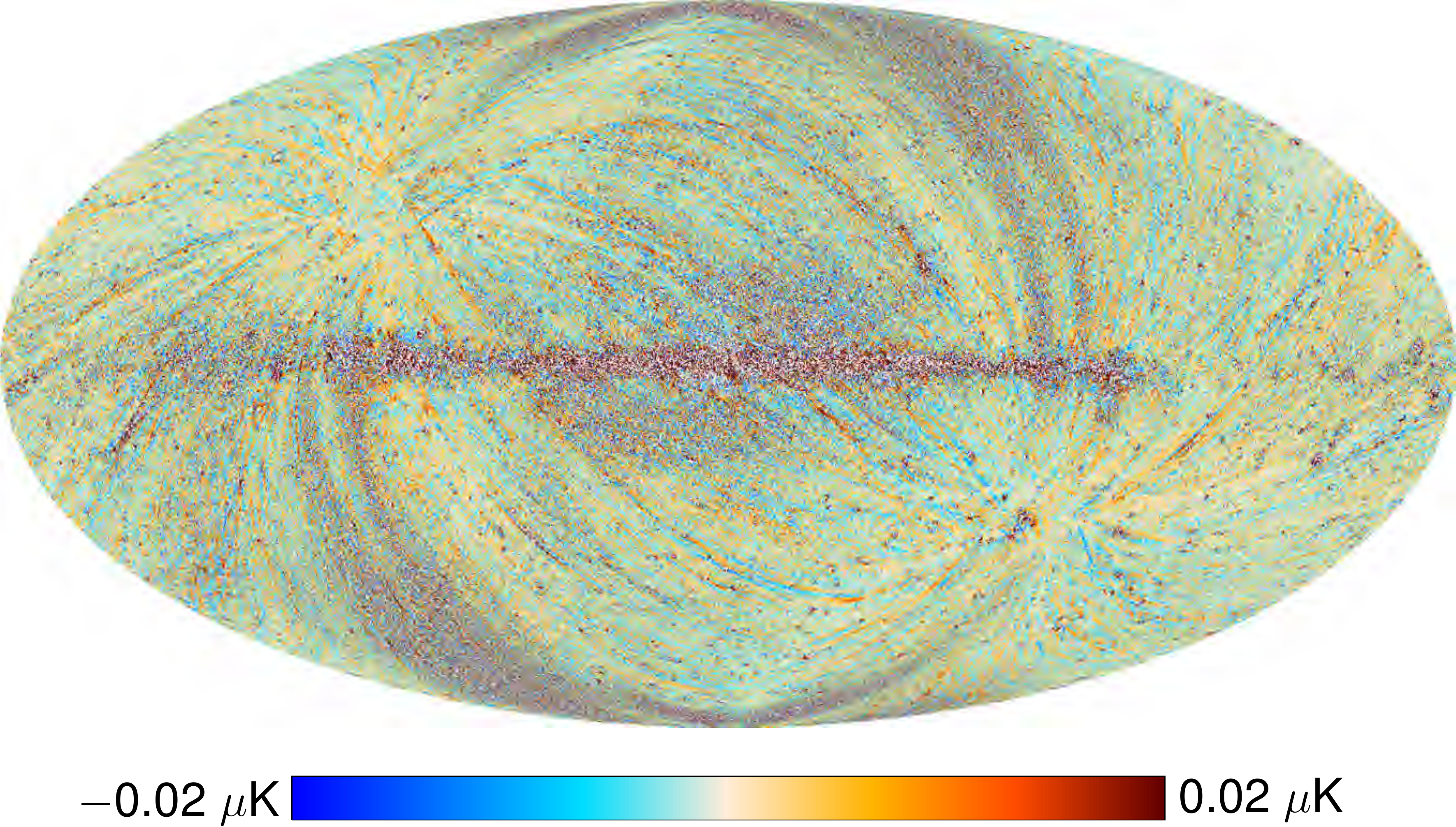}\\
	    &&&\\
	    44&
	    \includegraphics[width=56mm]{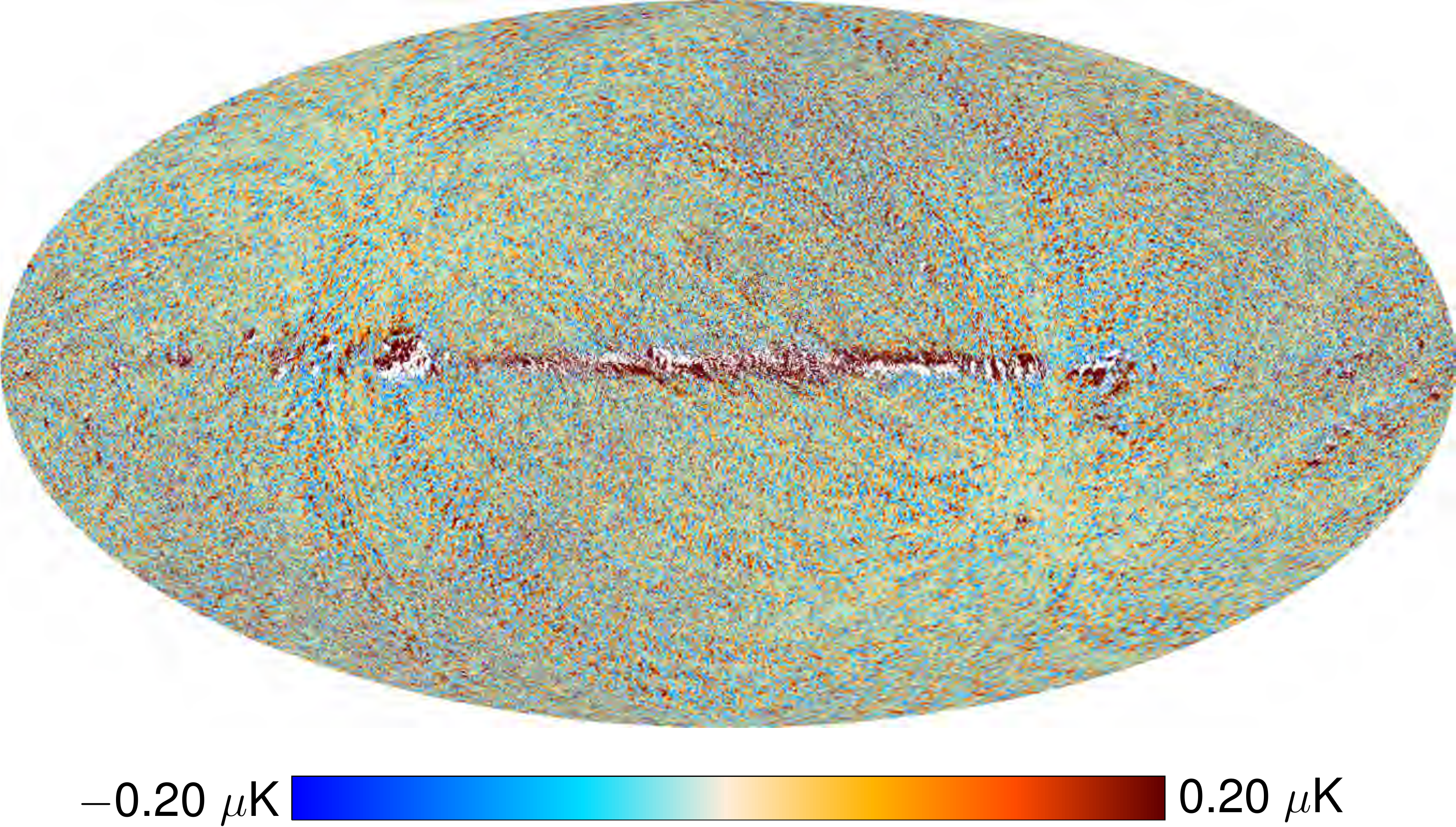}&
	    \includegraphics[width=56mm]{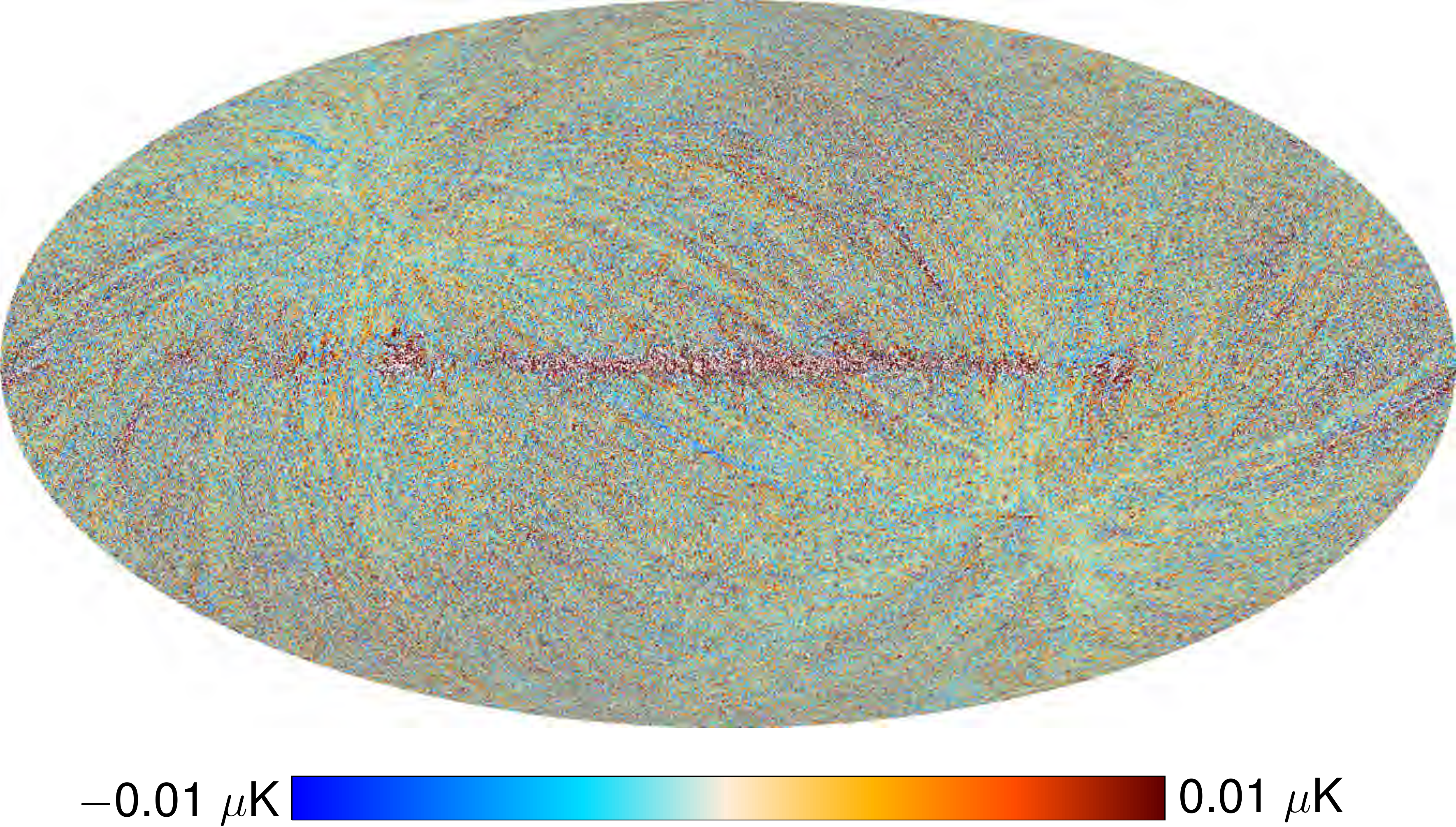}&
	    \includegraphics[width=56mm]{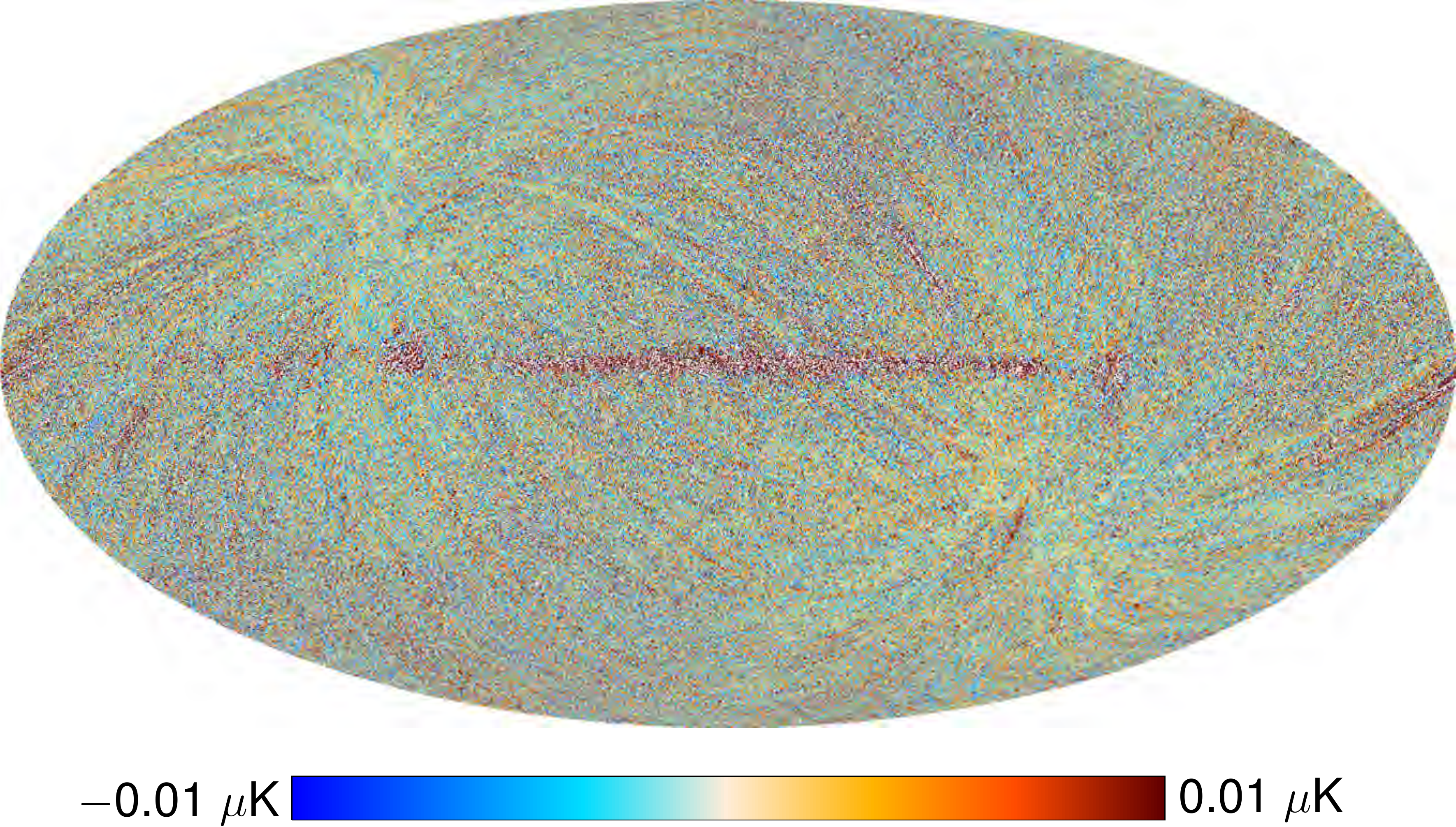}\\
	    &&&\\
	    70&
	    \includegraphics[width=56mm]{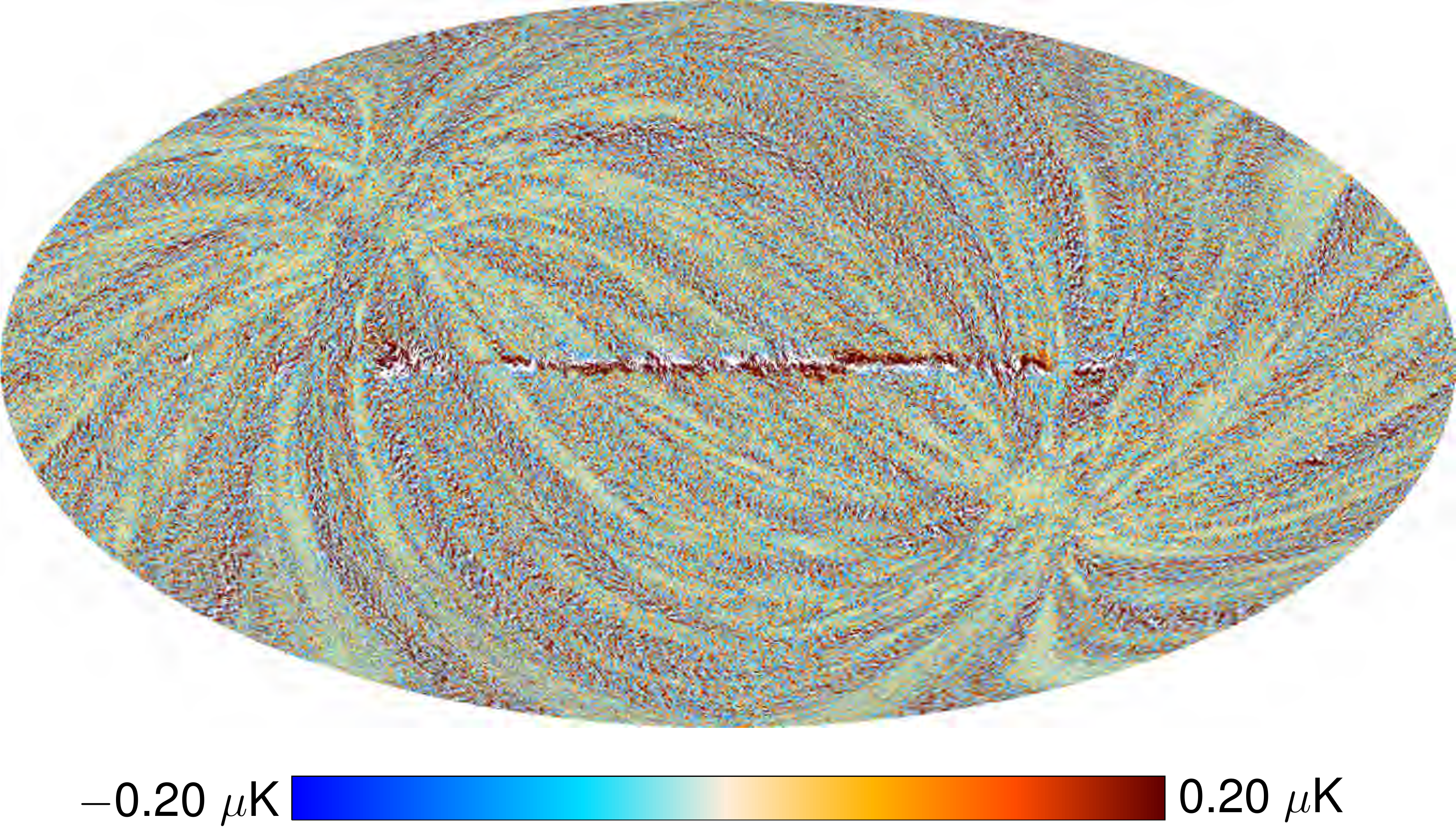}&
	    \includegraphics[width=56mm]{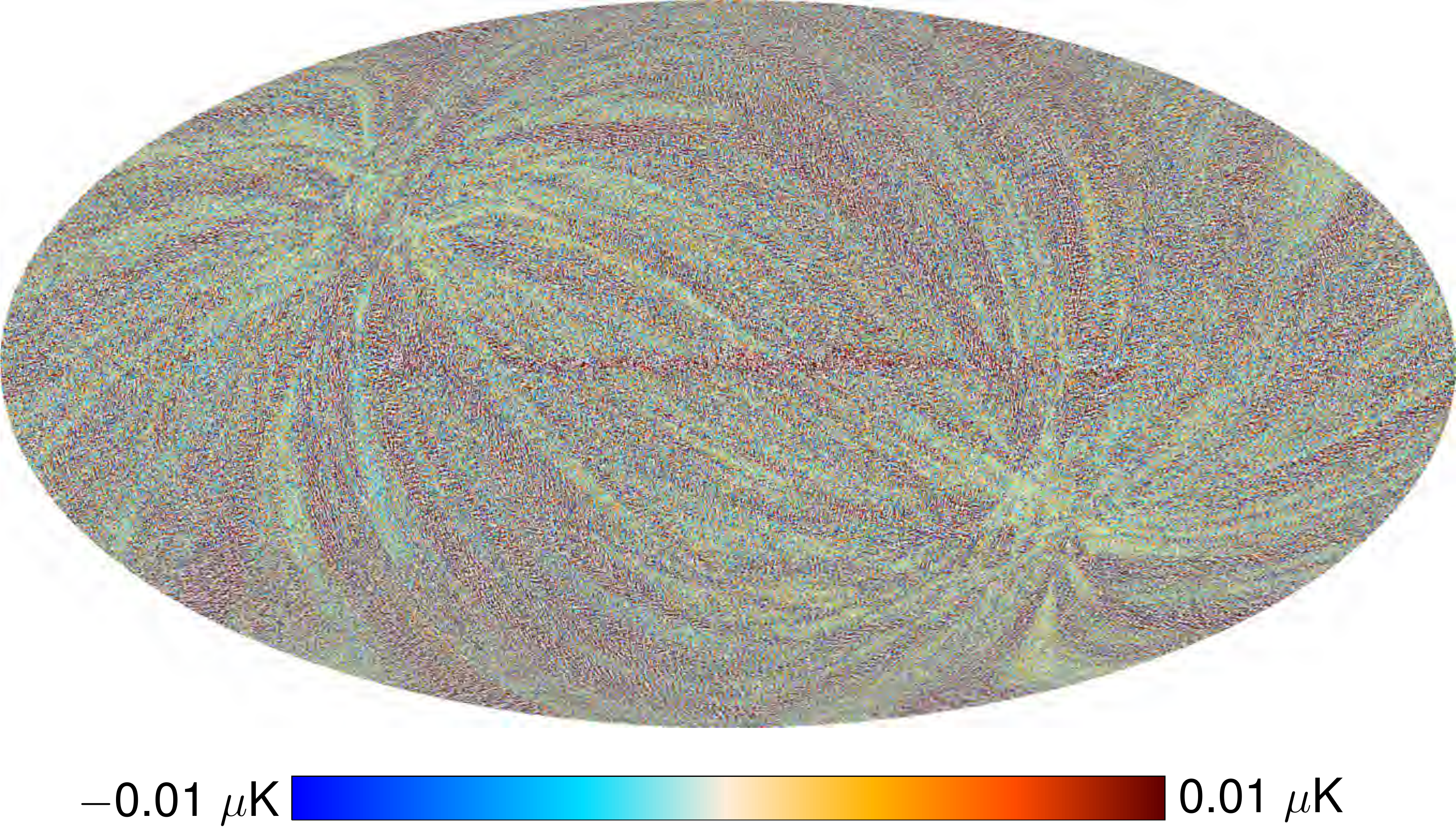}&
	    \includegraphics[width=56mm]{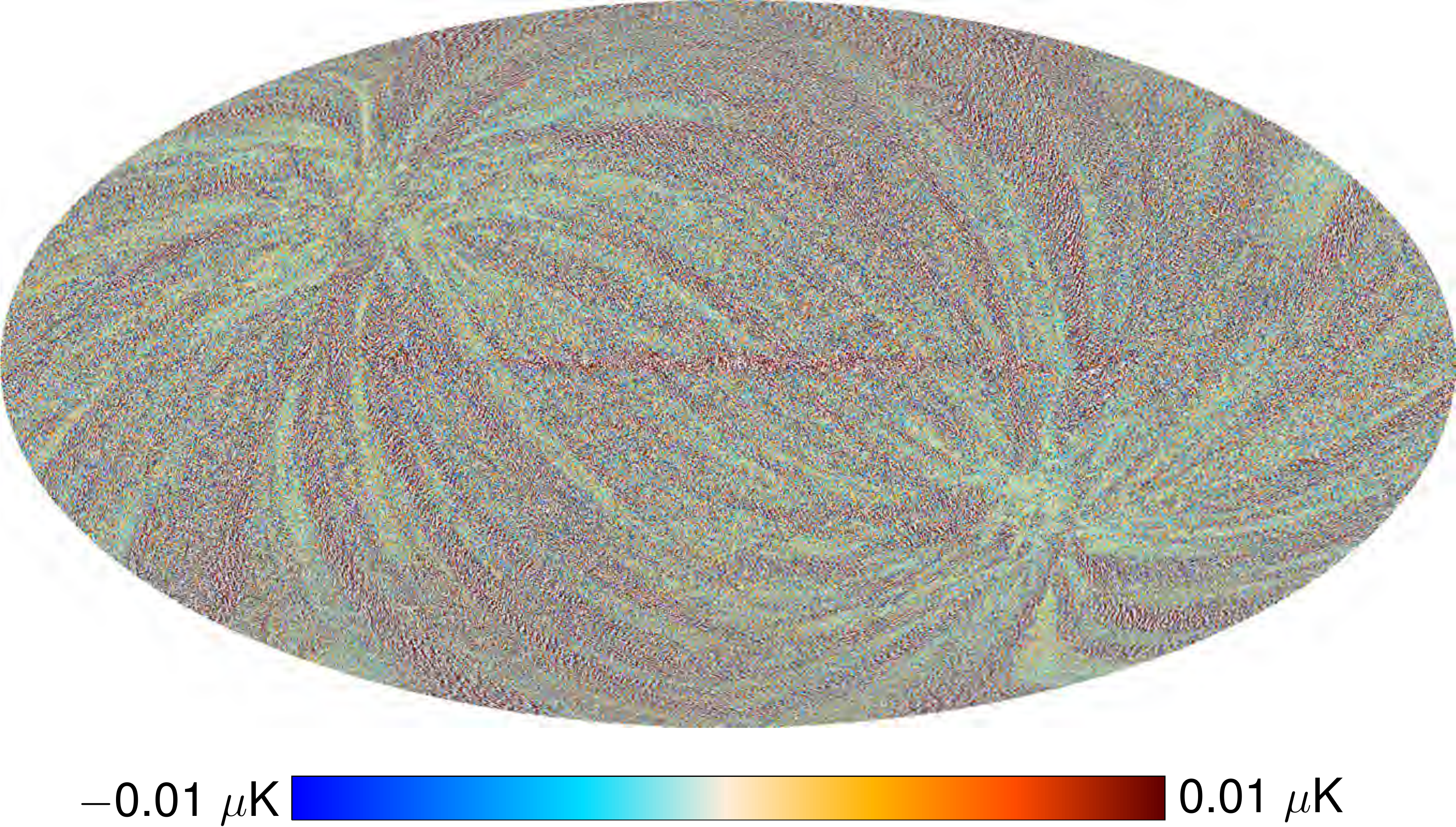}
	    \end{tabular}
	\end{center}

	\caption{Maps of the effect from pointing errors. Rows
        correspond to 30, 44, and 70\,GHz channels, while columns correspond to
        $I$, $Q$, and $U$.}
	\label{fig_whole_mission_intermediate_pointing_error_maps}
      \end{figure*}

  Figure~\ref{fig_whole_mission_intermediate_pointing_error_maps} shows full-sky maps of the estimated systematic effect caused by pointing uncertainty. 
      
  The level of the spurious residual is very small. In polarization it is much less than 1\,$\mu$K at all frequencies, whereas in total intensity it is larger, of the order of a few $\mu$K, being dominated by point sources along the Galactic plane. The 30\,GHz channel is the most affected by this effect; this is expected, because at 30\,GHz the emission from point sources is stronger than at higher frequencies, and the reconstruction of their positions on the sky is particularly sensitive to pointing accuracy.
      
      \subsubsection{Imperfect calibration}
\label{sec_assessment_simulations_imperfect_calibration}

  We assess the effect of uncertainties in the relative photometric calibration of each radiometer by differencing a model sky map and a second map obtained by applying the same calibration pipeline used in the data analysis. 
  
  We start by generating timelines from the measured sky maps at 30, 44, and 70\,GHz. We use these maps as our model sky, already convolved with the telescope beam pattern and radiometric bandpass response. These maps contain the CMB, the diffuse foreground emission, and a small residual of the instrument noise and systematic effects. We are interested in how well the calibration pipeline is able to reproduce this model sky, so that these residuals do not represent a limitation in our analysis.

  Then we add the following three components: (i) the dipole signal convolved with the model sidelobes \citep[see section~7.1 of][]{planck2014-a03}; (ii) the Galactic straylight estimated according to the procedure described in section~7.4 of \citet{planck2014-a03}; and (iii) the radiometer noise (white $+$ $1/f$) generated starting from the radiometer parameters in the instrument database. We report and discuss these parameters in table~10 of \citet{planck2014-a03}.

  The final step in data preparation is to convert the timelines into voltage units. We do this by multiplying each sample by the corresponding gain constant used in the data analysis. We have chosen this approach to have a variability that closely represents the actual measurements. We have also repeated this exercise with different choices of the fiducial gain solution and we have verified that the result does not depend on them at first order.
  
  At this point we apply the calibration pipeline to these timelines to recover the gain constants. These constants will not be identical to the input ones, because of the noise present in the data. The combination of the dipole and Galactic straylight variability with the instrument noise causes a difference between the input and recovered constants that varies with time. 
  
\begin{figure*}[!htpb]
  \begin{center}
    \begin{tabular}{m{.25cm} m{5.6cm} m{5.6cm} m{5.6cm}}
      & \begin{center}$I$\end{center} &\begin{center}$Q$\end{center}&\begin{center}$U$\end{center}\\    
      30&\includegraphics[width=56mm]{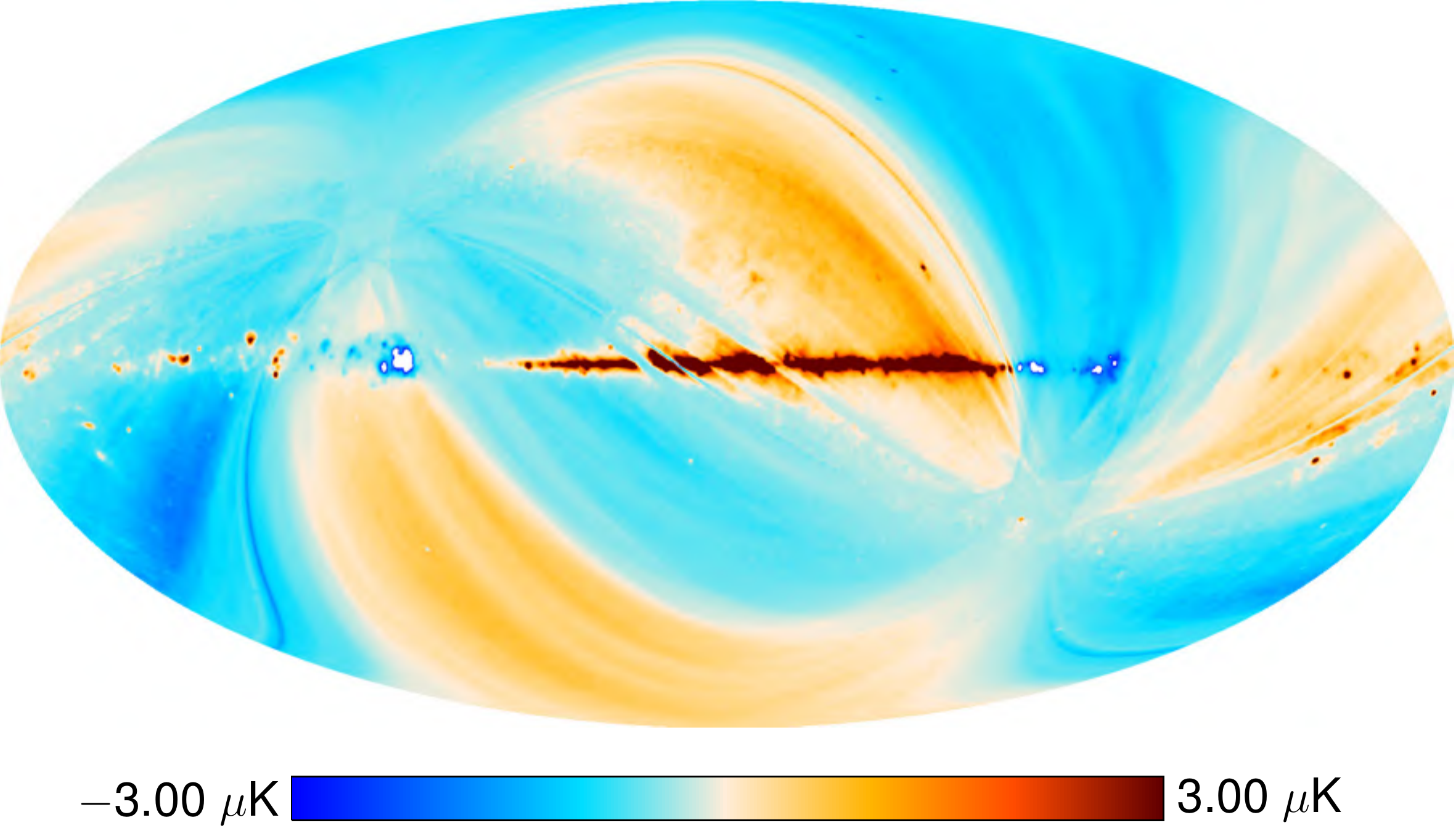}&
      \includegraphics[width=56mm]{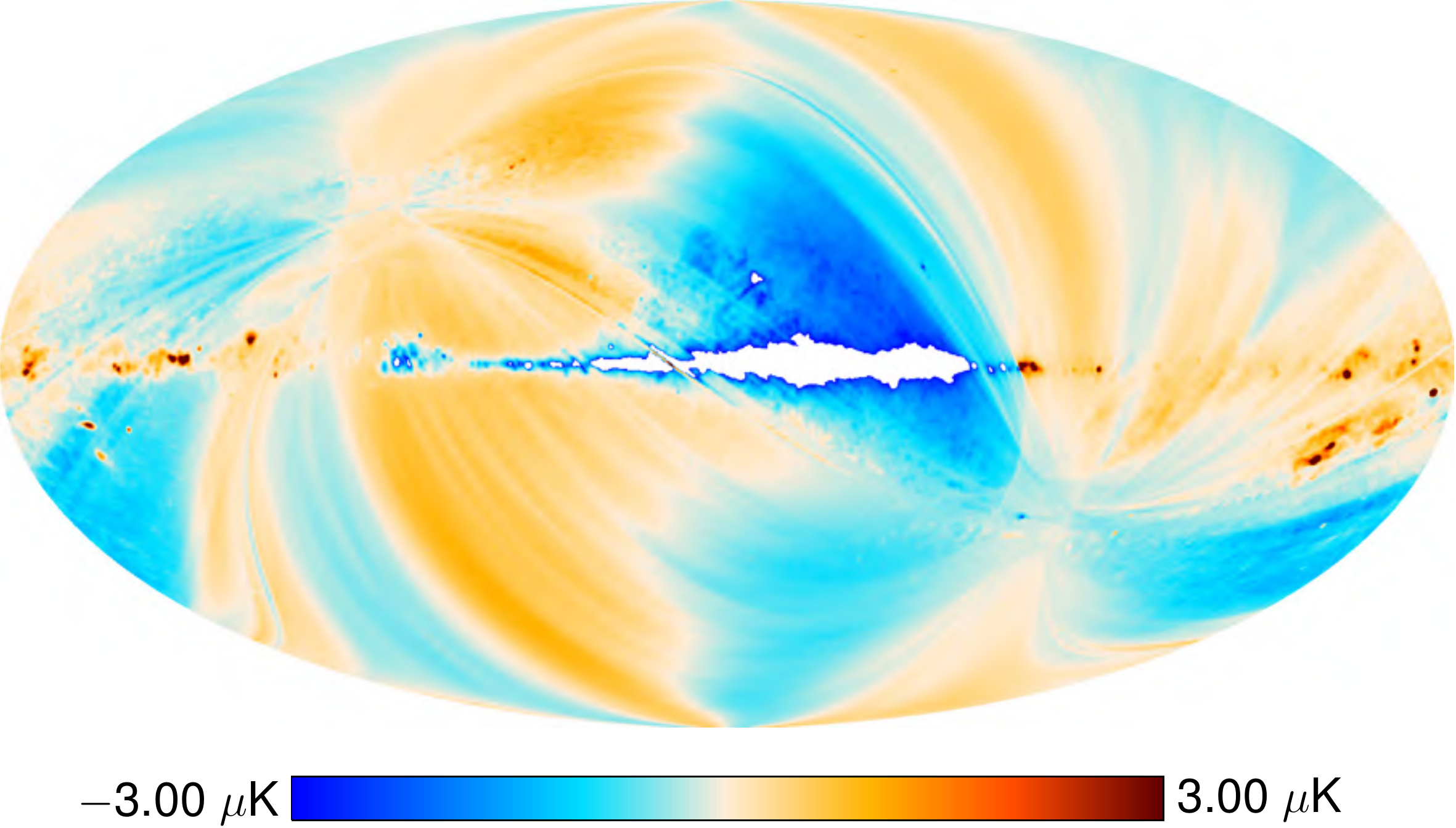}&
      \includegraphics[width=56mm]{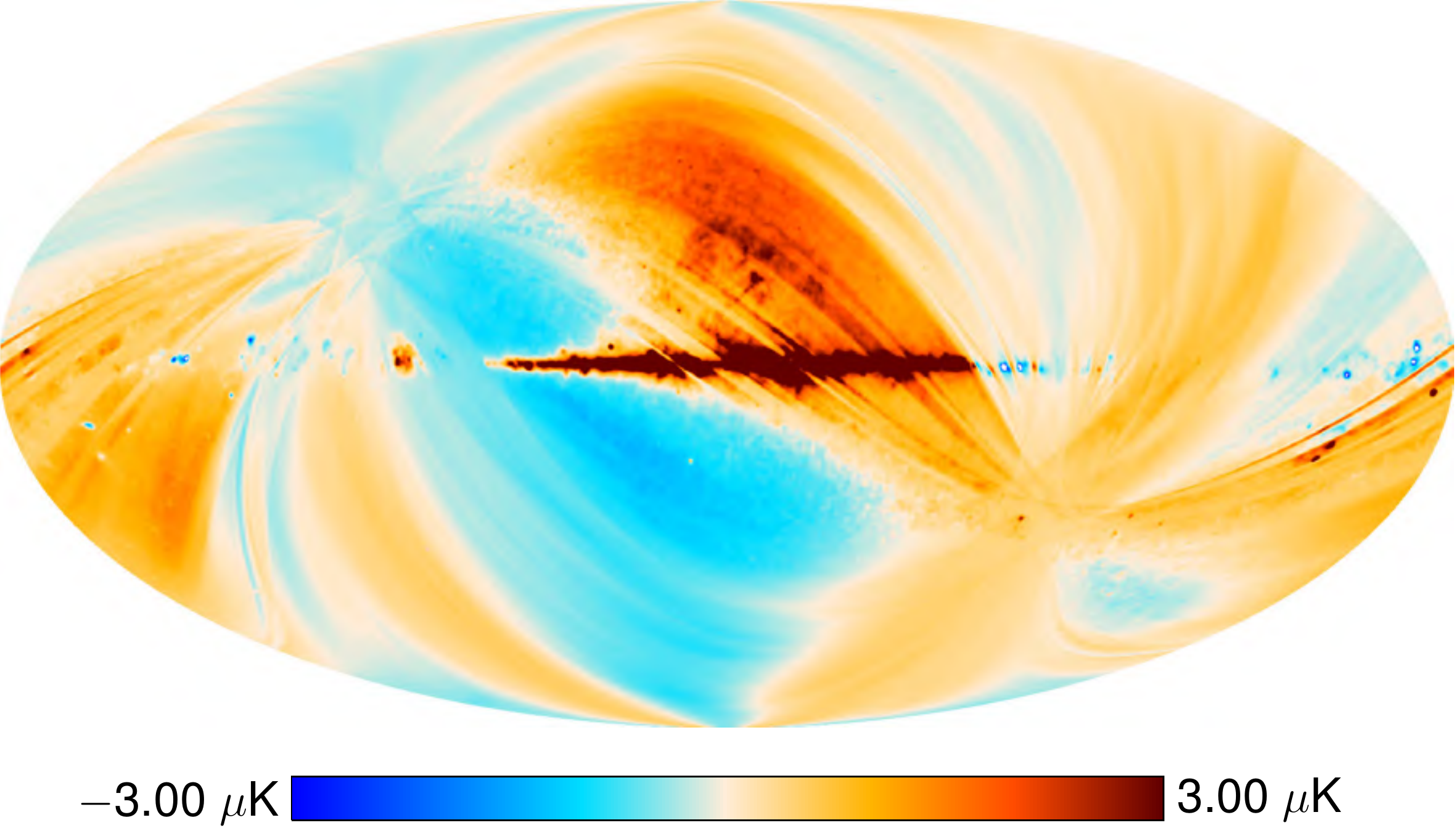}\\
      &&&\\
      44&\includegraphics[width=56mm]{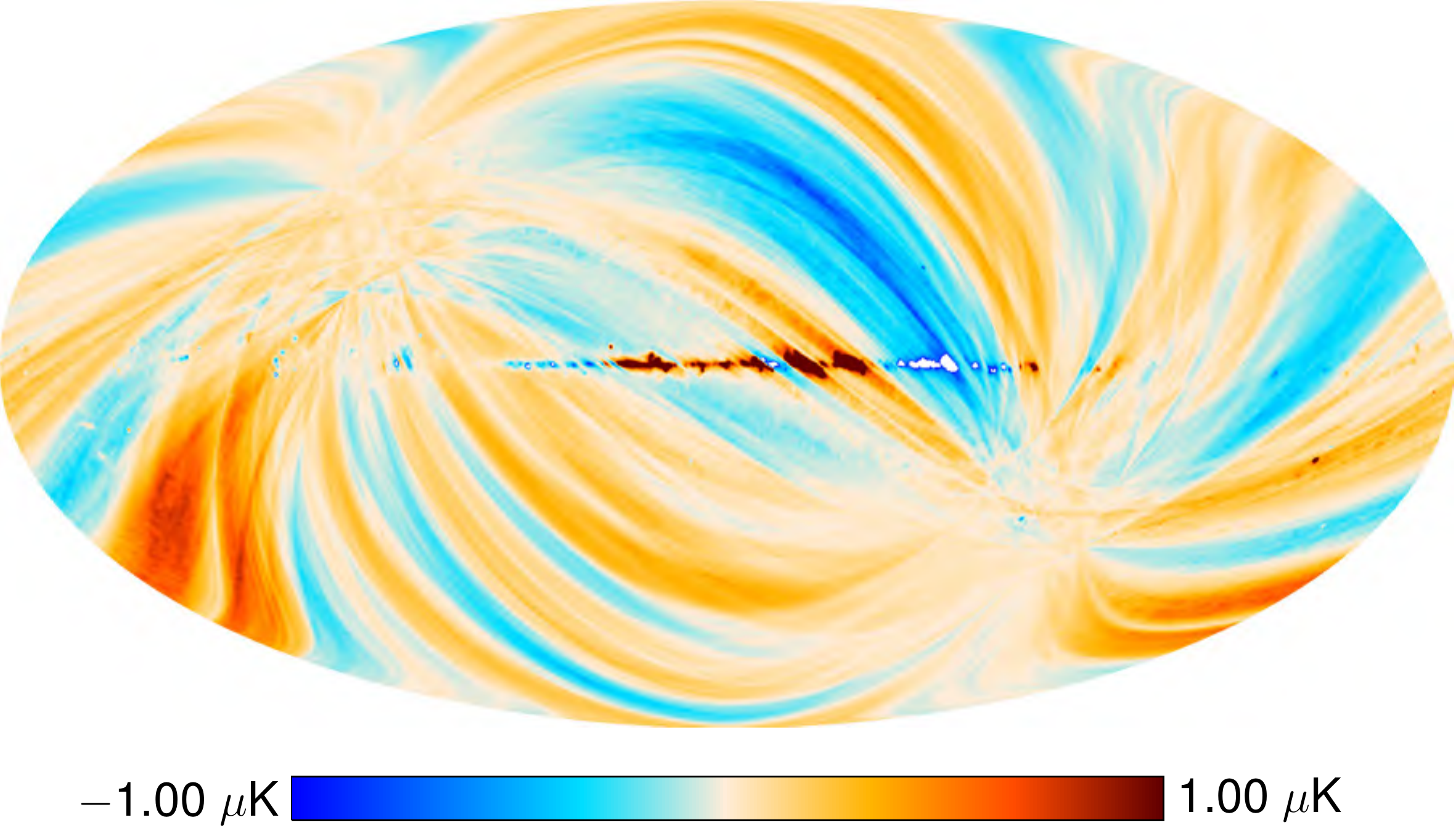}&
      \includegraphics[width=56mm]{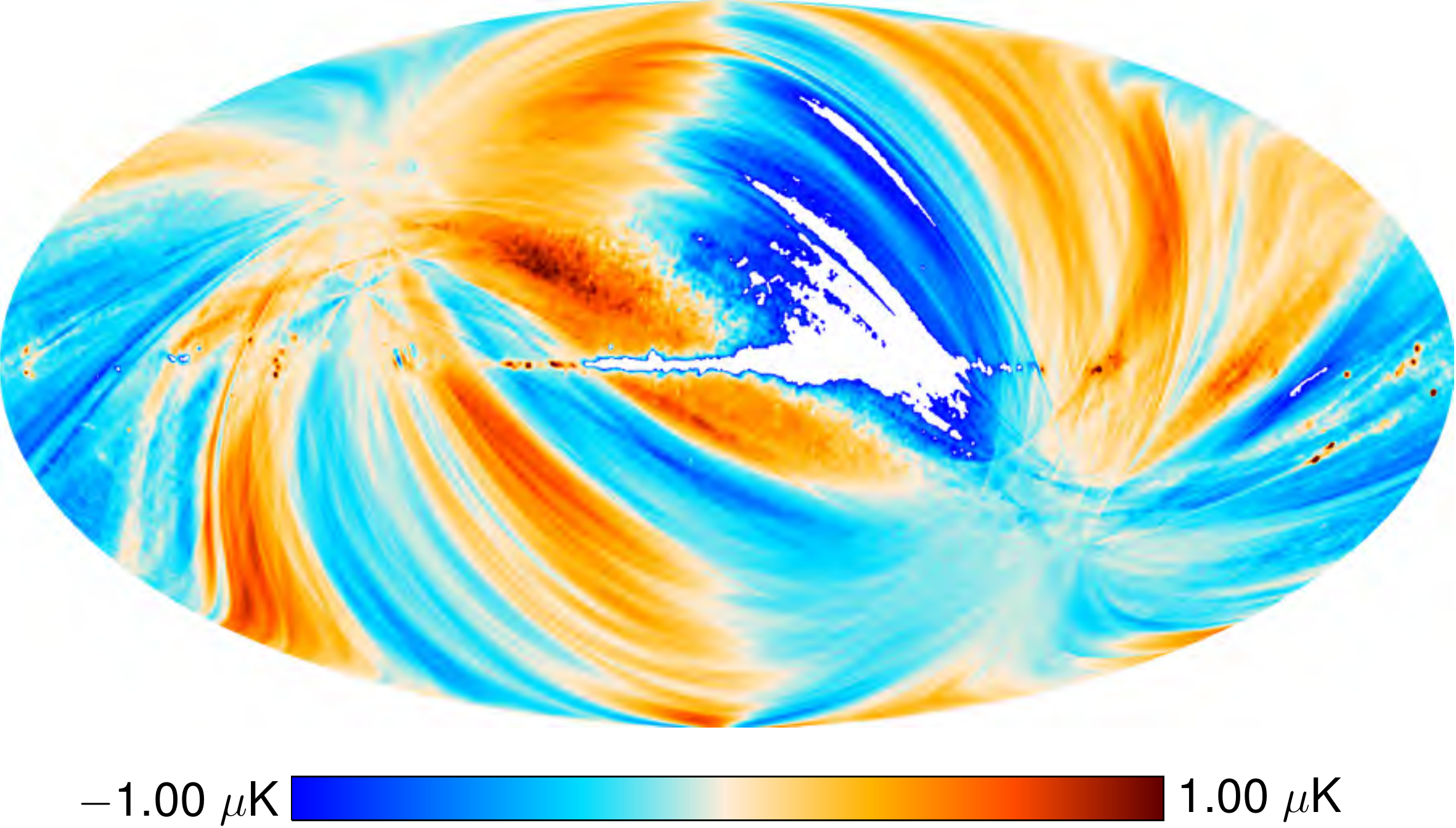}&
      \includegraphics[width=56mm]{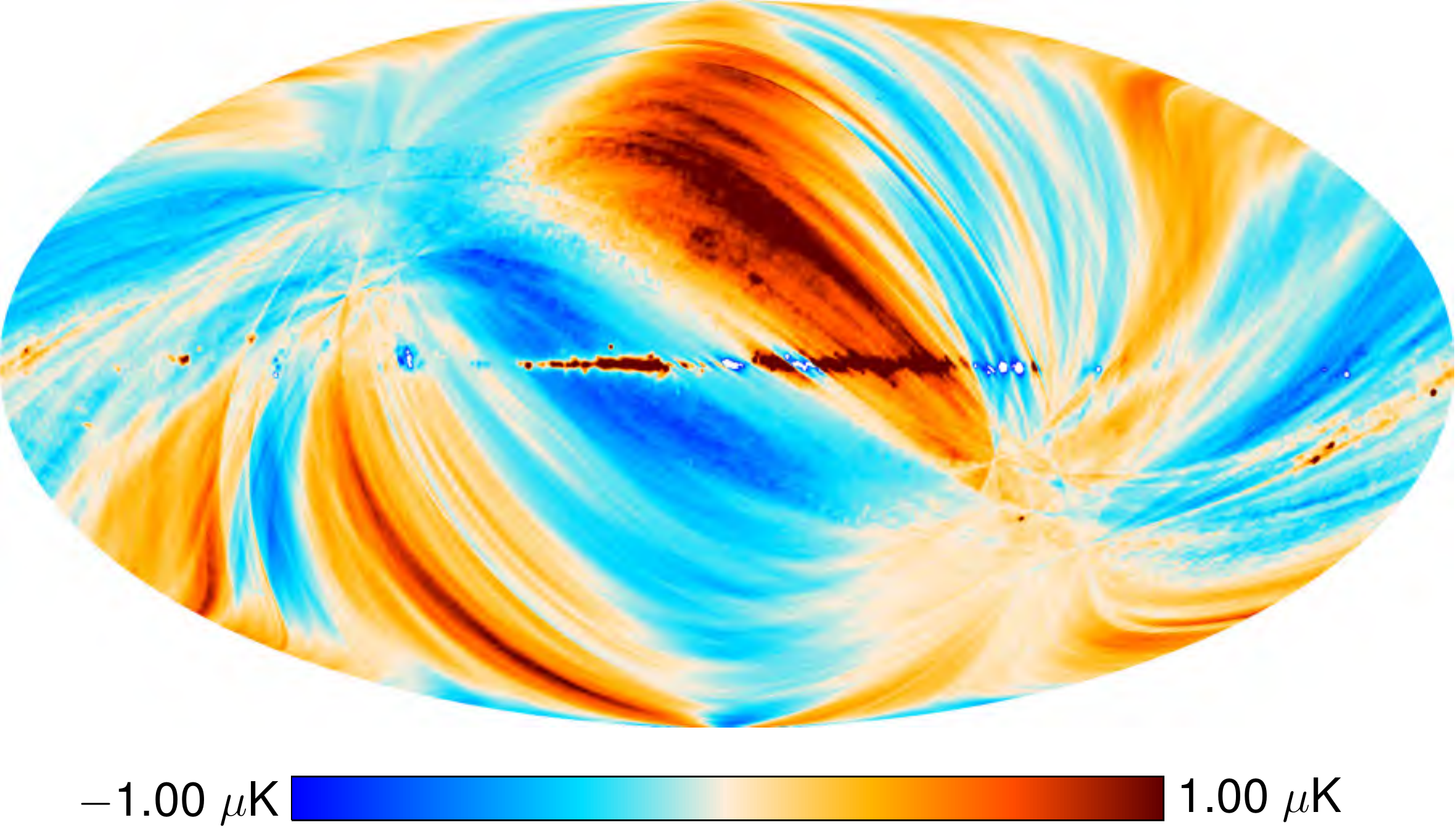}\\
      &&&\\
      70&\includegraphics[width=56mm]{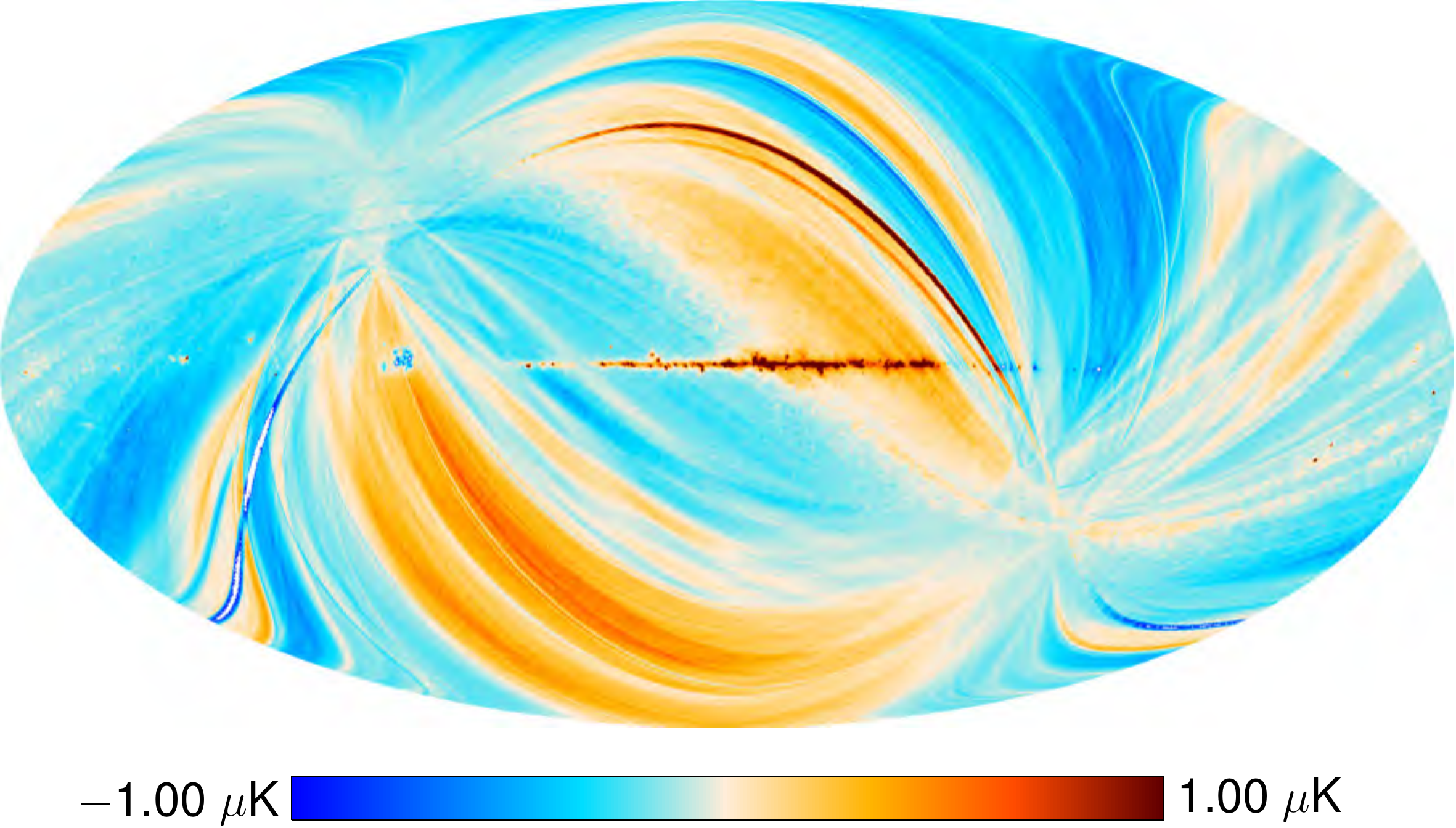}&
      \includegraphics[width=56mm]{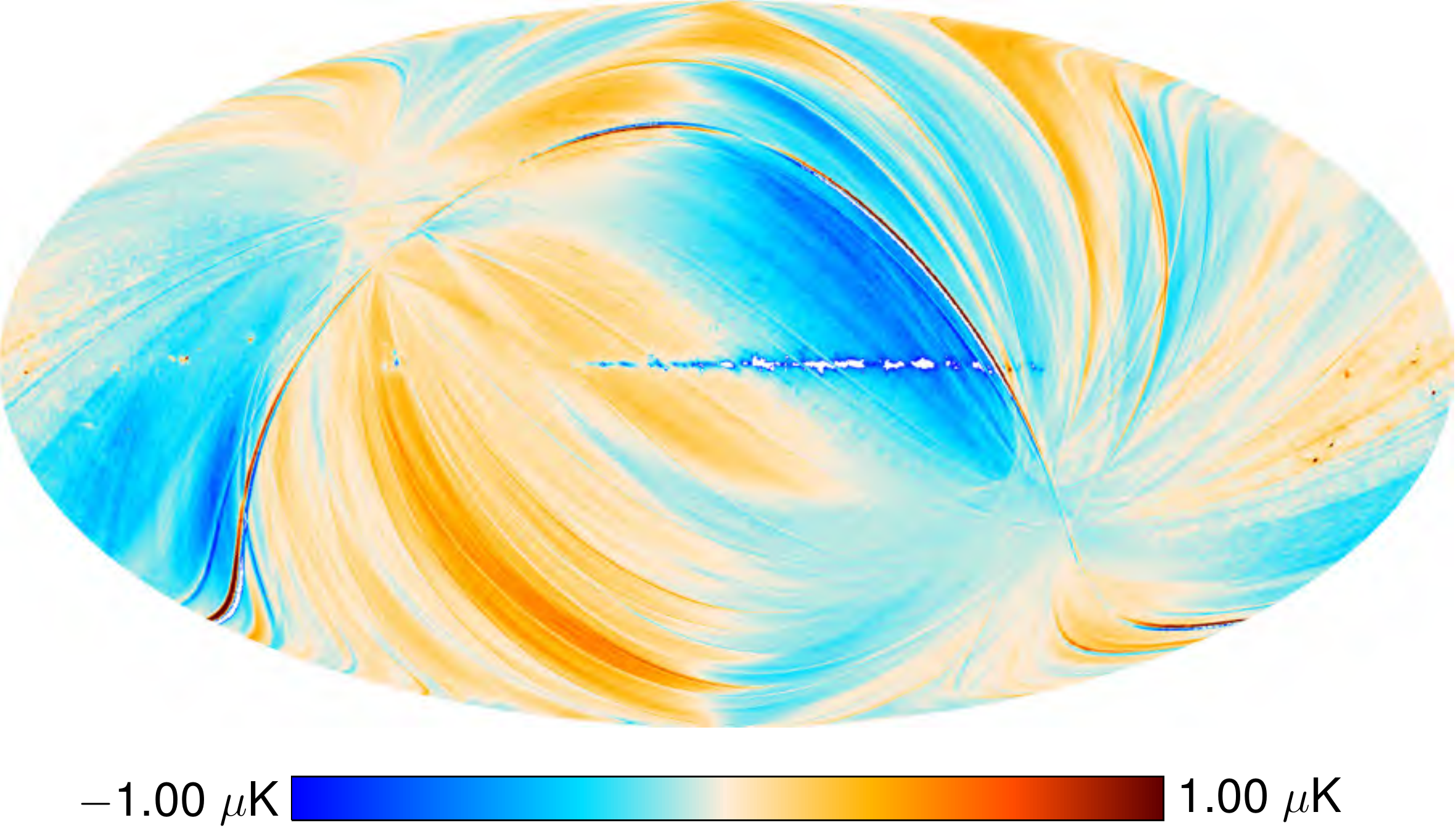}&
      \includegraphics[width=56mm]{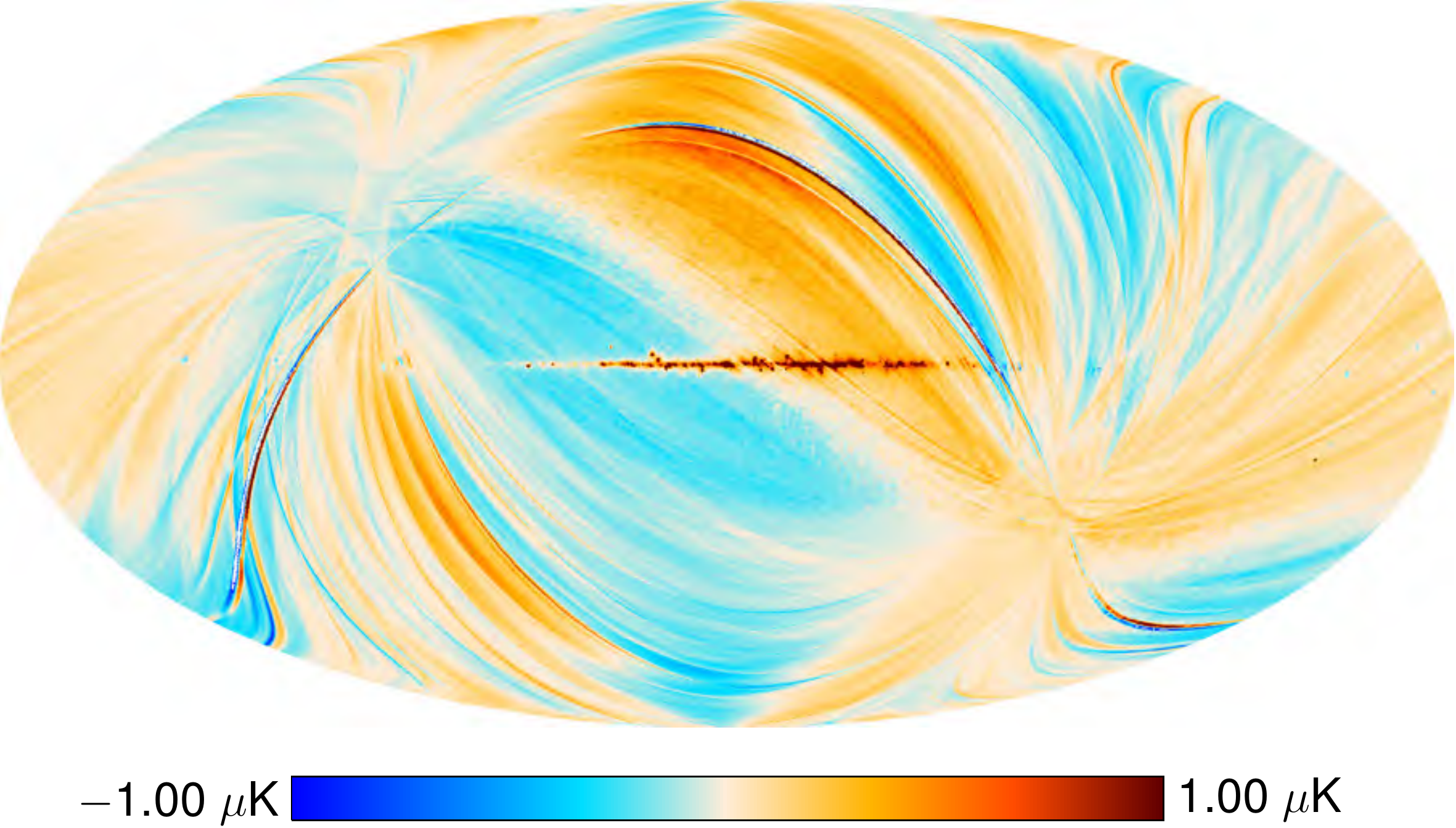}
      \end{tabular}
  \end{center}

  \caption{Maps of the effect from imperfect relative calibration. Rows correspond to 30, 44, and 70\,GHz channels, while columns correspond to $I$, $Q$, and $U$. Maps are smoothed to the beam optical resolution of each channel ($\theta_\mathrm{FWHM}=33\arcm$, $28\arcm$, and $13\arcm$, respectively).}
  \label{fig_whole_mission_CalibrationStatistical_maps}
  \end{figure*}

  We evaluate the impact of the difference between the input and recovered gain constants by differencing the input maps with those generate using the recovered gains. We show these maps in Fig.~\ref{fig_whole_mission_CalibrationStatistical_maps}. These maps show that the effect is of the order of $2\microK$ peak-to-peak at 44 and 70\,GHz, and $6\microK$ peak-to-peak at 30\,GHz.
  
  This is the effect with the largest impact on LFI polarization data and drives the systematic effect uncertainties. At 30\,GHz the residual  is about five orders of magnitudes less than the synchrotron emission in temperature and about ten times less in polarization. We expect, therefore, that we can use this channel as a foreground template with negligible impact from systematic uncertainties. We have verified this in the analysis presented in Sect.~\ref{sec_lowell}, where we assess the impact of systematic effects in the 30 and 70\,GHz channels on the reionization optical depth parameter, $\tau$. 
  
  At 70\,GHz the residual is less than the CMB $E$-mode power spectrum, apart from the range of multipoles 10--20. We evaluated the impact of these uncertainties on the reionization optical depth, $\tau$, and found them to be small. We discuss this assessment in Sect.~\ref{sec_lowell}. At 44\,GHz the residual effect is at the level of the $E$-mode power spectrum for $\ell<10$ and exceeds it in the multipoles range 10--30. In this release we did not use the 44\,GHz data in the extraction of $\tau$. We are currently evaluating strategies to improve the photometric calibration accuracy and we will report the results of this effort in the context of the next release.

      \subsubsection{ADC non linearity}
\label{sec_assessment_simulations_adc_non_linearity}

  We assess the effect of the ADC nonlinearity by taking differences between simulated maps containing the ADC effect and fiducial, clean maps. This is the same approach described in \citet{planck2013-p02a} and is based on the following steps.

  First we produce time-ordered data with and without a known ADC error for each detector. To do this we start from full-mission sky maps and rescan them into time-ordered data using the pointing information. We de-calibrate the data using a gain model based on the actual average of the calibration constant and time variations obtained from relative fluctuations of the reference voltage \citep[the ``4-K calibration'' method described in section~3 of][]{planck2013-p02b}. Then we add voltage offsets, drifts, and noise, in agreement with those observed in in the real, raw data.

  We add the ADC effect to the time-ordered data of each detector applying the inverse spline curve used to correct the data. For the channels where we do not apply any correction we add a conservative estimate of the effect using the estimator $\epsilon_\mathrm{ADC} = (1/\langle V_\mathrm{sky}\rangle) (\delta V_\mathrm{WN, sky}/\delta V_\mathrm{WN, ref})$, where $\langle V_\mathrm{sky}\rangle$ is the average sky voltage and $\delta V_\mathrm{WN, sky}$, $\delta V_\mathrm{WN, ref}$ are the estimates of the white noise in the sky and reference load voltages, respectively. We discuss the rationale behind this choice in appendix~B of \citet{planck2013-p02a}.

  We then determine ADC correction curves from these simulated time-ordered data and remove the estimated effect. A residual remains, though, since the reconstruction is not perfect because of the presence of noise in the data. The effect of this residual is what we estimate in our simulations.

\begin{figure*}[!htpb]
  \begin{center}
    \begin{tabular}{m{.25cm} m{5.6cm} m{5.6cm} m{5.6cm}}
      & \begin{center}$I$\end{center} &\begin{center}$Q$\end{center}&\begin{center}$U$\end{center}\\    
      30&\includegraphics[width=56mm]{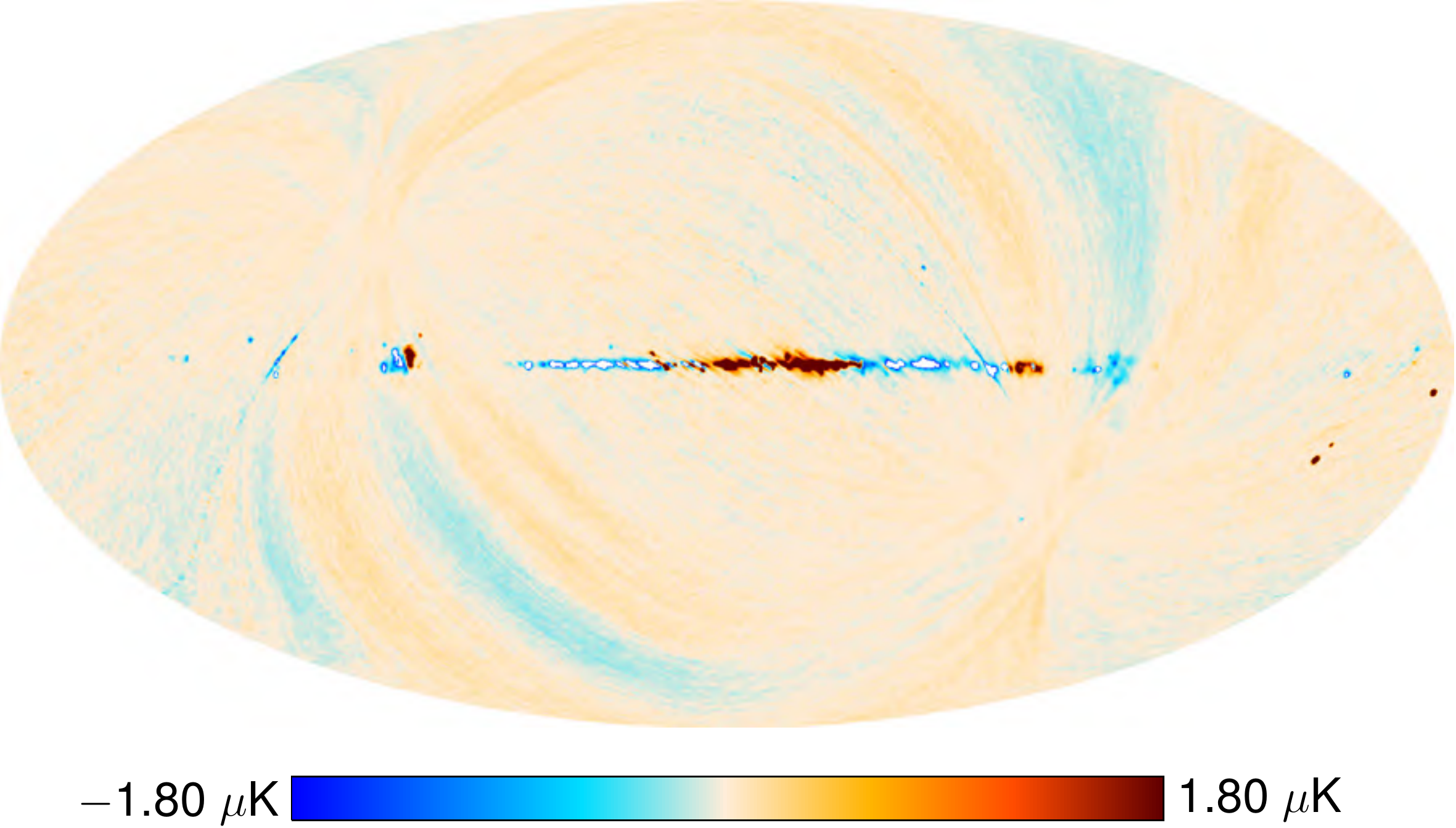}&
      \includegraphics[width=56mm]{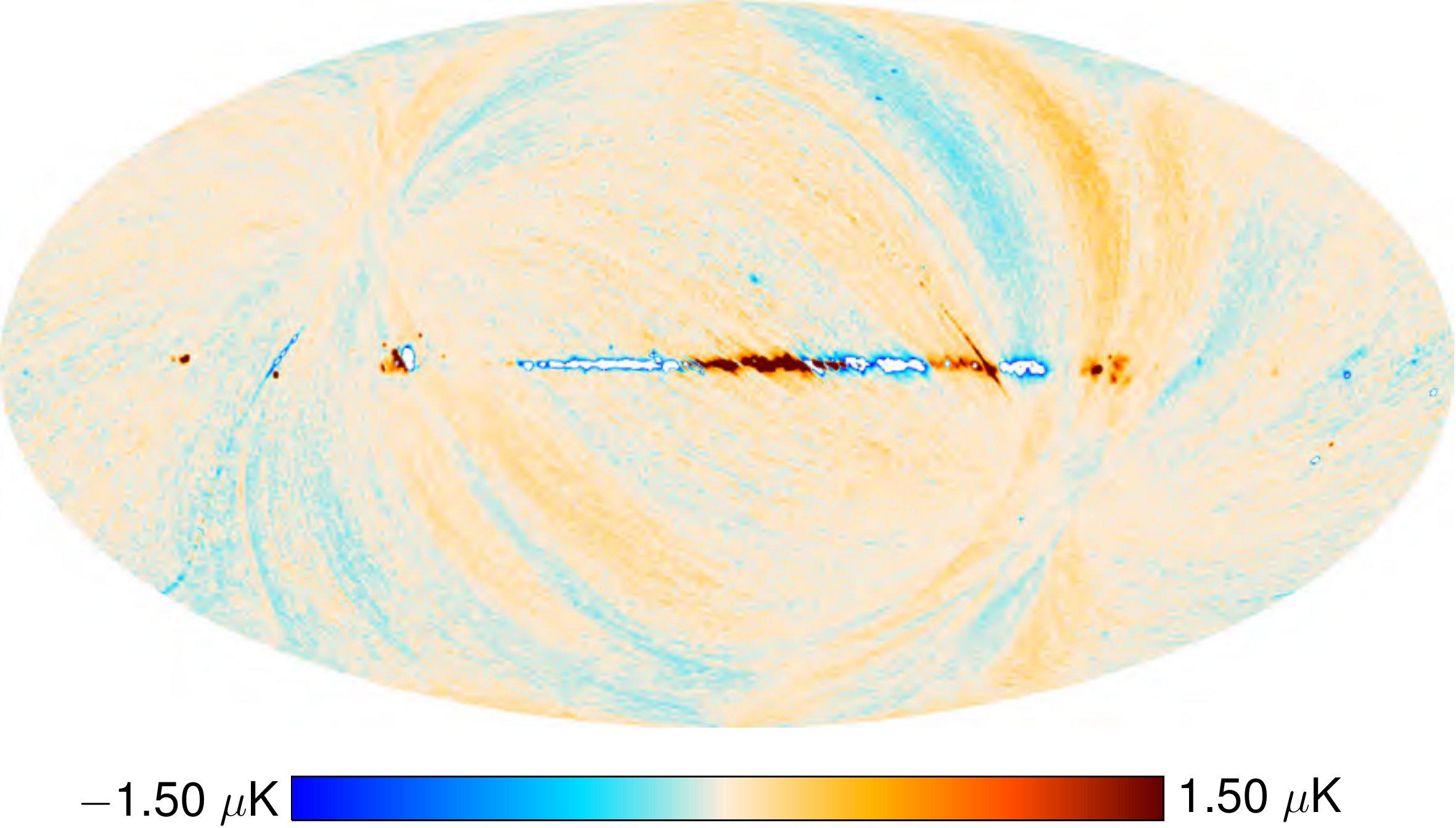}&
      \includegraphics[width=56mm]{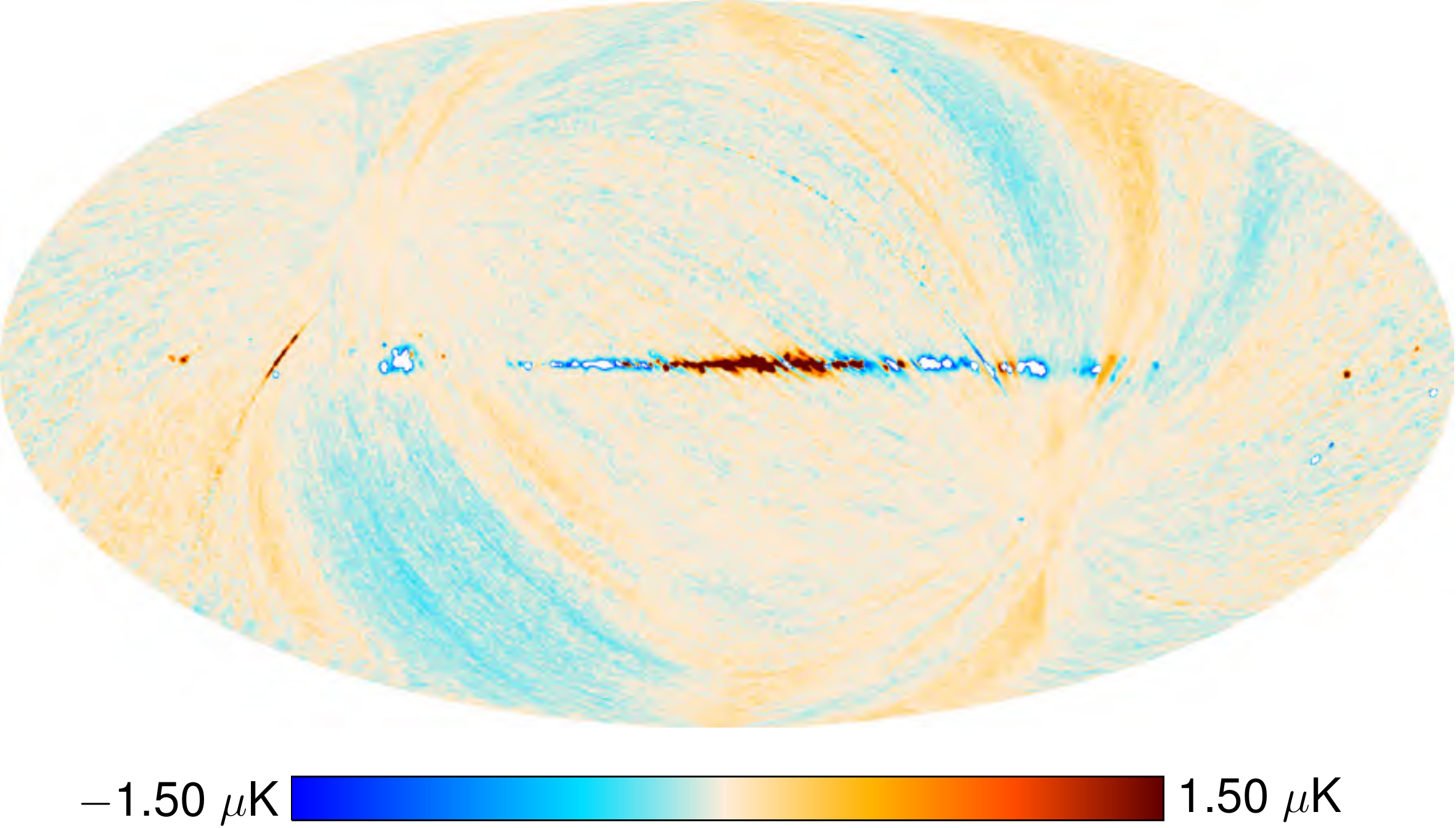}\\
      &&&\\
      44&\includegraphics[width=56mm]{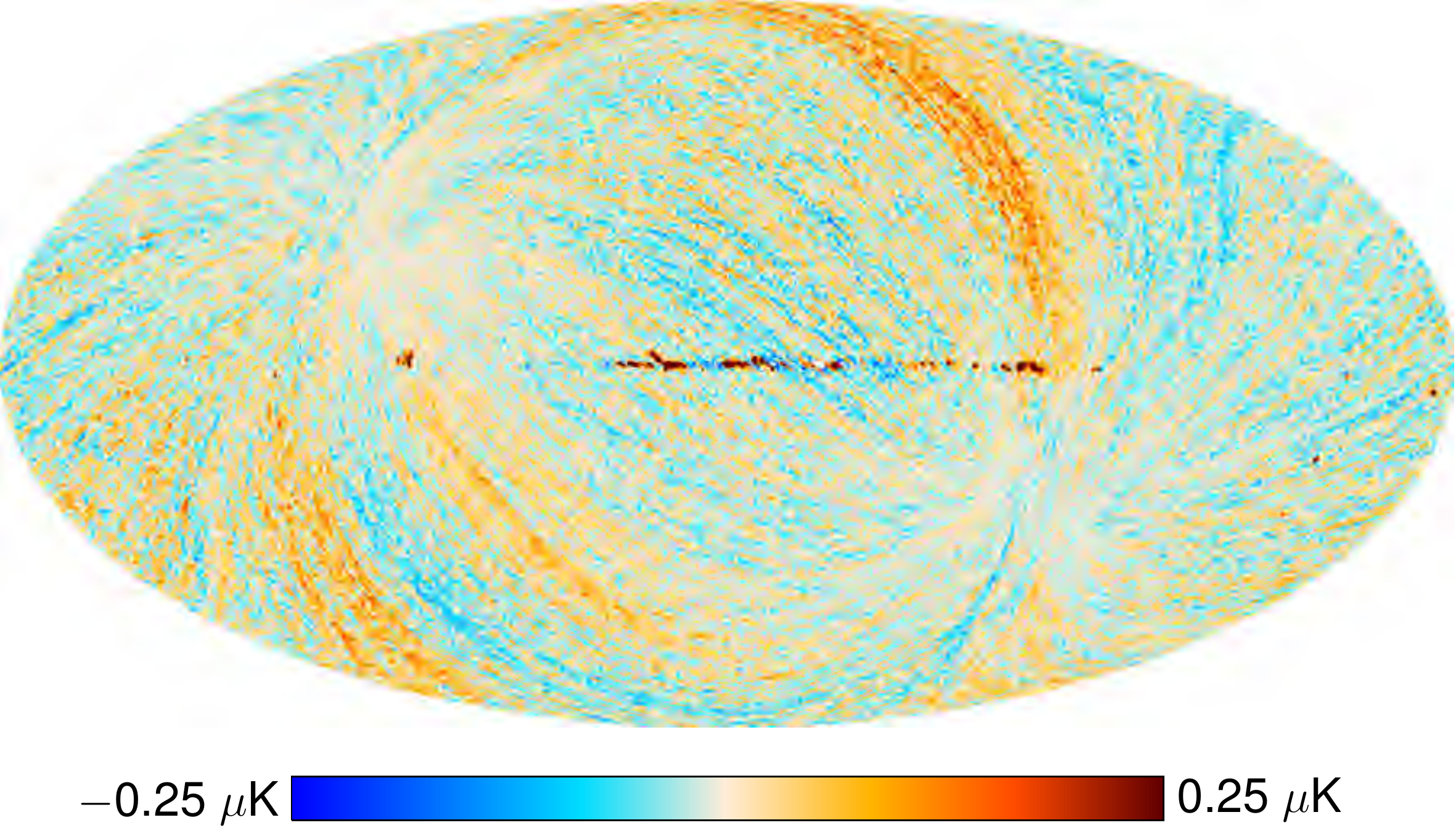}&
      \includegraphics[width=56mm]{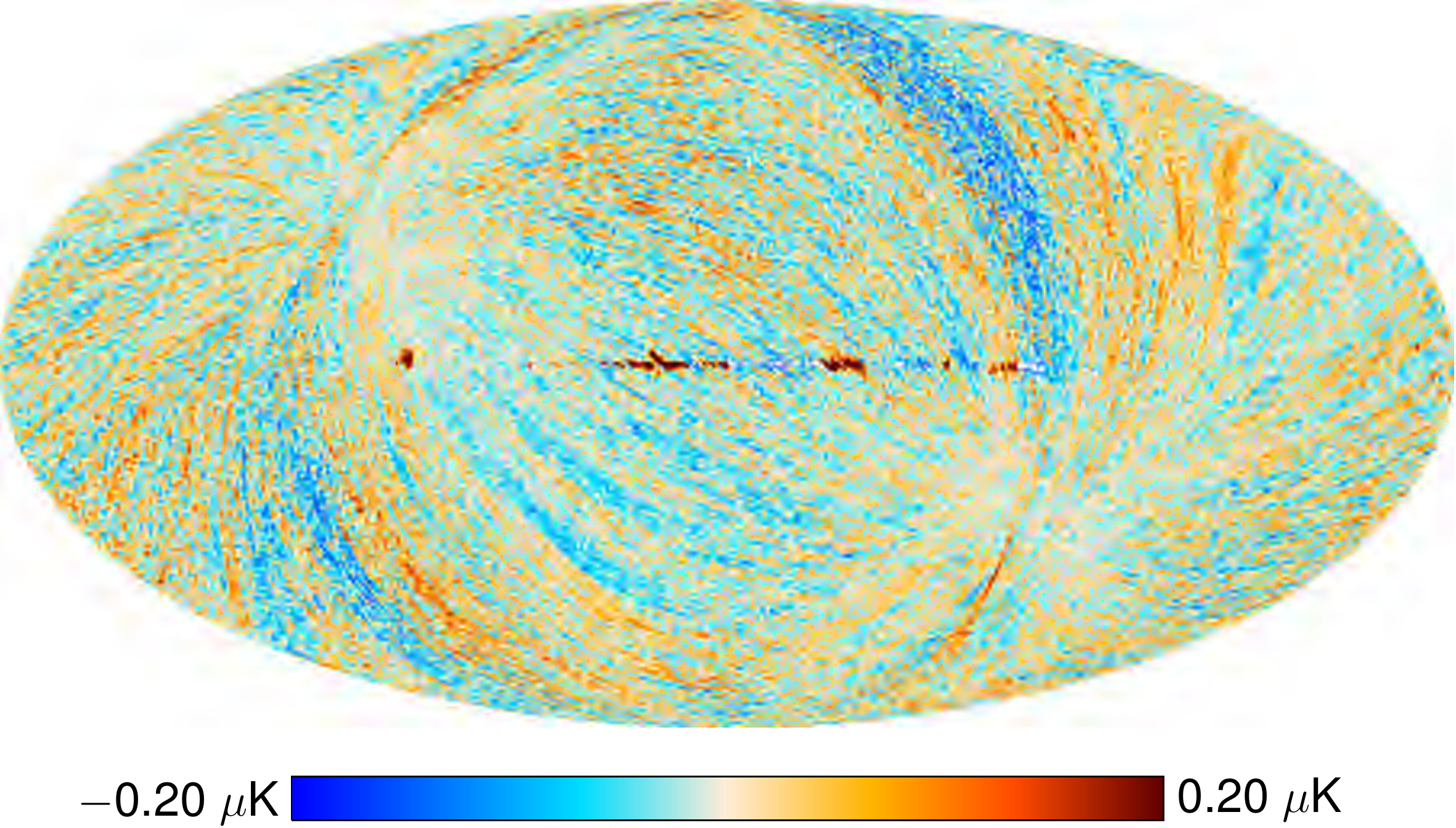}&
      \includegraphics[width=56mm]{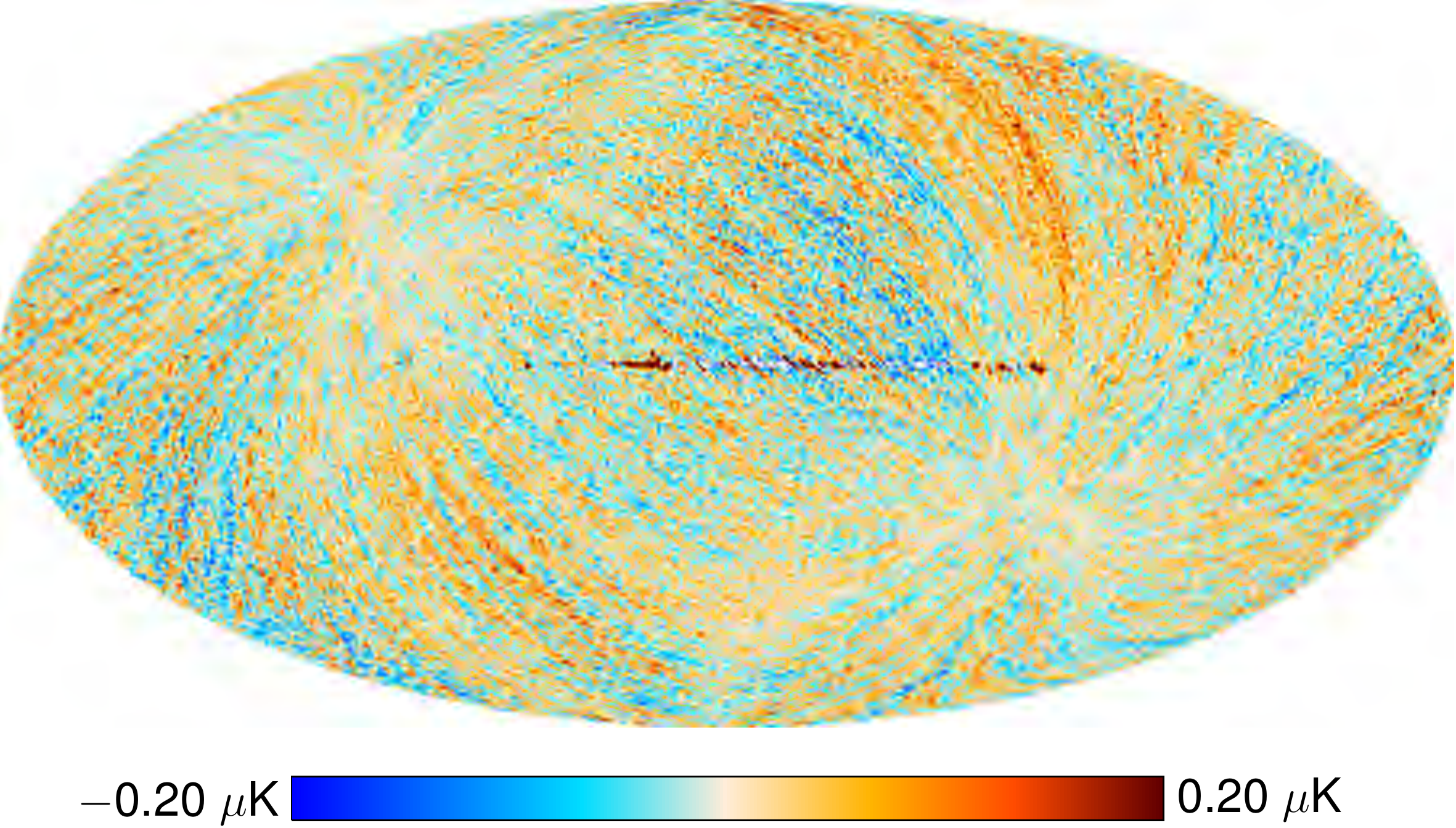}\\
      &&&\\
      70&\includegraphics[width=56mm]{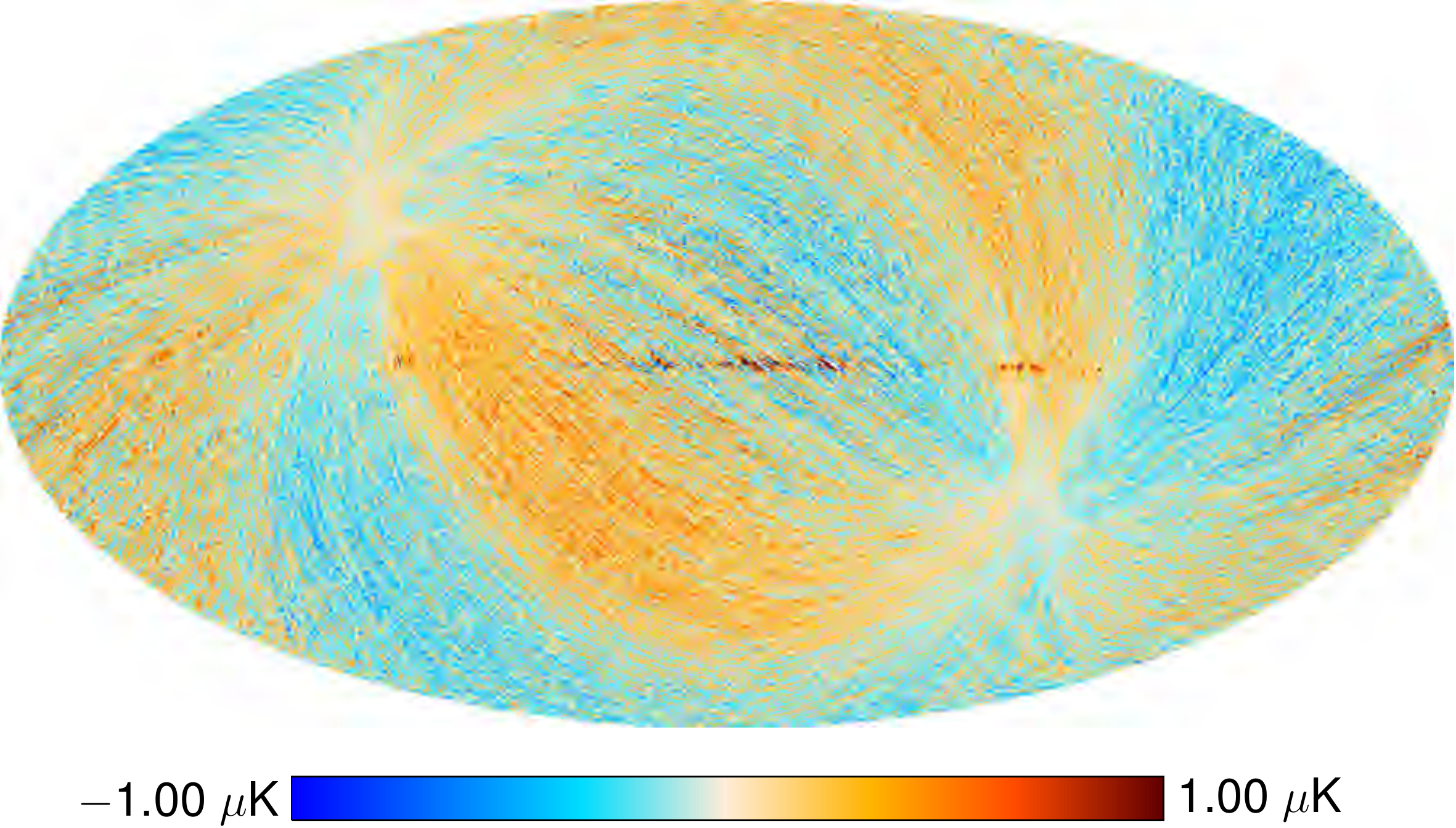}&
      \includegraphics[width=56mm]{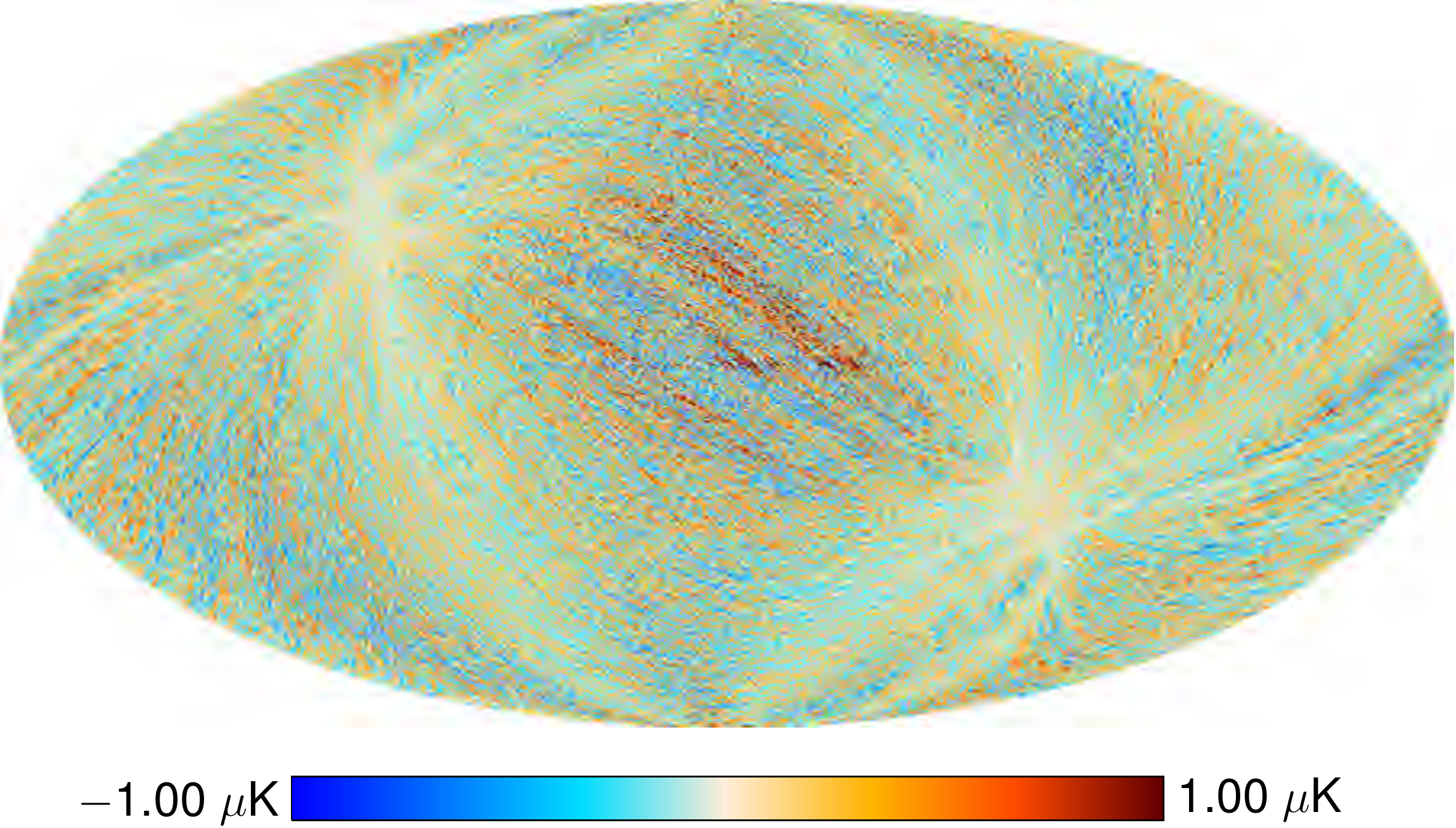}&
      \includegraphics[width=56mm]{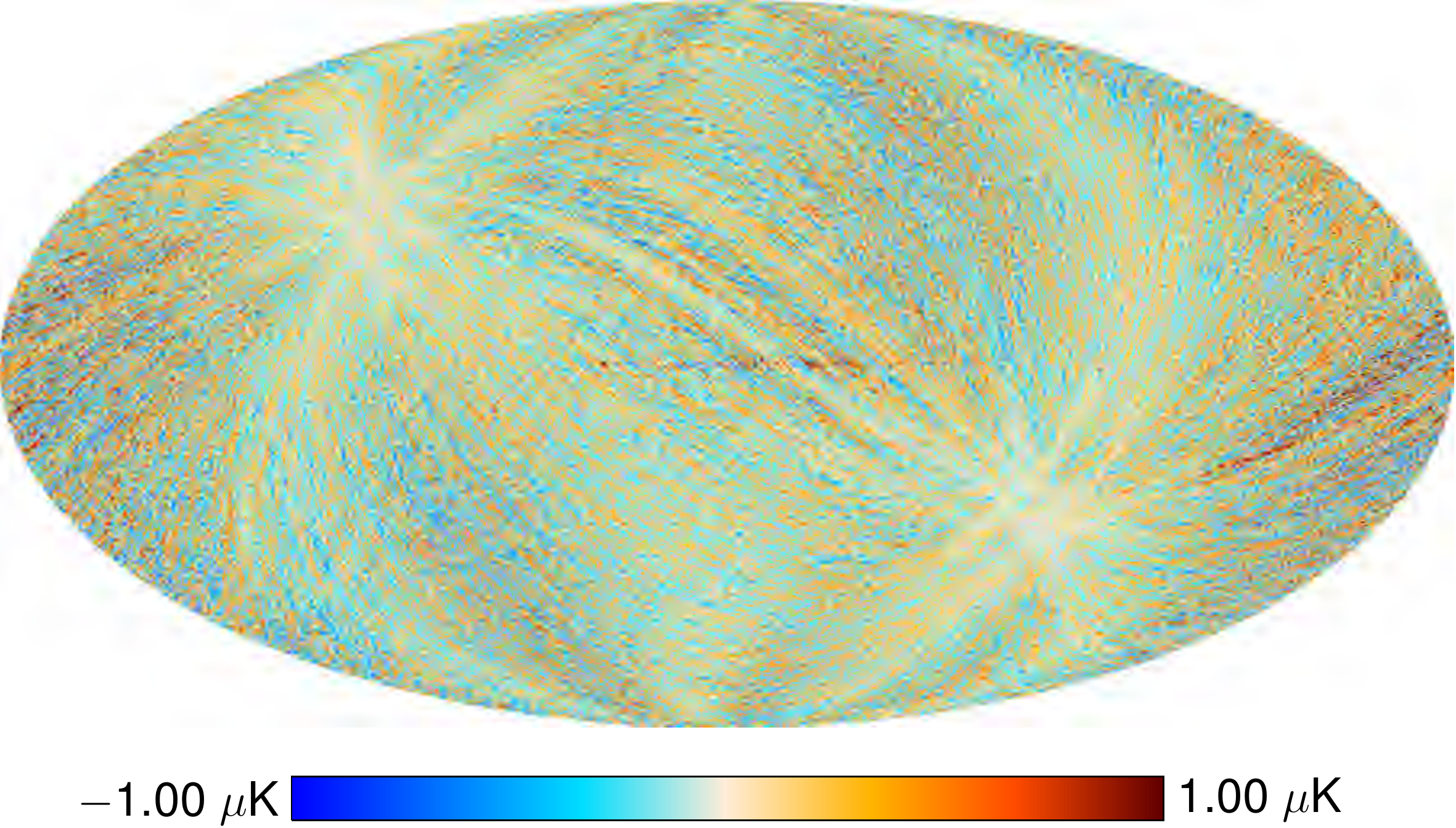}
      \end{tabular}
  \end{center}

  \caption{Maps of the ADC nonlinearity effect. Rows correspond to 30, 44, and 70\,GHz channels, while columns correspond to  $I$, $Q$, and $U$. Maps are smoothed to the beam optical resolution of each channel ($\theta_\mathrm{FWHM}=33\arcm$, $28\arcm$, and $13\arcm$, respectively).}
  \label{fig_whole_mission_ADC_maps}
  \end{figure*}

  Finally we produce difference maps from time-ordered data with and without the residual ADC effect (Fig.~\ref{fig_whole_mission_ADC_maps}). 
  
  In our previous results \citep{planck2013-p02a} the rms fluctuations away from the Galactic plane were about 1 and 0.3\,$\mu$K at 30 and 44\,GHz. Now they are 0.3 and 0.1\,$\mu$K, respectively. In these two frequency channels $Q$ and $U$ maps show ADC stripes at the level of 0.07 and 0.05$\muK$, respectively. In the Galactic plane we see features in both intensity and polarization at the level of a few $\muK$ at 30\,GHz and a fraction of a $\muK$ at 44\,GHz. 
  
  The case at 70\,GHz is more complicated, because the white noise is higher and also because the data from some of the diodes were not corrected. This leads to the appearance of a broad stripe with an amplitude of about $0.3\,\muK$. 
  
      \subsubsection{Thermal effects}
\label{sec_assessment_simulations_thermal}

  Temperature fluctuations of the 4-K reference loads, of the 20-K focal plane, and of the 300-K receiver back-end modules are a source of systematic variations in the measured signal. Fluctuations in the temperature of the 4-K reference loads couple directly with the differential measurements, while fluctuations in the 20-K and 300-K stages couple with the detected signal through thermal and radiometric transfer functions. Thermal transfer functions are described in section~3 of \citet{tomasi2010}, while a complete description of the radiometric coupling with temperature fluctuations can be found in \citet{terenzi2009b}. We also provide a general treatment of the susceptibility of the LFI receivers to systematic effects in section~3 of \citet{seiffert2002} (see, especially, equation~10).

\begin{figure*}[!htpb]
    \begin{center}
      \includegraphics[width=170mm]{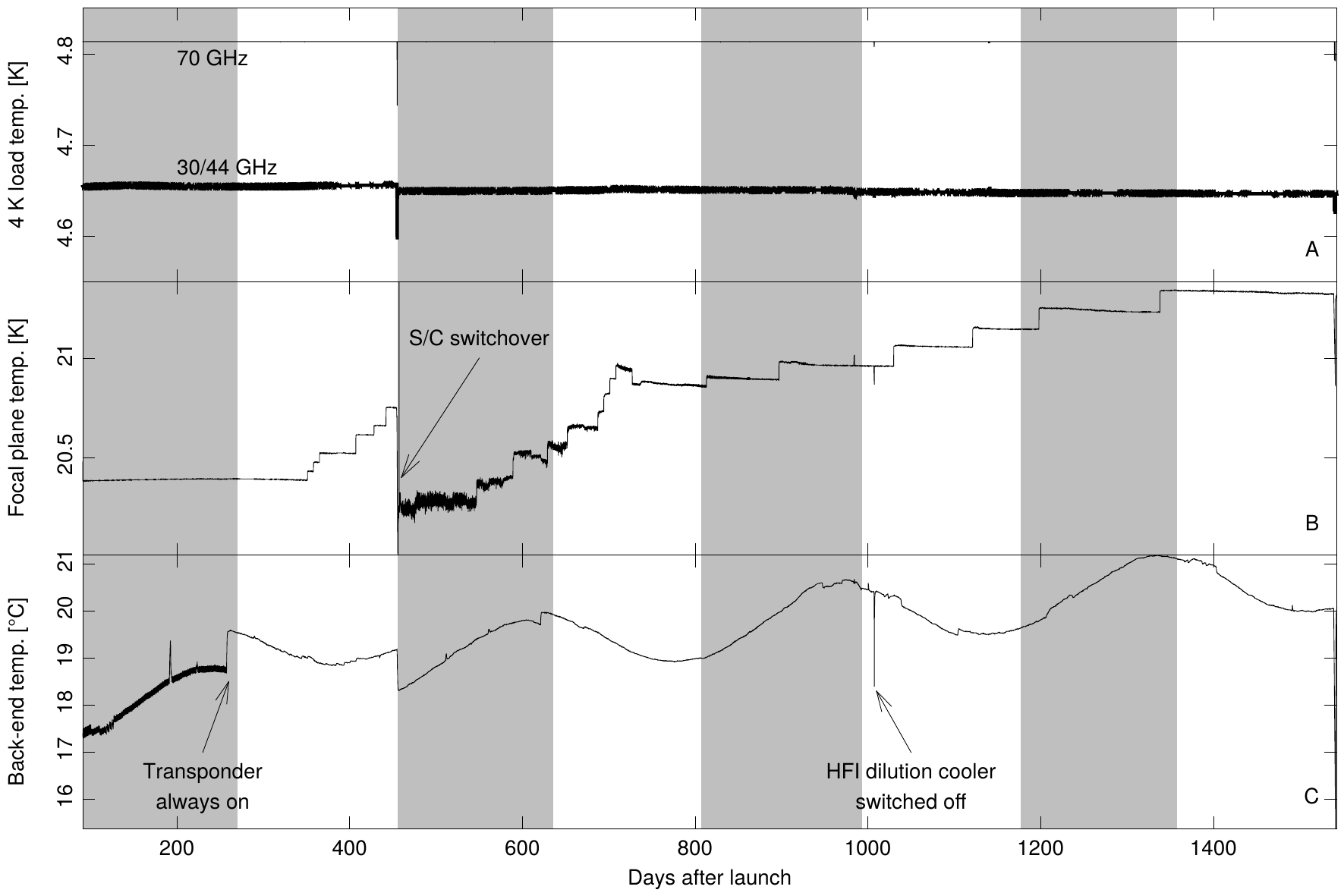}
    \end{center}
    \caption{
      \textit{Top}: temperature measured by two sensors mounted on the HFI 4-K shield at the level of 30 and 44\,GHz (bottom curve) and 70\,GHz (top curve) reference loads. The rms variation of this temperature over the full mission is $\sigma_\mathrm{30, 44} = 1.55$\,mK and $\sigma_\mathrm{70} = 80\,\mu$K. \textit{Middle}: temperature of the 20-K focal plane measured by a temperature sensor placed on the flange of the \texttt{LFI28} 30\,GHz feedhorn. The consecutive temperature steps after day 500 reflect changes to the set-point of the temperature control system. We applied these changes to control the level of temperature fluctuations. \textit{Bottom}: temperature of the 300-K back-end unit.
    }
    \label{fig_whole_mission_temperature_curves}
  \end{figure*}

  The three panels in Fig.~\ref{fig_whole_mission_temperature_curves} show the behavior of the relevant LFI temperature over the entire mission, with labels identifying relevant events that occurred during the mission. The grey bands identify the eight surveys. 
  
  The top panel shows the temperature of the 4-K loads at the level of the 30 and 44\,GHz (bottom curve) and 70\,GHz (top curve) channels. The rms variation of this temperature over the whole mission is $\sigma_\mathrm{30, 44} = 1.55$\,mK and $\sigma_\mathrm{70} = 80\,\mu$K.
  
  The middle panel displays the 20-K focal plane temperature measured by a sensor mounted on the flange of the 30\,GHz \texttt{LFI28} receiver feedhorn. There are some notable features. The most stable period is the one corresponding to the first survey. After the sorption cooler switchover we see a period of temperature instability that spanned about half of the third survey and that was later controlled by commanded temperature steps that we continued to apply until the end of the mission. The bottom panel shows the behavior of one of the the back-end temperature sensors. We notice a short period temperature fluctuation during the first survey, which was caused by the daily on-off switching of the transponder. This was later left on to reduce these 24-hour variations (see the temperature increase after day 200). We also see a temperature drop corresponding to the sorption cooler switchover and a seasonal periodic variation correlated with the yearly orbit around the Sun.
  
  We assess the effect of temperature variations following the procedure described in section~4.2.1 of \citet{planck2013-p02a}. We combine temperature measurements with thermal and radiometric transfer functions to obtain time-ordered data of the systematic effects. We then use these time-ordered data and pointing information to produce maps. For this release we used the same thermal and radiometric transfer functions that were applied in the analysis of the 2013 release.

\begin{figure*}[!htpb]
  \begin{center}
    \begin{tabular}{m{.25cm} m{5.6cm} m{5.6cm} m{5.6cm}}
      & \begin{center}$I$\end{center} &\begin{center}$Q$\end{center}&\begin{center}$U$\end{center}\\    
      30&\includegraphics[width=56mm]{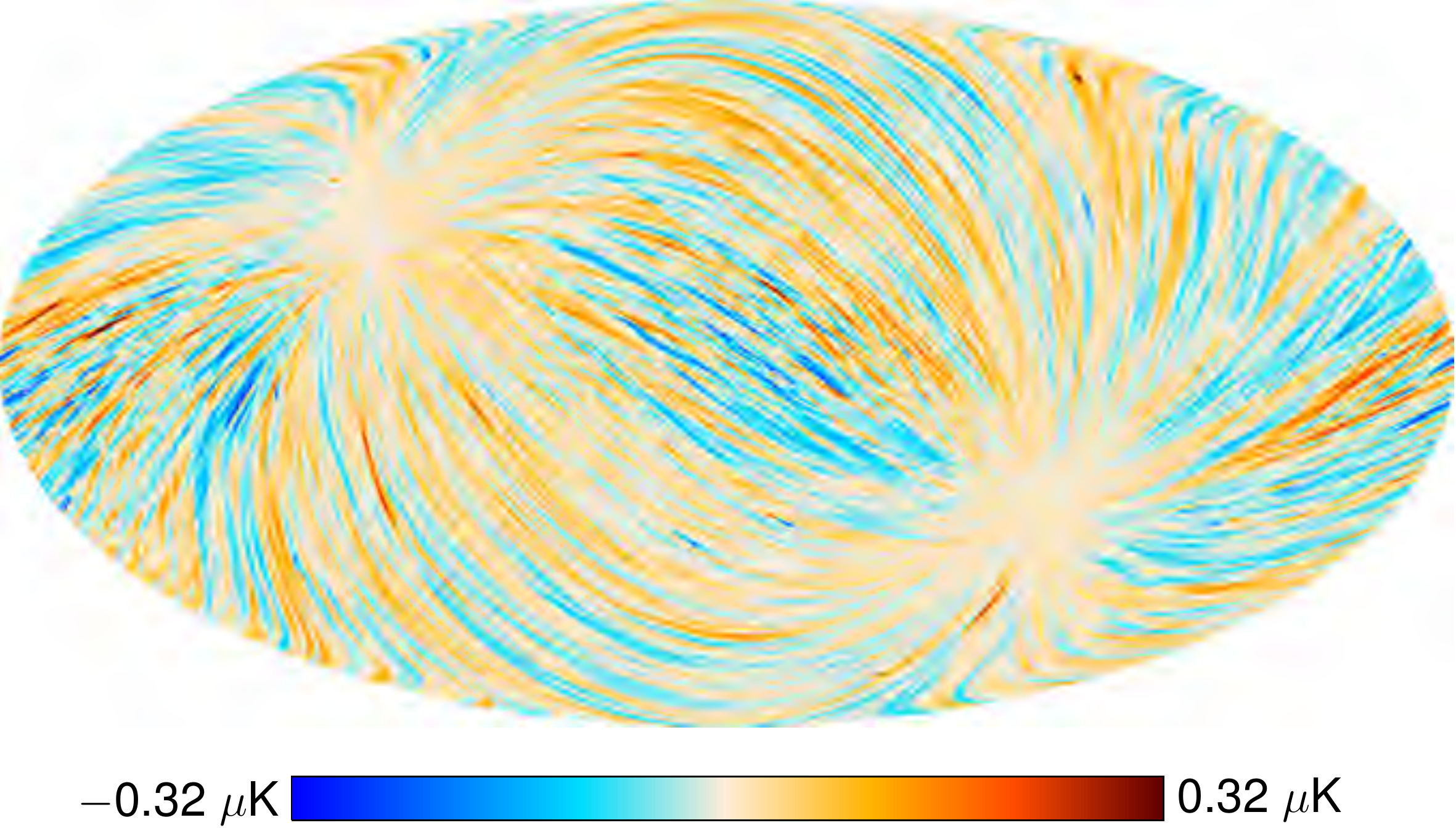}&
      \includegraphics[width=56mm]{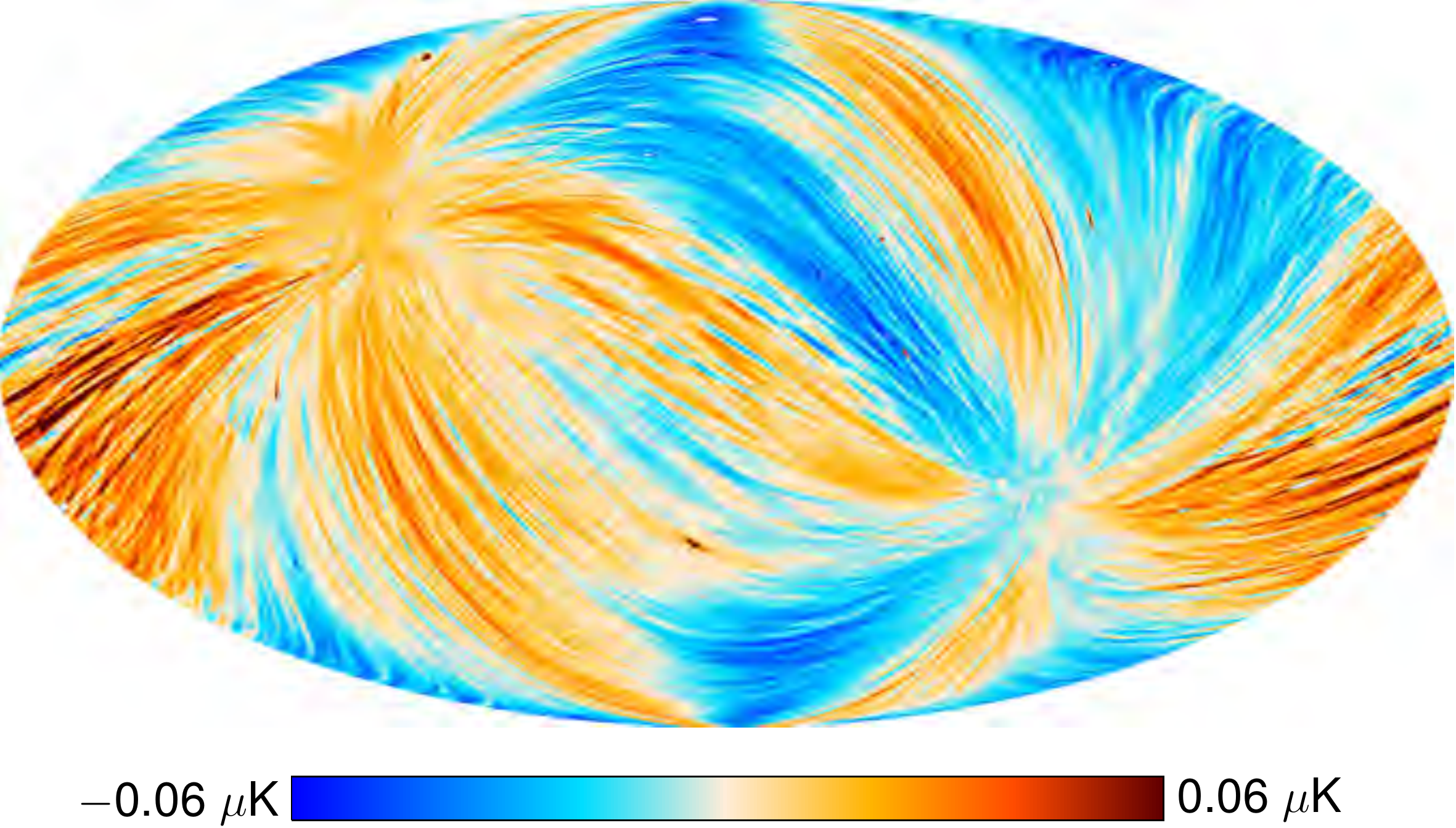}&
      \includegraphics[width=56mm]{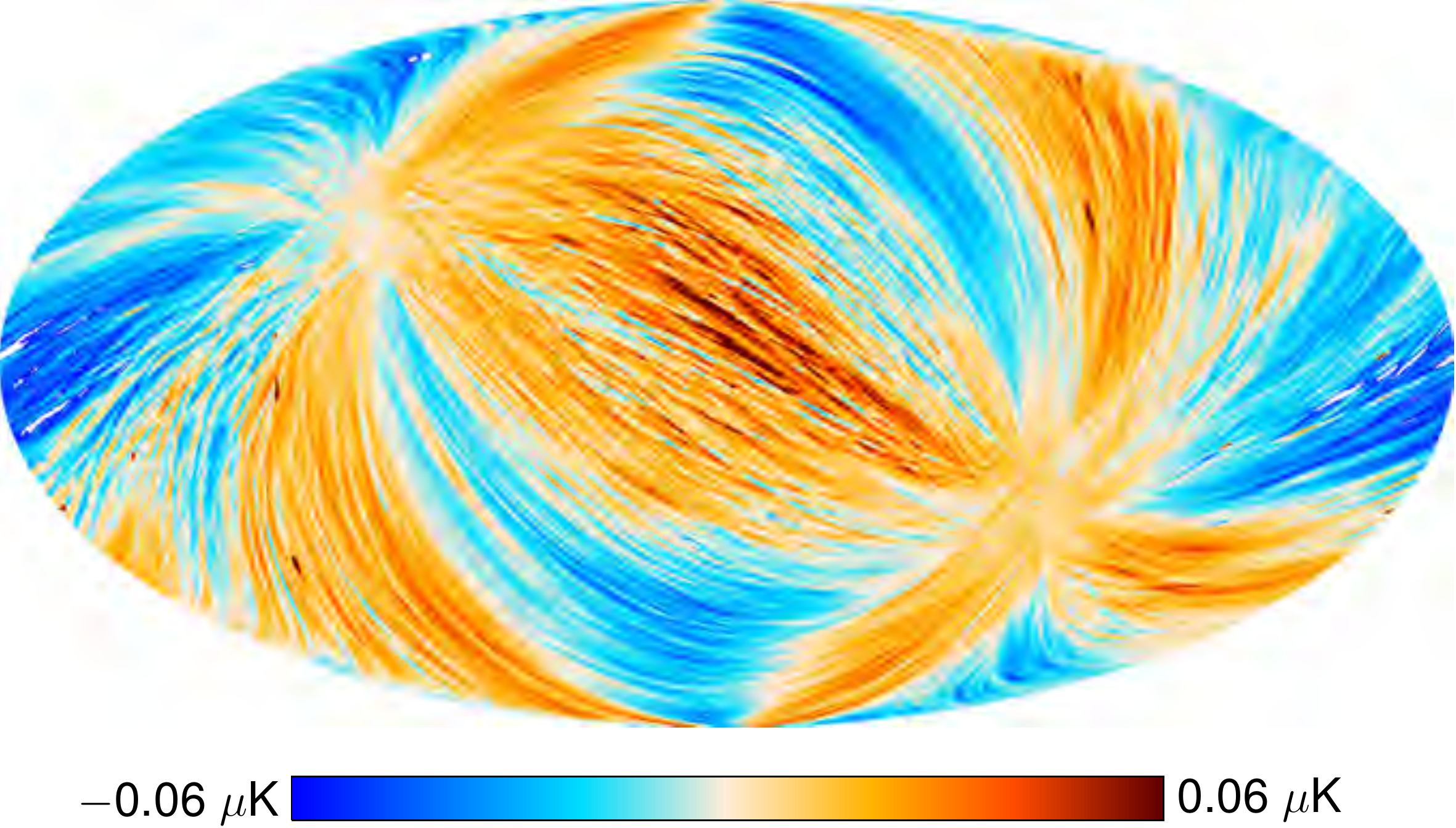}\\
      &&&\\
      44&\includegraphics[width=56mm]{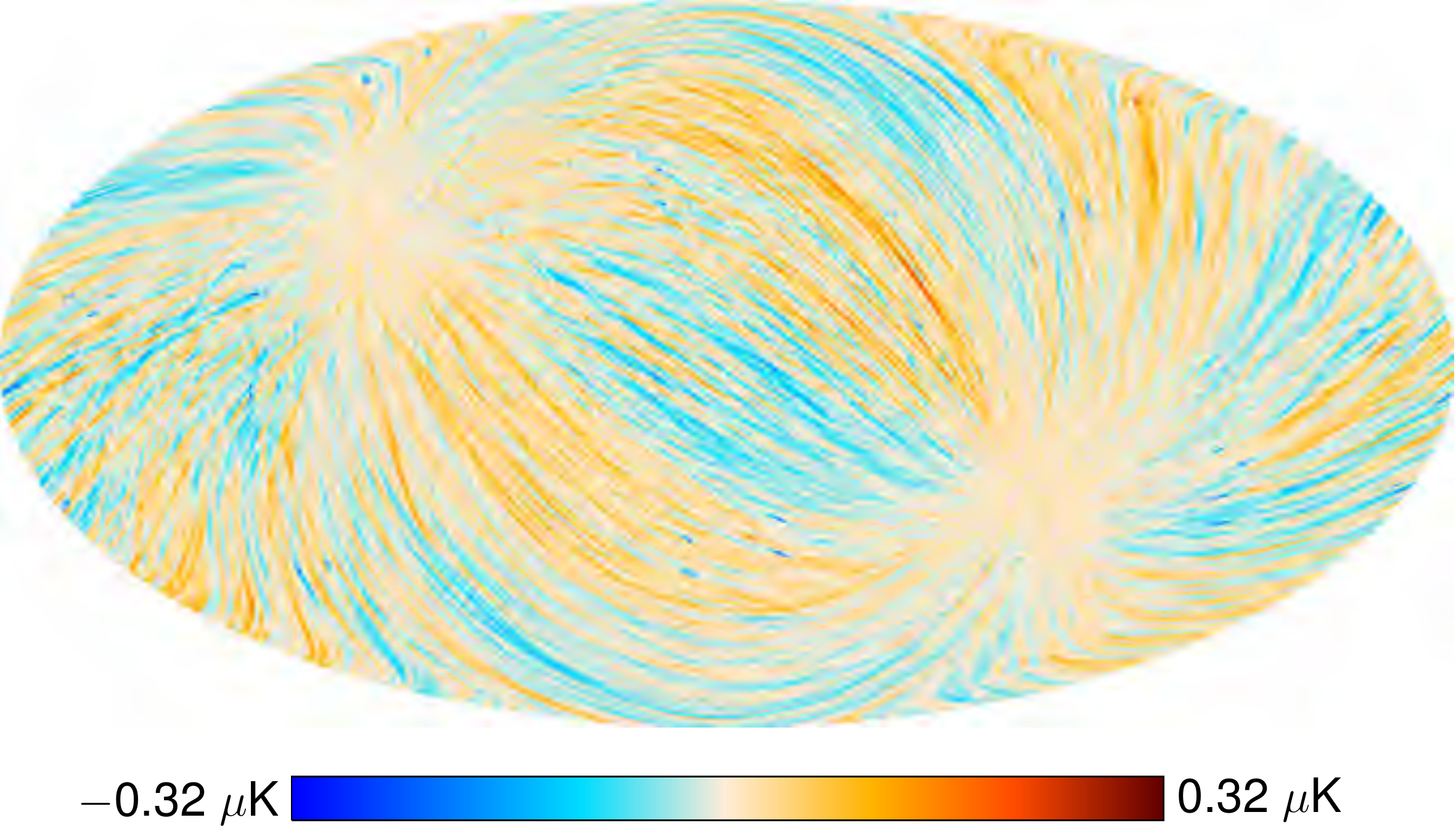}&
      \includegraphics[width=56mm]{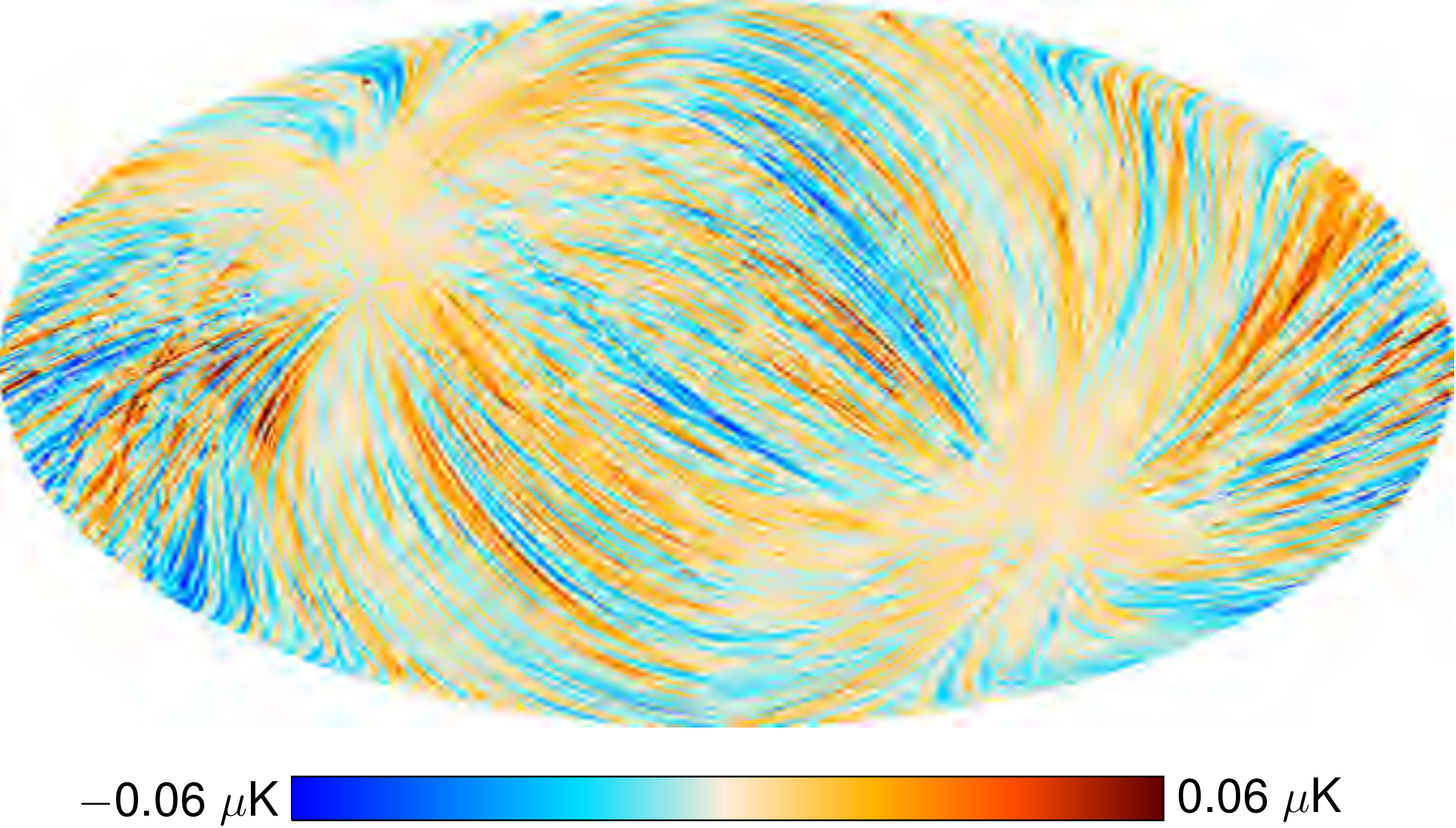}&
      \includegraphics[width=56mm]{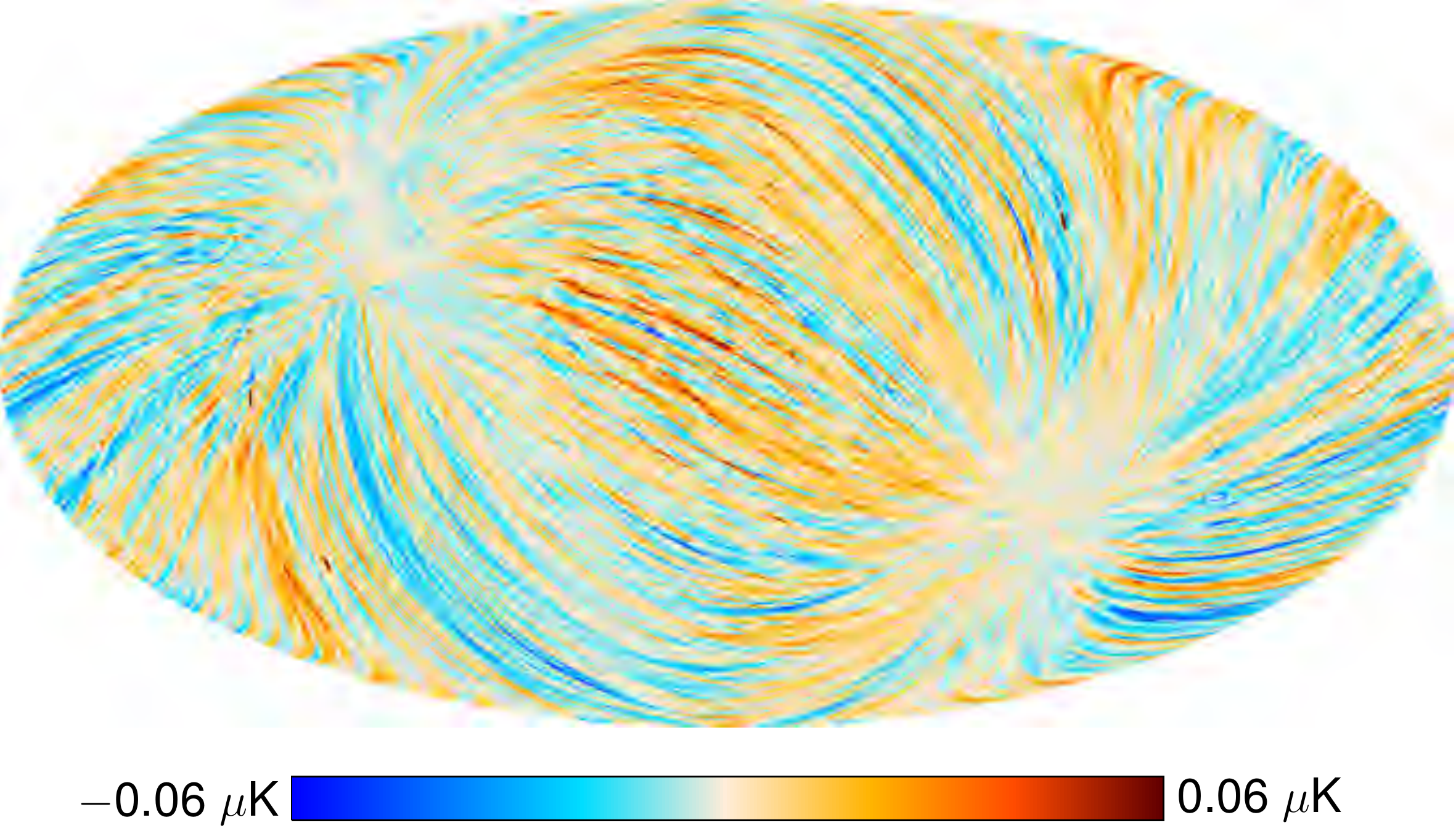}\\
      &&&\\
      70&\includegraphics[width=56mm]{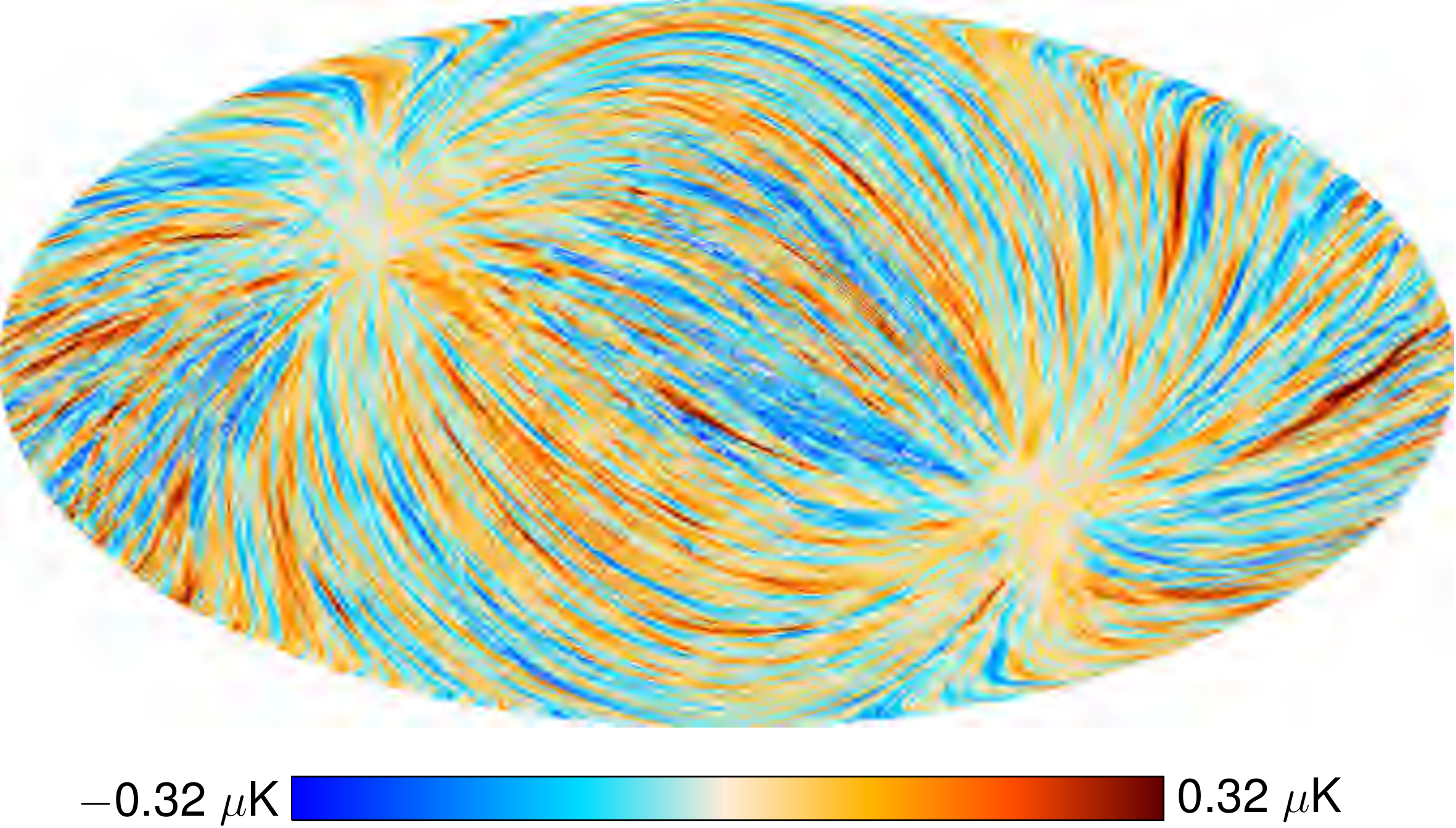}&
      \includegraphics[width=56mm]{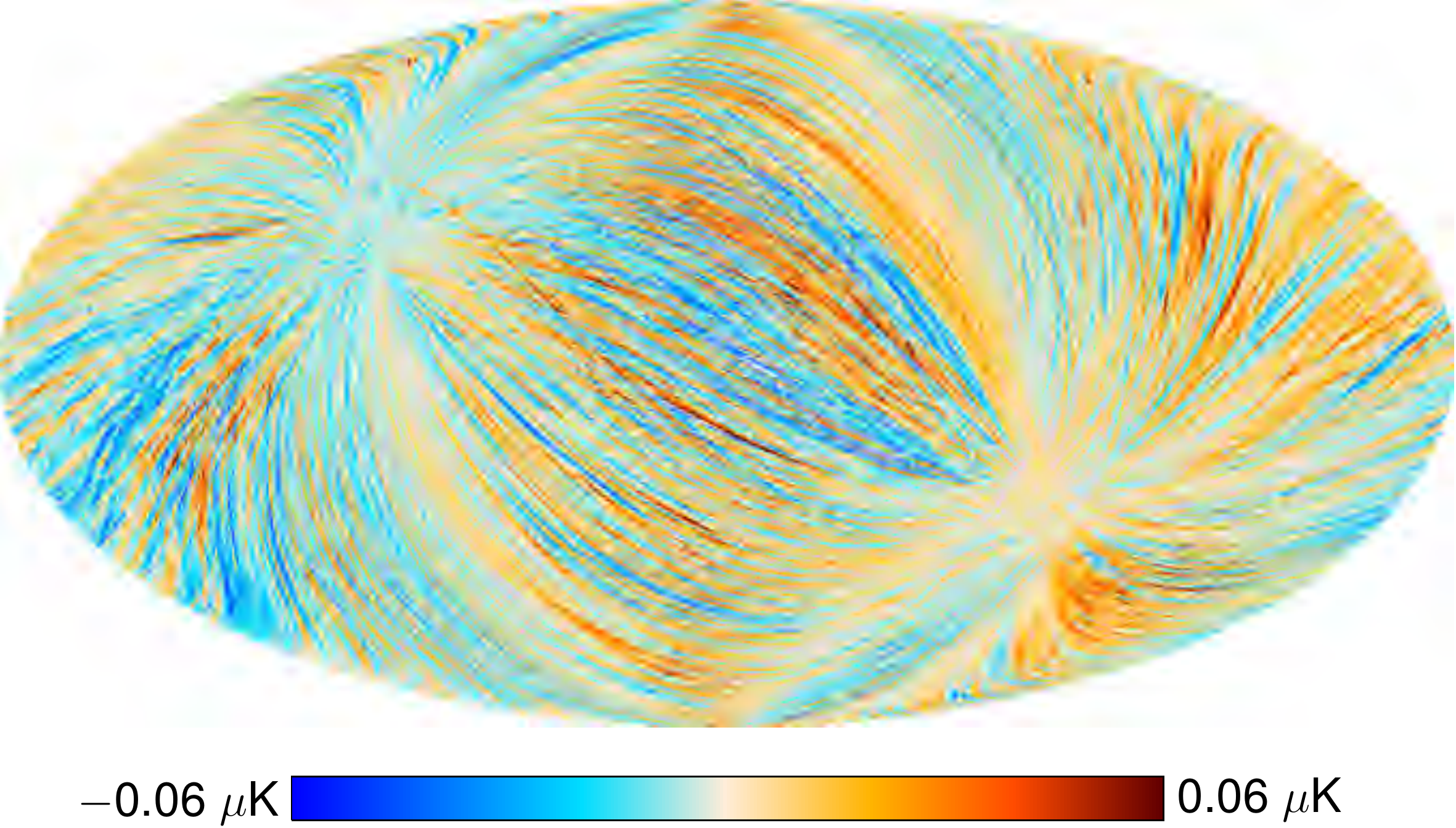}&
      \includegraphics[width=56mm]{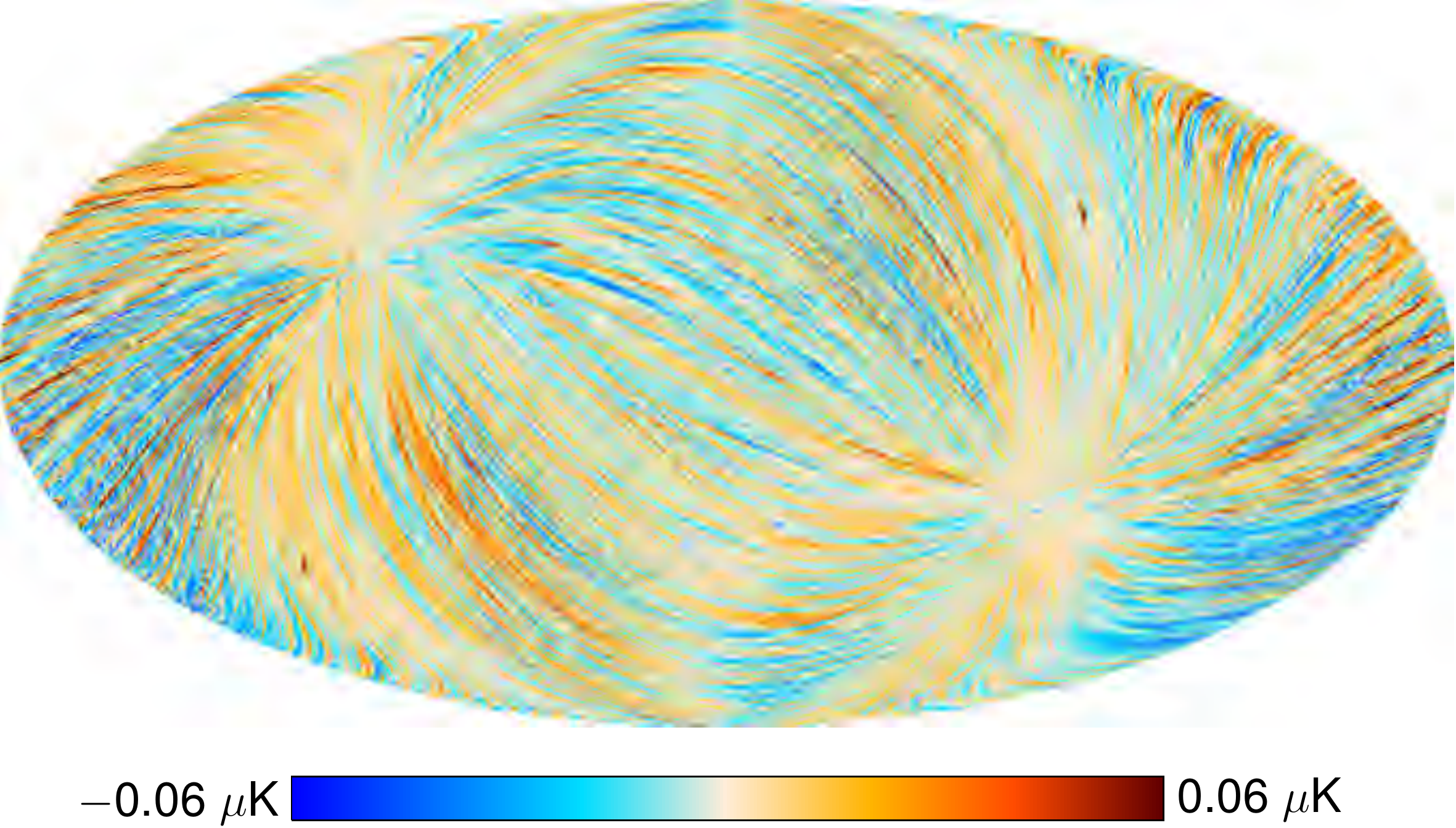}
      \end{tabular}
      
  \end{center}
  \caption{
    Maps of thermal systematic effects. Rows correspond to 30, 44, and 70\,GHz channels, while columns correspond to $I$, $Q$, and $U$. Maps are smoothed to the beam optical resolution of each channel ($\theta_\mathrm{FWHM}=33\arcm$, $28\arcm$, and $13\arcm$, respectively).
   }
  \label{fig_whole_mission_thermal_effects_T_maps}
  \end{figure*}

  In Fig.~\ref{fig_whole_mission_thermal_effects_T_maps} we show maps in total intensity and polarization of the combined thermal effects. In each of these maps the impact of thermal fluctuations is less than 1\,$\mu$K, which is negligible. 
	
      \subsubsection{Bias fluctuations}
\label{sec_assessment_simulations_bias}

  We assess the effect of bias fluctuations in the LFI front-end modules on temperature and polarization measurements, disentangling the various sources of electrical instabilities. They are:
  
  \begin{itemize}
    \item thermal fluctuations in the analog electronics;
    \item thermal fluctuations in the instrument front-end;
    \item electrical instabilities in analogue electronics; and
    \item electrical instabilities affecting the cold amplifiers that may be
    generated either inside or outside the device (e.g., electric line
    instabilities or cosmic rays).
  \end{itemize}

  We follow a procedure in which we correlate the radiometric signal with measured drain currents. For each radiometer we first remove fluctuations caused by thermal instabilities in the cold and warm units \citep[equation~1 in][]{planck2013-p02a}, and then we correlate the residual drain current fluctuations with the voltage output of both radiometer diodes \citep[equation~2 in][]{planck2013-p02a}.
  
  We obtain four coefficients: two of them ($\alpha_{20\,\mathrm{K}}$ and $\alpha_{300\,\mathrm{K}}$) are the correlation coefficients between the measured currents and temperatures, the other two ($\alpha_{\mathrm{sky}}$ and $\alpha_{\mathrm{ref}}$) link the total power sky and reference-load voltage outputs with drain current measurements corrected for thermal effects. 
  
  In this release we updated $\alpha_{\mathrm{sky}}$ and $\alpha_{\mathrm{ref}}$, exploiting the possibility of measuring drain currents every 6\,s instead of every 64\,s (the default time interval for this housekeeping parameter). To enable this faster acquisition we developed a dedicated telemetry procedure that was not available for the 2013 release.
  
  The thermal coefficients, instead, are the same as those used in the 2013 release. We measured them by exploiting controlled temperature variations during the in-flight calibration phase \citep{gregorio2013} and temperature fluctuations induced by the transponder being switched on and off every day during the first survey \citep[see figure~26 in][]{planck2011-1.3}. Since we did not perform other tests involving temperature changes during the mission we did not update $\alpha_{20\,\mathrm{K}}$ and $\alpha_{300\,\mathrm{K}}$ between the 2013 and 2015 data release.
  
  With these coefficients we use the drain current measurements to generate time-ordered-data that we project onto the sky. 
  
\begin{figure*}[!htpb]
  \begin{center}
    \begin{tabular}{m{.25cm} m{5.6cm} m{5.6cm} m{5.6cm}}
      & \begin{center}$I$\end{center} &\begin{center}$Q$\end{center}&\begin{center}$U$\end{center}\\    
      30&\includegraphics[width=56mm]{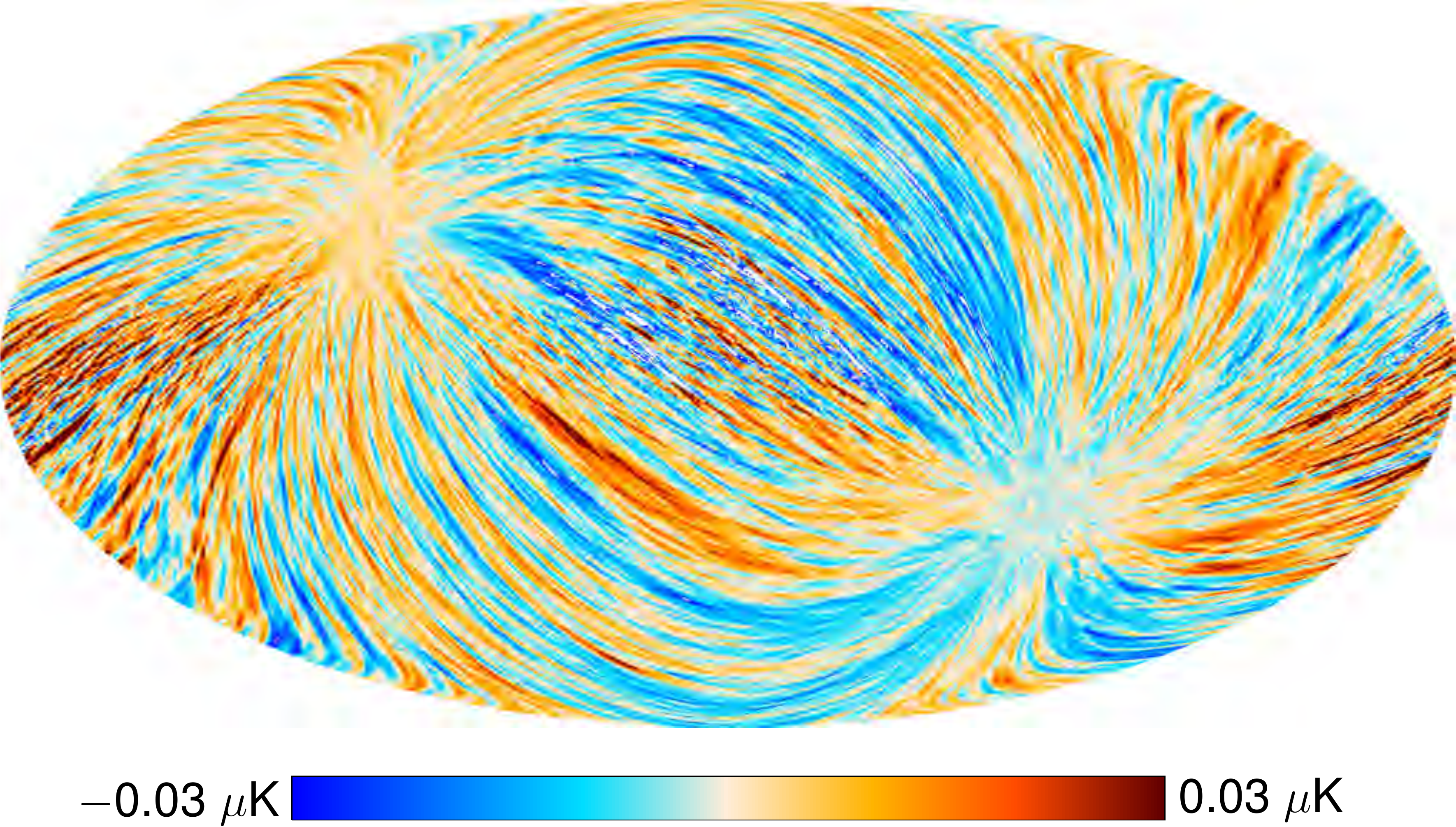}&
      \includegraphics[width=56mm]{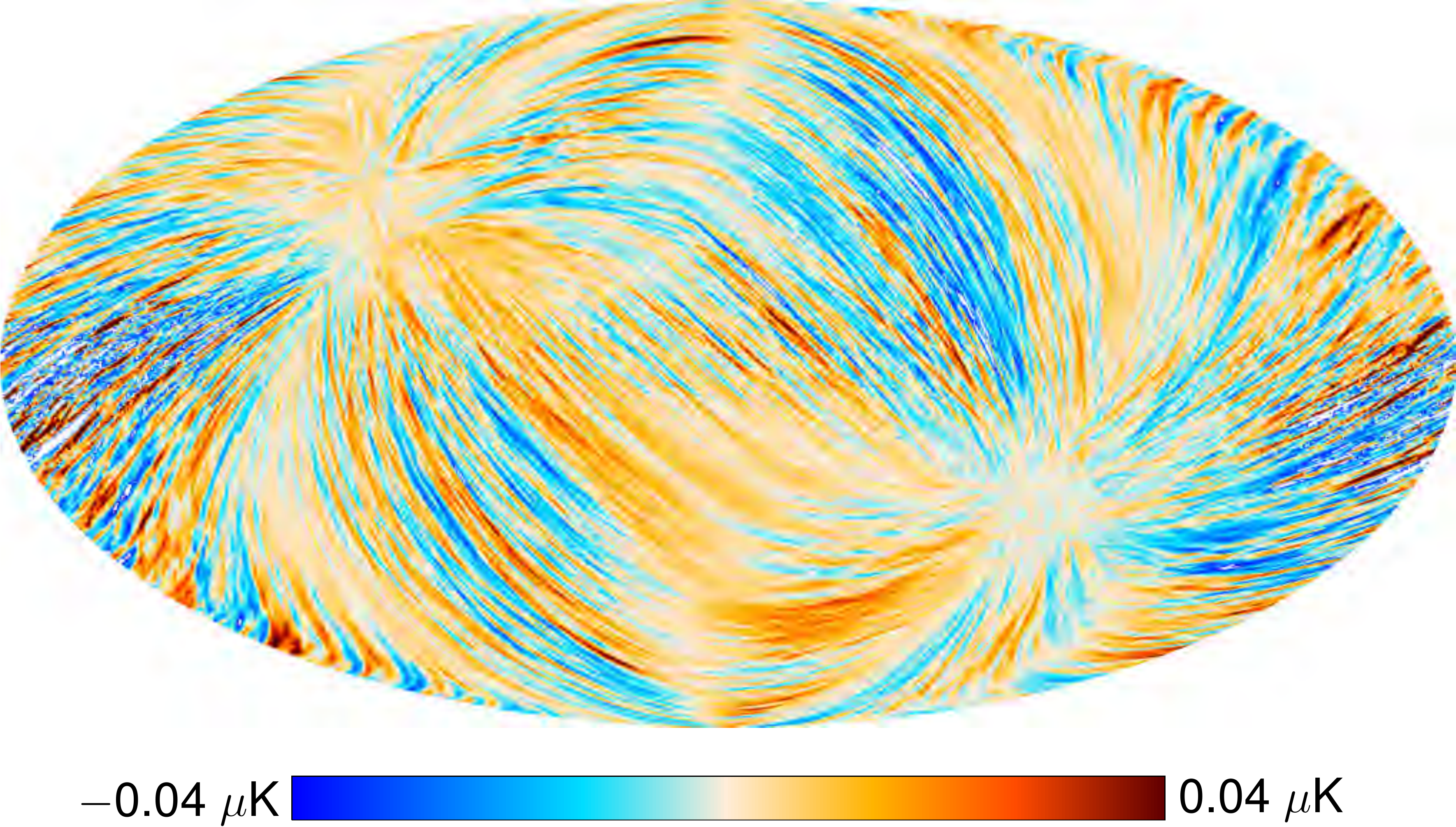}&
      \includegraphics[width=56mm]{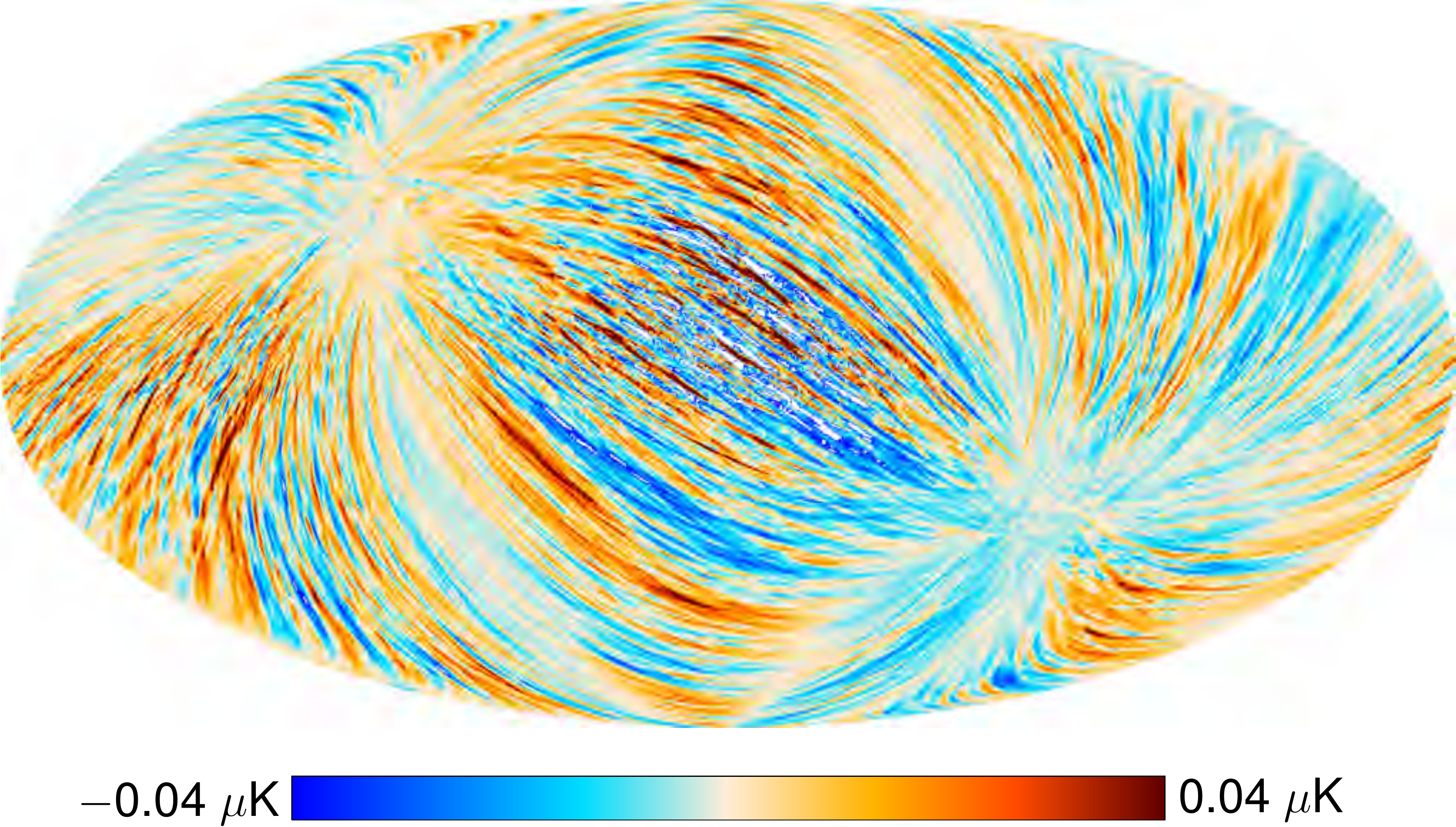}\\
      &&&\\
      44&\includegraphics[width=56mm]{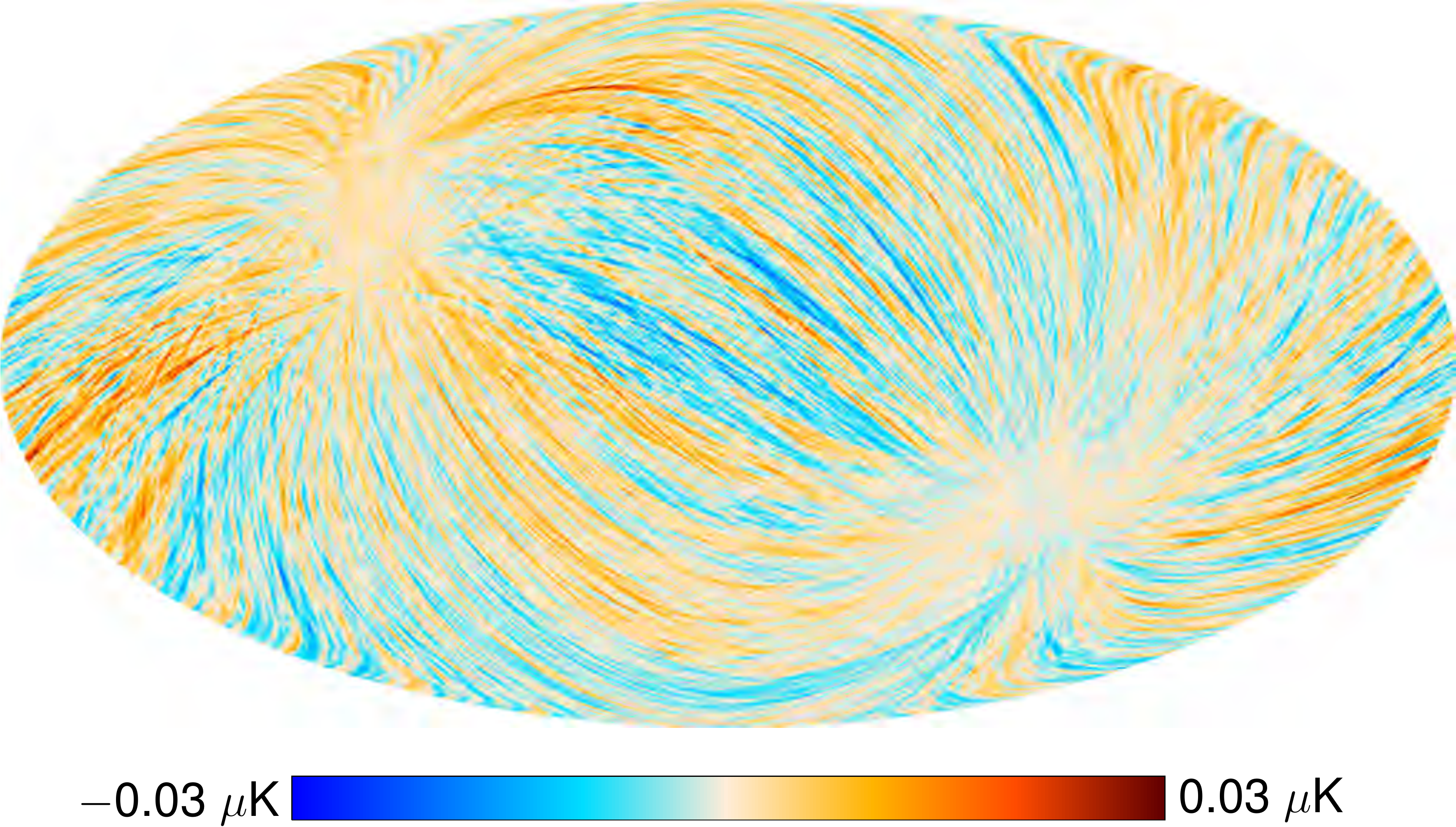}&
      \includegraphics[width=56mm]{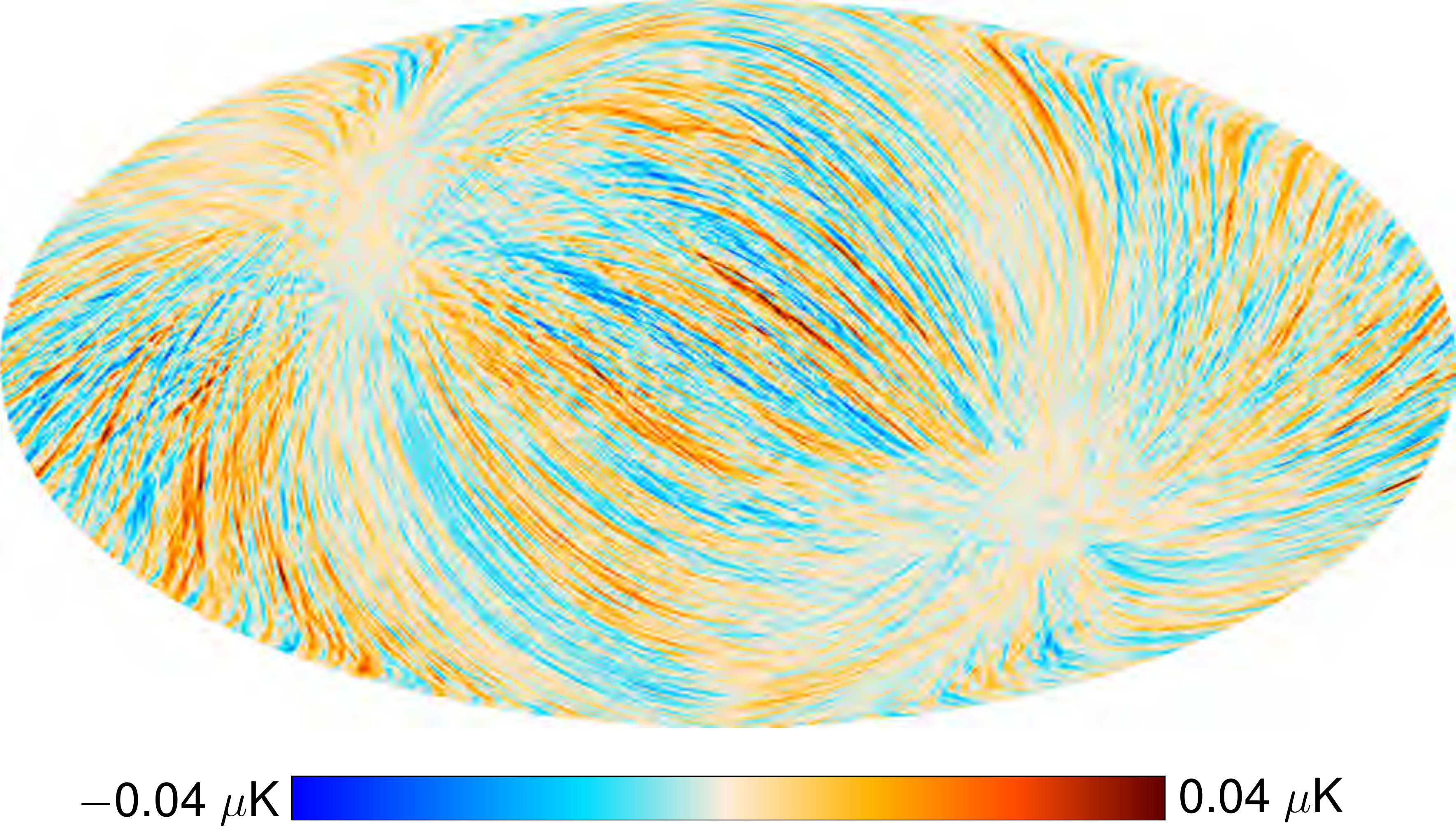}&
      \includegraphics[width=56mm]{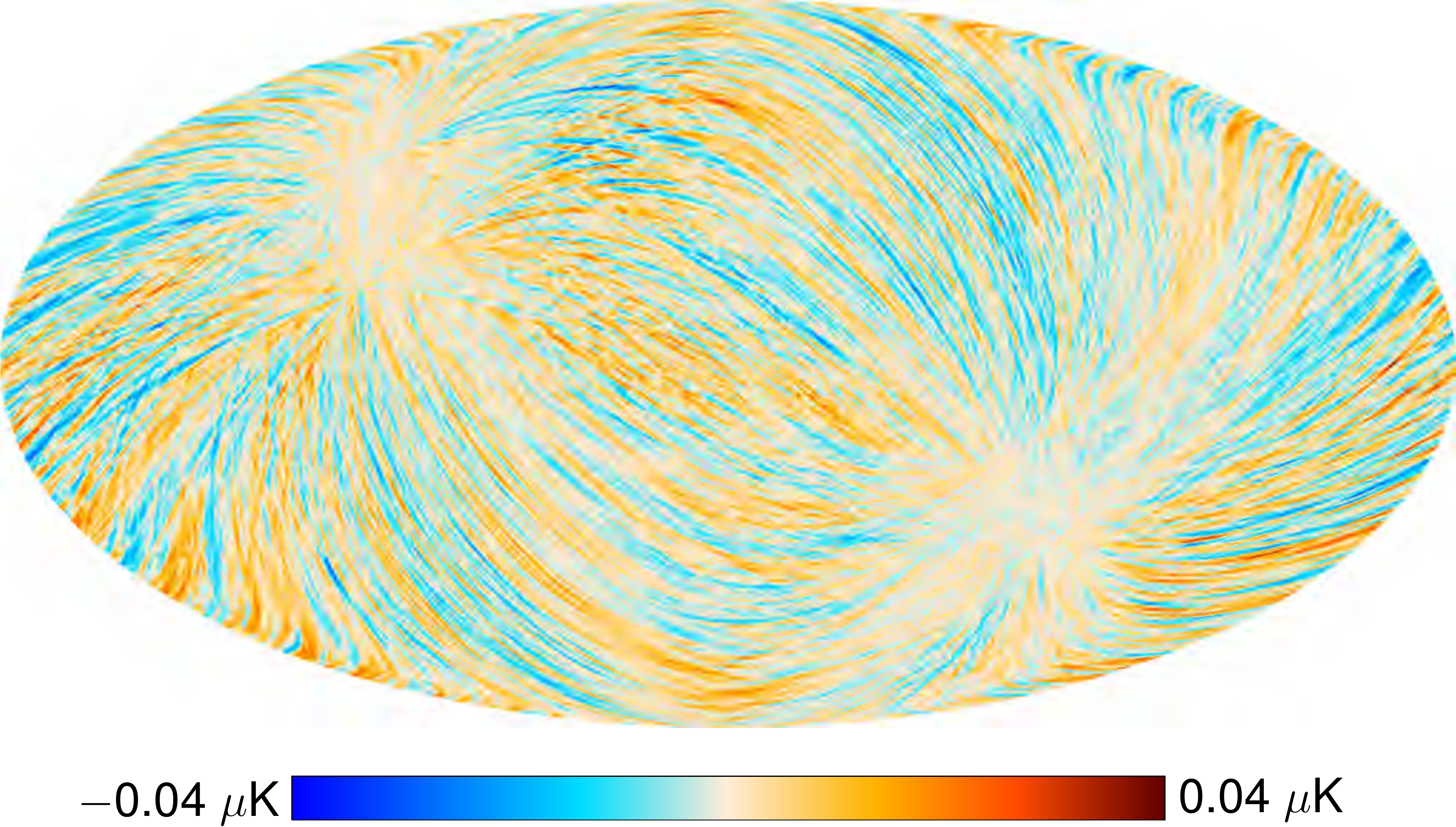}\\
      &&&\\
      70&\includegraphics[width=56mm]{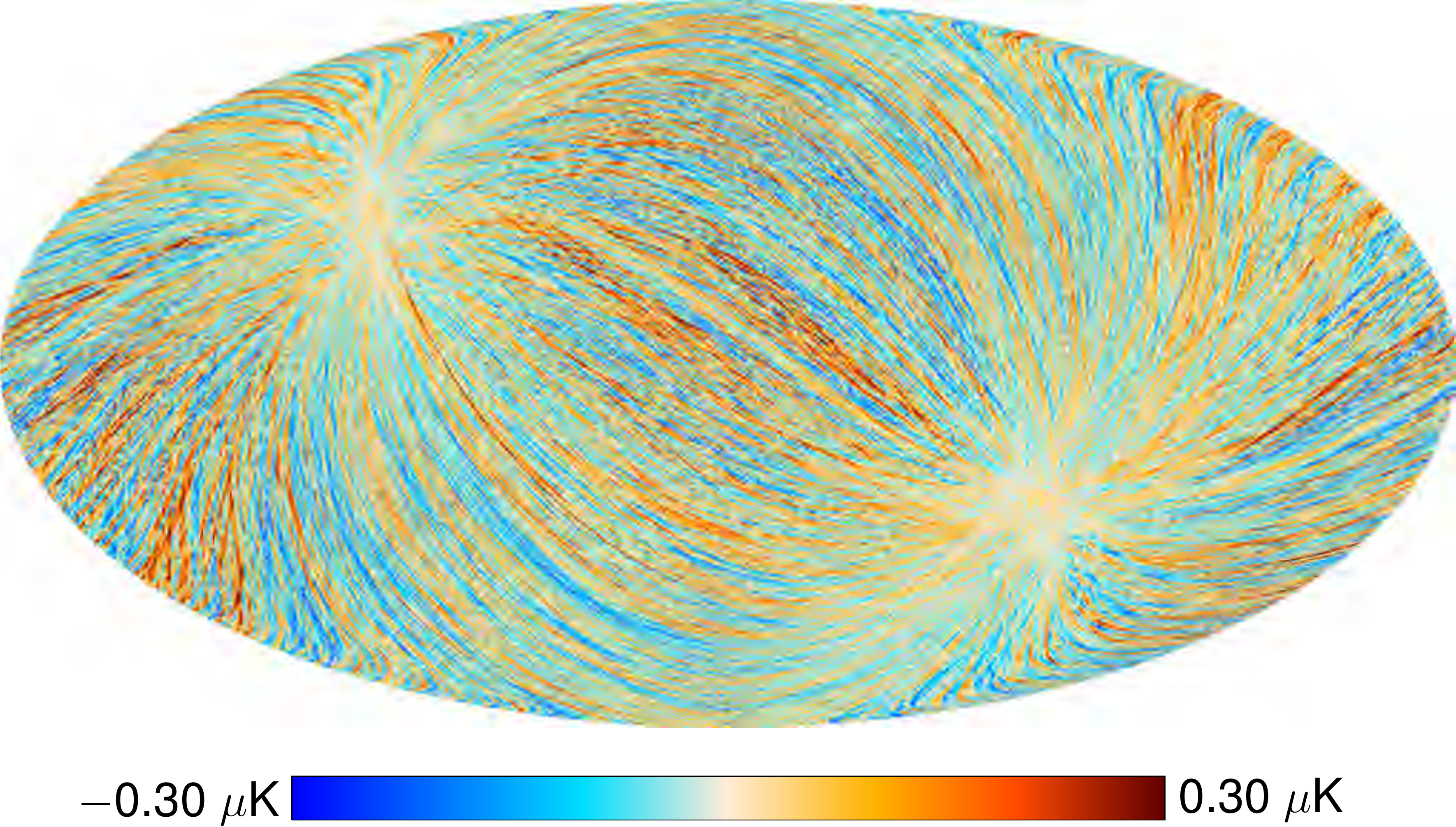}&
      \includegraphics[width=56mm]{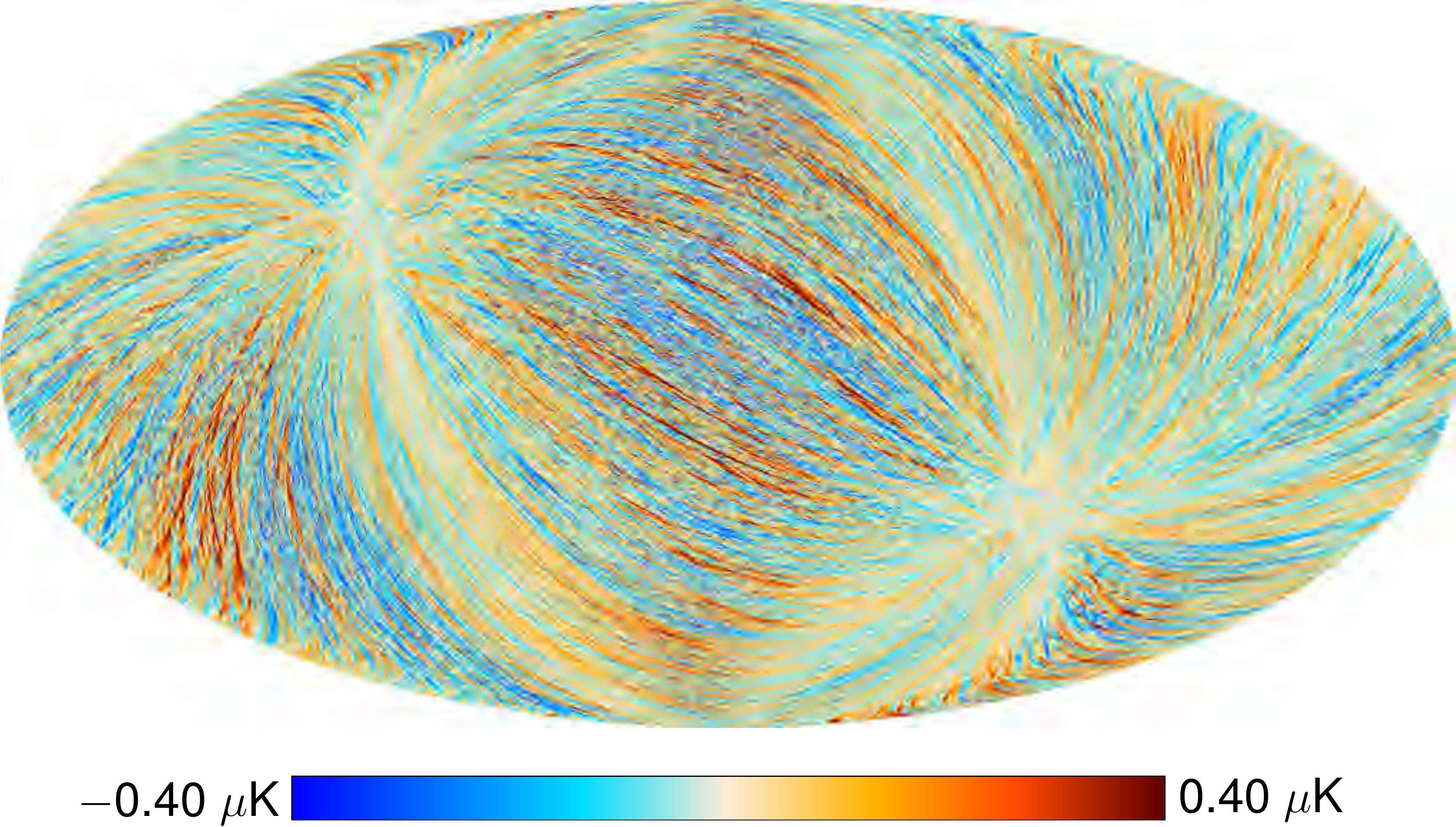}&
      \includegraphics[width=56mm]{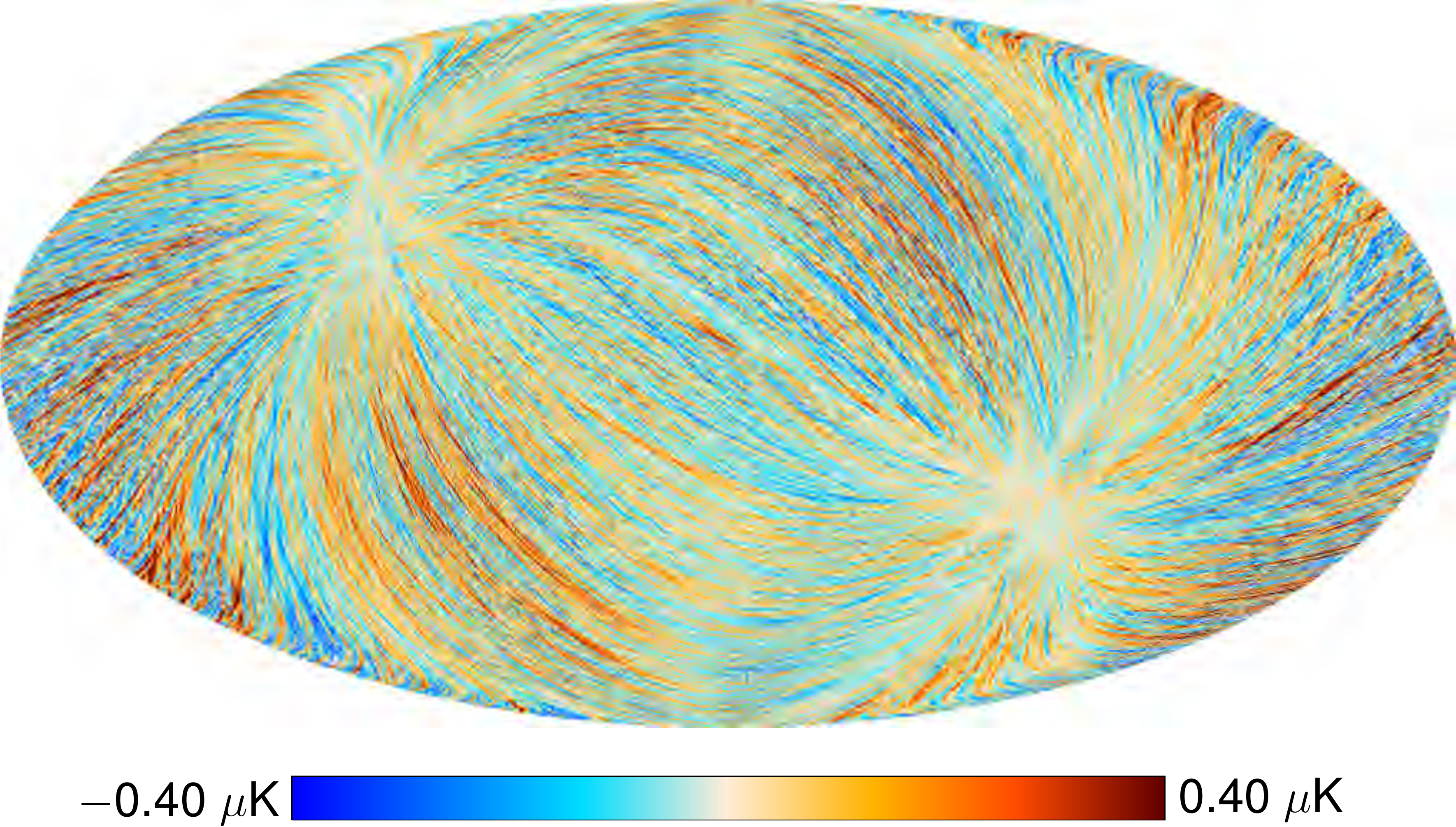}
      \end{tabular}
  \end{center}

  \caption{Maps of the effect from front-end bias fluctuations. Rows correspond to 30, 44, and 70\,GHz channels, while columns correspond to $I$, $Q$, and $U$. Maps are smoothed to the beam optical resolution of each channel ($\theta_\mathrm{FWHM}=33\arcm$, $28\arcm$, and $13\arcm$, respectively).}
  \label{fig_whole_mission_bias_maps}
  \end{figure*}

  
  Fig.~\ref{fig_whole_mission_bias_maps} shows maps in temperature and polarization of the systematic effect from bias fluctuations. This effect is less than 1\,$\mu$K both in temperature and polarization, which is negligible compared to the CMB at 44 and 70\,GHz and foregrounds at 30\,GHz. These results are also consistent with the 2013 analysis in temperature \citep{planck2013-p02a}.

      \subsubsection{1-Hz spikes}
\label{sec_assessment_simulations_spikes}

  This effect is caused by a well-known cross-talk between the housekeeping 1-Hz acquisition clock and the scientific data acquisition. We have described this effect in several previous papers. In section~5.2.5 of \citet{mennella2010} we discuss how we characterized it during ground tests, in section~4.1.2 of \citet{gregorio2013} we presented a similar characterization performed during flight calibration, in section~3.1 of \citet{planck2011-1.6} we explained how we build templates of this effect and remove them from the data, and in section~4.2.3 of \citet{planck2013-p02a} we show how we assess the impact of residual uncertainties on the LFI temperature results.

  Here we extend our analysis to temperature and polarization data using the full mission data set, using the same procedure explained in \citet{planck2013-p02a}.
  
\begin{figure*}[!htpb]
  \begin{center}
    \begin{tabular}{m{.25cm} m{5.6cm} m{5.6cm} m{5.6cm}}
      & \begin{center}$I$\end{center} &\begin{center}$Q$\end{center}&\begin{center}$U$\end{center}\\    
      30&\includegraphics[width=56mm]{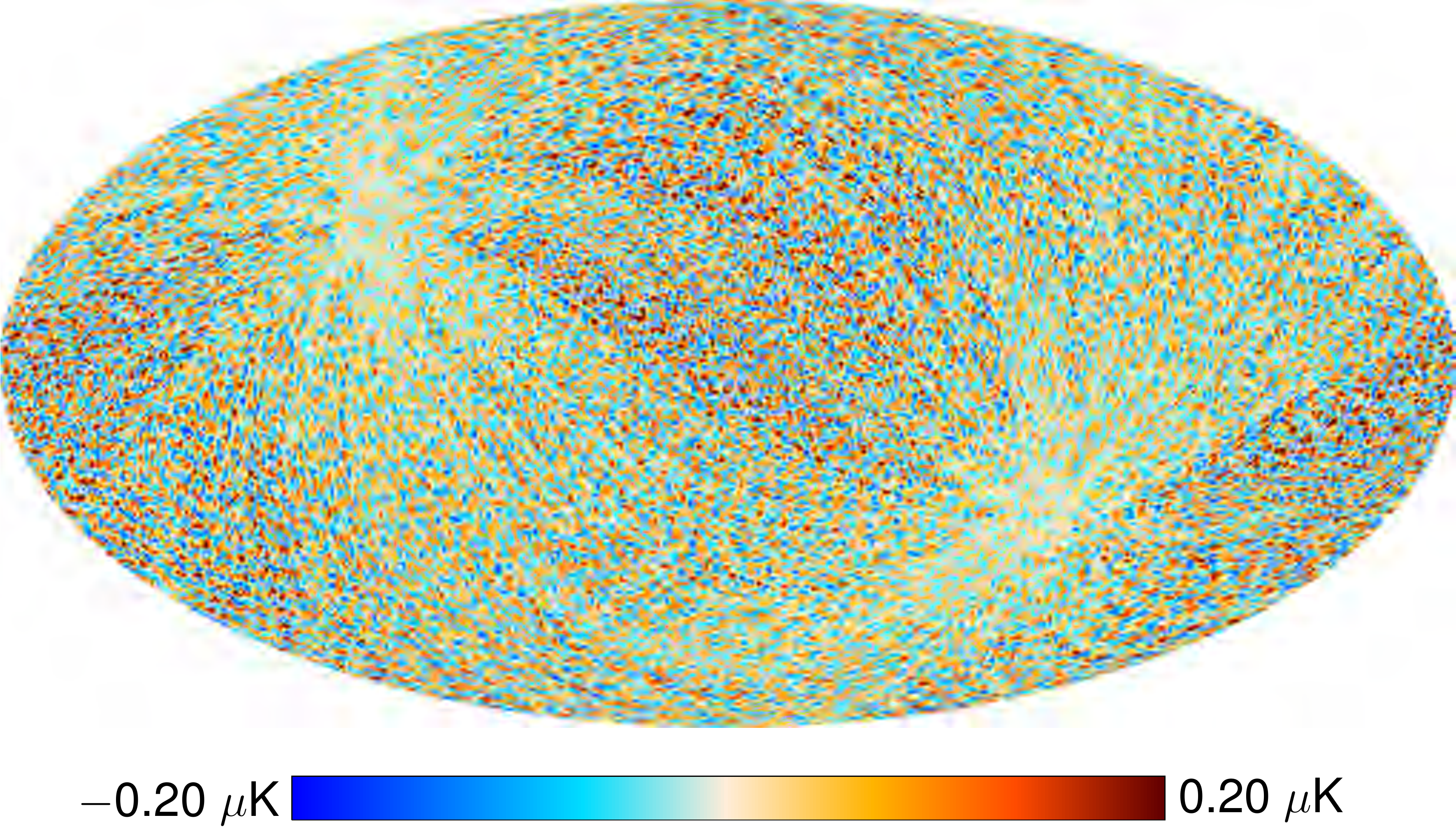}&
      \includegraphics[width=56mm]{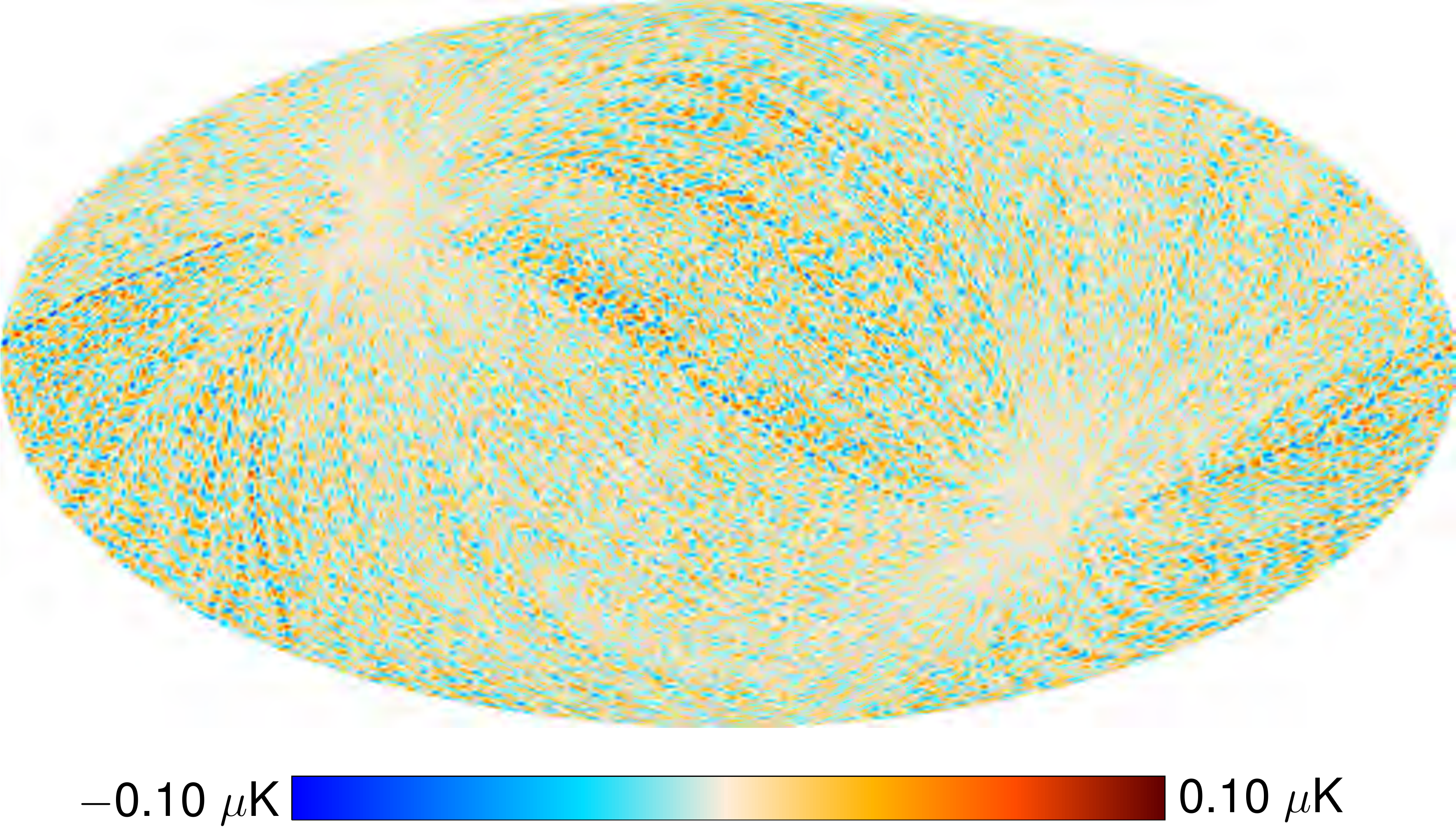}&
      \includegraphics[width=56mm]{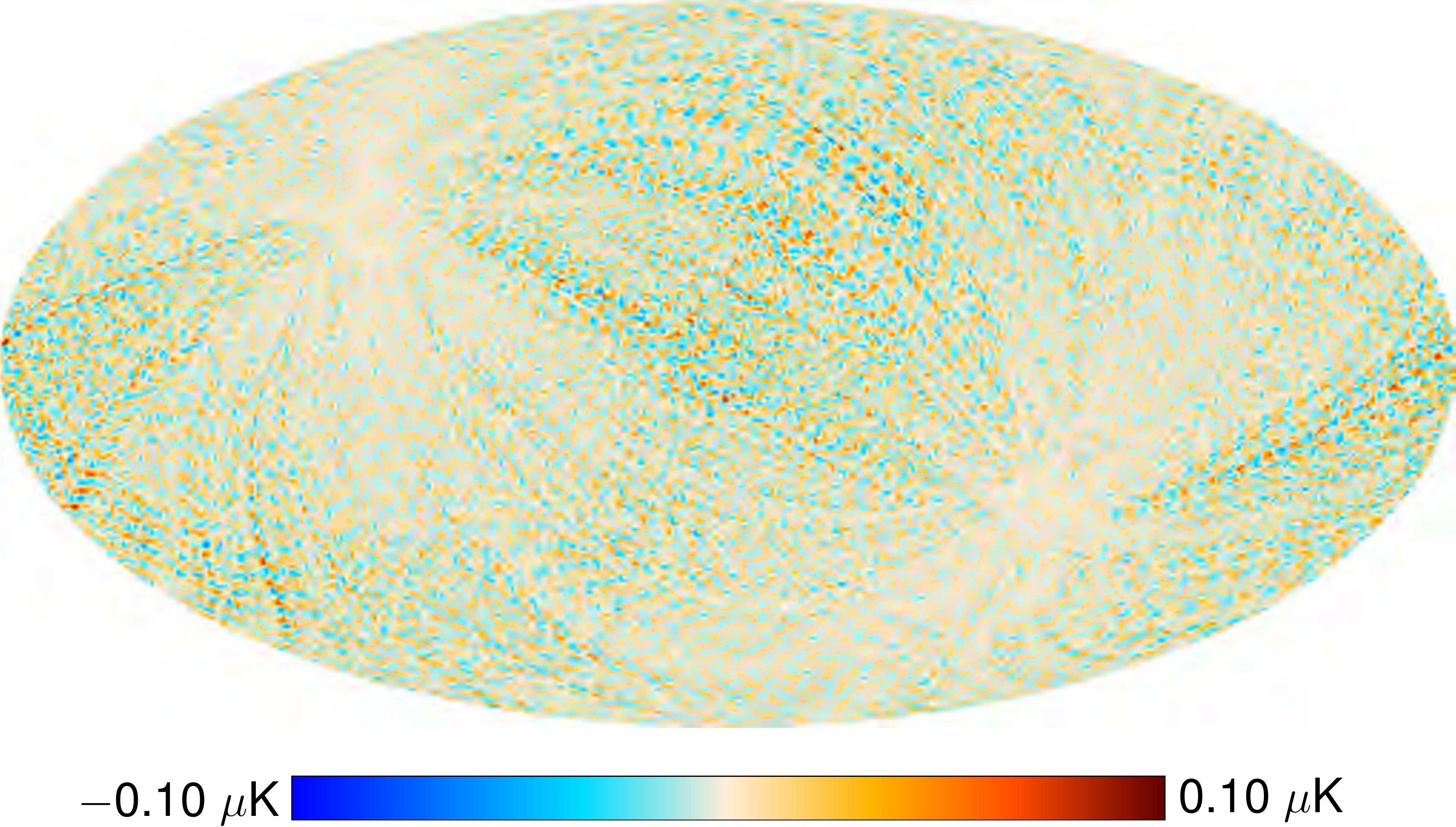}\\
      &&&\\
      44&\includegraphics[width=56mm]{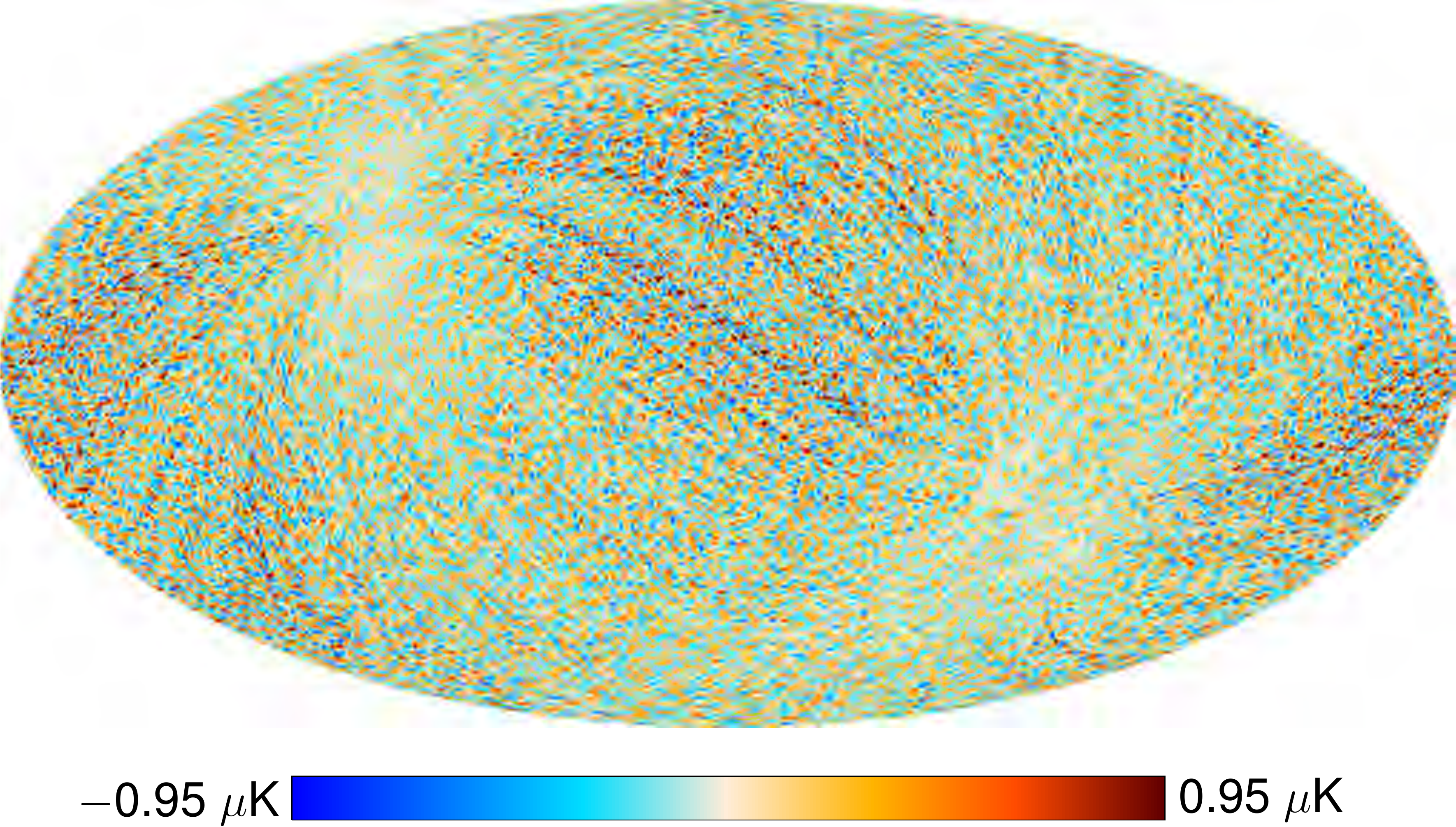}&
      \includegraphics[width=56mm]{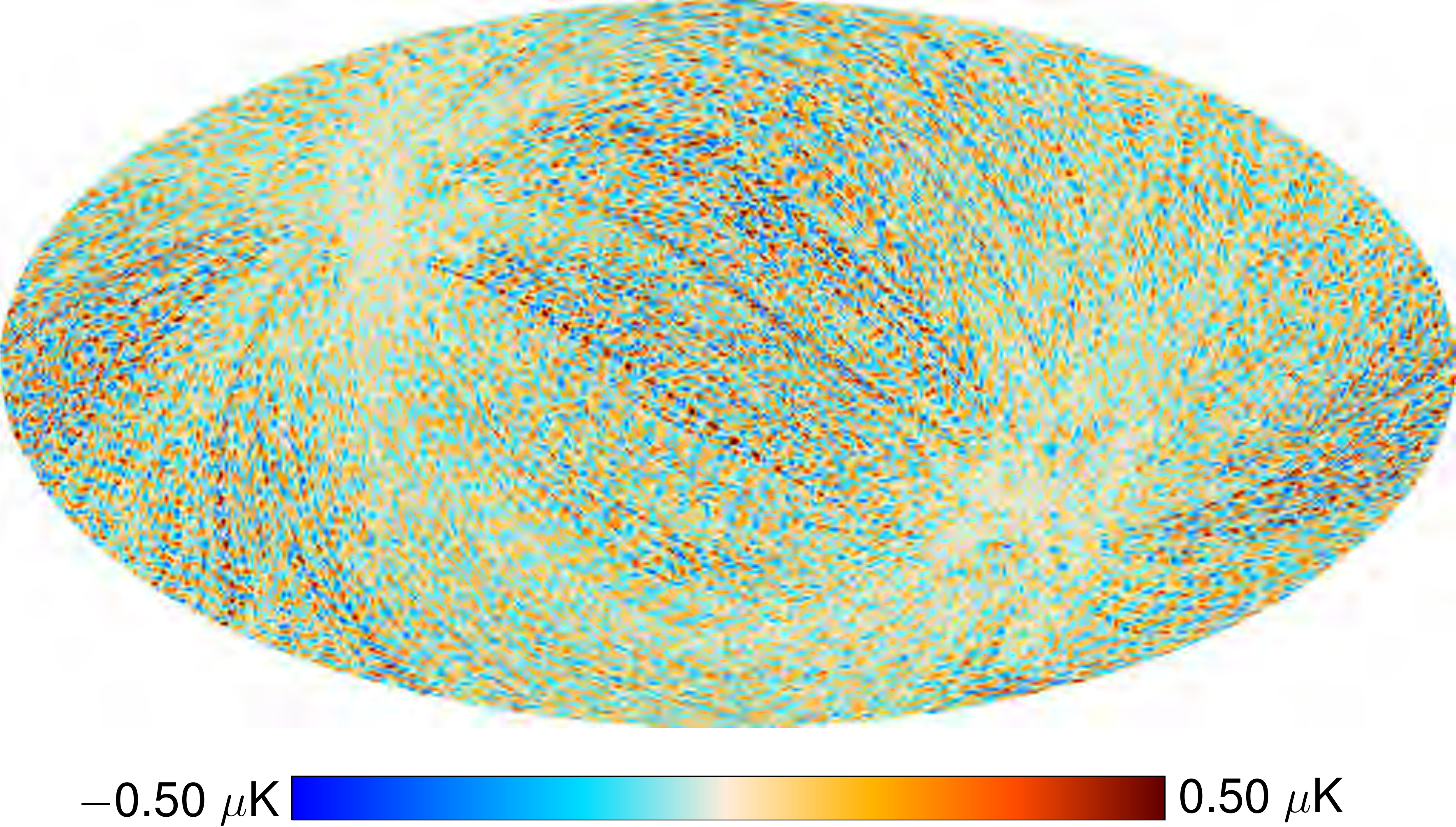}&
      \includegraphics[width=56mm]{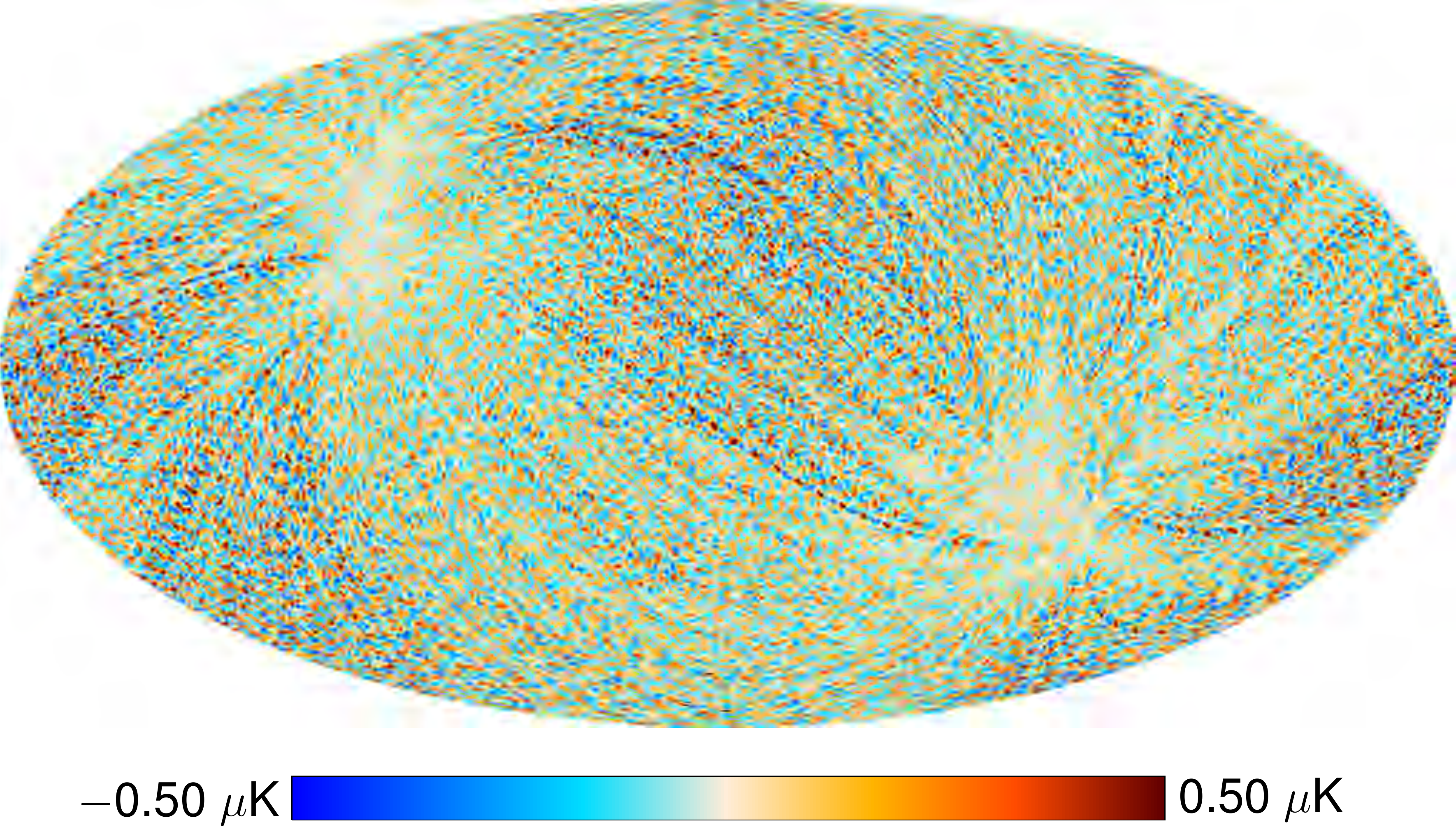}\\
      &&&\\
      70&\includegraphics[width=56mm]{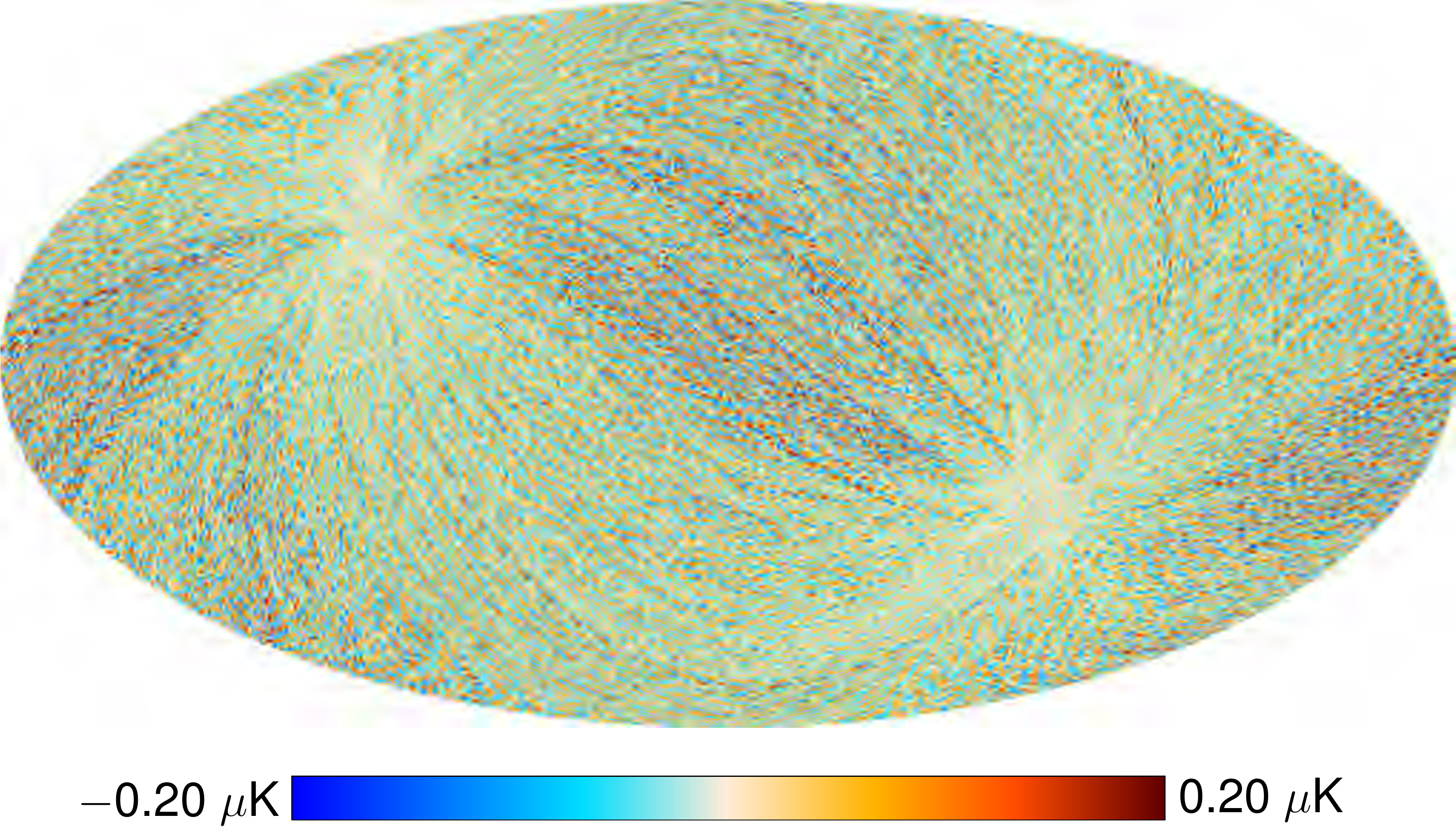}&
      \includegraphics[width=56mm]{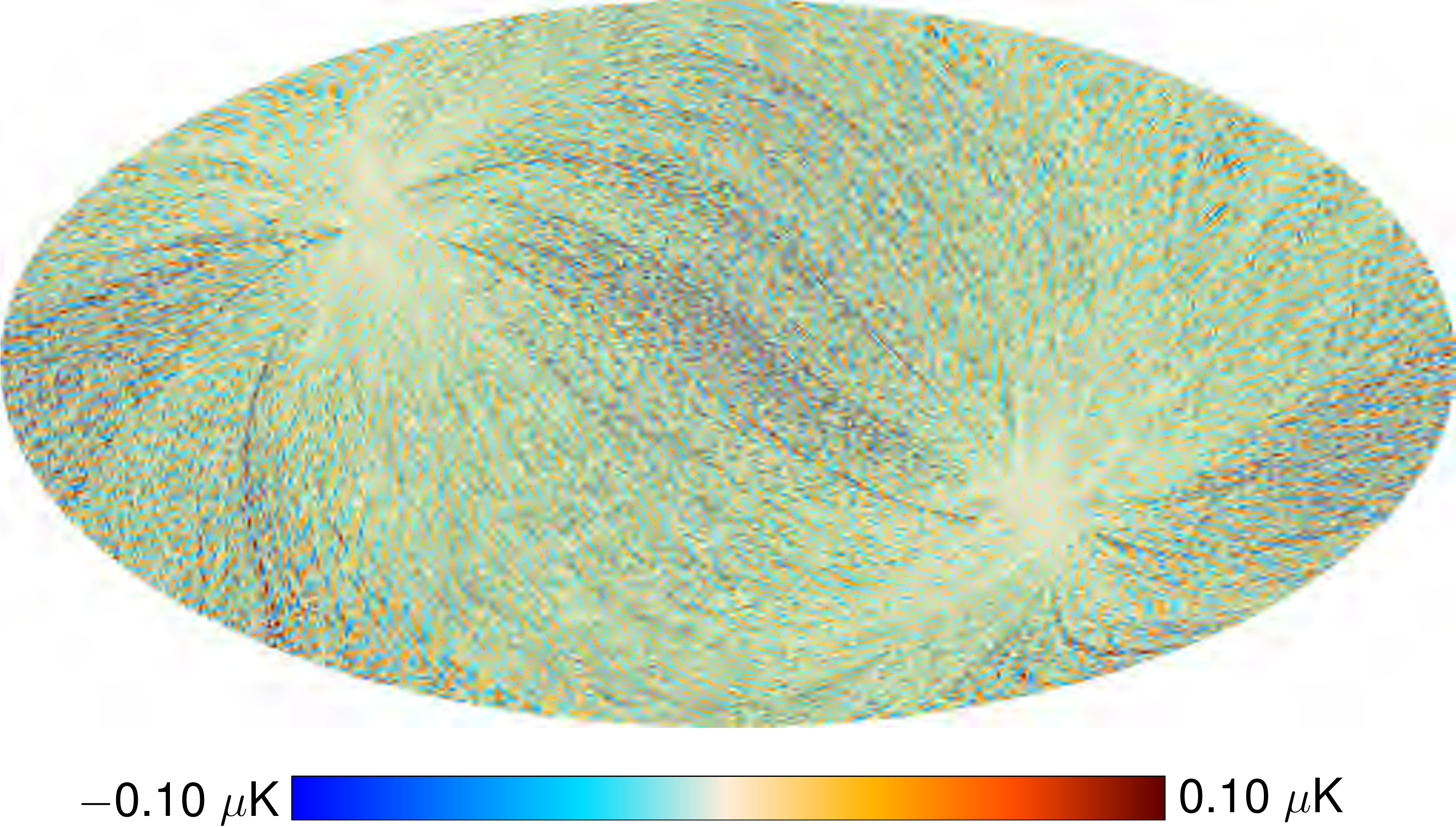}&
      \includegraphics[width=56mm]{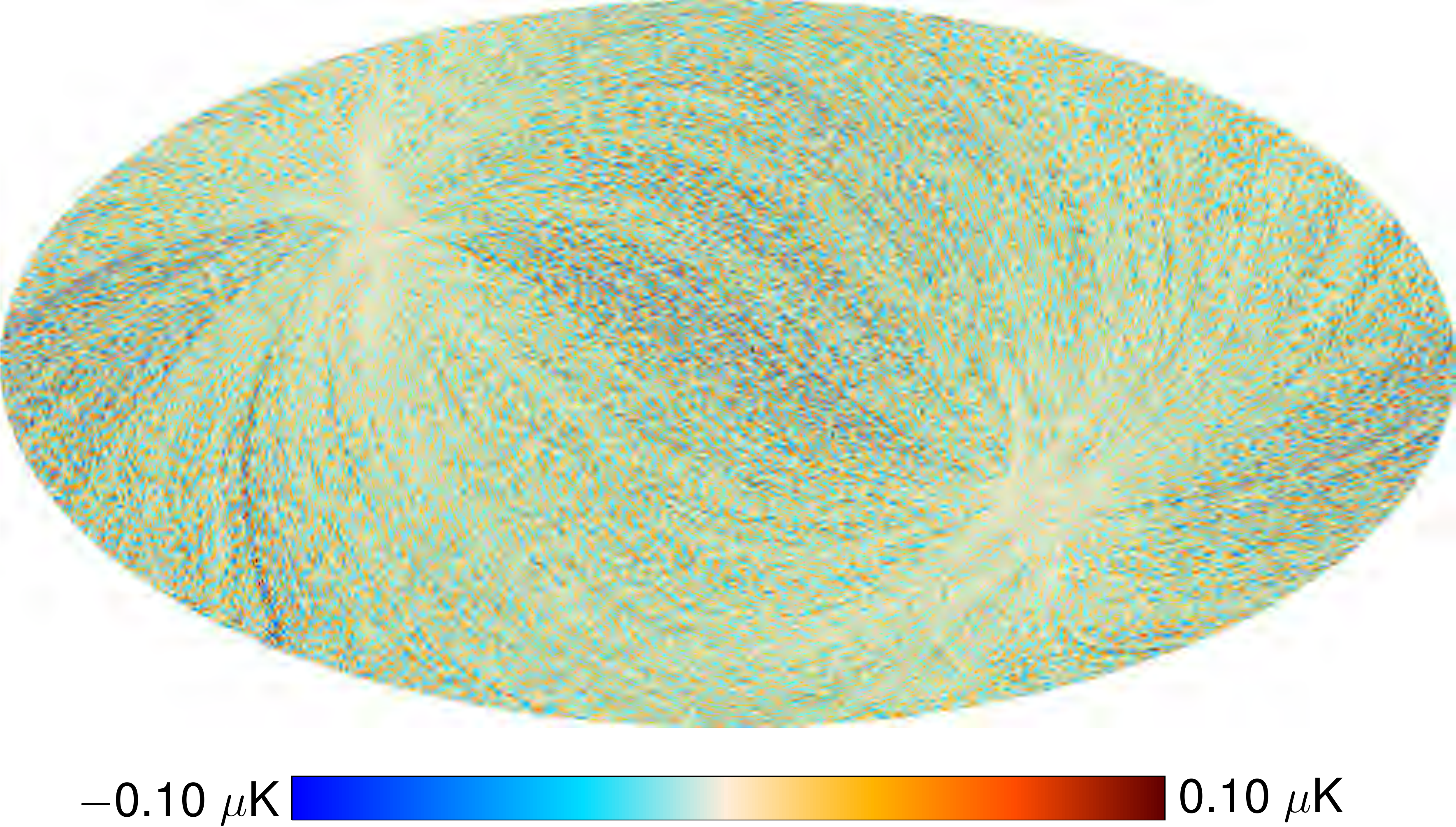}
      \end{tabular}
  \end{center}

  \caption{Maps of the effect from 1-Hz spikes. Rows correspond to 30, 44, and 70\,GHz channels, while columns correspond to  $I$, $Q$, and $U$. Maps are smoothed to the beam optical resolution of each channel ($\theta_\mathrm{FWHM}=33\arcm$, $28\arcm$, and $13\arcm$, respectively).}
  \label{fig_whole_mission_spikes_maps}
  \end{figure*}

  The maps in Fig.~\ref{fig_whole_mission_spikes_maps} show that this effect is less than 1\,$\mu$K at 30 and 70\,GHz and about 1--2\,$\mu$K at 44\,GHz. This channel, indeed, is the most affected by 1-Hz spikes and it is the only one that we correct by removing the signal template from the time-ordered data. The 30 and 70\,GHz data, instead, are only slightly affected by this spurious signal and we do not apply any correction. 
  
%

  
    \subsection{Assessing systematic effects via null tests (``top-down'')}
\label{sec_assessment_null_tests}

  We define ``null test''  as an  analysis based on differences between subsets of data (maps, timelines, power spectra, etc.) that, in principle, contain the same sky signal. The \Planck\ scanning strategy \citep[see section~4.1 of][]{planck2014-a01} together with the symmetric configuration of the focal plane (see figure~11 of \citealp{bersanelli2010} and figure~4 of \citealp{lamarre2010}) offer several useful null-test combinations, each with a different sensitivity to various kinds of systematic effects both in temperature and in polarization.

  We have routinely carried out null tests during the LFI data analysis. These tests played a key role in our understanding of the LFI systematic error budget, and led to improvements in the self-consistency of our data. These improvements have allowed us to use the LFI polarization results at the largest angular scales. 
  
  In this section we discuss the results of our null test analysis. First we discuss the general strategy and then we present the results obtained showing the robustness of the LFI data against several classes of systematic effects.

\subsubsection{Null tests strategy}

  We complement every \Planck-LFI data release with a suite of null tests combining data selected at various timescales. 
  
  The shortest timescale is that of a single pointing period ($\sim$40~minutes) that is split into two parts. We then difference the corresponding maps and obtain the so-called \textit{half-ring difference maps} that approximate the instrument noise and may contain systematic effects correlated on timescales $\lesssim 20$~minutes. 
  
  Then we have longer timescales: six months (a sky survey), one year, the full mission (four years). We can create a large number of tests by combining these timescales for single radiometers\footnote{We do not expect that single radiometer survey differences are strictly null. Indeed, the radiometers are polarized detectors that observe the sky with a different range of polarization angles for different surveys.  We use these tests to validate the radiometer stability, minimizing these effects by considering survey combinations with the same scanning patterns (survey 1 vs survey 3, survey 2 vs survey 4) or by combining radiometers to solve for $I$, $Q$ and $U$.}. We provide the detailed timing of each survey in the \Planck\ Explanatory Supplement\footnote{\burl{http://wiki.cosmos.esa.int/planckpla/index.php/Survey_scanning_and_performance}}.

  When we take a difference between two maps we apply a weighting to guarantee that we obtain the same level of white noise independently of the timescale considered. The weighting scheme is described by equations (30), (31), and (32) of \citet{planck2011-1.3}, where we normalize the white noise to the full mission (8 surveys) noise. This means that in equation (32) of \citet{planck2011-1.3} the term $\mathbf{hit}_\mathrm{full}(p)$ corresponds to the number of hits at each pixel, $p$, in the full map.  
  
  We assess the quality of the null tests by comparing null maps pseudo spectra  obtained from flight data with those coming from systematic effect simulations and  with noise-only  Monte Carlo realizations based on  the \Planck\ \textit{full focal plane} (FFP8) simulation \citep{planck2014-a14}.  For the systematic effect simulations we used \textit{global} maps by combining the effects listed in Table~\ref{tab_list_simulated_effects}. Monte Carlo realizations  include pointing, flagging, and a radiometer specific noise model based on the  measured noise power spectrum.  We create 1000 random realizations of such noise maps using the same destriping algorithm used for the real data, and compute null maps and pseudo-spectra in the same way. For each multipole, $\ell$, we calculate the mean $C_\ell$ and its dispersion by fitting the 1000 $C_\ell$s with an asymmetric Gaussian.
  
  Passing these null tests is a strong indication of self-consistency. Of course, some effects could be present, at a certain level, in the various timescales, so that they are canceled out in the difference and remain undetected. However, the combined set of map differences allows us to gain confidence of our data and noise model. 
  
  In the following part of this section we present the results of some null test analyses, that highlight the data consistency with respect to various classes of systematic effects. All the power spectra are pseudo-spectra computed on maps masked with the bottom mask shown in Fig.~\ref{fig_masks}.

  Our results show the level of consistency of the LFI data. At 70\,GHz, in particular, the data pass all our tests both in temperature and polarization. Small residuals exist at lower frequencies, especially at 30\,GHz. At this frequency we see the evidence of residuals probably due to a non perfect Galactic straylight removal from our data (see results in Sect.~\ref{sec_surveydiff}).

  None of the detected excess cases appear to be crucial for science analysis. Residuals in temperature (which are expected to be larger due to the much stronger sky signals) are still orders of magnitude below the power of the signal from the sky (see left-hand panels of Figs.~\ref{fig_systematic_effects_power_spectrum_30}-\ref{fig_systematic_effects_power_spectrum_70}). We use the 30\,GHz channel in polarization (middle and right-hand panels of Fig.~\ref{fig_systematic_effects_power_spectrum_30}) as a synchrotron monitor for the \Planck\ CMB channels, and the observed deviations are small for foreground analysis or component separation.

\subsubsection{Highlighting residuals in subsets of data with ``full $-$ year'' difference maps}
\label{sec_fullyeardiff}

    We check the consistency of data acquired during each year using the full-mission map of the corresponding frequency as a reference to identify particular years that appear anomalous compared to others. For example, we test for spurious residuals in the first year of data of the 70\,GHz channel by taking the difference between the year-1 and full-mission maps at 70\,GHz. 
   
    The full-mission maps may contain some residual systematic effects, but this is not a problem in the context of this test, which aims at highlighting relative differences among the various year-long datasets. Furthermore systematic effects average out more efficiently in full-mission maps than in single-year or single-survey maps. We verified this by calculating the peak-to-peak variations in simulated maps containing the sum of all systematic effects.  Then we took the ratio between this peak-to-peak value calculated for year-maps and that for full-mission maps, and obtained values that from 1.2 at 30\,GHz to 2.1 at 70\,GHz.
   
\begin{sidewaysfigure*}
    \begin{center}
	\hspace{1.35cm} $TT$ \hspace{5.75cm} $EE$ \hspace{5.9cm} $BB$\\
	\begin{tabular}{m{.2cm} m{5.3cm} m{.25cm} m{5.3cm} m{.25cm} m{5.3cm}}
	\begin{turn}{90}30\,GHz\end{turn}&
	\includegraphics[width=6.0cm]{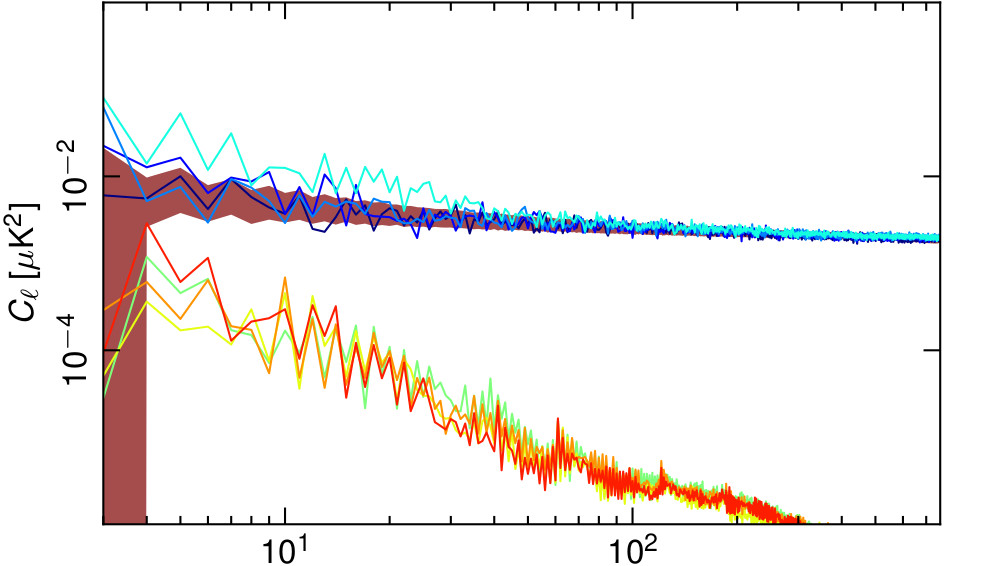} &
	&
	\includegraphics[width=6.0cm]{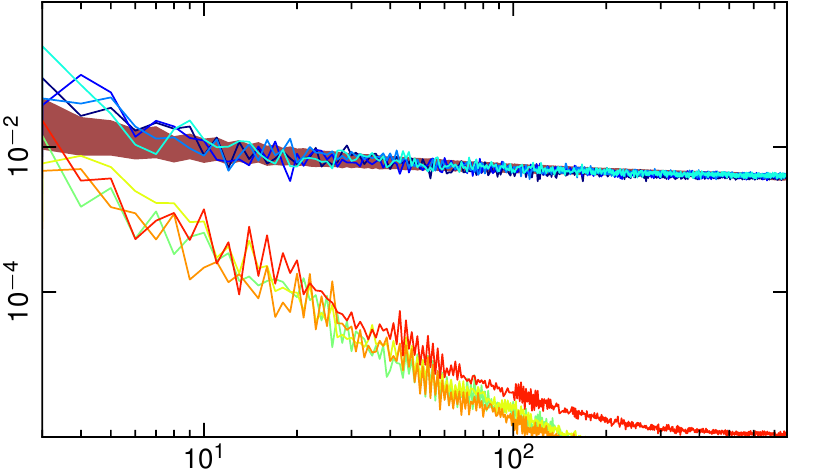} &
	&
	\includegraphics[width=6.0cm]{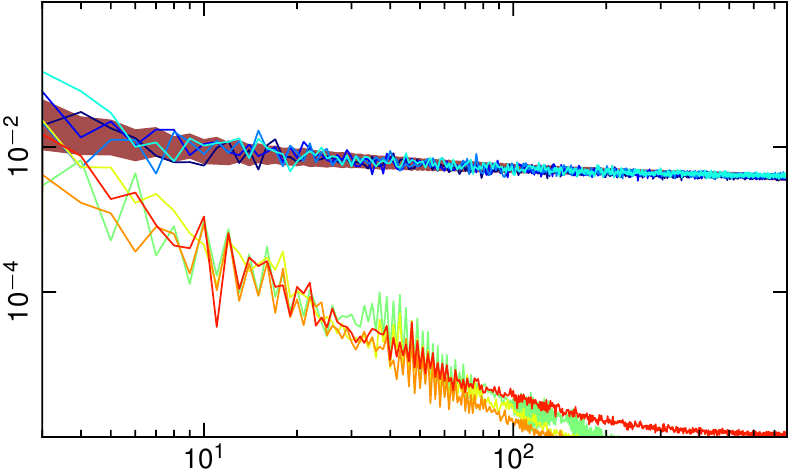}\\
	\begin{turn}{90}44\,GHz\end{turn}&
	\includegraphics[width=6.0cm]{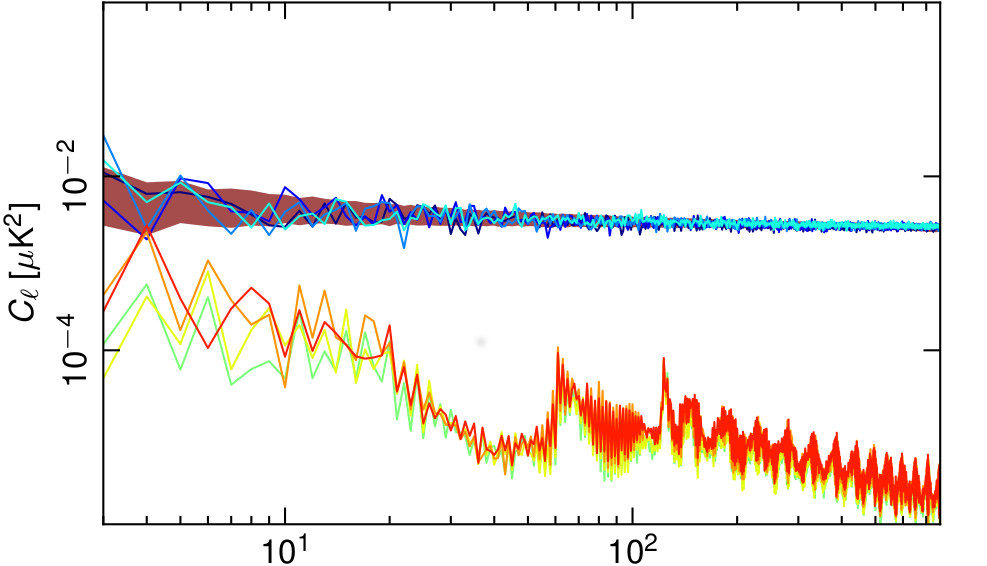} &
	&
	\includegraphics[width=6.0cm]{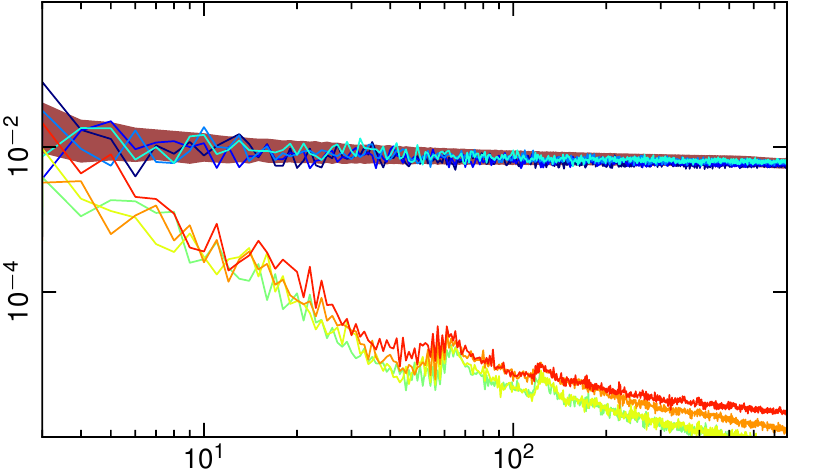} &
	&
	\includegraphics[width=6.0cm]{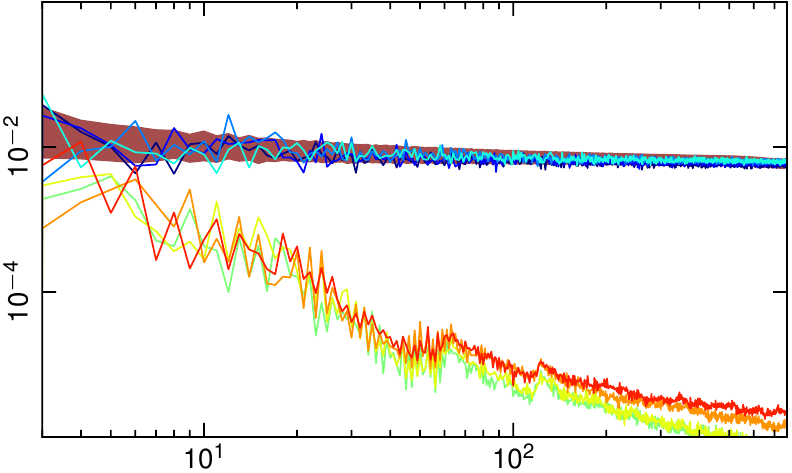}\\
	\begin{turn}{90}70\,GHz\end{turn}&
	\includegraphics[width=6.0cm]{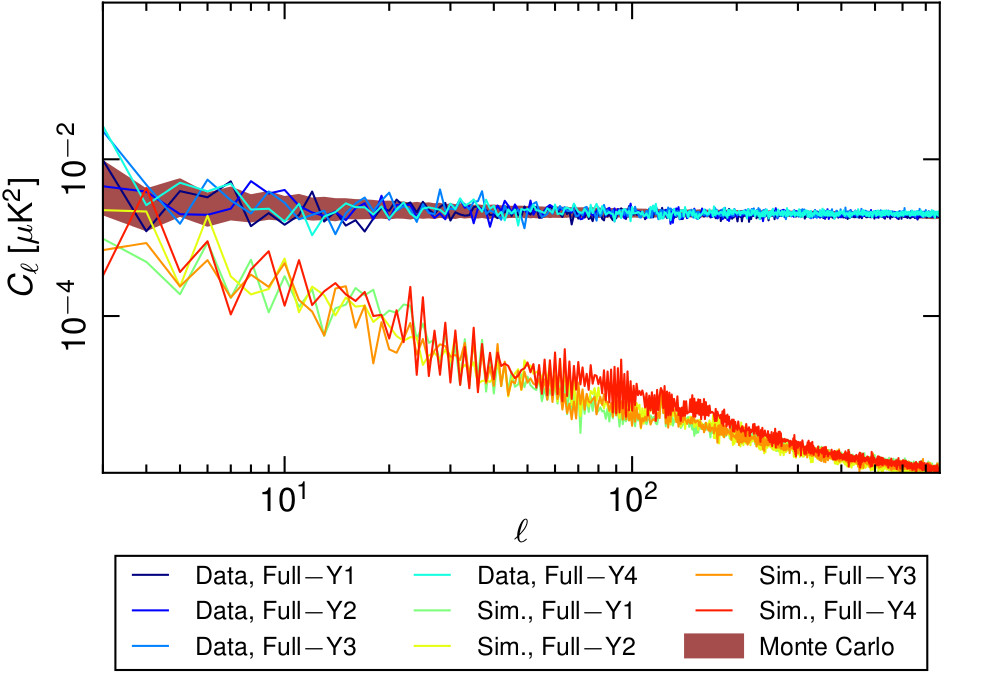} &
	&
	\includegraphics[width=6.0cm]{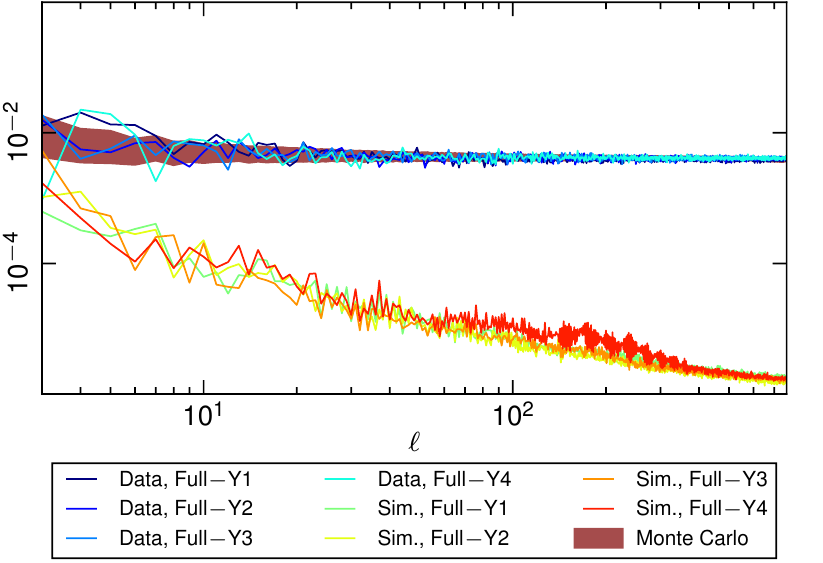} &
	&
	\includegraphics[width=6.0cm]{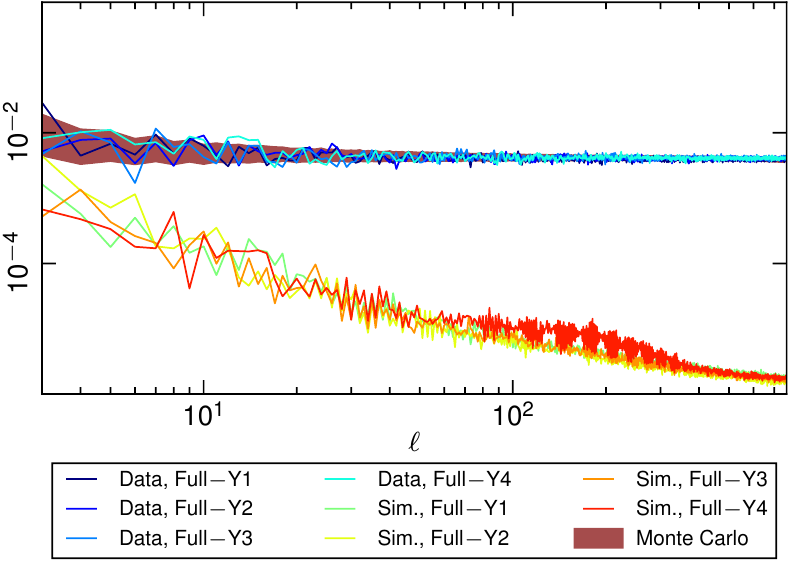}\\
	\end{tabular}
	\caption{
	\label{fig_full-year-null}
	Angular power spectra of \textit{full $-$ year} maps. Colored lines represent the four nulls obtained from data and simulations {of the effects listed in Table~\ref{tab_list_simulated_effects}}. The purple colored area represents the envelope from Monte Carlo simulations as described in the text. {In this figure we can appreciate that the power spectra of the data from the null maps are well matched by noise, except for a few low-$\ell$ points at 30\,GHz.  Furthermore, the power spectra of the simulated systematic effects fall well below the data at all but the lowest $\ell$ values.}}
    \end{center}

  \end{sidewaysfigure*}

  In Fig.~\ref{fig_full-year-null} we show the results of pseudo-spectra from {\it full $-$ year} difference maps obtained from data and simulations compared to the range of spectra obtained from the 1000 noise-only Monte Carlo simulations. This range is indicated by the colored area, which represents the rms spread of the simulated spectra. 

  Our analysis shows that at 44 and 70\,GHz the null power spectra are well explained by noise, while at 30\,GHz there are some residuals slightly exceeding the $1\sigma$ region of the Monte Carlo simulations. The null spectra from simulated systematic effects are below the noise level apart from the first multipoles where the simulations in some cases reach the noise level.
  
\subsubsection{Checking for time varying effects with ``odd $-$ even'' year difference maps}
\label{sec_oddevenyeardiff}

  We check for time varying effects considering a small change that was implemented in the \Planck\ scanning strategy after the first two years. At the beginning of the third year the precession phase angle of \Planck\ spin axis was shifted by 90$^\circ$ \citep[see section~4.2 of ][]{planck2013-p01}.
  
  This shift produced a slight symmetry-break between the scanning strategy of the first two and second two years of LFI observations. As a consequence, an exact repetition of the same configuration of the beams and sidelobes relative to the sky only occurs for the survey pairs $S_1$ and $S_3$, $S_2$ and $S_4$, $S_5$ and $S_7$, $S_6$ and $S_8$. Thus differences between these survey pairs contain only time-variable effects (neglecting pointing errors and secular changes in the optics, which are known to be small), while any beam asymmetry will cancel out exactly. Similarly, bandpass effects in a null test involving a given radiometer of a given quadruplet are removed. Since these difference maps contain only stochastic residuals, the combination at each frequency:
  
  \begin{equation}
  (S_1 - S_3) + (S_2 - S_4) + (S_5 - S_7) + (S_6 - S_8) = Y_{1 + 3} - Y_{2 + 4},
  \end{equation}
  gives a high signal-to-noise monitor of residuals dominated by time-variable effects, such as relative calibration, ADC non-linearity, thermal effects.  

\begin{sidewaysfigure*}
    \begin{center}
	\hspace{1.35cm} $TT$ \hspace{5.75cm} $EE$ \hspace{5.9cm} $BB$\\
	\begin{tabular}{m{.2cm} m{5.3cm} m{.25cm} m{5.3cm} m{.25cm} m{5.3cm}}
	\begin{turn}{90}30\,GHz\end{turn}&
	\includegraphics[width=6.0cm]{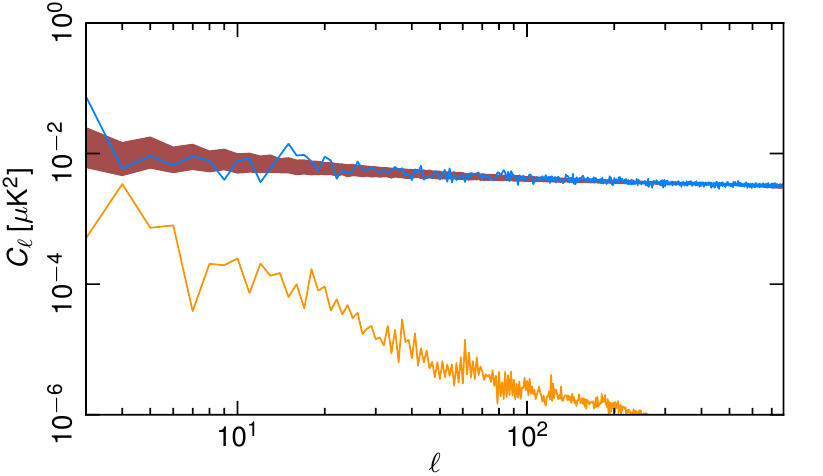} &
	&
	\includegraphics[width=6.0cm]{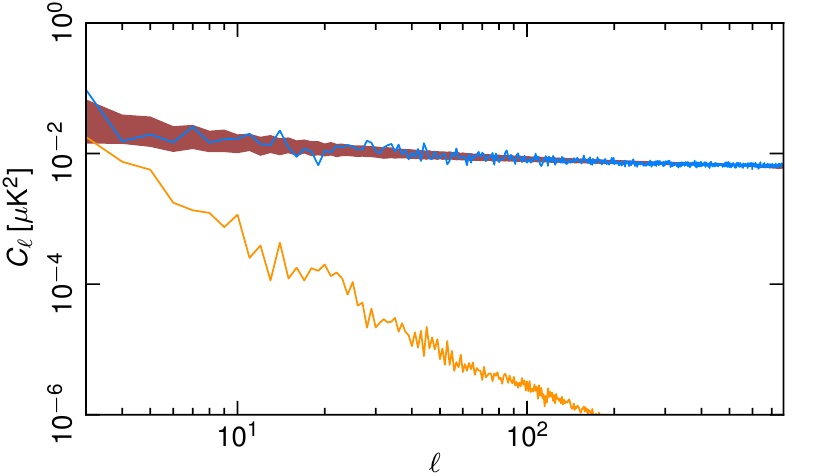} &
	&
	\includegraphics[width=6.0cm]{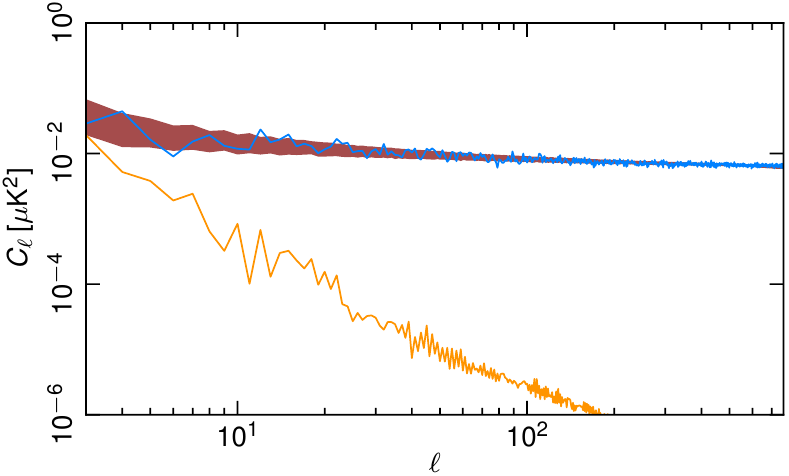}\\
	\begin{turn}{90}44\,GHz\end{turn}&
	\includegraphics[width=6.0cm]{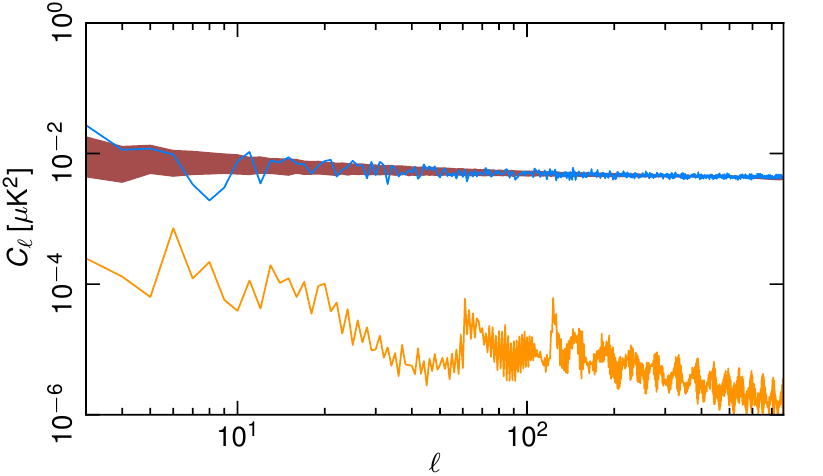} &
	&
	\includegraphics[width=6.0cm]{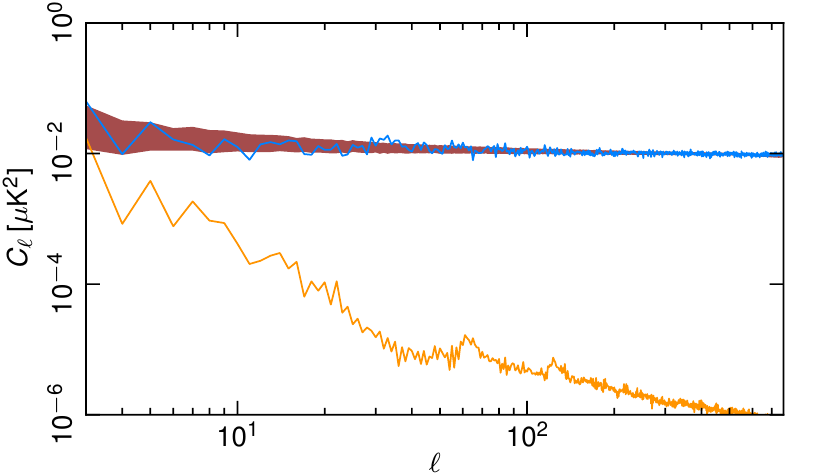} &
	&
	\includegraphics[width=6.0cm]{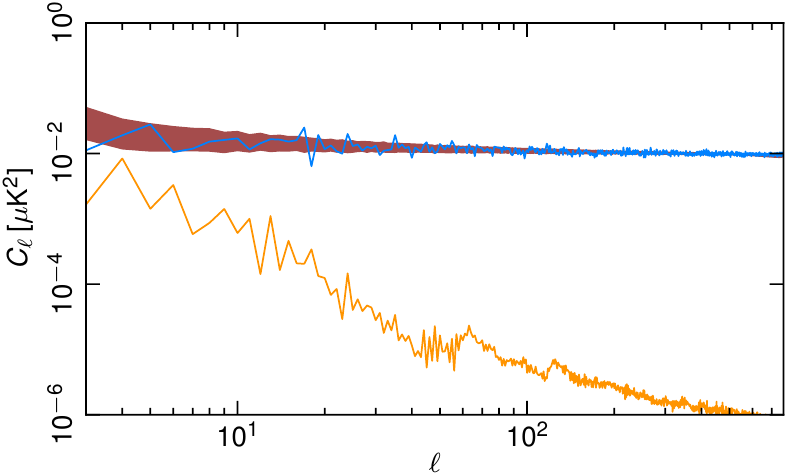}\\
	\begin{turn}{90}70\,GHz\end{turn}&
	\includegraphics[width=6.0cm]{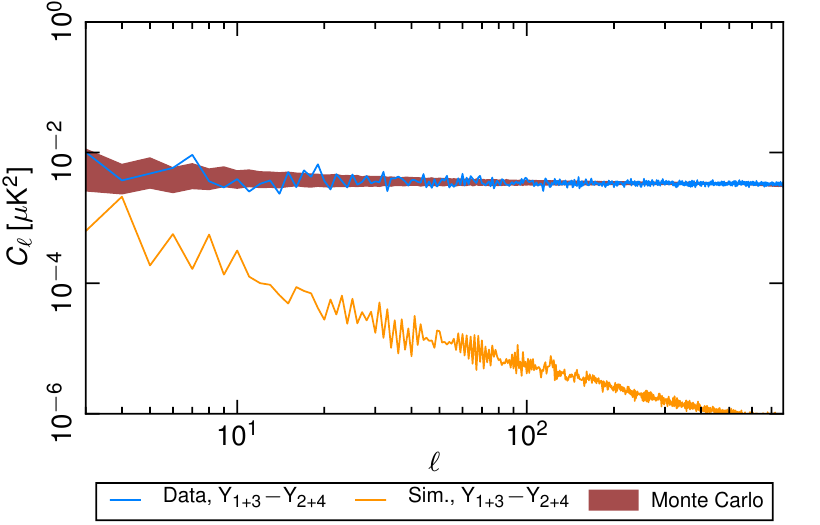} &
	&
	\includegraphics[width=6.0cm]{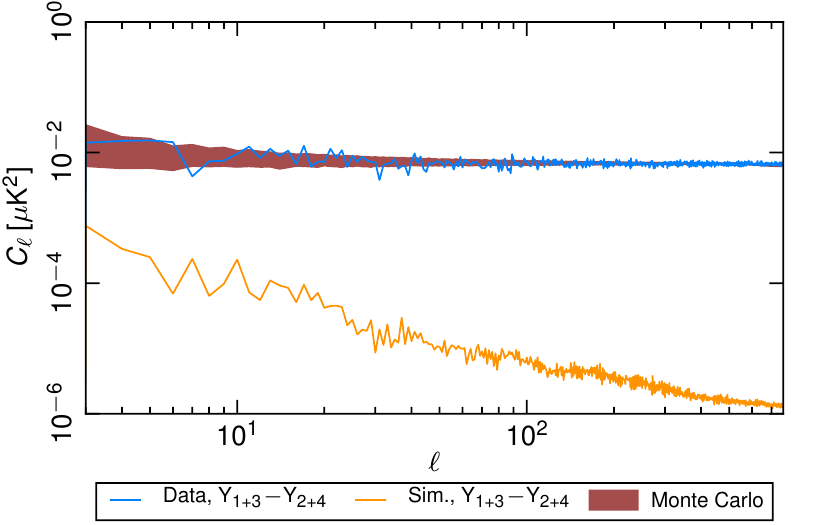} &
	&
	\includegraphics[width=6.0cm]{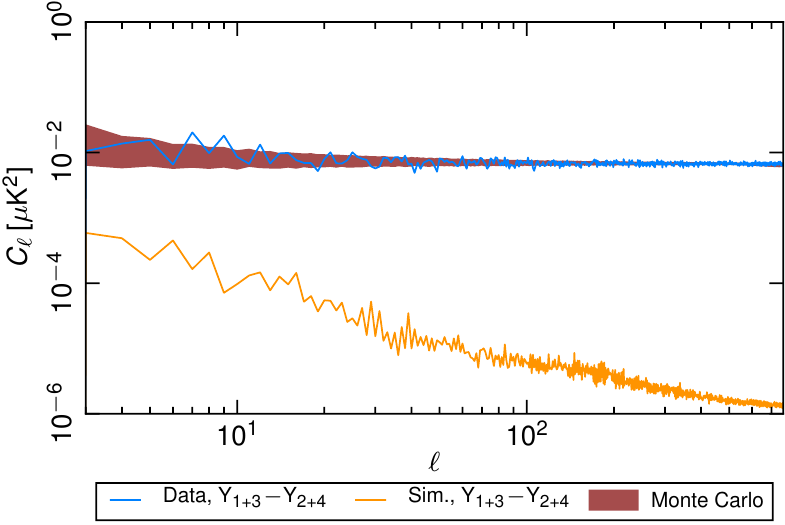}\\
	\end{tabular}
      \caption{
      \label{fig_odd-even-null}
      Angular power spectra of \textit{odd $-$ even} year maps. Colored lines represent the four nulls obtained from data and simulations {of the effects listed in Table~\ref{tab_list_simulated_effects}}. The colored area represents the envelope from Monte Carlo simulations as described in the text.}
    \end{center}

  \end{sidewaysfigure*}

  In Fig.~\ref{fig_odd-even-null} we show the result of null tests for such year combination for all LFI frequency bands, both in temperature and polarization. Again we compare the results with Monte Carlo noise simulations, as well as with the level of contamination predicted by our systematic effect simulations. In all cases we find very good consistency  between data from the null maps and the noise.  We also find that the systematic effects run well below both .

\subsubsection{Highlighting straylight residuals with consecutive survey difference maps}
\label{sec_surveydiff}

  We looked for residual effects in null maps constructed from differences between consecutive surveys, which we expect to be dominated by Galactic straylight. Indeed, when the spacecraft changes from an odd to an even survey it re-visits the same sky patch with its orientation reversed by about 180$^\circ$. As a consequence, the coupling of the beams with the sky is reversed, so that differences between odd and even surveys highlight the effect of main beam asymmetries and straylight from sidelobes. This implies that consecutive survey differences are among the most demanding null tests.

\begin{sidewaysfigure*}
    \begin{center}
	\hspace{1.35cm} $TT$ \hspace{5.75cm} $EE$ \hspace{5.9cm} $BB$\\
	\begin{tabular}{m{.2cm} m{5.3cm} m{.25cm} m{5.3cm} m{.25cm} m{5.3cm}}
	\begin{turn}{90}30\,GHz\end{turn}&
	\includegraphics[width=6.0cm]{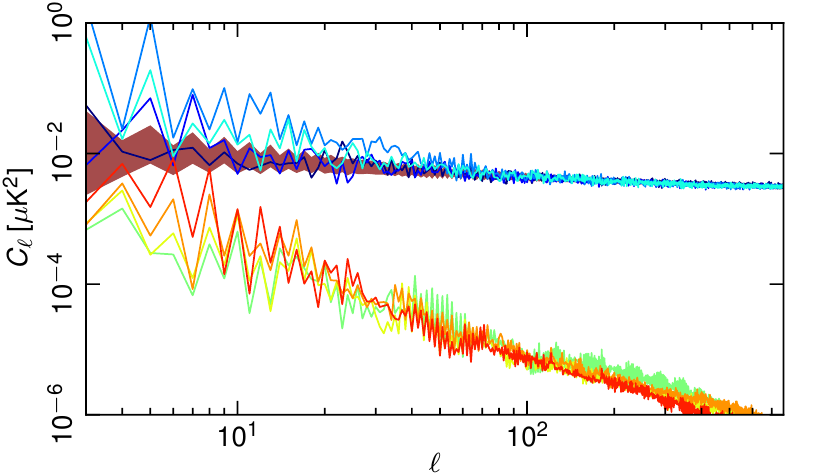} &
	&
	\includegraphics[width=6.0cm]{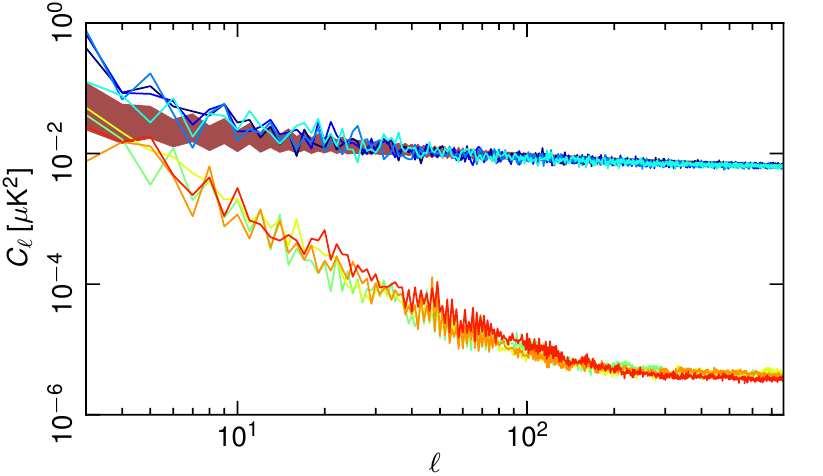} &
	&
	\includegraphics[width=6.0cm]{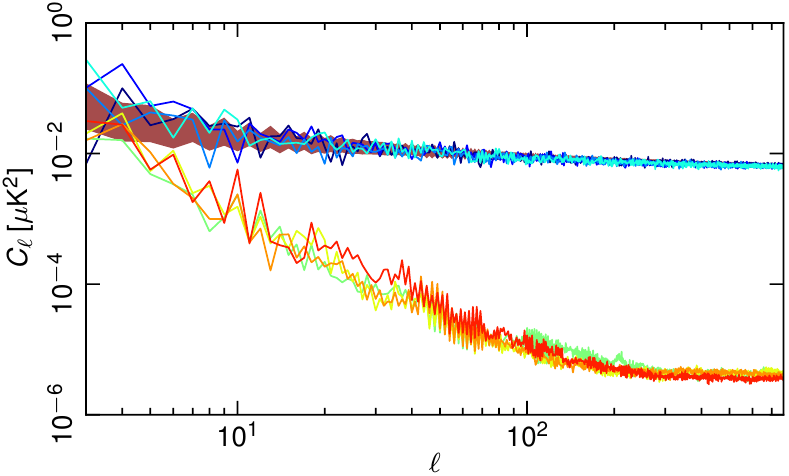}\\
	\begin{turn}{90}44\,GHz\end{turn}&
	\includegraphics[width=6.0cm]{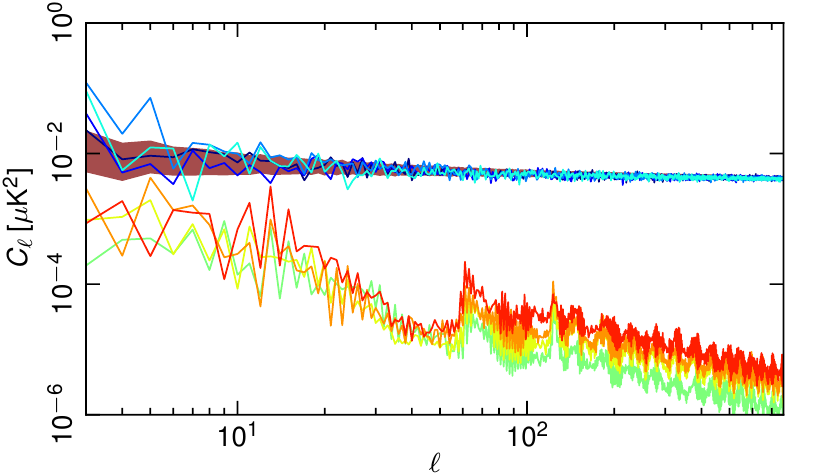} &
	&
	\includegraphics[width=6.0cm]{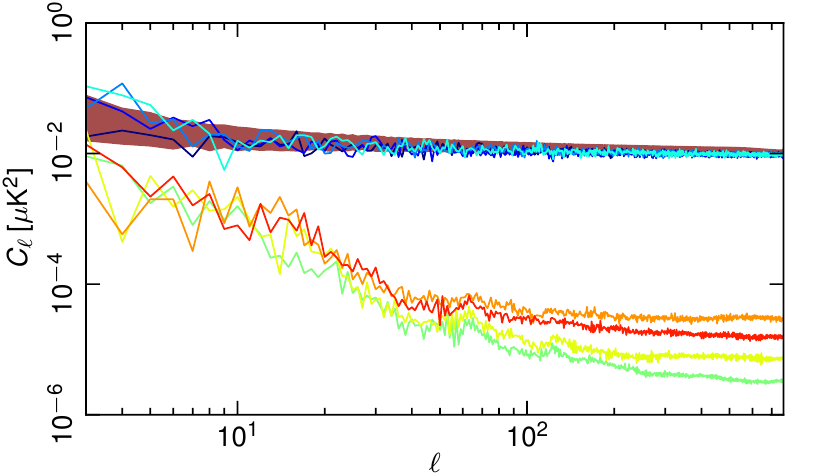} &
	&
	\includegraphics[width=6.0cm]{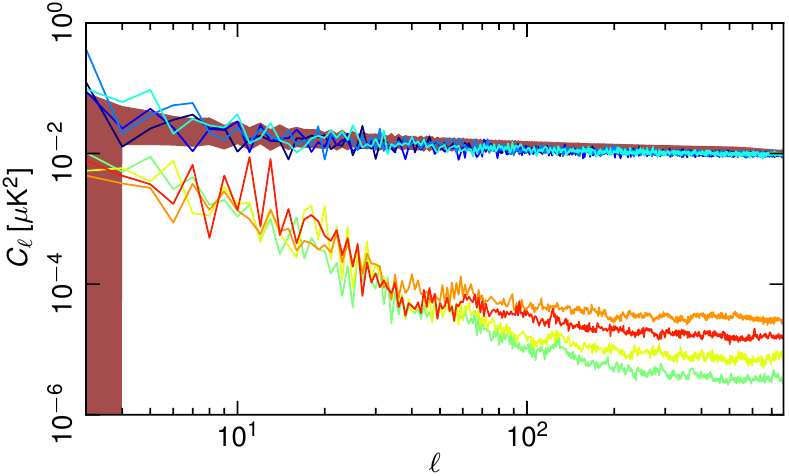}\\
	\begin{turn}{90}70\,GHz\end{turn}&
	\includegraphics[width=6.0cm]{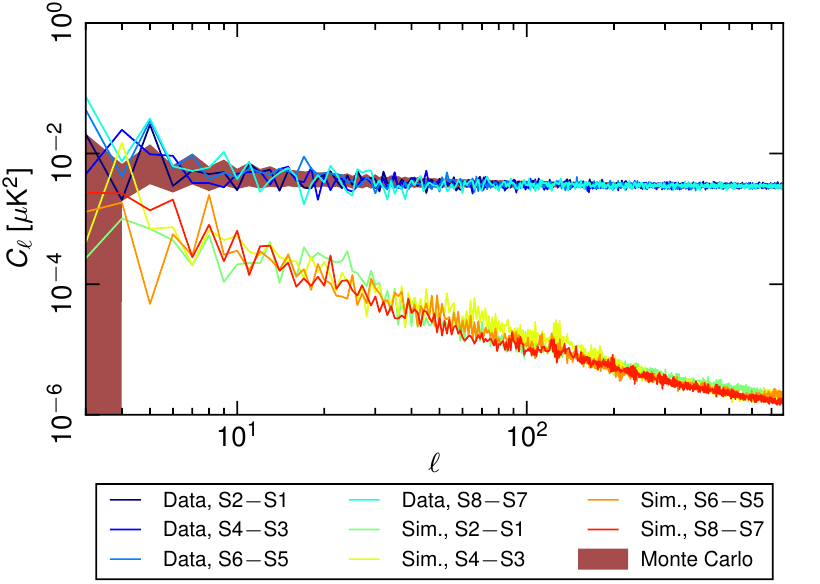} &
	&
	\includegraphics[width=6.0cm]{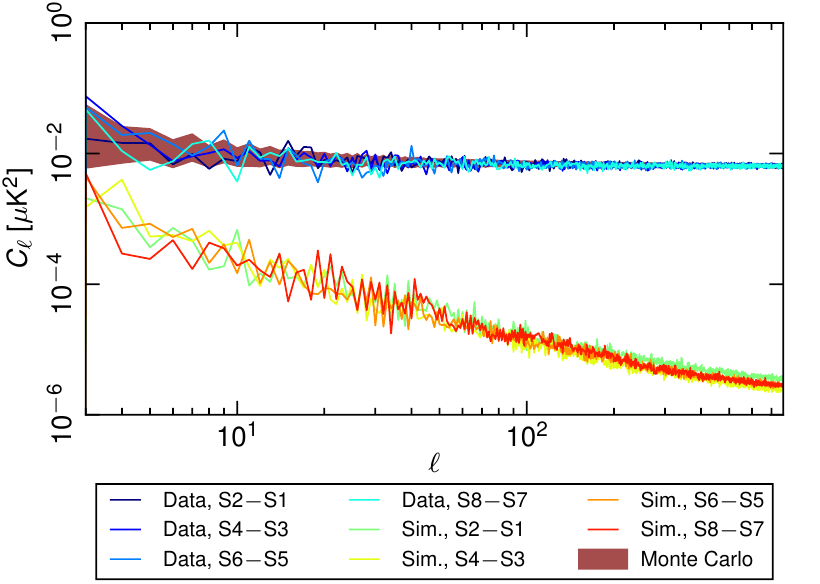} &
	&
	\includegraphics[width=6.0cm]{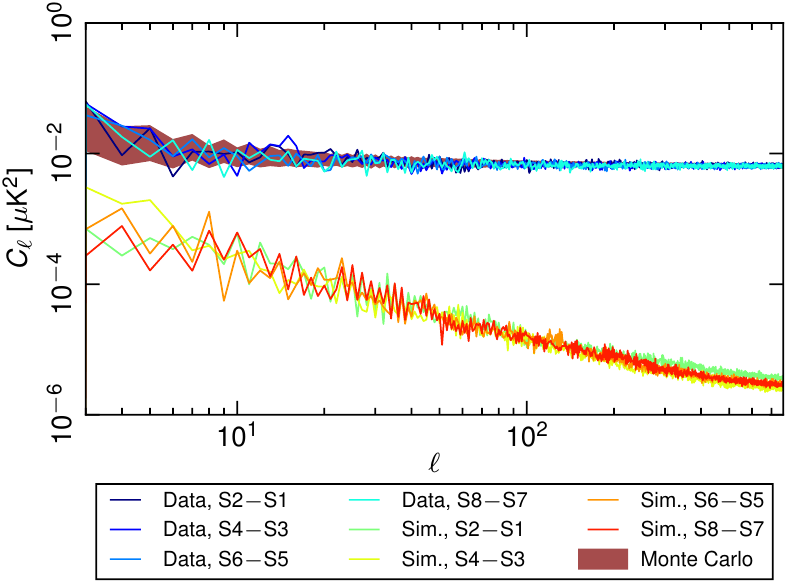}\\
	\end{tabular}
      \caption{
      \label{fig_ssodd-sseven-null}
      Angular power spectra of consecutive survey difference maps. Colored lines represent the four nulls obtained from data and simulations {of the effects listed in Table~\ref{tab_list_simulated_effects}}. The colored area represents the envelope from Monte Carlo simulations as described in the text.}
    \end{center}

  \end{sidewaysfigure*}

  In Fig.~\ref{fig_ssodd-sseven-null} we display the angular power spectra of a wide sample of consecutive survey difference maps and compare them with noise and systematic effects simulations.

  In this case, at 30\,GHz the null spectra from the data exceed the predictions from both the noise Monte Carlo and the systematic effects simulations that assume a perfect subtraction of straylight effects. This is true particularly for the spectra in temperature. Some excess in temperature is present also at 44 and 70\,GHz.  
  
  These results give useful hints on the accuracy of the optical model of the LFI sidelobes used to estimate the straylight contribution. In this respect we are planning further studies to improve this model exploiting the information provided by these null tests. The outcome of these studies will be reported in the context of the next \Planck\ release.

    \subsection{Impact of systematic effects at large angular scales}
\label{sec_lowell}
  
  In this section we describe the assessment of systematic effect uncertainties on the detection of optical depth, $\tau$, from LFI data at large angular scales. For the \Planck\ 2015 release the extraction of the $\tau$ parameter is based on the LFI 70\,GHz data, using the LFI 30\,GHz channel for removing polarized foreground synchrotron and the HFI 353\,GHz channel to clean polarized dust emission \citep{planck2014-a13,planck2014-a15}. 

  To quantify the impact of residual effects on $\tau$, we carried out an end-to-end analysis, propagating the simulated effects to maps, power spectra, and parameters, following the same processing steps adopted in the data analysis. We start from the map containing the sum of all the systematic effects at 70\,GHz in polarization, add to it a realization of white noise and $1/f$ noise derived from FFP8 simulations for the full mission, and finally add a CMB realization. The corresponding map at 30\,GHz is used in the template-fitting procedure to quantify its impact on the synchrotron removal process. Here we neglect the propagation of systematic effects from the 353\,GHz channel. However, we expect a very small contribution from this channel considering the expected level of systematic effects \citep[see section~7 of][]{planck2014-a09} and the scaling coefficient for dust between 353 and 70\,GHz, $\beta \approx 0.0077$ \citep[][section~2.3]{planck2014-a13}. 

  In more detail, we consider the linear combination $\vec{m}_{\rm clean} = \vec{m}_{70} -\alpha \,\vec{m}_{30}$, where $\alpha = 0.063$ is the effective synchrotron scaling ratio between the 30 and 70\,GHz channels, and adjust the effective noise covariance matrix accordingly \citep[see][for a more detailed discussion]{planck2014-a13}. In practice, this is equivalent to assuming that our cleaning procedure leaves no foreground residual in the final map, and for the scope of this analysis we only need to consider the impact of the rescaled 30\,GHz noise and residual effects on the foreground-cleaned map. 

  From the resulting foreground-cleaned map, we extract the power spectra for the temperature and polarization components at multipoles $\ell \leq 29$. We calculate the spectra over the sky region used to derive $\tau$, shown in the bottom panel of Fig.~\ref{fig_masks}.

  To quantify the result, we calculate the bias introduced by the systematic effects on the three parameters that are most sensitive to low multipoles, i.e., $\tau$, $r$, and the amplitude of scalar perturbations, $A_s$. We produce 1000 Monte Carlo FFP8 simulations of the CMB polarized sky plus white and $1/f$ noise with systematic effects, and 1000 similar simulations but containing only CMB and noise. For each realization we then calculate the marginalized distributions for each of the three parameters $X = \tau, r, A_s$ and calculate the differences $\Delta X = X_{\rm syst.} - X_\mathrm{no-syst.}$, which represent the bias introduced in the estimates of $X$ by the combination of all systematic effects. 

  For $\log(A_\mathrm{s})$ and $r$ we find median bias values of $-0.026$ and $0.11$, respectively, which would correspond to a 0.2\,$\sigma$ effect on the amplitude parameter and an increase of 15\,\% on the upper limit on $r$ (95\% CL). However, the dominant \Planck\ constraints on these two parameters come effectively from temperature power spectrum at high multipoles, so the actual impact on the \Planck\ results is very small. 

  For the optical depth, we find a mean bias $\langle \Delta \tau \rangle = 0.005$, or 0.2--0.25 times the standard deviation of the value of $\tau$ measured by LFI \citep{planck2014-a15}. This result shows that the impact of all systematic effects on the measurement of $\tau$ is within 1$\sigma$. The measured $\langle \Delta \tau \rangle$ is compatible with a positive but sub-dominant bias by residual systematics, with an impact on $\tau$ well within the statistical uncertainty. 

  We emphasize that this result is based on our bottom-up approach, and therefore it relies on the accuracy and completeness of our model of all known instrumental systematic effects. As we have shown, at large angular scales systematics residuals from our model are only marginally dominated by the $EE$ polarized CMB signal. For this reason we plan to produce a further independent tests on these data based both on null tests and on cross-spectra between the 70\,GHz map and the HFI 100 and 143\,GHz maps. Such a cross-instrument approach may prove particularly effective, because we expect that systematic effects between the two \Planck\ instruments are largely uncorrelated. We will discuss these analyses in a forthcoming paper in combination with the release of the low-ell HFI polarization data at 100--217\,GHz and in the final 2016 Planck release.
  
  
    \subsection{Propagation of systematic effects through component separation}
\label{sec_assessment_compsep}

  In this section we discuss how we assess the impact of residual systematic effects in the LFI data on the CMB power spectra after component separation (see Fig.~\ref{fig_component_separation_global_compsep} in Sect.~\ref{sec_summary_table}). 

  \Planck\ component separation exploits a set of algorithms to derive each individual sky emission component. They are minimum variance in the needlet domain ({\tt NILC}) or use foreground templates generally based on differences between two \Planck\ maps that are close in frequency ({\tt SEVEM}), as well as parametric fitting conducted in the pixel ({\tt Commander}) and harmonic ({\tt SMICA}) domains. We describe them in detail in \citet{planck2014-a11}. 

  To assess residuals after component separation we use LFI systematic effect maps as the input for a given algorithm, setting the HFI channels to zero. This means that the output represents only the LFI systematic uncertainty in the corresponding CMB reconstruction. In \citet{planck2013-p02a} we exploited a global minimum-variance component-separation implementation, {\tt AltICA}, to derive weights used to combine the LFI systematic effect maps. Here we generalize the same procedure using {\tt NILC} and {\tt SEVEM}. Both are based on minimum-variance estimation of the weights, but in localized spatial and harmonic domains, and so optimally subtract foregrounds where they are most relevant ({\tt NILC}), and exploit foreground templates generally constructed by diffferencing two nearby \Planck\ frequency channels ({\tt SEVEM}). 

\begin{figure*}[!htpb]
 \begin{center}
   \begin{tabular}{m{.25cm} m{5.6cm} m{5.6cm} m{5.6cm}}
    & \begin{center}$I$\end{center} &\begin{center}$Q$\end{center}&\begin{center}$U$\end{center}\\    
    \begin{turn}{90}NILC\end{turn}&\includegraphics[width=56mm]{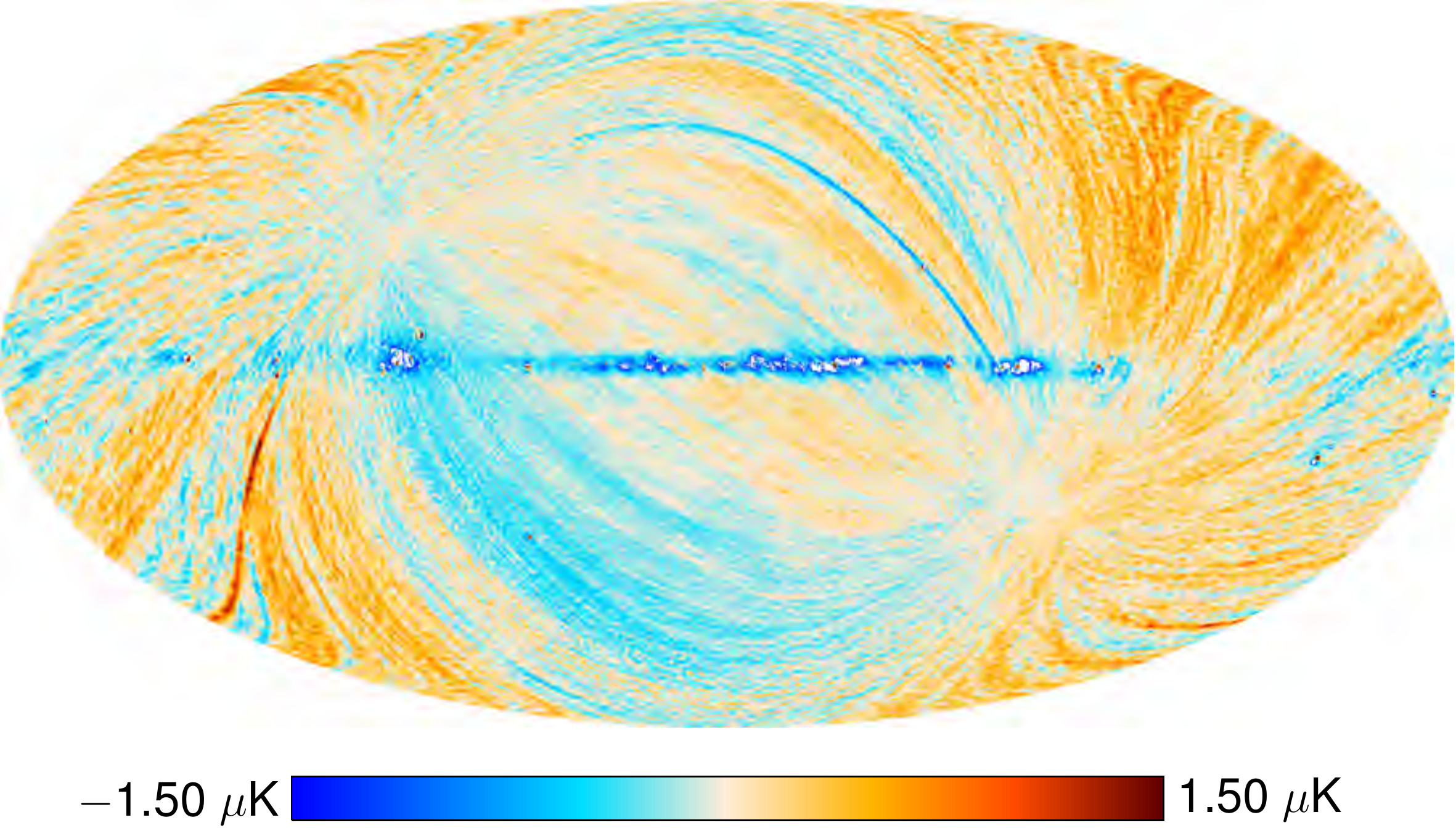}&
    \includegraphics[width=56mm]{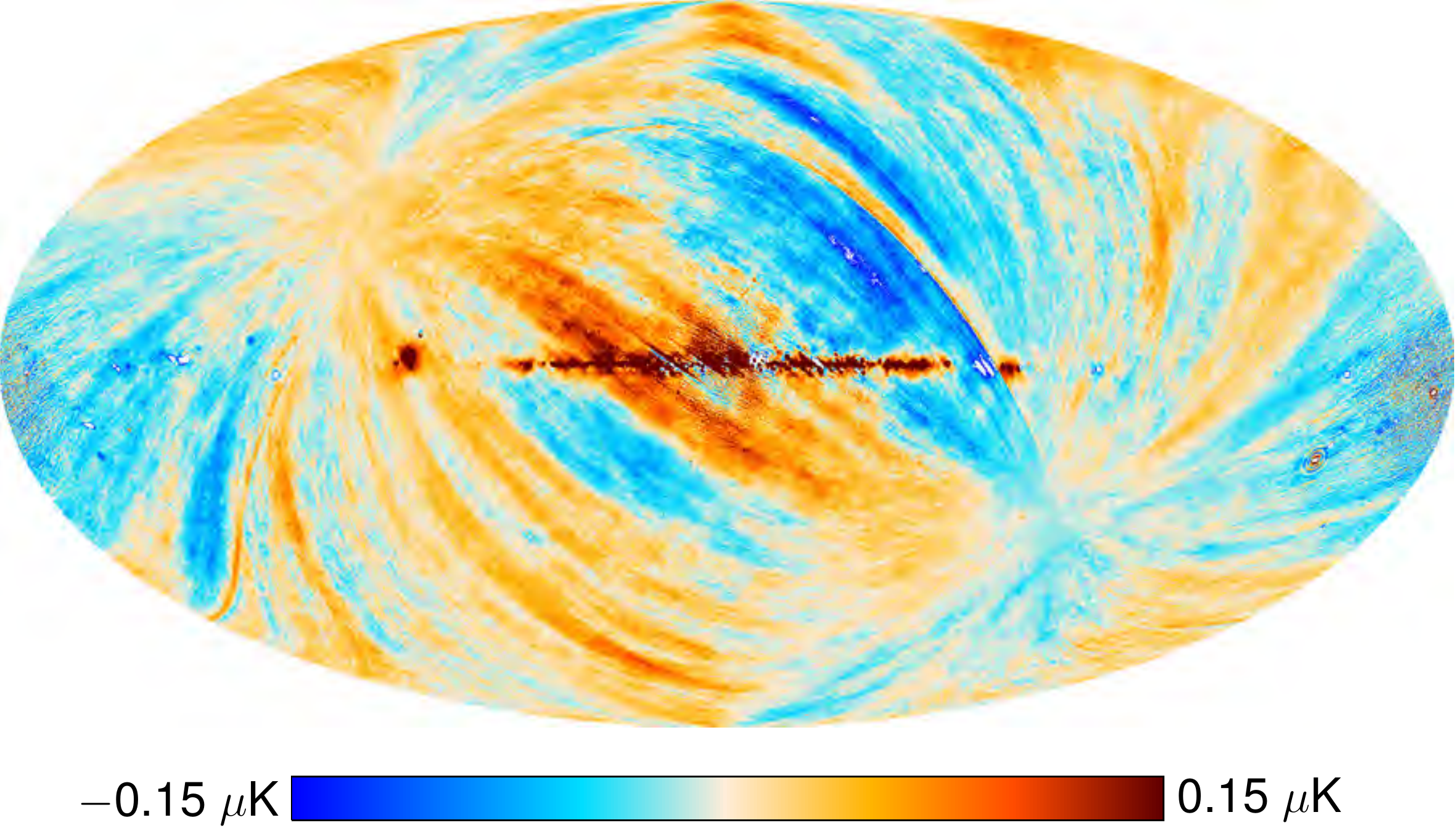}&
    \includegraphics[width=56mm]{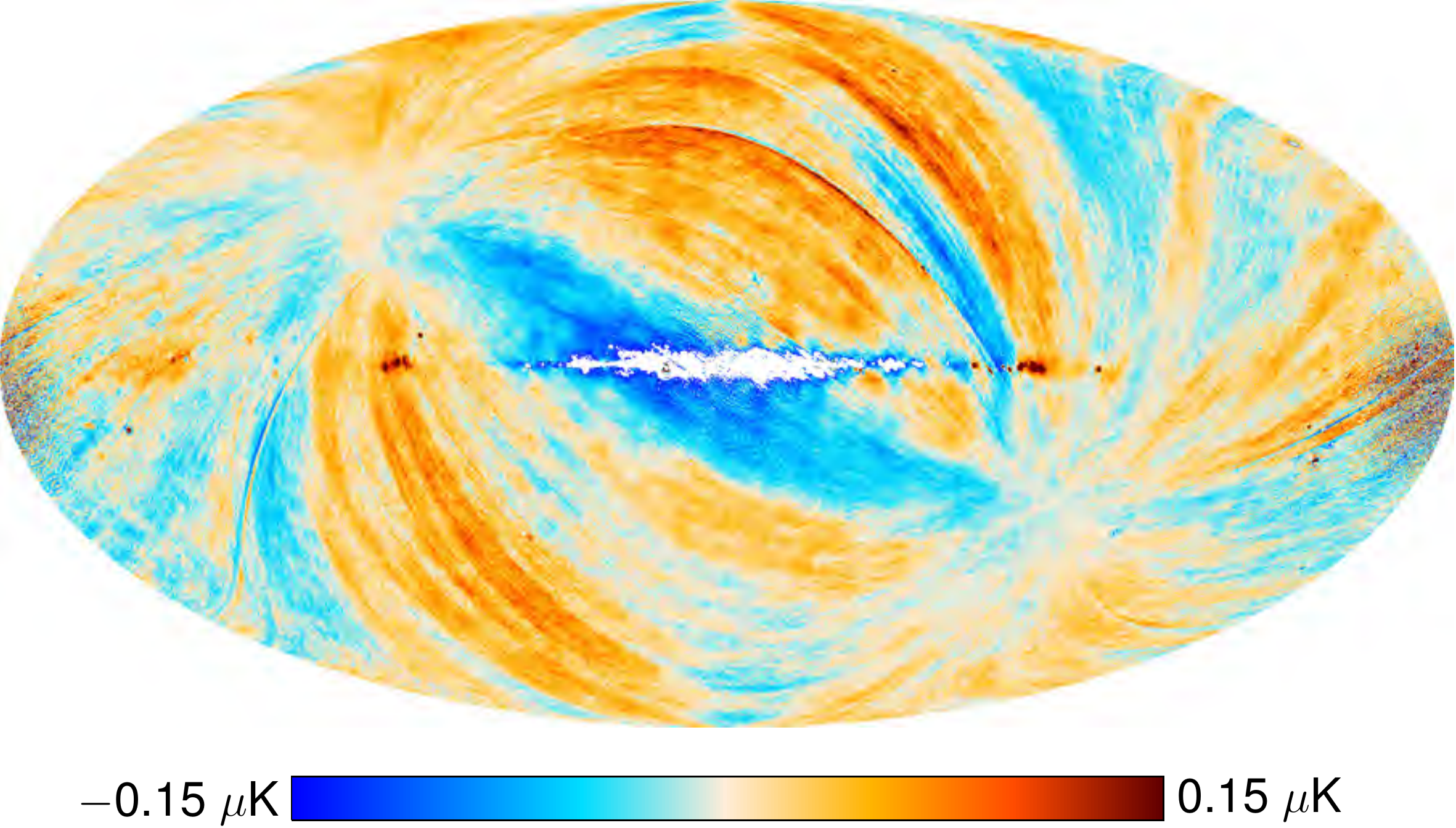}\\
    &&&\\
    \begin{turn}{90}SEVEM\end{turn}&\includegraphics[width=56mm]{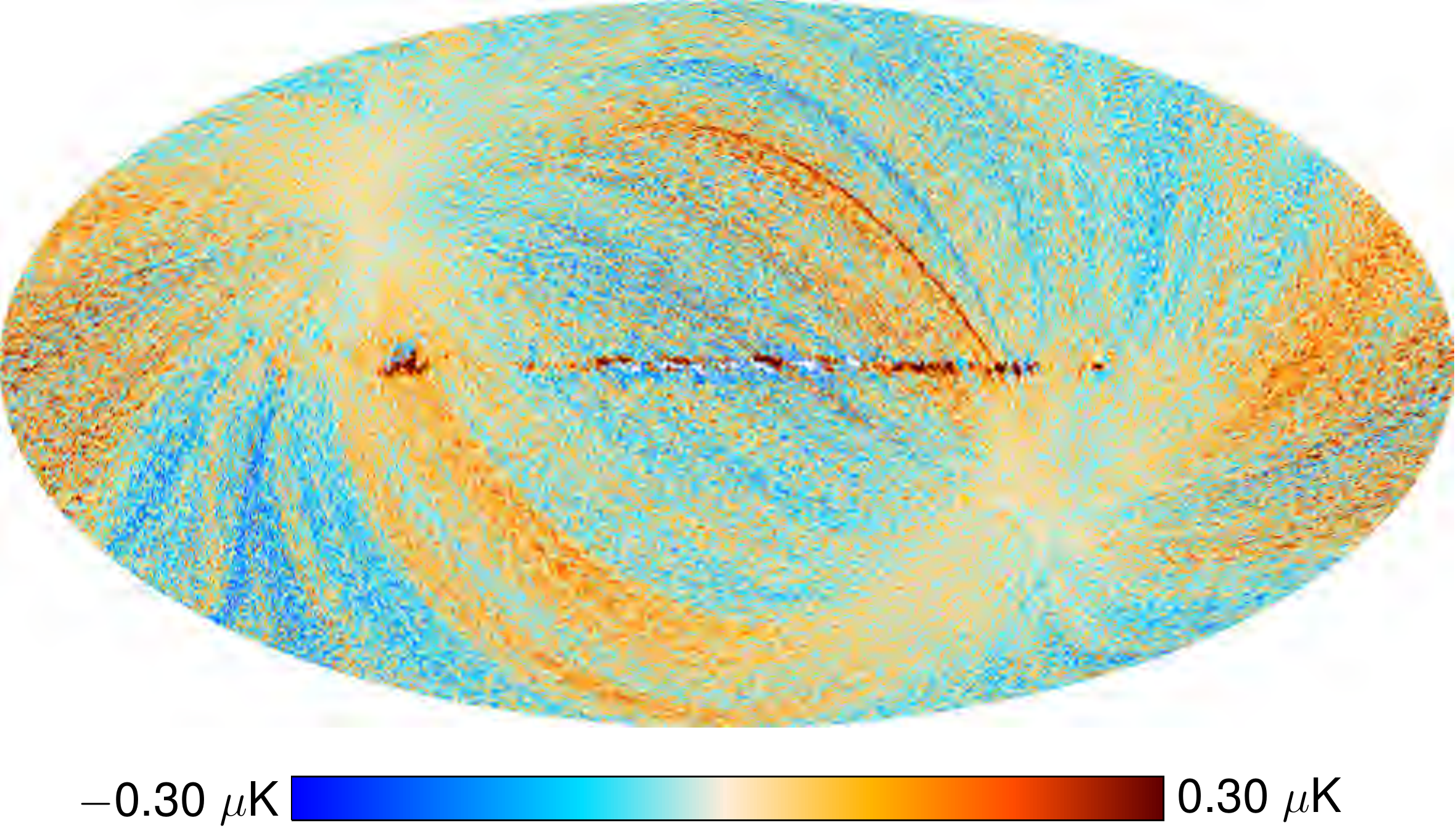}&
    \includegraphics[width=56mm]{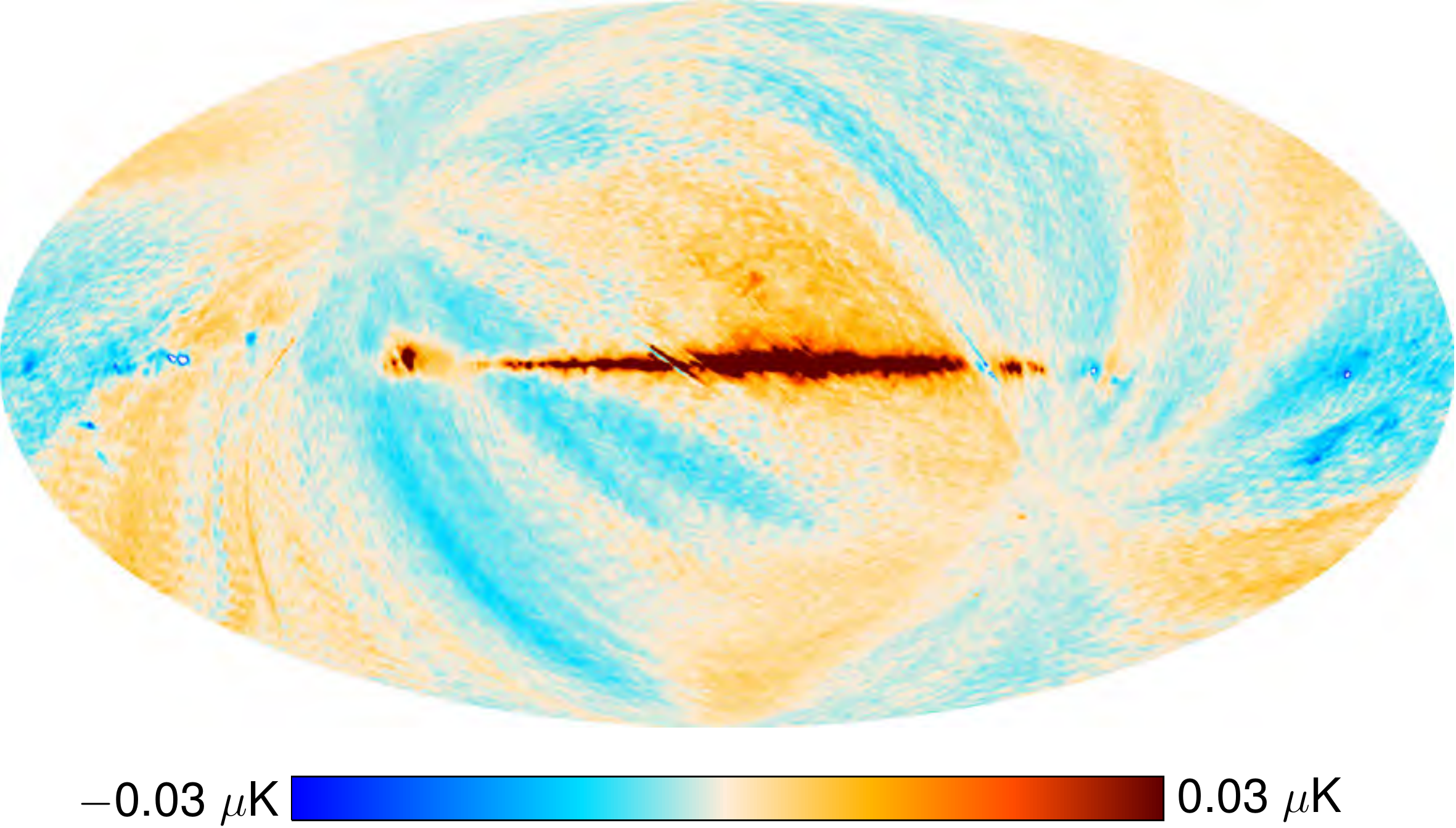}&
    \includegraphics[width=56mm]{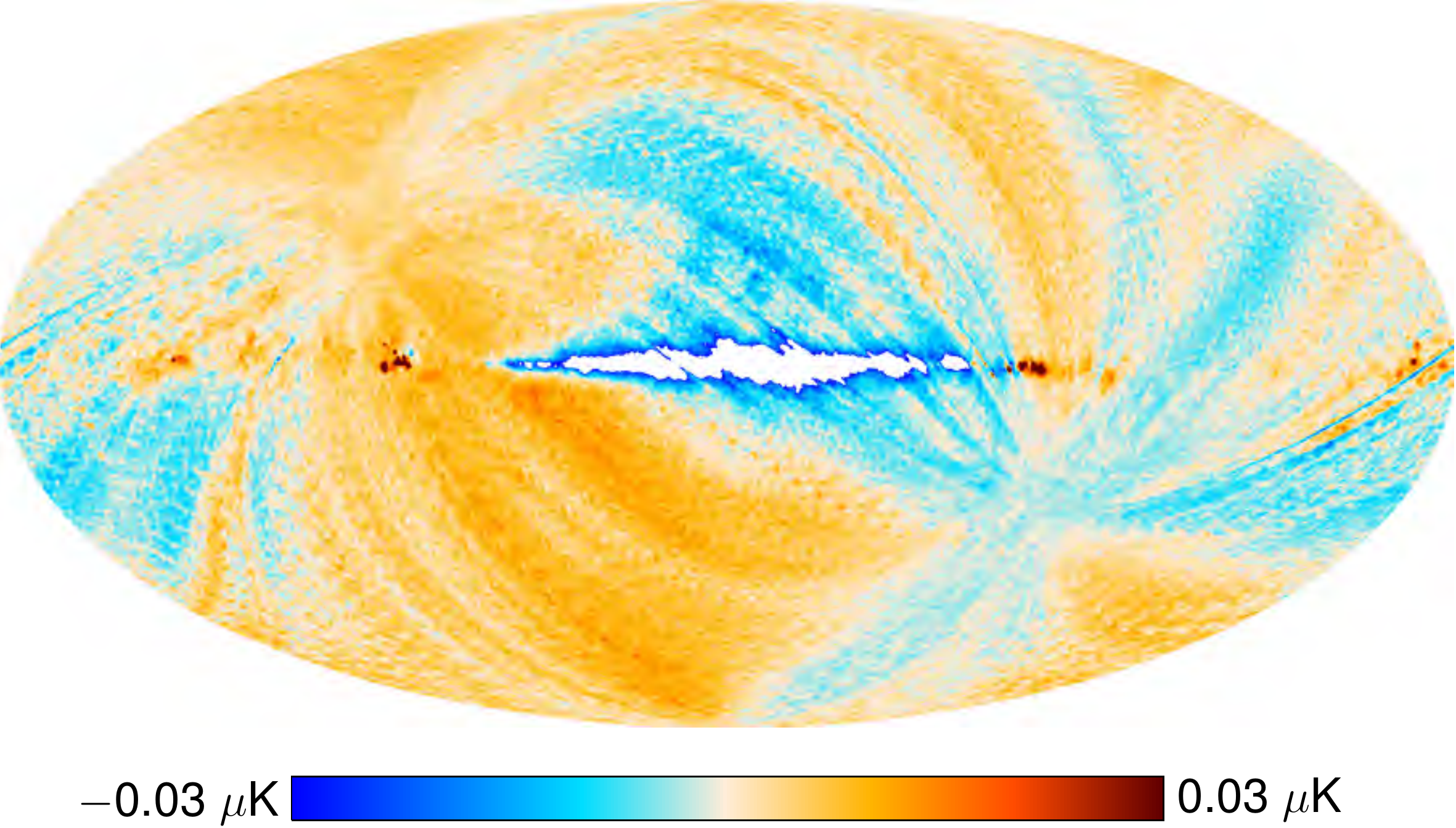}
    \end{tabular}
 \end{center}

 \caption{Maps in total intensity and polarization of the LFI systematic
 effects after component separation. {\it Top}: maps extracted with
 \texttt{NILC}. {\it Bottom}: maps extracted with \texttt{SEVEM}. Notice that
 the colour scale of \texttt{SEVEM} maps is 5 times smaller than that of
 \texttt{NILC} maps.}
 \label{fig_compsep_maps}
\end{figure*}

  Fig.~\ref{fig_compsep_maps} shows maps in total intensity and polarization of the LFI systematic effects after component separation. Maps extracted with \texttt{NILC} appear in the top row, while maps extracted with \texttt{SEVEM} appear in the bottom row. The structures that are most prominent outside the Galactic residuals appear to be associated with the scan strategy. Residuals are about 5 times larger for {\tt NILC} than for \texttt{SEVEM}, for the reasons described in Sect.~\ref{sec_summary_table}. It is important to stress that component separation does not alter the relative strength of the various systmatic effects treated in this paper, but simply filters them through the given foreground-cleaning pipeline. 

  Fig.~\ref{fig_component_separation_global_compsep} in Sect.~\ref{sec_summary_table} shows power spectra obtained from these maps compared with the best-fit \Planck\ 2015 $\Lambda$CDM cosmology.  To calculate these spectra we have first applied the masks shown in Fig.~\ref{fig_masks} and then computed pseudo-spectra corrected for the unseen sky fraction.

  In total intensity we confirm the results presented in \citet{planck2013-p02a}: the impact of known LFI systematic effects is at least two orders of magnitude less than the CMB. In polarization we observe a different residual level, depending on the algorithm used. The weighting strategy of {\tt NILC} at large angular scales performed in the needlet domain yields a residual effect that is larger by about 1.5 orders of magnitude compared to \texttt{SEVEM}. We further discuss this discrepancy in Sect.~\ref{sec_summary_table}. The 70-GHz channel is one with less foreground contamination and higher angular resolution. Hence, \texttt{NILC} weights this channel more compared to the others. This ultimately causes a larger level of residual systematic effects.

  The residual effect in polarization after processing with \texttt{NILC} is comparable to interesting levels of cosmological $B$-modes at large scales. This particular point needs further attention for the next \Planck\ data release, where component-separation solutions will be relevant for characterizing polarization accurately at large angular scales.

  
    \subsection{Gaussianity statitical tests}
\label{sec_assessment_gaussianity}

  In this section we present the results of statistical tests assessing the impact of known systematic effects in the LFI data on non-Gaussianity studies.

  The presence of systematic effect residuals can bias the statistical isotropy properties of the \Planck\ maps \citep{planck2014-a18} or the constraints on primordial non-Gaussianity \citep{planck2014-a19}. Therefore it is important to understand the impact of known systematic effects on the most relevant non-Gaussianity studies carried out within this release.

  In the \Planck\ 2013 release the non-Gaussianity studies were carried out using temperature data in two steps \citep{planck2013-p02a}. Firstly, we estimated an upper limit on the ``detectability level'' of all the known effects summed into a single ``global'' map. This level was defined as the factor we must multiply the global map by to generate a significant non-Gaussian deviation. Secondly, we measured the bias that these systematic effects could introduce on the local nonlinear coupling $f_{\rm NL}$ parameter. 

  In the current release we follow the same approach, considering, additionally, the polarization signal at low $\ell$. We have also considered the three usual cases (namely local, equilateral, and orthogonal) for the bispectrum shape when defining $f_{\rm NL}$.

  We characterize the level of detectability of the non-Gaussian contamination by comparing simulations that contain the systematic effect map added and rescaled by a global factor, $f_\mathrm{sys}$, with the null hypothesis (i.e., no systematic effects). We consider two scenarios, measuring the level of detectability of the systematic effects over: (i) the CMB + noise background; and (ii) the noise background only. These can be written
  \begin{eqnarray}
    \Delta T_\mathrm{(i)} \left( \hatn \right) & = &  \Delta T_\mathrm{CMB}
    \left( \hatn \right)  +  \Delta T_\mathrm{noise} \left( \hatn \right)
    + f^\mathrm{(i)}_\mathrm{sys} \Delta T_\mathrm{sys} \left( \hatn \right),   \\
    \Delta T_\mathrm{(ii)} \left( \hatn \right) & = & \Delta T_\mathrm{noise}
    \left( \hatn \right)  + f^\mathrm{(ii)}_\mathrm{sys} \Delta T_\mathrm{sys}
    \left( \hatn \right).   
  \end{eqnarray}
  
  For each case we calculate the detectability factor using a $\chi^2$ test on Monte Carlo simulations. First we produce two sets of 1000 simulations with and without the systematic effects added and then we define $\chi^2 = V \tens{C}^{-1}V^\textsf{T}$. The vector $V$ can be $V=[T]$, $V=[Q,U]$, or $V=[T,Q,U]$, while $\tens{C}$ is the corresponding covariance matrix. Under the assumption of normality, this statistics follow a $\chi^2$ distribution with $N_{T}$, $N_{Q}+N_{U}$ or $N_{T}+N_{Q}+N_{U}$ degrees of freedom, respectively.\footnote{$N_{T}$, $N_{Q}$, and $N_{U}$ are the total number of pixels available for the $T$, $Q$, and $U$ maps, respectively.} This is one of the estimators used in the Gaussianity analyses of \citet{planck2014-a18}.

  To define the ``detectability factors'' $f^\mathrm{(i)}_\mathrm{sys}$ and $f^\mathrm{(ii)}_\mathrm{sys}$ we consider a method to discriminate two $\chi^2$ histograms corresponding to simulations with and without systematic contamination. We use the common choice based on the significance level of the two histograms, defined as the fraction of cases of the null hypothesis (no systematic effects added) with a $\chi^2$ larger than the median of the alternative hypothesis (systematic effects added). Both distributions are considered different when this number is lower than 5\,\%. Therefore $f^\mathrm{(i)}_\mathrm{sys}$ and $f^\mathrm{(ii)}_\mathrm{sys}$ are defined as the global scaling factor of the systematic map that is necessary to detect non-Gaussianity deviations with 95\,\% confidence.


\begin{table}[tmb]               
  \begingroup
  \newdimen\tblskip \tblskip=5pt
  \caption{Level of detectability of non-Gaussianity caused by simulated
  systematic effects in the LFI maps. Numbers without parenthesis correspond
  to the case of CMB plus noise background, $f^{(i)}_{\rm sys}$, and
  numbers in brackets correspond to the case of the noise background,
  $\big(f^{(ii)}_{\rm sys}\big)$.}            
  \label{tab_gaussianity_detectability}                     
  \nointerlineskip
  \vskip 2mm
  \footnotesize
  \setbox\tablebox=\vbox{
    \newdimen\digitwidth 
    \setbox0=\hbox{\rm 0} 
    \digitwidth=\wd0 
    \catcode`*=\active 
    \def*{\kern\digitwidth}
    \newdimen\signwidth 
    \setbox0=\hbox{+} 
    \signwidth=\wd0 
    \catcode`!=\active 
    \def!{\kern\signwidth}
  {\tabskip=0pt
  \halign{ 
  \hbox to 1.0in{#\leaderfil}\tabskip=0em& 
  \hfil#\hfil\tabskip=1.0em& 
  \hfil#\hfil& 
  \hfil#\hfil\tabskip=0pt 
  \cr                       
  \noalign{\doubleline}
  \omit&
  \hfil 30\,GHz\hfil& 
  \hfil 44\,GHz\hfil& 
  \hfil 70\,GHz\hfil\cr 
  \noalign{\vskip -3pt}
  \omit&\multispan3\hrulefill\cr
  \noalign{\vskip 2pt}
    $I$&	16.34~(0.52)& 9.06~(1.72)& 12.98~(0.78)\cr
    \noalign{\vskip 4pt}
    $Q,U$&	*0.73~(0.55)& 1.25~(1.00)& *1.89~(1.54)\cr
    \noalign{\vskip 4pt}
    $I,Q,U$&	*0.74~(0.44)& 1.30~(0.98)& *1.92~(0.91)\cr
    \noalign{\vskip 5pt\hrule\vskip 3pt}
  }
  }}
  \endPlancktable          

  \endgroup
\end{table}   


  %
  We measure the levels of detectability, reported in Table~\ref{tab_gaussianity_detectability}, for temperature and polarization maps at low resolution ($\ell_\mathrm{max} = 95$). For temperature-only we obtain $f^\mathrm{(i)}_\mathrm{sys} = $  16.34, 9.06, and 12.98 for 30, 44, and 70\,GHz, respectively. These values are consistent with those obtained in the previous release \citep{planck2013-p02a}. Including polarization results in lower levels being detectable for both cases (CMB plus noise and noise-only backgrounds). The values found are $f^\mathrm{(i)}_\mathrm{sys} >$ 0.73, 1.25, and 1.81 for 30, 44, and 70\,GHz, respectively. If we consider only the noise background the values decrease for all the cases (see Table~\ref{tab_gaussianity_detectability}). The level of detectability for the 70\,GHz channel in the CMB plus noise background case is always larger than the critical limit of $f^\mathrm{(i)}_\mathrm{sys} = 1$. This is the case that is particularly relevant for non-Gaussianity tests \citep{planck2014-a18,planck2014-a19}.


\begin{table}[tmb]               
  \begingroup
  \newdimen\tblskip \tblskip=5pt
  \caption{Relative variation $\Delta f_{\mathrm{NL}}/\sigma(f_{\mathrm{NL}})$
  as a percentage, caused by simulated systematic effects in the LFI data.}
  \label{tab_gaussianity_fnl}                     
  \nointerlineskip
  \vskip 2mm
  \footnotesize
  \setbox\tablebox=\vbox{
    \newdimen\digitwidth 
    \setbox0=\hbox{\rm 0} 
    \digitwidth=\wd0 
    \catcode`*=\active 
    \def*{\kern\digitwidth}
    \newdimen\signwidth 
    \setbox0=\hbox{+} 
    \signwidth=\wd0 
    \catcode`!=\active 
    \def!{\kern\signwidth}
  {\tabskip=0pt
  \halign{ 
  \hbox to 1.0in{#\leaderfil}\tabskip=0em& 
  \hfil#\hfil\tabskip=1.5em& 
  \hfil#\hfil& 
  \hfil#\hfil\tabskip=0pt 
  \cr                       
  \noalign{\doubleline}
  \omit&
  \hfil 30\,GHz\hfil& 
  \hfil 44\,GHz\hfil& 
  \hfil 70\,GHz\hfil\cr 
  \noalign{\vskip -3pt}
  \omit&\multispan3\hrulefill\cr
  \noalign{\vskip 2pt}
    Local&	        $-0.90$& $-0.01$& $-0.05$\cr
    \noalign{\vskip 4pt}
    Equilateral&	$*1.80$& $*0.02$& $*0.02$\cr
    \noalign{\vskip 4pt}
    Orthogonal&	        $*2.22$& $*0.02$& $*0.06$\cr
    \noalign{\vskip 5pt\hrule\vskip 3pt}
  }
  }}
  \endPlancktable          

  \endgroup
\end{table}   


  The second aspect on non-Gaussianity we have considered is the impact of systematic effects on the primordial non-Gaussianity $f_{\rm NL}$ parameter. We define the bias on this parameter, $\Delta f_{\rm NL}$, as the mean difference between the two $f_{\rm NL}$ values measured in maps with and without systematic effects, i.e., $\Delta f_{\rm NL} \equiv  f^\mathrm{sys}_\mathrm{NL} - f^\mathrm{clean}_\mathrm{NL}$. 
  
  To obtain a limit on this bias, we have first computed the full-sky bispectrum of the global systematic effect maps, following the formalism of \citet{komatsu2002}, and then we have cross-correlated it with the primordial bispectrum. We removed the bias generated by extragalactic point sources or the CIB-lensing, following the procedure described, e.g., in \citet{curto2013,curto2015}. 
  
  Table~\ref{tab_gaussianity_fnl} shows the values of the bias $\Delta f_{\rm NL}$ calculated at high resolution ($\ell_{\rm max} = 1024$) for the LFI channels. The bias is normalized to the corresponding dispersion of $f_{\rm NL}$ to estimate the relative impact on the measurement of this parameter. For the three LFI channels, the impact of systematic effects on $f_{\rm NL}$ is negligible, being lower than 0.90\,\% for the local shape, 1.80\,\% for the equilateral shape and 2.22\,\% for the orthogonal shape. The 30\,GHz channel has the highest amplitude for this bias, whereas the 44 and 70\,GHz channels have maximum amplitudes of 0.02\,\% and 0.03\,\%, respectively.

  
  \section{Summary of uncertainties due to systematic effects}
\label{sec_summary_table}

  This section provides a top-level overview of the residual\footnote{We use the word ``residual'' to refer to the spurious signal remaining in the final LFI maps due to a systematic effect, that is after any removal steps applied by the data analysis pipeline.} uncertainties in the \Planck-LFI CMB maps and power spectra, introduced by systematic effects. We list these effects in Table~\ref{tab_list_systematic_effects} and summarize the main results of our analysis, which are discussed in Sect.~\ref{sec_assessment} and corresponding subsections.

  Tables~\ref{tab_summary_systematic_effects_maps_30}, \ref{tab_summary_systematic_effects_maps_44}, and \ref{tab_summary_systematic_effects_maps_70} report the peak-to-peak\footnote{In this paper we call ``peak-to-peak'' the difference between the 99\% and the 1\% quantiles of the pixel value distributions.} and rms systematic effect uncertainties in LFI maps. To calculate these uncertainties we have used {\tt HEALPix} \citep{gorski2005} maps with simulated systematic effects degraded to $N_\mathrm{side} = 128$ (corresponding to a pixel size of around 28\arcm) at 30 and 44\,GHz, and $N_\mathrm{side} = 256$ (corresponding to a pixel size of about 14\arcmin) at 70\,GHz. This pixel sizes approximate the optical beam angular resolution. Maps were masked with the top and middle masks shown in Fig.~\ref{fig_masks}, also used for power spectra estimation. 

      \begin{table}[tmb]               
  \begingroup
  \newdimen\tblskip \tblskip=5pt
  \caption{Summary of systematic effect uncertainties on 30\,GHz maps$^{\rm a}$
  in \muKCMB.  Columns give the peak-to-peak (``p-p'') and rms levels
  for Stokes $I$, $Q$, and $U$ maps.}
  \label{tab_summary_systematic_effects_maps_30}                     
  \nointerlineskip
  \footnotesize
  \setbox\tablebox=\vbox{
    \newdimen\digitwidth 
    \setbox0=\hbox{\rm 0} 
    \digitwidth=\wd0 
    \catcode`*=\active 
    \def*{\kern\digitwidth}
    \newdimen\signwidth 
    \setbox0=\hbox{+} 
    \signwidth=\wd0 
    \catcode`!=\active 
    \def!{\kern\signwidth}
  {\tabskip=0pt
  \halign{ 
  \hbox to 1.3in{#\leaderfil}\tabskip=0em& 
  \hfil#\hfil\tabskip=1.0em& 
  \hfil#\hfil& 
  \hfil#\hfil& 
  \hfil#\hfil& 
  \hfil#\hfil& 
  \hfil#\hfil\tabskip=0pt\cr                       
  \noalign{\doubleline}
  \omit&
  \multispan2\hfil $I$\hfil& 
  \multispan2\hfil $Q$\hfil&
  \multispan2\hfil $U$\hfil\cr   
  \noalign{\vskip -3pt}
  \omit&\multispan6\hrulefill\cr
  \noalign{\vskip 2pt}
  \omit& p-p& rms& p-p& rms& p-p& rms\cr
    \noalign{\vskip 4pt}
    Near sidelobes&          0.72& 0.13& 0.05& 0.01& 0.05& 0.01\cr            
    \noalign{\vskip 4pt}                                                      
    Pointing&                0.37& 0.07& 0.02& 0.01& 0.02& 0.00\cr            
    \noalign{\vskip 4pt}                                                      
    Polarization angle&      0.02& 0.00& 0.53& 0.11& 0.64& 0.15\cr            
    \noalign{\vskip 4pt}                                                      
    1-Hz spikes&             0.54& 0.11& 0.11& 0.02& 0.09& 0.02\cr            
    \noalign{\vskip 4pt}                                                      
    Bias fluctuations&       0.07& 0.01& 0.07& 0.01& 0.06& 0.01\cr            
    \noalign{\vskip 4pt}                                                      
    ADC nonlinearity&        0.42& 0.09& 0.54& 0.11& 0.56& 0.11\cr            
    \noalign{\vskip 4pt}                                                      
    Calibration&             2.43& 0.55& 2.53& 0.46& 2.34& 0.43\cr            
    \noalign{\vskip 4pt}                                                      
    Thermal fluct. (300\,K)& 0.00& 0.00& 0.00& 0.00& 0.00& 0.00\cr            
    \noalign{\vskip 4pt}                                                      
    Thermal fluct. (20\,K)&  0.12& 0.03& 0.06& 0.02& 0.06& 0.02\cr            
    \noalign{\vskip 4pt}                                                      
    Thermal fluct. (4\,K)&   0.29& 0.06& 0.06& 0.01& 0.05& 0.01\cr            
  \omit&\multispan6\hrulefill\cr                                              
  \noalign{\vskip 2pt}
    Total$^{\rm b}$&	            2.72& 0.61& 2.79& 0.52& 2.42& 0.49\cr
  \noalign{\vskip 5pt\hrule\vskip 3pt}
  }
  }}
  \endPlancktable          
  \tablenote {{\rm a}} Calculated for a pixel size approximately equal to the
  average beam FWHM. A null value indicates a residual $<10^{-2}$\,\muKCMB.\par
  \tablenote {{\rm b}} The total has been computed on maps resulting from the
  sum of individual systematic effect maps.\par
  \endgroup
\end{table}   


\begin{table}[tmb]               
  \begingroup
  \newdimen\tblskip \tblskip=5pt
  \caption{Summary of systematic effect uncertainties on 44\,GHz maps in
  \muKCMB.  Columns give the peak-to-peak (``p-p'') and rms levels
  for Stokes $I$, $Q$, and $U$ maps.}            
  \label{tab_summary_systematic_effects_maps_44}                     
  \nointerlineskip
  \footnotesize
  \setbox\tablebox=\vbox{
    \newdimen\digitwidth 
    \setbox0=\hbox{\rm 0} 
    \digitwidth=\wd0 
    \catcode`*=\active 
    \def*{\kern\digitwidth}
    \newdimen\signwidth 
    \setbox0=\hbox{+} 
    \signwidth=\wd0 
    \catcode`!=\active 
    \def!{\kern\signwidth}
  {\tabskip=0pt
  \halign{ 
  \hbox to 1.3in{#\leaderfil}\tabskip=0em& 
  \hfil#\hfil\tabskip=1.0em& 
  \hfil#\hfil& 
  \hfil#\hfil& 
  \hfil#\hfil& 
  \hfil#\hfil& 
  \hfil#\hfil\tabskip=0pt\cr                       
  \noalign{\doubleline}
  \omit&
  \multispan2\hfil $I$\hfil& 
  \multispan2\hfil $Q$\hfil&
  \multispan2\hfil $U$\hfil\cr   
  \noalign{\vskip -3pt}
  \omit&\multispan6\hrulefill\cr
  \noalign{\vskip 2pt}
  \omit& p-p& rms& p-p& rms& p-p& rms\cr
    \noalign{\vskip 4pt}
    Near sidelobes&          0.09& 0.02& 0.00& 0.00& 0.00& 0.00\cr            
    \noalign{\vskip 4pt}                                                      
    Pointing&                0.30& 0.06& 0.01& 0.00& 0.01& 0.00\cr            
    \noalign{\vskip 4pt}                                                      
    Polarization angle&      0.04& 0.01& 0.35& 0.07& 0.38& 0.10\cr            
    \noalign{\vskip 4pt}                                                      
    1-Hz spikes&             1.99& 0.40& 0.88& 0.18& 1.04& 0.21\cr            
    \noalign{\vskip 4pt}                                                      
    Bias fluctuations&       0.04& 0.01& 0.05& 0.01& 0.05& 0.01\cr            
    \noalign{\vskip 4pt}                                                      
    ADC nonlinearity&        0.30& 0.06& 0.36& 0.07& 0.34& 0.07\cr            
    \noalign{\vskip 4pt}                                                      
    Calibration&             1.05& 0.18& 1.57& 0.29& 1.31& 0.26\cr            
    \noalign{\vskip 4pt}                                                      
    Thermal fluct. (300\,K)& 0.00& 0.00& 0.00& 0.00& 0.00& 0.00\cr            
    \noalign{\vskip 4pt}                                                      
    Thermal fluct. (20\,K)&  0.04& 0.02& 0.06& 0.01& 0.05& 0.01\cr            
    \noalign{\vskip 4pt}                                                      
    Thermal fluct. (4\,K)&   0.23& 0.05& 0.05& 0.01& 0.06& 0.01\cr            
  \omit&\multispan6\hrulefill\cr                                              
  \noalign{\vskip 2pt}
    Total&                    2.29& 0.45& 1.95& 0.37& 1.76& 0.37\cr
    \noalign{\vskip 4pt}
  \noalign{\vskip 5pt\hrule\vskip 3pt}
  }
  }}
  \endPlancktable          

  \endgroup
\end{table}   


\begin{table}[tmb]               
  \begingroup
  \newdimen\tblskip \tblskip=5pt
  \caption{Summary of systematic effect uncertainties on 70\,GHz maps in
  \muKCMB.  Columns give the peak-to-peak (``p-p'') and rms levels
  for Stokes $I$, $Q$, and $U$ maps.}
  \label{tab_summary_systematic_effects_maps_70}                     
  \nointerlineskip
  \footnotesize
  \setbox\tablebox=\vbox{
    \newdimen\digitwidth 
    \setbox0=\hbox{\rm 0} 
    \digitwidth=\wd0 
    \catcode`*=\active 
    \def*{\kern\digitwidth}
    \newdimen\signwidth 
    \setbox0=\hbox{+} 
    \signwidth=\wd0 
    \catcode`!=\active 
    \def!{\kern\signwidth}
  {\tabskip=0pt
  \halign{ 
  \hbox to 1.3in{#\leaderfil}\tabskip=0em& 
  \hfil#\hfil\tabskip=1.0em& 
  \hfil#\hfil& 
  \hfil#\hfil& 
  \hfil#\hfil& 
  \hfil#\hfil& 
  \hfil#\hfil\tabskip=0pt\cr                       
  \noalign{\doubleline}
  \omit&
  \multispan2\hfil $I$\hfil& 
  \multispan2\hfil $Q$\hfil&
  \multispan2\hfil $U$\hfil\cr   
  \noalign{\vskip -3pt}
  \omit&\multispan6\hrulefill\cr
  \noalign{\vskip 2pt}
  \omit& p-p& rms& p-p& rms& p-p& rms\cr
    \noalign{\vskip 4pt}
    Near sidelobes&          0.30& 0.07& 0.01& 0.00& 0.01& 0.00\cr          
    \noalign{\vskip 4pt}                                                    
    Pointing&                0.60& 0.11& 0.03& 0.01& 0.03& 0.01\cr          
    \noalign{\vskip 4pt}                                                    
    Polarization angle&      0.02& 0.00& 0.08& 0.02& 0.08& 0.02\cr          
    \noalign{\vskip 4pt}                                                    
    1-Hz spikes&             0.39& 0.08& 0.17& 0.03& 0.15& 0.03\cr          
    \noalign{\vskip 4pt}                                                    
    Bias fluctuations&       0.68& 0.14& 0.84& 0.17& 0.95& 0.18\cr          
    \noalign{\vskip 4pt}                                                    
    ADC nonlinearity&        1.56& 0.33& 1.92& 0.39& 2.05& 0.41\cr          
    \noalign{\vskip 4pt}                                                    
    Calibration&             1.06& 0.23& 0.98& 0.18& 0.77& 0.16\cr          
    \noalign{\vskip 4pt}                                                    
    Thermal fluct. (300\,K)& 0.00& 0.00& 0.00& 0.00& 0.00& 0.00\cr          
    \noalign{\vskip 4pt}                                                    
    Thermal fluct. (20\,K)&  0.44& 0.08& 0.07& 0.01& 0.08& 0.02\cr          
    \noalign{\vskip 4pt}                                                    
    Thermal fluct. (4\,K)&   0.38& 0.08& 0.04& 0.01& 0.05& 0.01\cr          
  \omit&\multispan6\hrulefill\cr                                            
  \noalign{\vskip 2pt}
    Total&                   2.24& 0.47& 2.27& 0.46& 2.38& 0.48\cr
    \noalign{\vskip 4pt}
  \noalign{\vskip 5pt\hrule\vskip 3pt}
  }
  }}
  \endPlancktable          

  \endgroup
\end{table}   


  The rms uncertainty in LFI maps from known systematic effects is $\la0.5\,\mu$K in polarization and $\la 1\,\mu$K in temperature. The improvements\footnote{See Sects.~4, 6, and 7 of \citet{planck2014-a03}} introduced into the LFI pipeline have allowed us to reduce the peak-to-peak uncertainty by a factor ranging from 3.5 at 70\,GHz to 7.7 at 30\,GHz, compared to the 2013 analysis \citep{planck2013-p02a}. At 30 and 70\,GHz calibration and analogue-to-digital converter (ADC) nonlinearity are the prevailing effects, while at 44\,GHz calibration and 1-Hz spikes dominate. 
    
  In our assessment we have not included the residual effects from far sidelobes, because we remove Galactic straylight directly from the timelines. This removal is based on optical simulations, which implies that a residual effect may be present in the data. Estimating this remaining signal is complex and computationally demanding, since it requires us to generate Monte Carlo simulations of the far sidelobes. For the present analysis we have used the following approach regarding far sidelobes: we have assessed the impact of systematic effects assuming the perfect removal of Galactic straylight; and additionally we have quantified how much the far sidelobes would affect our results if they were not removed at all. 
      
\begin{sidewaysfigure*}
      \hspace{5.cm} $TT$ \hspace{7.4cm} $EE$ \hspace{7.4cm} $BB$\\
      \begin{tabular}{m{.2cm} m{25cm}}
	\begin{turn}{90}30\,GHz\end{turn}&\includegraphics[width=24cm]{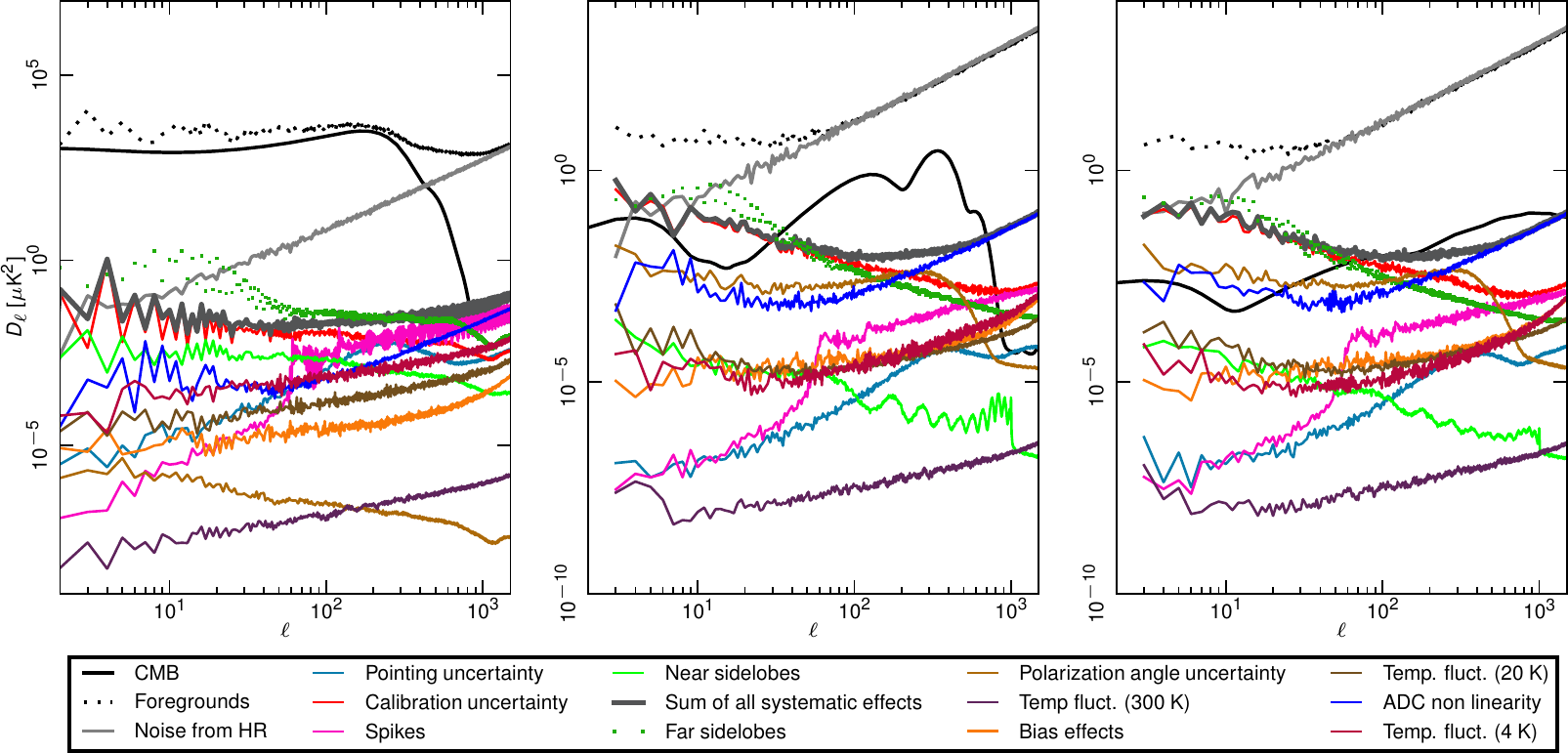}\\
      \end{tabular}
      \caption{
	\label{fig_systematic_effects_power_spectrum_30}
	Angular power spectra of the various systematic effects at 30\,GHz, compared to the CMB and foreground temperature and polarization spectra and to the instrumental noise from half-ring (HR) difference maps.  The CMB $TT$ and $EE$ spectra are best fits to the \Planck\ cosmological parameters \citep[see figures~9 and 10 in][]{planck2014-a01} filtered by the LFI window functions. The example CMB $B$-mode spectrum is based on \Planck-derived cosmological parameters and assumes a tensor-to-scalar ratio $r=0.1$, a tensor spectral index $n_\mathrm{T}=0$, and no beam-filtering. The thick dark-grey line represents the total contribution. The dotted dark-green line is the contribution from the far sidelobes that has been removed from the data and is therefore not considered in the total.
	}
    \end{sidewaysfigure*}

\begin{sidewaysfigure*}
      \hspace{5.cm} $TT$ \hspace{7.4cm} $EE$ \hspace{7.4cm} $BB$\\
      \begin{tabular}{m{.2cm} m{25cm}}
	\begin{turn}{90}44\,GHz\end{turn}&\includegraphics[width=24cm]{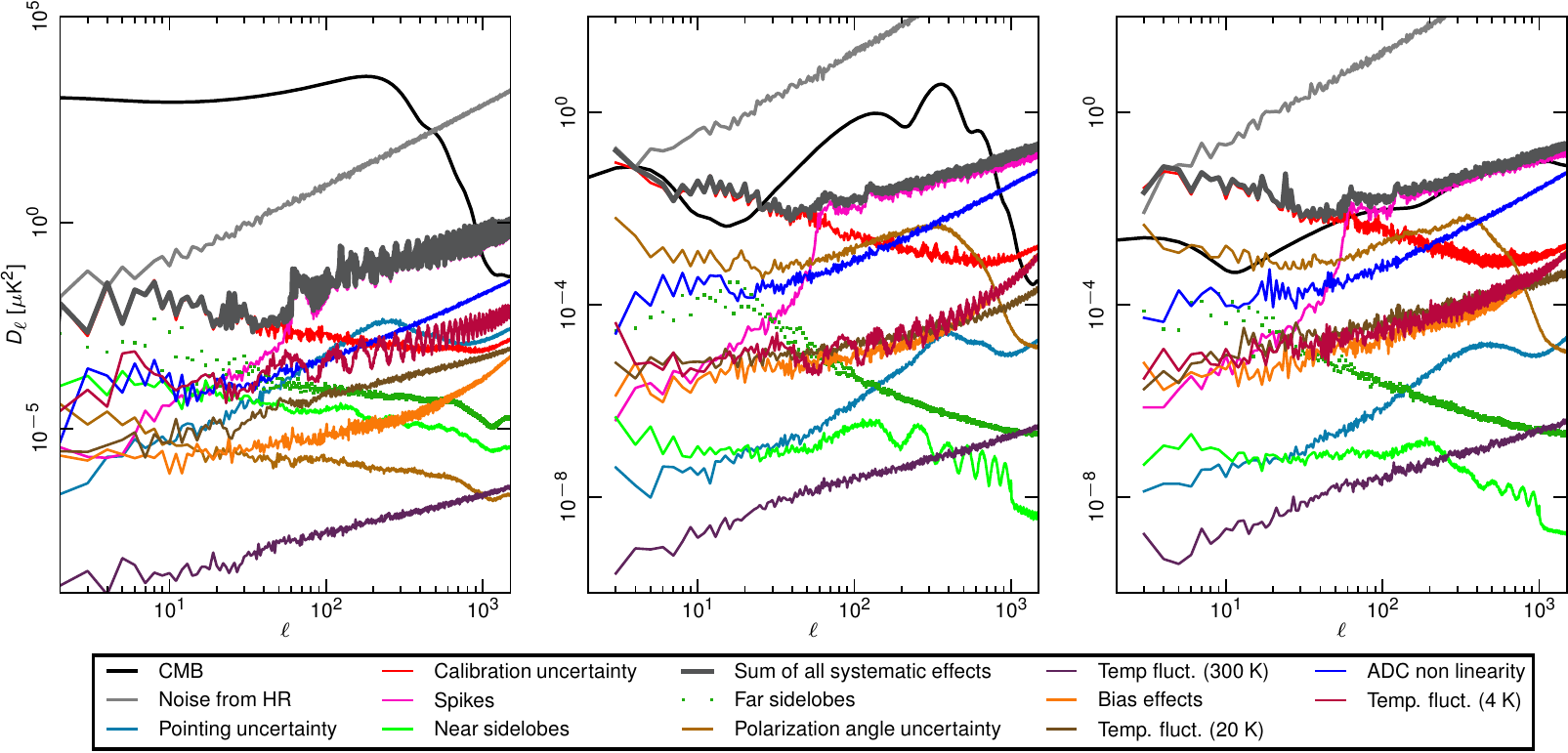}\\
      \end{tabular}
      \caption{
	\label{fig_systematic_effects_power_spectrum_44}
	Angular power spectra of the various systematic effects at 44\,GHz, compared to the CMB temperature and polarization spectra and to the instrumental noise from half-ring (HR) difference maps.  The CMB $TT$ and $EE$ spectra are best fits to the \Planck\ cosmological parameters \citep[see figures~9 and 10 in][]{planck2014-a01} filtered by the LFI window functions. The example CMB $B$-mode spectrum is based on \Planck-derived cosmological parameters and assumes a tensor-to-scalar ratio $r=0.1$, a tensor spectral index $n_\mathrm{T}=0$, and no beam-filtering. The thick dark-grey line represents the total contribution. The dotted dark-green line is the contribution from far the sidelobes that has been removed from the data and is therefore not considered in the total.
	}

    \end{sidewaysfigure*}

\begin{sidewaysfigure*}
      \hspace{5.cm} $TT$ \hspace{7.4cm} $EE$ \hspace{7.4cm} $BB$\\
      \begin{tabular}{m{.2cm} m{25cm}}
	\begin{turn}{90}70\,GHz\end{turn}&\includegraphics[width=24cm]{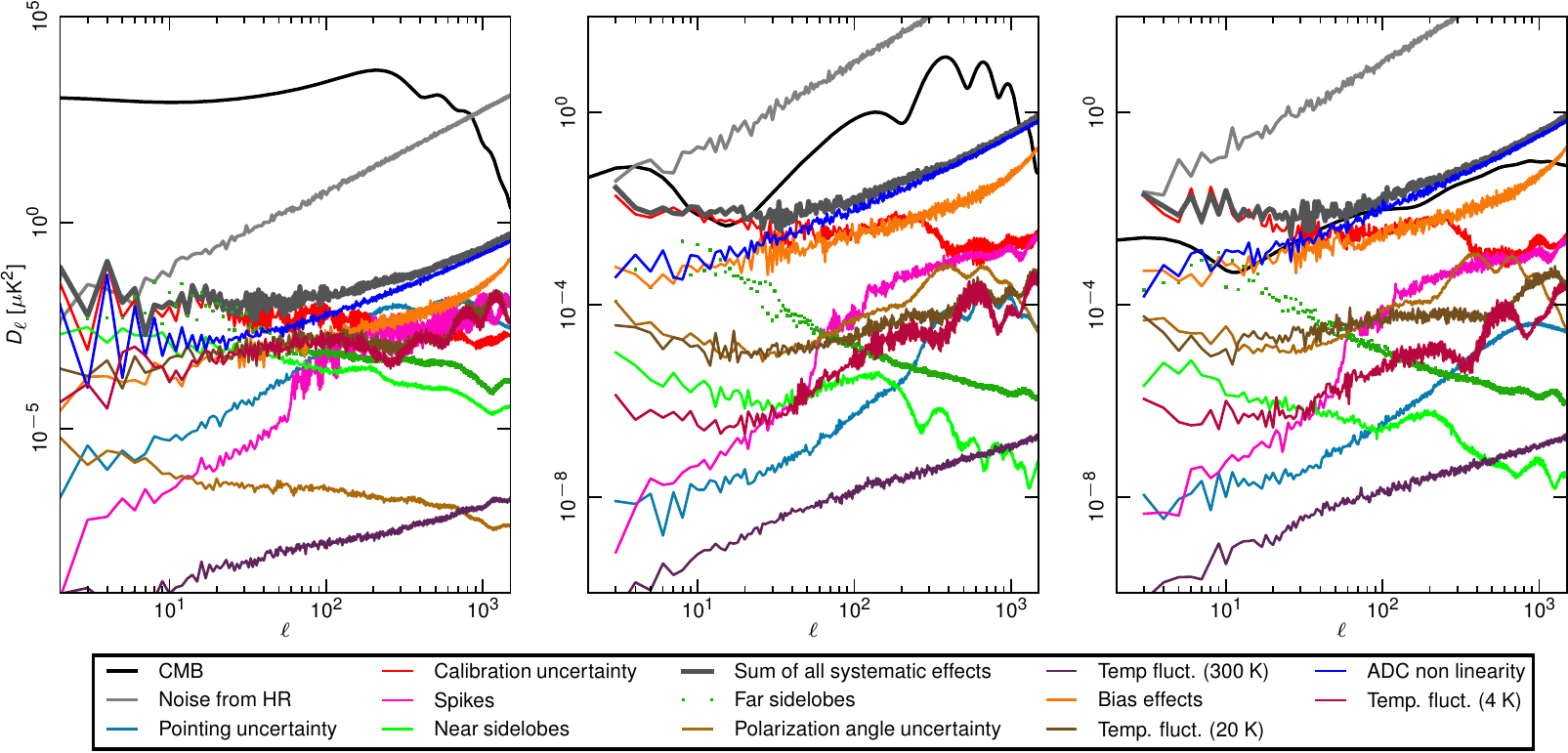}\\
      \end{tabular}
      \caption{
	\label{fig_systematic_effects_power_spectrum_70}
	Angular power spectra of the various systematic effects at 70\,GHz, compared to the CMB temperature and polarization spectra and to the instrumental noise from half-ring (HR) difference maps.  The CMB $TT$ and $EE$ spectra are best fits to the \Planck\ cosmological parameters \citep[see figures~9 and 10 in][]{planck2014-a01} filtered by the LFI window functions. The example CMB $B$-mode spectrum is based on \Planck-derived cosmological parameters and assumes a tensor-to-scalar ratio $r=0.1$, a tensor spectral index $n_\mathrm{T}=0$, and no beam-filtering. The thick dark-grey line represents the total contribution. The dotted dark-green line is the contribution from far the sidelobes that has been removed from the data and is therefore not considered in the total.}
    \end{sidewaysfigure*}

  Figures~\ref{fig_systematic_effects_power_spectrum_30}, \ref{fig_systematic_effects_power_spectrum_44}, and \ref{fig_systematic_effects_power_spectrum_70} provide an overview of the power spectra in temperature and polarization for each systematic effect, compared to the foreground levels at 30\,GHz and to the cosmological signal at 44 and 70\,GHz. At 30\,GHz we use the spectrum obtained from measured data as an approximation of the foreground spectrum at this frequency. At 44 and 70\,GHz we use the power spectrum coming from the best fits to the \Planck\ cosmological parameters \citep[see figures~9 and 10 in][]{planck2014-a01} filtered by the LFI window functions. The example CMB $B$-mode spectrum is based on \Planck-derived cosmological parameters and assumes a tensor-to-scalar ratio $r=0.1$, a tensor spectral index $n_\mathrm{T}=0$, and no beam-filtering. Instrumental noise here is based on ``half-ring'' difference maps, as described in sections~12.1 and 12.2 of \citet{planck2014-a03}.
    
  In the same figure we also show the power spectra of Galactic straylight detected by the far sidelobes (the dotted green lines), which indicate the level of the effect that we expect to have removed from the data.
    
  At 30\,GHz the systematic effects are all lower than the foreground signal. The Galactic straylight is higher than the noise level at $\ell\la 20$. For this reason we removed an estimate of Galactic straylight from the timelines, based on our best knowledge of the far sidelobes. These results show that the 30\,GHz channel gives a reliable foreground template, with uncertainties set by the instrumental noise.

  At 44 and 70\,GHz the level of Galactic straylight is lower than the CMB. It is reasonable to assume that any residual that could be present in the data must be less than the total effect reported here and, therefore, negligible compared to the CMB. 
    
  The power spectrum of the sum of all systematic effects (dark-grey line) is higher than the $E$-mode spectrum in the $\ell$ range 10--15 and is marginally below that for multipoles $< 10$, at both 44 and 70\,GHz. This could have an impact on the extraction of the optical depth, $\tau$, which is strongly dependent on the $C_\ell^{EE}$ spectrum at very low $\ell$s.
    
  We have evaluated the impact of the simulated effects on $\tau$ (see Sect.~\ref{sec_lowell}) and found a bias that is about 0.2 times the standard deviation, showing that the uncertainty on this parameter is dominated by statistics and the contribution from systematic effects is only of marginal importance.
    
  We have also assessed the uncertainty caused by LFI systematic effects on the CMB power spectra estimated by \Planck\ after component separation. 
    
  In our procedure (described in Sect.~\ref{sec_assessment_compsep}) we set the HFI channels to zero to evaluate the systematic uncertainty of LFI only in the CMB reconstruction. It is a generalization of the approach described in \citet{planck2013-p02a}, based on component-separation weights calculated via minimum variance over the whole sky area considered. In this test we first input maps with the sum of all systematic effects into the component separation pipeline, then we apply the top and middle mask in Fig.~\ref{fig_masks} to the resulting maps and, finally, we calculate the pseudo-spectra. 
    
\begin{figure*}
    \begin{center}
      \includegraphics[width=17cm]{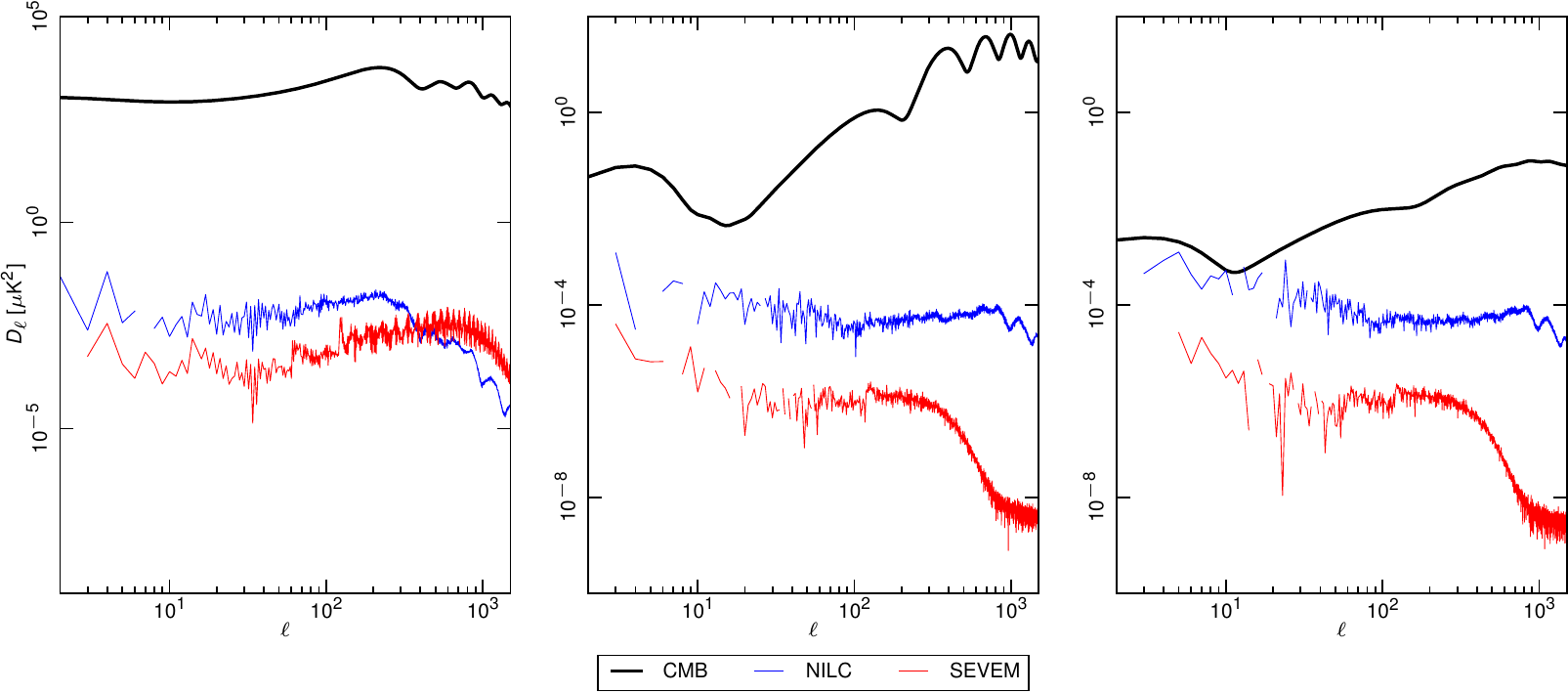}
    \end{center}
    \caption{Angular power spectra for the combined LFI systematic effects
    templates in the CMB $TT$, $EE$, and $BB$ reconstructions of the {\tt NILC} and
    {\tt SEVEM} pipelines, compared with the \Planck\ fiducial cosmology.}
    \label{fig_component_separation_global_compsep}
    \end{figure*}

  Figure~\ref{fig_component_separation_global_compsep} shows the angular power spectra of the sum of all known LFI systematic effects in the component-separation outputs of the {\tt NILC} and {\tt SEVEM} algorithms described in \citet{planck2014-a11}. These plots highlight the level of the residual effects compared with the \Planck\ 2015 best-fit cosmology. 
    
  The results in total intensity confirm the findings of our previous data release. The residual systematic effects are several orders of magnitude lower than the CMB power spectrum at all angular scales.
    
  The results in polarization show that the residual effects resulting after the application of the {\tt SEVEM} algorithm are about 1.5--2 orders of magnitude lower than those resulting from {\tt NILC}, at all angular scales. This means that the residual effects obtained with \texttt{NILC} have an amplitude comparable to cosmological $B$-modes with $r\sim 0.1$.
    
  The reason for this discrepancy in the component-separated outputs is the different weighting that the two codes apply to the LFI channels. In \texttt{NILC} the LFI channels are weighted more than in \texttt{SEVEM}, which also implies a larger impact of the systematic effects. Let us recall the reasons for this different weighting. 
    
  {\tt NILC} implements a minimum variance approach in the needlet domain, and produces a set of weights for each $\ell$-band in which it is applied. For this reason, in the LFI channels the weights are particularly relevant at large angular scales, where foregrounds are most important. 
    
  \texttt{SEVEM}, on the other hand, applies a smoothing to the LFI channels and then calculates the minimum variance coefficients over the entire range of multipoles, which eventually results in smaller weights for the LFI channels and, therefore, a smaller contribution of their systematic effects. 
  
  \section{Conclusions}
\label{sec_conclusions}

This is the era of precision cosmology. The advances in detector and space technology in the last 20 years now allow us to test theories describing the evolution of the Universe with statistical uncertainties that were unimaginable at the time the CMB was discovered, more than 50 years ago. 

\Planck\ has produced the most sensitive full-sky maps of the microwave sky to date. We have exploited its unprecedented statistical power to obtain the most precise angular temperature power spectrum of the CMB \citep{planck2014-a13}, as well as cosmological parameters with relative errors below the percent level in some cases \citep{planck2014-a15}. 

In the last ten years several experiments from the ground and the stratosphere have successfully tested new technologies that are further increasing sensitivity and opening new frontiers for cosmology by exploiting measurements of the CMB anisotropy polarization.

However, precision is nothing without accuracy. Understanding and controlling systematic uncertainties is one of the greatest challenges for present and future measurements of the CMB. The control of systematic effects has indeed been a challenge for \Planck, both in the development phase and during data analysis.

In this paper we have discussed the systematic effect uncertainties of the \Planck\ Low Frequency Instrument data in the context of the second cosmological data release. This is the result of work begun almost 20 years ago, when we started developing the instrument with systematic effects control as one of the main drivers for the instrument and data-analysis pipeline designs.

Our approach follows two complementary paths.

 \begin{itemize}
  \item The first uses measured data and exploits the redundancy in the scanning strategy to divide the observations into periods of various length in which the observed sky is the same. We used the analysis of difference maps constructed on such periods (``null tests'') to highlight possible spurious residual signals exceeding the instrumental noise.\vspace{.2cm}

  \item the second uses our knowledge of the instrument to build physical models of the various known systematic effects that are simulated from timelines to maps. Here we exploit, as much as possible, actual flight measurements, such as pointing, temperatures, and radiometric data.
 \end{itemize}

We use simulations to quantify the uncertainties introduced by systematic effects in the maps and power spectra, and compare our predictions with null-test results to identify residuals that are not accounted for by our model. We also use our simulations to assess the impact of these effects on cosmological parameters (like the reionization optical depth, $\tau$) on the measurements of the CMB statistical
properties, and on component separation.

Our results for temperature data confirm the findings of the first \Planck\ release \citep{planck2013-p02a}: the measurements are limited by instrumental noise and at all relevant angular scales the systematic effects are several orders of magnitude below the power spectrum of the CMB itself.

Our analysis for polarization demonstrates the robustness of the LFI data for scientific analysis, in particular regarding the measurement of $\tau$ and the statistical analysis of CMB maps. Systematic effects, however, are more challenging in polarization than in temperature and their level is close to the $E$-mode signal, especially at large angular scales.
 
Uncertainties in the relative photometric calibrations dominate the LFI systematic effects budget, especially at large angular scales. This is an area in the data analysis pipeline that is still being improved in preparation for the next \Planck\ release.
 
Our data could also contain residual Galactic straylight caused by an imperfect knowledge of the beam sidelobes. We do not consider this residual in our budget, but null spectra from consecutive surveys indicate a possible presence of such a spurious signal at 30\,GHz. 

At 70\,GHz the systematic effects compete with the CMB $E$-modes for multipoles in the range 10--20. This does not preclude an accurate measurement of $\tau$, which depends mainly on multipoles $\ell<10$ \citep{planck2014-a13}. Using systematic effects simulations we have shown that the bias introduced on $\tau$ is less than $0.25$ times the standard deviation of the measured parameter. Forthcoming analyses will include independent estimations, based on null tests and on cross-correlation between the LFI 70\,GHz map and the HFI 100 and 143\,GHz maps.

We have also evaluated the impact on the scalar perturbations amplitude, $\ln(A_{\rm s})$, and on the upper limit to the tensor-to-scalar ratio, $r$, derived with large-scale polarization data. In this case the effect on $\ln(A_{\rm s})$ is approximately $0.2\,\sigma$, while the upper limit on $r$ is increased by the systematic effects by around $15\,\%$. For these two parameters, however, the main \Planck\ constraint comes from the temperature power spectrum at high multipoles, so that the actual impact is negligible.

At 30\,GHz the systematic effects are much smaller than the Galactic emission at all multipoles. We use this channel as a foreground monitor, which implies that we are not limited by systematic effects at this frequency for any angular scale, in either temperature or polarization.

The 44\,GHz channel displays residuals that compete with the $E$-mode polarization for $\ell\leq 10$ and dominate the signal for multipoles in the range 10--20. We do not use this channel in the current polarization analysis, so these effects do not play a role in the measurement of $\tau$. We use the 44\,GHz data, however, in the component separation analysis. 
  
The contribution of LFI systematic effects on CMB maps and power spectra after component separation is smaller than the CMB signal at all scales, both in temperature and polarization. We have assessed this using two component separation codes, namely \texttt{NILC} (a minimum variance code in the needlet domain) and \texttt{SEVEM} (a code based on foreground templates). With both
codes the LFI systematic uncertainties do not limit accurate measurement of the CMB temperature and polarization spectra. As expected, we find that the use of \texttt{SEVEM} results in a lower level of residuals compared to \texttt{NILC}, because of the different weighting of the LFI data applied by the two codes.

The presence of known systematic effects in the LFI data does not significantly impact non-Gaussianity studies. We have used maps with the simulated effects combined with CMB and noise maps and found that, at 70\,GHz, the amplitude of these effects must be at least a factor of 2 larger to detect a significant non-Gaussianity. We have also assessed the bias on the $f_\mathrm{NL}$
parameter and found that it is less than $0.1\,\%$ at 44 and 70\,GHz and $<2.2\,\%$ at 30\,GHz. 

Finally, we comment about the systematic uncertainties on the $B$-mode polarization measurements. Our analysis shows that at 70\,GHz the level of systematic effects is smaller than the instrumental noise, but larger than a $B$-mode power spectrum for $r=0.1$. This does not impact our polarization analysis, based on $E$-mode polarization data, but shows, once again, the importance of understanding and controlling systematic effects in future experiments aiming at the detection of this elusive signal.
  
Understanding and controlling systematic effects in the LFI data has been a challenge from which we have gained even deeper knowledge of our instrument and learned several valuable lessons for the future. This is a future destined to be one of even more precise and accurate cosmology, but also one of increasing challenge to control systematics effects,

\begin{acknowledgements}

  The Planck Collaboration acknowledges the support of: ESA; CNES, and
CNRS/INSU-IN2P3-INP (France); ASI, CNR, and INAF (Italy); NASA and DoE (USA);
STFC and UKSA (UK); CSIC, MINECO, JA, and RES (Spain); Tekes, AoF, and
CSC (Finland); DLR and MPG (Germany); CSA (Canada); DTU Space (Denmark);
SER/SSO (Switzerland); RCN (Norway); SFI (Ireland); FCT/MCTES (Portugal);
ERC and PRACE (EU). A description of the Planck Collaboration and a list of
its members, indicating which technical or scientific activities they have
been involved in, can be found at
\burl{http://www.cosmos.esa.int/web/planck/planck_collaboration}.
Some of the results in this paper have been derived using the
{\tt HEALPix} package.

\end{acknowledgements}

\bibliographystyle{aat}
\bibliography{Planck_bib,custom}

\raggedright

\end{document}